\documentclass[aps,prd,reprint,superscriptaddress,nofootinbib,longbibliography,floatfix]{revtex4-2}

\usepackage{amsmath,amssymb,amsfonts,mathtools}
\usepackage{mathrsfs}
\usepackage{bm}
\usepackage{booktabs}
\usepackage{graphicx}
\usepackage{hyperref}
\usepackage{slashed}
\usepackage{enumitem}
\usepackage{capt-of}

\begin{document}

\title{Dark matter glueball candidate from a $G(2)$--$E_6$--$E_7$ exceptional grand unified theory: \\$G$-parity, spin, mass and stability}

\author{Nicol\`o Masi}
\email{masin@bo.infn.it}
\affiliation{
	INFN, Physics Department of Bologna University, Via Irnerio 46 Bologna, 40126, Italy}

\begin{abstract}
We determine the gauge-invariant identity, mass scale and ultraviolet stability of the dark matter candidate arising from \(G(2)\to SU(3)_C\) in the exceptional \(G(2)\!-\!E_6\!-\!E_7\) construction. The broken \(G(2)\) sector contains odd \(1^{+-}\) and \(0^{--}\) channels, whereas the scalar \(0^{++}\sim X\bar X\) state is even and unprotected. For \(m_X\simeq5.7\times10^{13}\,\mathrm{GeV}\), weak-binding reference masses are \(M_{2X}\simeq1.14\times10^{14}\,\mathrm{GeV}\) and \(M_{3X}\simeq1.71\times10^{14}\,\mathrm{GeV}\), while the exact pole masses remain nonperturbative. Gauge-invariant Fr\"ohlich--Morchio--Strocchi (FMS) operators, Hall--Post bounds, \(Y\)-junction arguments and the pure-\(SU(3)\) glue spectrum favor \(1^{+-}\) in their controlled regimes without excluding a deeply bound \(0^{--}\) state.

We then test whether this dark grading survives the full chiral exceptional embedding. An on-shell analysis finds no independent purely dark odd operator through dimension seven: the first nonvanishing basis appears at dimension nine. So the minimal one-copy exceptional embedding does not provide an exact ultraviolet dark \(G\)-parity. Moving the dark Higgs from the common \(\mathbf{1463}_H\) parent to a separated \(\mathbf{1539}_H\) removes the scalar-parent obstruction and the relevant dimension-nine exceptional parents vanish on the selected pure-dark component in the undressed limit. A genuinely gauged or geometric \(\mathbb Z_{2,D}\) would therefore leave a bosonic parity after dark Higgsing. Its extension to the mirror-free theory nevertheless fails because the required \(G(2)\) conjugation also conjugates color, \(\mathbf3_C\leftrightarrow\bar{\mathbf3}_C\). The remaining obstruction to exact dark matter stability is therefore ultraviolet and chiral, rather than low-energy or purely scalar.
\end{abstract}

\maketitle

\section{Introduction and motivations}

The $G(2)$ dark sector was originally introduced as a minimal exceptional
enlargement of the strong sector and was subsequently embedded in the Class-B
$G(2)$--$E_6$--$E_7$ unification framework
\cite{MasiG2SciRep2021,MasiG2,MasiE6E7}.  In the present work we return to the
microscopic dark matter (DM) candidate generated by the breaking
$G(2)\to SU(3)_C$ and ask three connected questions: i) which
 gauge-invariant confined state is actually protected and what is its spin; ii)
 what mass scale characterizes its physical pole; iii) can the dark $G$-parity
 of the isolated gauge--Higgs theory survive the complete chiral grand unified theory (GUT)
 embedding?  The last question is especially important, because the absence
of an obvious decay channel is not enough to guarantee an exact stabilizing
symmetry in the ultraviolet theory.

We keep the expression ``dark matter glueball candidate'', but the
gauge-invariant problem is more general than this constituent name suggests.
The physical pole can contain broken-vector and pure-glue components and need
not coincide with a conventional pure glueball.  The isolated $G(2)$
gauge--Higgs theory possesses a protected odd sector and admits the usual
Higgs/confinement continuity interpretation
\cite{ButtazzoG2Higgs,ButtazzoScalarGauge2020,FradkinShenker1979,OsterwalderSeiler1978,HayashiHiggsConfinement2024}.
We therefore begin from the confined spectrum and its total-spin, parity and charge-conjugation quantum numbers $J^{PC}$ and
only afterwards move to the exceptional embedding, where the same protecting
$G$-parity must coexist with the mirror-free chiral Standard Model (SM).

\paragraph{Dark $G$-parity along the construction.}
Throughout this paper, \emph{dark $G$-parity} denotes the discrete
$\mathbb Z_2$ grading that distinguishes the potentially protected odd
sector of the broken $G(2)$ theory.  The terminology is used as a convenient
name for the same physical selection rule followed through the successive
effective and ultraviolet descriptions of the model; it should not be
confused with the familiar isospin $G$-parity of ordinary quantum chromodynamics (QCD), although the name is inspired by meson physics. Its defining
feature here is that the nontrivial $G(2)$ transformation acts as charge
conjugation on the daughter $SU(3)$ gauge sector and is combined, when
necessary, with a sign acting on the field that selects the broken vacuum.

In the isolated bosonic $G(2)$ gauge--Higgs theory, the vacuum-preserving
transformation is
\begin{equation}
	G_B\equiv P_B=Z_S U_B,
	\label{eq:GBdictionary}
\end{equation}
where
\begin{equation}
	Z_S:\ S\to-S
\end{equation}
is the accidental scalar sign and $U_B\in G(2)$ implements the relevant
conjugation on the gauge sector.  In the most general renormalizable isolated
$G(2)$ gauge--Higgs action this is an exact accidental dark $G$-parity, as in
Ref.~\cite{ButtazzoG2Higgs}: it is not a fundamental symmetry of an arbitrary
Wilsonian effective field theory (EFT).  The on-shell operator analysis below sharpens this statement: no independent
pure-dark odd local operator survives through dimension seven, while the first
nonzero basis appears at dimension nine.  Since $U_B$ is an inner gauge transformation, it acts
trivially on a fully $G(2)$-gauge-invariant local observable; the physical
$G_B$ eigenvalue is therefore the parity of the number of $S$ insertions in
its gauge-invariant interpolator.  After expanding around the Higgs vacuum,
the same $U_B$ action appears as charge conjugation on the daughter $SU(3)$
gauge components.  Thus the $C$-odd Fr\"ohlich--Morchio--Strocchi (FMS) channels identified below are exactly
the $G_B$-odd channels, without attributing a physical global charge to the
gauge transformation $U_B$ itself.

When the same question is formulated inside the original one-copy
Class-B effective theory, the selected dark coordinate carries instead the
accidental pure-dark grading
\begin{equation}
	G_D=Z_\chi U_B,
	\label{eq:GDdictionaryIntro}
\end{equation}
with $Z_\chi$ the corresponding sign of that dark coordinate.  This should
not be interpreted as a new independent symmetry: $G_D$ is the EFT
realization of the same protecting conjugation after the exceptional
projection has been imposed.

Finally, in the separated-$\mathbf{1539}_H$ construction, if an additional
gauged or geometric $\mathbb Z_{2,D}$ is present, the dark Higgs vacuum
preserves the diagonal bosonic transformation
\begin{equation}
	P_D=z_D U_B .
	\label{eq:PDdictionaryIntro}
\end{equation}
The sequence
\begin{equation}
	G_B
	\quad\longrightarrow\quad
	G_D
	\quad\longrightarrow\quad
	P_D
\end{equation}
therefore represents three successive realizations of the same underlying
dark-$G$-parity idea: exact accidentally at the renormalizable isolated
bosonic level, accidental in the selected Class-B EFT, and potentially gauged
in the separated-parent
ultraviolet construction.  A central purpose of this paper is to determine
how far this selection rule can actually be promoted.  In particular, a
bosonic realization is not sufficient: an exact ultraviolet symmetry must
also act consistently on the exceptional matter parents and, ultimately,
within the mirror-free chiral Standard Model spectrum.

This symmetry-first viewpoint is also essential for identifying the dark
particle itself.  Labels such as $X\bar X$, $XXX$, or ``glueball'' describe
dominant operator or constituent content but do not by themselves define a
physical state.  After confinement, the observables are gauge-invariant
eigenstates, and different interpolating operators can overlap with the same
pole whenever they share the appropriate quantum numbers.  The Class-B
construction of
\begin{equation}
	E_6\to G(2)\times SU(3)_A
\end{equation}
and its $E_7$ uplift
\cite{MasiG2,MasiG2SciRep2021,MasiE6E7}
must therefore satisfy two requirements simultaneously: it must retain a
$G$-odd confined state suitable for the dark sector and project the
exceptional parent content onto the observed mirror-free chiral
SM spectrum without destroying the symmetry that would protect
that state.

The isolated-theory analysis of Ref.~\cite{ButtazzoG2Higgs} already
identifies a $C$-odd sector and baryon-like gauge-invariant operators.
The gauge-invariant FMS construction developed below sharpens this picture:
a baryonic interpolator can have a leading $X\bar X$ component, so
``two-vector'' does not imply $G$ even and ``three-vector'' does not by
itself define the protected particle.  The physical problem is instead to
identify the lowest gauge-invariant $G$-odd pole and determine its spin,
mass and ultraviolet stability.  These three questions must therefore be
treated together.

\paragraph{From the earlier $0^{++}$ assignment to the $G$-odd spectrum.}
The earlier Class-B GUT analysis \cite{MasiE6E7}, following the usual
secluded-Yang--Mills intuition and the phenomenological treatment of the
$G(2)$-origin confined sector, described the lightest scalar
$J^{PC}=0^{++}$ glueball as the natural late-time DM relic and
mentioned an optional constituent sign $X_\mu\to-X_\mu$ as an additional
protection.  Recent studies of confining dark sectors likewise emphasize that
relic abundance, stability and phenomenology depend on the full glueball
spectrum, portal structure and cosmological history rather than on a scalar
mass ordering alone \cite{CarenzaGlueball2022,CarenzaGlueball2023,
McKeenGlueball2025,YamadaYonekura2023,BiondiniNonAbelian2024}.
The present gauge-invariant analysis refines that working identification.  The
relevant question is not which channel is the lightest
scalar confined state, but which gauge-invariant state is odd under the
actual dark $G$-parity.  We show below that the scalar
$X_\mu\bar X^\mu$ channel is $G$ even and that the $G(2)/SU(3)$ algebra
does not admit an independent $X\to-X$ $\mathbb Z_2$.  The symmetry-protected
dark candidates therefore belong instead to the $G$-odd $1^{+-}$ and
$0^{--}$ sectors.  The subsequent $E_6/E_7$ analysis addresses the separate
question of whether this genuine dark $G$-parity can survive ultraviolet
completion.

\paragraph{Ultraviolet construction and fixed Class-B input.}
The ultraviolet setting is the previously developed Class-B exceptional
construction, whose relevant symmetry-breaking chain begins as
\begin{equation}
	E_7\longrightarrow E_6\times U(1)_X,
	\qquad
	E_6\longrightarrow G(2)\times SU(3)_A .
	\label{eq:UVchainIntro}
\end{equation}
Here $G(2)$ contains the dark/strong sector discussed above, whereas
$SU(3)_A$ is the ancestor of the electroweak gauge group.  The $E_7$
uplift is required in particular by the abelian sector: the additional
$U(1)_X$ permits the hypercharge embedding
\begin{equation}
	Y=\frac13(A+X),
	\label{eq:YIntro}
\end{equation}
where the diagonal $SU(3)_A$ generator is normalized as
\begin{equation}
	A=\sqrt3\,t_A^8,
	\qquad
	q_A=2A,
\end{equation}
or equivalently
\begin{equation}
	Y=\frac16(q_A+2X).
\end{equation}
The same $X$ charge has a second role: besides completing the hypercharge
assignment, it distinguishes the exceptional matter branches and therefore
provides the charge on which the Class-B chirality selector acts.  The
$E_7$ uplift is consequently relevant here not merely as an enlargement of
the gauge group, but as the structure that simultaneously fixes the
abelian embedding and makes the component-resolved visible-sector
projection possible.

Throughout the branch tables we use left-handed Weyl notation:
\(Q_L\) and \(L\) denote the Standard Model quark and lepton
\(SU(2)_L\) doublets, respectively, while \(u^c\), \(d^c\), \(e^c\)
and \(N^c\) denote the left-handed charge-conjugate fields associated
with the right-handed up quark, down quark, charged lepton and neutrino.
Thus the superscript \(c\) labels the usual conjugate SM
representation and is not an additional Class-B quantum number.
For orientation, the selected light Class-B matter branches may be summarized
as
\begin{table}[htbp]
	\centering
	\small
	\setlength{\tabcolsep}{5pt}
	\renewcommand{\arraystretch}{1.16}
	\caption{Compact Class-B matter assignment used in the present stability
		analysis.  The complete parity-adapted table, including the $A$ charges,
		hypercharge and scalar directions, is given in Table~\ref{tab:ClassB}.}
	\label{tab:ClassBIntro}
	\begin{tabular}{@{}c c c c@{}}
		\toprule
		field & $E_6\times U(1)_X$ branch & $G(2)\times SU(3)_A$ & $X$\\
		\midrule
		$Q_L$ & $\overline{\mathbf{351}}_{+1}$
		& $(\mathbf{7},\overline{\mathbf{3}}_A)$ & $+1$\\
		$u^c$ & $\mathbf{351}_{-1}$
		& $(\mathbf{7},\mathbf{15}_A)$ & $-1$\\
		$d^c$ & $\mathbf{351}_{-1}$
		& $(\mathbf{7},\overline{\mathbf{6}}_A)$ & $-1$\\
		$L$ & $\mathbf{2925}_{-3}$
		& $(\mathbf{1},\mathbf{8}_A)$ & $-3$\\
		$e^c$ & $\mathbf{2925}_{+3}$
		& $(\mathbf{1},\mathbf{8}_A)$ & $+3$\\
		$N^c$ & $\mathbf{2925}_{+3}$
		& $(\mathbf{1},\mathbf{10}_A)$ & $+3$\\
		\bottomrule
	\end{tabular}
\end{table}

The charged branches are descendants of the single real $E_7$ matter
macro-parent
\begin{equation}
	\overline{\mathbf{351}}_{+1}\oplus
	\mathbf{351}_{-1}\oplus
	\mathbf{2925}_{-3}\oplus
	\mathbf{2925}_{+3}
	\subset\mathbf{27664}_{E_7}.
	\label{eq:ClassBIntroMacroParent}
\end{equation}
On the selected Standard Model family including $N^c$, the same $U(1)_X$
charges obey
\begin{equation}
	X=3(B-L),
	\label{eq:ClassBIntroXBL}
\end{equation}
a relation used later in the comparison between dark $G$-parity and
baryon/lepton-number symmetries.

The symbol \(\chi\) used hereafter denotes the dark-breaking scalar direction,
not the DM particle itself.  More precisely,
\(\chi\sim(\mathbf{7},\mathbf{1}_A)_0\) is the selected
\(G(2)\)-fundamental Higgs component whose vacuum expectation value
\(\langle\chi\rangle=v_\chi e_7\) breaks
\(G(2)\to SU(3)_C\).  
%The six broken gauge bosons then transform as
%\(\mathbf{3}_C\oplus\overline{\mathbf{3}}_C\) and are denoted
%schematically by \(X_\mu\) and \(\bar X_\mu\).  The dark-matter candidates
%studied in this work are gauge-invariant confined \(G\)-odd states in the
%\(1^{+-}\) and \(0^{--}\) sectors built from these broken-vector and
%gluonic degrees of freedom; 
\(\chi\) itself is therefore the dark Higgs
field that creates the broken phase, not the DM state.  In the
separated-\(\mathbf{1539}_H\) completion discussed later, the same physical
dark-breaking direction is realized as the selected component
\(\Xi_D\subset\mathbf{1539}_H\).

For the present stability analysis we do not construct a new grand-unified
model: we keep the published Class-B breaking pattern and matter
assignment fixed and ask whether the dark $G$-parity identified in the
isolated $G(2)$ theory can survive inside them. 

A central result inherited from the special
\(E_6\to G(2)\times SU(3)_A\) embedding is the renormalizable seclusion of
the broken \(G(2)\) sector from the selected light SM fields.
In the Class-B assignment the direct light--light--\(\chi\) vertex vanishes
exactly, so ordinary renormalizable decays of the heavy
\(G(2)/SU(3)_C\) sector into the selected light fermions are absent.
At this level the remaining ordinary gauge-sector decay channels belong to
the daughter \(SU(3)_C\), identified here with QCD, and hence ultimately
lead to gluonic or hadronic states.  This statement concerns the
renormalizable selected-light theory only: higher-dimensional exceptional
connectors can reopen visible channels, and whether the protecting grading
extends to an exact symmetry of the complete mirror-free chiral theory is
precisely the ultraviolet question addressed below.
 
The visible-sector
construction is based on the compact $dP_2$ geometry with selector flux \cite{MasiE6E7}
\begin{equation}
	F_X=E_1-E_2
\end{equation}
and the six Class-B matter supports described in
Secs.~\ref{sec:compact-base} and \ref{sec:selector-flux}.  For the dark-state
problem only three outputs of that construction are essential:
(i) the resulting mirror-free chiral light spectrum,
(ii) the vanishing of the renormalizable light--light--$\chi$ kernel, and
(iii) the component projector whose compatibility with dark $G$-parity must
be tested.  Only the selector ingredients needed to define the chiral projector are retained below.

For the present paper these compactification ingredients are used only to define
the published Class-B selector and the resulting chiral projector; the detailed
string-geometric completion is not part of the dark-matter analysis.

For numerical orientation we retain the Class-B benchmark of
Ref.~\cite{MasiE6E7},
\begin{widetext}
	\begin{equation}
		M_\Omega\simeq M_\Theta\simeq1.03\times10^{13}~{\rm GeV},
		\qquad
		M_\chi\simeq10^{14}~{\rm GeV},
		\qquad
		M_{E_6}\simeq6.9\times10^{15}~{\rm GeV},
		\label{eq:BenchmarkScales}
	\end{equation}
\end{widetext}
with
%\begin{widetext}
\begin{equation}
		g_{G(2)}(M_\chi)\simeq0.565,
		\qquad
		m_X\simeq5.7\times10^{13}~{\rm GeV}.
		\label{eq:BenchmarkMasses}
\end{equation}
%\end{widetext}

The vector mass follows the normalization used explicitly in
Ref.~\cite{MasiE6E7}.  In a weak-binding constituent estimate these numbers
correspond to the reference scales
\begin{equation}
\begin{aligned}
M_{2X}&\simeq2m_X\simeq1.14\times10^{14}~{\rm GeV},\\
M_{3X}&\simeq3m_X\simeq1.71\times10^{14}~{\rm GeV}.
\end{aligned}
\label{eq:BenchmarkCompositeMasses}
\end{equation}
These are constituent reference masses, not predictions for physical
confined poles.  Once confinement and operator mixing are included, the
DM mass is determined by the lowest gauge-invariant state in the
relevant $G$-odd spectrum.  This is why the $1^{+-}$ two-vector/glue and
$0^{--}$ three-vector sectors are treated as separate variational problems
below.

\paragraph{Literal Class-B benchmark and the spectroscopic deformation.}
A distinction between the literal Class-B benchmark and the confining
spectroscopy problem studied below is essential.  In the fixed Class-B
construction the daughter subgroup
\begin{equation}
	SU(3)_C\subset G(2)
\end{equation}
is identified with ordinary QCD color.  Consequently the confinement scale
associated with the unbroken daughter theory is parametrically tiny compared
with the mass of the broken $G(2)/SU(3)_C$ vectors.  Using, only for numerical
orientation, the conventional pure-Yang--Mills normalization \cite{TeperGlueball1998}
$\sqrt{\sigma}\simeq0.44\,{\rm GeV}$ gives
\begin{equation}
	x_{\rm QCD}\equiv\frac{\sqrt{\sigma}}{m_X}
	\sim 8\times10^{-15},
	\qquad
	k_1\sqrt{\sigma}\sim2.7~{\rm GeV},
	\label{eq:LiteralClassBx}
\end{equation}
for the benchmark $m_X\simeq5.7\times10^{13}\,$GeV.  The precise numerical
conversion is not important for the arguments below.  For example, the
$r_0$-based scale setting of Ref.~\cite{AthenodorouTeper2020} corresponds to
$\sqrt{\sigma}\simeq0.485\,{\rm GeV}$ and hence
$k_1\sqrt{\sigma}\simeq2.9\,{\rm GeV}$.  Light dynamical quarks further modify
the literal QCD spectrum, while none of the dimensionless ordering arguments
developed below depends on choosing either of these scale conventions.

This enormous hierarchy has an immediate physical consequence.  If one
temporarily considers only the bosonic conjugation grading inherited from
$G(2)$, the daughter theory already contains $C$-odd gauge-sector
excitations at the hadronic scale.  An ultraheavy state dominated by
$X\bar X$ or $XXX$, with mass of order $10^{14}\,$GeV, therefore cannot be
the lowest odd excitation of the \emph{literal} Class-B benchmark.  Moreover,
in the complete chiral Class-B theory even this bosonic grading is not an
exact symmetry: the element $U_B$ that implements the conjugation does not
act within the mirror-free light-matter projector, as proved in
Sec.~\ref{sec:chiralColorParityNoGo}.  The fixed benchmark thus presents two
logically distinct obstructions to interpreting the ultraheavy level as an
exactly protected dark relic: it is not the lightest state in the bosonic
odd spectrum, and the corresponding grading does not extend to an exact
symmetry of the complete chiral theory.

The calculations performed below therefore address a deliberately broader
spectroscopic question.  We continuously vary
\begin{equation}
	x\equiv\frac{\sqrt{\sigma}}{m_X}
\end{equation}
away from its literal QCD value and study the ordering of the independent
$1^{+-}$ and $0^{--}$ gauge-invariant sectors as the confinement scale
becomes non-negligible relative to the heavy-vector mass.  The adopted interval
$x\simeq0.05$--$0.47$ should consequently be interpreted as a
\emph{deformed or isolated confining regime} used to determine when a
two-vector-, three-vector-, or glue-dominated odd pole could become the
lowest relevant state.  It is not the value of $x$ realized by the literal
Class-B QCD benchmark, nor is it being used here as a hidden assumption that
QCD itself confines at the ultraheavy scale.

\paragraph{Cosmological abundance.}
The mass scale relevant here, $m_{\rm DM}\sim10^{14}\,$GeV, lies far above
the partial-wave unitarity range of a standard thermal relic produced by
freeze-out from an equilibrium bath \cite{GriestKamionkowski1990}.
Consequently, if one of the $G$-odd confined states identified below is both
sufficiently long lived and cosmologically populated, its primordial abundance
cannot arise from conventional weakly interacting massive particle (WIMP)-like thermal freeze-out.  It must instead
be seeded by a nonthermal or nonstandard production mechanism, for example
inflaton or reheating decays, production through the $G(2)$-breaking dark-Higgs
sector, or gravitational particle creation
\cite{ChungKolbRiotto1998,ChungKolbRiotto1999,
	ChungCrottyKolbRiotto2001}.  Related possibilities were already considered
in the earlier $G(2)$ and Class-B cosmological analyses
\cite{MasiG2,MasiE6E7}.

Production and subsequent dark sector evolution should be distinguished.
Once some energy density has been transferred to the secluded
$G(2)$-origin sector, its strong interactions may establish an internal dark
bath even when that bath was never in thermal equilibrium with the Standard
Model.  Confinement, number-changing reactions and a later dark freeze-out
can then redistribute the initially produced abundance among the physical
confined states.  Such processes determine the final yield only after the
initial population mechanism and the dark-to-visible entropy or temperature
ratio have been specified.
The present work does not attempt this cosmological calculation.  
%We do not
%fix the inflaton branching fractions, reheating temperature, dark-Higgs
%abundance, initial dark-sector entropy, or nonperturbative number-changing
%rates, and therefore do not predict $\Omega_{\rm DM}h^2$.  
Our purpose is
instead prior and microscopic: to determine which gauge-invariant
$G$-odd state could in principle survive long enough to constitute such a
population.  Accordingly, all lifetime requirements quoted below are
conditional on that state having acquired a cosmologically relevant
abundance.

\paragraph{Main results and organization.}
The discussion leads to four separate but closely connected results.
First, the renormalizable isolated $G(2)$ gauge--Higgs theory has an exact
accidental bosonic dark $G$-parity,
\begin{equation}
	G_B=P_B=Z_SU_B ,
\end{equation}
and the renormalizable spectrum separates into two independent $G$-odd
$J^{PC}$ sectors.  The $1^{+-}$ sector contains a gauge-invariant
two-vector/glue channel, while the natural three-vector scalar belongs to
the $0^{--}$ sector.  Heavy-particle inequalities, the $Y$-junction
comparison and the decoupled pure-$SU(3)$ spectrum independently favor the
$1^{+-}$ channel in their respective controlled limits, although none of
these arguments excludes a deeply bound or near-threshold nonperturbative
$0^{--}$ pole.

Second, the Class-B chiral embedding has a nontrivial positive selection
property.  After projection onto the light Standard Model sector, the
renormalizable light--light--$\chi$ coupling vanishes exactly.  This removes
the most immediate decay source, but it is not by itself an ultraviolet
stability theorem: higher-dimensional parent operators are allowed and must
be analyzed separately.  The purely dark operator basis can, however, be sharpened
considerably.  The scalar $\chi F^3$ family vanishes exactly by a mismatch between
the Lorentz and internal permutation symmetries.  The other apparent
dimension-seven representative,
$O_{abc}(F_{\mu\nu}\chi)^a(D^\mu\chi)^b(D^\nu\chi)^c$, is algebraically
nonzero but redundant modulo integration by parts and the leading equations of
motion.  A complete on-shell contact-term classification then gives no independent
pure-dark odd operator at $d=3,5,7$.  The first nonzero basis occurs at $d=9$;
in particular a mixed-helicity $\chi F^4$ operator survives in the unique
$[3,1]$ permutation channel.

Third, the scalar-sector obstruction of the original common-parent
construction can be repaired.  Placing the dark breaking field in a
separated $\mathbf{1539}_H$ reverses the relevant carrier sign, removes the
common-$\mathbf{1463}_H$ obstruction and admits an aligned broken-phase
vacuum.  If an additional gauged or geometric $\mathbb Z_{2,D}$ exists, the
dark Higgsing leaves the diagonal bosonic remnant
\begin{equation}
	P_D=z_DU_B ,
\end{equation}
providing a candidate ultraviolet realization of the same dark $G$-parity.
This construction is conditional on the existence of that discrete factor
and, phenomenologically, on a strongly split $\mathbf{1539}_H$ threshold
spectrum compatible with the published Class-B running.  The distinction is important: the $\mathbf{1539}_H$ representation
repairs the scalar-parent problem by itself, but it does not remove the first
physical $d=9$ odd operator by representation theory alone.  The exact hook
plethysm below gives one $\mathbf{1463}_H$ and two $\mathbf{1539}_H$ parent channels for the
$[3,1]$ $\Xi F^4$ structure.  The additional $\mathbb Z_{2,D}$ is therefore
load-bearing if the complete odd-$\Xi$ tower is to be removed.

The analysis of the even-$\Xi$ operators further sharpens this result.  Even though
dimension-six operators containing two $\Xi$ fields are not forbidden, the
potentially direct selected-light contribution obtained by replacing one
$\Xi$ by the dark vacuum expectation value (VEV) and leaving one transverse
$\mathbf{3}_C\oplus\bar{\mathbf{3}}_C$ fluctuation vanishes exactly by unbroken
$SU(3)_C$ triality, to arbitrary order in additional VEV insertions.
Radial and two-transverse portals are not removed by this theorem and require
state-dependent matching to the physical odd poles.

Fourth, the remaining obstruction to a complete exact dark
$G$-parity is chiral rather than scalar.  The nontrivial
$SU(3)_C\subset G(2)$ normalizer action exchanges
$\mathbf{3}_C\leftrightarrow\bar{\mathbf{3}}_C$, whereas the selected mirror-free
Standard Model projector is not equivariant under that transformation.
Consequently the bosonic $P_D$ construction cannot be extended to an exact
internal dark parity of the complete mirror-free Class-B spectrum under the
fixed breaking chain.  The precise symmetry class excluded by this
fixed-architecture no-go theorem is stated in
Sec.~\ref{sec:chiralColorParityNoGo}.

These conclusions concern the identity and microscopic protection of the
candidate and should be kept separate from its cosmological abundance, which
is not computed here.  The paper is therefore organized in the same order
as the physics.  We first determine the gauge-invariant odd spectrum,
exclude the scalar $0^{++}\sim X\bar X$ meson as the protected state and
construct the $1^{+-}$ and $0^{--}$ variational sectors.  We then embed
those channels into the Class-B exceptional parent, derive the light-field
component zeros and analyze the massive Higgs/sheaf structure.  The
subsequent sections test the separated-$\mathbf{1539}_H$ bosonic
$G$-parity construction, its threshold and vacuum consistency, the surviving
higher-dimensional operators, and finally the chiral obstruction to
promoting that symmetry to the complete theory.  The closing
discussion summarizes separately the exact results, the conditional bosonic
completion, and the remaining nonperturbative and ultraviolet questions.

\section{The DM candidate and dark $G$-parity in the isolated $G(2)$ theory: mass and spin hierarchies}
\label{sec:isolated-g2-stability}

\subsection{$G(2)\to SU(3)$, the broken vectors, and the confined spectrum}
\label{sec:G2SU3brokenvectors}

The starting point is the spontaneous breaking of the exceptional gauge group
$G(2)$ to its maximal $SU(3)$ subgroup, which is the basic mechanism used in
the earlier $G(2)$ constructions and in the present Class-B embedding
\cite{MasiG2SciRep2021,MasiG2,MasiE6E7,ButtazzoG2Higgs}.  In the fundamental
representation one has
\begin{equation}
	\mathbf{7}\longrightarrow
	\mathbf{1}\oplus\mathbf{3}\oplus\bar{\mathbf{3}},
\end{equation}
and a vacuum expectation value aligned with the singlet direction leaves
$SU(3)$ unbroken.  Equivalently, the vacuum manifold is
\begin{equation}
	G(2)/SU(3)\simeq S^6 .
\end{equation}
At the level of gauge fields the adjoint decomposes as
\begin{equation}
	\mathbf{14}\longrightarrow
	\mathbf{8}\oplus\mathbf{3}\oplus\bar{\mathbf{3}}.
	\label{eq:G2adjSU3}
\end{equation}
The eight generators in the $\mathbf{8}$ form the unbroken $SU(3)$ gauge
sector, whereas the six generators in the coset acquire a mass from the
$G(2)$-breaking vacuum.

We denote the corresponding massive vector fields by
\begin{equation}
	X_\mu^\alpha\sim\mathbf{3},
	\qquad
	\bar X_{\mu\alpha}\sim\bar{\mathbf{3}},
	\qquad \alpha=1,2,3 .
	\label{eq:Xtransformations}
\end{equation}
Thus $X$ and $\bar X$ should not be viewed as additional colorless vector
particles appended to the theory: they are the massive gauge bosons of the
six broken $G(2)/SU(3)$ directions and remain charged under the unbroken
non-Abelian subgroup.  Consequently they cannot occur as asymptotic
one-particle states once the daughter $SU(3)$ confines.  Only
$SU(3)$-singlet gauge-invariant combinations can appear in the physical
spectrum.

The simplest constituent descriptions then include a meson-like sector
\begin{equation}
	X\bar X ,
\end{equation}
three-vector baryon-like sectors
\begin{equation}
	XXX,\qquad
	\bar X\bar X\bar X ,
\end{equation}
and states dominated by operators built solely from the unbroken
$SU(3)$ field strength.  Ref.~\cite{ButtazzoG2Higgs} exhibits precisely this
structure in the broken $G(2)$ gauge--Higgs theory.  The terminology
``meson'', ``baryon'', and ``glueball'' is useful for identifying dominant
operator content, but it must not be confused with an exact classification
of the physical particles: after confinement, the observable states are
poles of gauge-invariant correlation functions, and operators with the same
quantum numbers may mix.

This point is particularly important for the dark matter problem.  A
constituent count alone does not determine whether a state is protected.
The scalar contraction $X_\mu\bar X^\mu$, the derivative two-vector
operators, the three-vector operators and the pure-glue operators transform
differently under the conjugation inherited from $G(2)$.  The relevant
classification is therefore by the quantum numbers and dark $G$-parity of
the corresponding \emph{gauge-invariant} interpolating operators, rather
than by the number of massive vectors appearing in a particular Higgs-phase
expansion.  We turn to this symmetry next.

\subsection{Dark $G$-parity in the isolated $G(2)$ theory}
\label{sec:isolated-Gparity}

The next point is the symmetry that can protect the confined state.  In the
broken theory this symmetry belongs to the full gauge--Higgs system and cannot
be introduced as an independent sign of the massive vectors
\cite{ButtazzoG2Higgs}.  Let the
$G(2)$-breaking scalar be written as
\begin{equation}
	S(x)=w\,e_7+\delta S(x),
\end{equation}
where the unit vector $e_7$ selects the vacuum direction whose stabilizer is
the daughter $SU(3)$.  There exists an element $U_B\in G(2)$ belonging to
the nontrivial normalizer class of this $SU(3)$ embedding whose action on
the vacuum direction is
\begin{equation}
	U_B e_7=-e_7 .
	\label{eq:UBe7}
\end{equation}
On the unbroken and broken gauge generators, this same transformation
implements the conjugation which exchanges the
$\mathbf{3}$ and $\bar{\mathbf{3}}$ coset directions.  In the Higgs-phase
description it therefore acts on the massive vectors as the charge
conjugation
\begin{equation}
	X_\mu^\alpha
	\longleftrightarrow
	-\,\bar X_{\mu\alpha},
	\label{eq:UBXconjugation}
\end{equation}
up to the basis conventions specified below.

Equation~\eqref{eq:UBe7}, however, also shows why $U_B$ by itself is
\emph{not} an unbroken physical symmetry of the chosen vacuum.  Acting only
with $U_B$ maps
\begin{equation}
	\langle S\rangle=w e_7
	\quad\longrightarrow\quad
	-w e_7 .
\end{equation}
The isolated scalar theory possesses the accidental sign symmetry
\begin{equation}
	Z_S:\qquad S\longrightarrow-S ,
\end{equation}
so the vacuum is instead left invariant by the combined transformation
\begin{equation}
	G_B\equiv P_B=Z_SU_B ,
	\qquad
	G_B\langle S\rangle=\langle S\rangle,
	\qquad
	G_B^2=1 .
	\label{eq:IsolatedGparity}
\end{equation}
It is this diagonal transformation, rather than $U_B$ alone, that defines
the exact accidental discrete symmetry of the renormalizable isolated bosonic
Higgs action.  This qualification is essential: higher-dimensional
$G(2)$-invariant operators need not respect $Z_S$.  The completeness analysis in
Sec.~\ref{sec:pureDarkOperatorBasis} nevertheless shows that the accidental
protection is stronger than a naive tensor count suggests: the physical
pure-dark odd basis is empty through dimension seven and begins at dimension
nine.  There is also
an important gauge-theory qualification.  Since $U_B$ is an inner $G(2)$ gauge
transformation, it acts trivially on a fully $G(2)$-gauge-invariant local operator.
On such operators the physical action of $G_B=Z_SU_B$ therefore reduces exactly to
the parity of the number of $S$ insertions.  The charge-conjugation language used
below refers to the Higgs-phase/FMS components after the vacuum direction has been
chosen; it is not an additional global action of $U_B$ on a gauge-invariant state.

We refer to $G_B$ as the \emph{dark $G$-parity}.  The terminology is meant
to emphasize its physical role rather than an identity with the familiar
isospin $G$-parity of QCD.  Here the construction combines a discrete
transformation of the Higgs sector with the $G(2)$ transformation that acts
as charge conjugation on the daughter gauge sector.  Once one restricts to
gauge-sector states after
\begin{equation}
	G(2)\longrightarrow SU(3),
\end{equation}
the scalar-sign factor has no independent action on the gauge fields and
the $G_B$ eigenvalue reduces to the charge-conjugation grading.  Therefore
\begin{equation}
	C=-1
	\quad\Longleftrightarrow\quad
	G_B=-1
\end{equation}
for the confined gauge-sector channels relevant here.  The $C$-odd
channels are precisely the $G$-odd sector of the isolated bosonic theory,
whereas $C$-even states belong to the unprotected $G$-even sector.

This observation changes the formulation of the DM problem.
The protecting quantum number is not ``the number of $X$ constituents''
and there is no independent $X\to-X$ particle-number parity.  What can be
protected is instead the lightest \emph{gauge-invariant $G$-odd
	eigenstate}.  A physical state may have a leading $X\bar X$, three-vector,
or glue component in a particular operator basis while its stability is
fixed by its $G_B$ quantum number.  In particular, the scalar
$0^{++}\sim X_\mu\bar X^\mu$ channel considered below is $G$ even, whereas
the $1^{+-}$ two-vector/glue and $0^{--}$ three-vector sectors provide
$G$-odd candidates.

The remainder of the paper asks whether this renormalizable accidental
symmetry of the isolated $G(2)$ gauge--Higgs system can be promoted to an exact
selection rule when the theory is embedded into the
Class-B exceptional construction.  This is a substantially stronger
requirement than the absence of renormalizable portals: the transformation
must first lift consistently to the $E_6/E_7$ scalar and matter parents and
must ultimately act within the mirror-free chiral Standard Model
subspace.  The successive realizations denoted below by $G_D$ and $P_D$
should therefore be understood as attempts to reproduce, at the EFT and
ultraviolet levels respectively, the same protecting dark $G$-parity
identified here.

\subsection{Why the scalar $X\bar X$ channel is not the protected relic}
\label{sec:xxbar-nogo}
\paragraph{Relation to the earlier GUT treatment.}
The earlier Class-B analysis \cite{MasiE6E7} used the standard secluded
Yang--Mills intuition that the lightest confined state is typically the
scalar $0^{++}$ glueball and adopted that channel as a convenient
representative of the late dark relic.  That was a phenomenological working
identification, not a derivation of the state protected by the specific
discrete symmetry of the broken $G(2)$ theory.  The distinction becomes
important once the confined spectrum is classified by gauge-invariant
operators rather than by constituent labels alone.

There are in fact two conceptually different scalar objects that can carry
$J^{PC}=0^{++}$: a state dominated by the unbroken $SU(3)$ field strength and
the heavy meson-like contraction $X_\mu\bar X^\mu$.  Their microscopic
composition is different, but both are even under the conjugation relevant
for dark $G$-parity.  The present analysis therefore refines, rather than
merely relabels, the earlier picture: the protected relic cannot be selected
simply by choosing the lightest scalar channel.  Moreover, the optional
constituent sign $X_\mu\to-X_\mu$ mentioned in Ref.~\cite{MasiE6E7} is not an
independent symmetry of the $G(2)/SU(3)$ algebra.  The $E_6/E_7$ uplift is
thus not what makes the $0^{++}$ state unprotected: that conclusion already
holds in the isolated gauge--Higgs theory.  The ultraviolet problem studied
later is instead whether the genuinely $G$-odd channels can retain their
protecting symmetry after the chiral exceptional embedding is restored.

The scalar two-vector color singlet
\begin{equation}
M_0=X_\mu^\alpha\bar X^\mu_\alpha
\end{equation}
is not protected by the $G(2)$ conjugation.  Ref.~\cite{ButtazzoG2Higgs} explicitly identifies the corresponding
$W\bar W$ meson as a state that decays into glueballs.  Their charge
conjugation acts as
\begin{equation}
\resizebox{0.98\columnwidth}{!}{$\displaystyle
C:\qquad
X_\mu^\alpha\longmapsto-\bar X_{\mu\alpha},
\qquad
A_{\rm imag}\longmapsto A_{\rm imag},
\qquad
A_{\rm real}\longmapsto-A_{\rm real},
$}
\label{eq:ButtazzoCfull}
\end{equation}
and therefore
\begin{equation}
C(M_0)=M_0.
\end{equation}
The scalar meson is therefore $C$ even and, in the isolated gauge sector,
$G_B$ even.  It is not protected by the dark $G$-parity identified above.

Nor can this conclusion be reversed by assigning an independent constituent
sign $X\to-X$.  The
$G(2)/SU(3)$ algebra contains
\begin{equation}
[X,X]\sim\bar X,\qquad
[\bar X,\bar X]\sim X,
\end{equation}
so a phase
$X\to e^{i\alpha}X$, $\bar X\to e^{-i\alpha}\bar X$
must obey
\begin{equation}
e^{3i\alpha}=1.
\end{equation}
The allowed phases are consequently cubic roots of unity.  The coset
supports the familiar cubic grading associated with semi-annihilation, but
not a $\mathbb Z_2$ particle-number symmetry for a single $X$ constituent
\cite{ButtazzoG2Higgs}.

%This
%statement does not constitute a classification of exotic non-invertible
%symmetries.

The direct conclusion is that
$M_0=X_\mu\bar X^\mu$ is not the protected DM state of the isolated
$G(2)$ theory.  The symmetry statement alone is sufficient to exclude this
channel as the protected relic.  As an independent dynamical check, the same scalar
channel is also extremely short lived in the weak-confinement benchmark.  With
$|\psi(0)|^2\simeq(C_F\alpha_s m_X)^3/(8\pi)$ and
$\sigma v(X\bar X\to gg)\sim\pi\alpha_s^2/m_X^2$, one obtains, up to
spin/color coefficients of order unity,
\begin{equation}
\resizebox{0.98\columnwidth}{!}{$\displaystyle
\Gamma_{0,2g}\sim\frac{C_F^3}{8}\alpha_s^5m_X
\simeq1.79\times10^5\ {\rm GeV},
\qquad
{\tau_{0,2g}\sim3.68\times10^{-30}\ {\rm s}.}
$}
\label{eq:XXbar2glifetime}
\end{equation}
This estimate is quoted only to illustrate the fate of the unprotected
scalar and is not used for the odd tensor channel below.

The no-go is therefore specific: it excludes the scalar contraction
$X_\mu\bar X^\mu$, not every gauge-invariant state whose FMS expansion
contains two heavy vectors.  A different Lorentz and internal contraction can
be $G$ odd.  Constructing that operator is the next step.

\subsection{Gauge-invariant origin of the $1^{+-}$ two-vector channel}
\label{sec:codd-two-vector}

This exclusion leaves open an important possibility: a
\emph{gauge-invariant} odd operator may still project onto a two-vector state
without being the scalar contraction $X_\mu\bar X^\mu$.  The natural place
to look is the octonionic baryon operator of Ref.~\cite{ButtazzoG2Higgs},
written there schematically as $O_{ijk}S^iS^jS^k$ with derivatives suppressed.
Those derivatives are not a cosmetic detail.  For three commuting scalar
fields evaluated at the same spacetime point,
\begin{equation}
O_{ijk}S^iS^jS^k=0,
\end{equation}
and the undifferentiated local contraction vanishes identically.  A nonzero
local interpolator must therefore distinguish at least two scalar factors.
The lowest-derivative gauge-invariant completion is
\begin{equation}
{
\mathcal B_{\mu\nu}^{(2)}
=
O_{ijk}\,
S^i(D_\mu S)^j(D_\nu S)^k .
}
\label{eq:B2FMSoperator}
\end{equation}
This operator is manifestly $G(2)$ gauge invariant.  Its Lorentz indices are
antisymmetric,
\begin{equation}
\mathcal B_{\mu\nu}^{(2)}
=
-\mathcal B_{\nu\mu}^{(2)}.
\end{equation}
The symmetry assignment is immediate.  Each factor $S$ or $D_\mu S$ changes
sign under the accidental scalar reflection, so the complete operator is odd
under $Z_S$ and hence under the vacuum-preserving dark $G$-parity
$G_B=P_B=Z_SU_B$ of Eq.~\eqref{eq:IsolatedGparity}.  On the daughter gauge
sector the $U_B$ factor acts as charge conjugation.  This is precisely the
combination needed for a gauge-invariant operator that is odd even though its
leading Higgs-phase content will contain one $X$ and one $\bar X$.

To expose that content, expand about the Higgs vacuum
\begin{equation}
S=w\,n+\cdots,\qquad n=e_7.
\end{equation}
At leading order
\begin{equation}
D_\mu S=-igw\,G_\mu^aT_a n+\cdots ,
\end{equation}
and therefore
\begin{equation}
\resizebox{0.98\columnwidth}{!}{$\displaystyle
\mathcal B_{\mu\nu}^{(2)}
=
-g^2w^3 K_{ab}\,G_\mu^aG_\nu^b+\cdots ,
\qquad
K_{ab}
=
O_{7jk}(T_an)^j(T_bn)^k.
$}
\label{eq:B2Kdef}
\end{equation}

Using the exact seven-dimensional generators given in Appendix A.4 of
Ref.~\cite{ButtazzoG2Higgs}, and ordering the six broken generators as
\begin{equation}
(8,9,11,12,13,14),
\end{equation}
we find the exact rational matrix
\begin{equation}
{
K=\frac13
\begin{pmatrix}
0&-1&0&0&0&0\\
1&0&0&0&0&0\\
0&0&0&1&0&0\\
0&0&-1&0&0&0\\
0&0&0&0&0&-1\\
0&0&0&0&1&0
\end{pmatrix}.
}
\label{eq:B2Kmatrix}
\end{equation}
Ref.~\cite{ButtazzoG2Higgs} uses
\begin{align}
X+\bar X&=(G_8,G_{11},G_{13}),\\
i(X-\bar X)&=(G_9,-G_{12},G_{14}).
\end{align}
Substitution into Eq.~\eqref{eq:B2Kdef} gives, up to the common
covariant-derivative sign convention,
\begin{equation}
{
\mathcal B_{\mu\nu}^{(2)}
=
-\frac{2i}{3}g^2w^3
\sum_{\alpha=1}^{3}
\left(
\bar X_{\mu\alpha}X_\nu^\alpha
-\bar X_{\nu\alpha}X_\mu^\alpha
\right)
+\cdots .
}
\label{eq:B2FMSleading}
\end{equation}
Equation~\eqref{eq:B2FMSleading} is the key result.  Its leading term is a
color-singlet $X\bar X$ bilinear, but with the antisymmetric Lorentz structure
required to make it odd under Eq.~\eqref{eq:ButtazzoCfull}.  Thus the
statement ``two vectors imply $C=+$'' is false: it applies to the scalar
contraction, not to the full two-vector sector.

The Lorentz structure also fixes the lowest spin assignment.  Dualizing the
purely spatial components,
\begin{equation}
\mathcal V_i
=
\frac12\epsilon_{ijk}\mathcal B_{jk}^{(2)}.
\end{equation}
one obtains an axial vector.  In the nonrelativistic $S$-wave limit the
leading channel therefore has
\begin{equation}
{J^{PC}=1^{+-}.}
\label{eq:B2JPC}
\end{equation}
The mixed components $\mathcal B_{0i}$ furnish the opposite-parity vector
projection and are velocity suppressed in the simplest nonrelativistic ground
state.  The lowest protected two-vector candidate is therefore the
$1^{+-}$ channel, not a scalar.

\paragraph{Quantum-number classification of the two-vector sector.}
The explicit operator above can be embedded in the complete nonrelativistic
classification of a neutral $X\bar X$ pair.  Charge conjugation exchanges the
triplet and antitriplet constituents.  For two spin-one constituents with
orbital angular momentum $L$ and total spin $S=0,1,2$, the standard assignments
are
\begin{equation}
P=(-1)^L,
\qquad
C=(-1)^{L+S}.
\label{eq:XXbarPCrules}
\end{equation}
The exchange phase is $(-1)^L(-1)^{2-S}=(-1)^{L+S}$.  Unlike the familiar
fermion--antifermion derivation, no additional anticommutation sign appears \cite{Varshalovich1988,LandauLifshitzQM,MessiahQM}.
Table~\ref{tab:XXbarJPC} then makes the symmetry content transparent.  In the
isolated Higgs theory $G_B=P_B$ acts as charge conjugation on gauge-sector
states, so every $C=-1$ row is a dark-$G$-odd channel.

\begin{table}[htbp]
\centering
\small
\caption{Lowest nonrelativistic $X\bar X$ channels.  ``Odd'' means eigenvalue
$P_B=-1$ under the protecting discrete transformation, not negative spatial
parity.  The explicitly constructed FMS operator in
Eq.~\eqref{eq:B2FMSoperator} selects the lowest $1^{+-}$ row.}
\label{tab:XXbarJPC}
\resizebox{\columnwidth}{!}{%
\begin{tabular}{ccclcc}
\toprule
$L$ & $S$ & allowed $J$ & $J^{PC}$ & $P_B$ & status \\
\midrule
0 & 0 & 0 & $0^{++}$ & $+$ & unprotected scalar \\
0 & 1 & 1 & $1^{+-}$ & $-$ & protected-sector candidate \\
0 & 2 & 2 & $2^{++}$ & $+$ & unprotected \\
1 & 0 & 1 & $1^{--}$ & $-$ & odd excitation \\
1 & 1 & $0,1,2$ & $0^{-+},1^{-+},2^{-+}$ & $+$ & even sector \\
1 & 2 & $1,2,3$ & $1^{--},2^{--},3^{--}$ & $-$ & odd excitations \\
2 & 1 & $1,2,3$ & $1^{+-},2^{+-},3^{+-}$ & $-$ & odd excitations \\
\bottomrule
\end{tabular}%
}
\end{table}

%\paragraph{No $G$-odd scalar can be built from a pure two-vector component.}
The table has an immediate corollary that is important whenever the
microscopic state is modeled by an effective scalar.  For two spin-one
constituents, total $J=0$ requires $L=S$; hence
\begin{equation}
C=(-1)^{L+S}=(-1)^{2L}=+1.
\end{equation}
Hence it follows that a pure two-vector $X\bar X$ component cannot have either
$0^{+-}$ or $0^{--}$ quantum numbers.
A $G$-odd scalar can therefore not arise from a pure two-vector component.
Nor can mixing turn the $1^{+-}$ two-vector pole into a $0^{+-}$ or $0^{--}$
scalar: different $J^{PC}$ sectors belong to distinct correlator blocks.  Any
protected scalar must instead have a different leading microscopic structure,
which will be identified below with the three-vector sector.

\paragraph{Gauge-invariant state versus constituent language.}
It is useful to keep the two statements separate.  The scalar meson
$X_\mu\bar X^\mu$ remains unprotected and may decay.  What
Eq.~\eqref{eq:B2FMSleading} establishes is that the same pair of heavy fields
can appear as the leading term of a \emph{different}, gauge-invariant and
$G$-odd operator.  A state that is baryon-like in the exact $G(2)$ operator
language may therefore be two-vector dominated in a Higgs-phase FMS
expansion.  This is precisely the distinction emphasized by the original FMS
construction and by modern studies of gauge-invariant spectra
\cite{FrohlichMorchioStrocchi1980,FrohlichMorchioStrocchi1981,MaasFMSReview,
MaasMarklMuller2023,MaasTorekSU3,MaasSondenheimerTorek2017,
MaasSondenheimerTorekSUN2018,MaasPedro2016}: physical particles are associated with poles of local
gauge-invariant correlators, whereas elementary-field ``constituents'' are
terms in a gauge-dependent vacuum expansion.

\subsection{Why two massive vectors are the minimal FMS content}
\label{sec:minimal-fms-two}

The appearance of two massive vectors in Eq.~\eqref{eq:B2FMSleading} is not
an artifact of a specially chosen derivative structure.  Representation theory
shows that two is the minimum possible number of broken-vector fields in a
local $G$-odd interpolator descending from the octonionic scalar-baryon
family.

First consider operators with one scalar and field strengths.  One scalar and
one adjoint cannot form a $G(2)$ singlet because
\begin{equation}
\mathbf{7}\otimes\mathbf{14}=\mathbf{7}\oplus\mathbf{27}\oplus\mathbf{64}
\end{equation}
contains no $\mathbf{1}$.  Adding a second adjoint does not help: a singlet would require a
$\mathbf{7}$ inside $\mathbf{14}\otimes\mathbf{14}$, but
\begin{equation}
\mathbf{14}\otimes\mathbf{14}
=
\mathbf{1}\oplus\mathbf{14}\oplus\mathbf{27}\oplus\mathbf{77}\oplus\mathbf{77}'
\end{equation}
contains no $\mathbf{7}$ \cite{Slansky1981,AdamsExceptional}.  Thus the first internal $G(2)$ contraction containing only one scalar insertion
appears with three field strengths, schematically $\epsilon SGGG$.  This statement
concerns the internal representation only.  Its Lorentz projections must still be
checked separately: the $1^{+-}$ vector projection used below is nonzero, whereas a
derivative-free Lorentz-scalar three-field-strength projection vanishes because the
internal tensor is symmetric while the scalar Lorentz contraction is antisymmetric.
A scalar $0^{--}$ glue interpolator therefore requires derivative dressing (or an
extended Wilson-loop construction), as is also familiar from QCD oddball-current
analyses \cite{PimikovEtAl2017Oddballs}.

The scalar-baryon family leads to the complementary conclusion.  The
invariant tensor $O_{ijk}$ requires three fundamentals, but three undifferentiated
commuting scalars vanish when contracted antisymmetrically.  At least two
factors must therefore be distinguished by derivatives.  After expanding
around the Higgs vacuum, the remaining undifferentiated scalar can be replaced
by $wn$, while each $D_\mu S$ contributes one broken gauge field.  Therefore
\begin{equation}
{
N_X^{\rm FMS,min}=2
}
\label{eq:minimalFMSvalence}
\end{equation}
for the odd $O\,S^3$ operator family.  In particular, there is no local
one-$X$ gauge-invariant interpolator in this family.  The two-vector channel is
therefore the \emph{minimal} Higgs-phase realization of an exact odd
$G(2)$ baryonic operator, not a contrived higher excitation.

\subsection{From operator overlap to a physical two-vector pole}
\label{sec:B2pole}

An interpolating operator and a physical bound-state pole are not the same
statement.  Equation~\eqref{eq:B2FMSleading} proves nonzero overlap with a
color-singlet $X\bar X$ configuration, but the associated correlator could in
principle contain only the two-particle continuum.  In the weakly coupled Higgs
regime there is nevertheless a simple dynamical reason to expect a bound-state
pole: the singlet interaction is attractive,
\begin{equation}
V(r)=-\frac{C_F\alpha_s}{r},
\qquad C_F=\frac43,
\qquad
\mu=\frac{m_X}{2}.
\end{equation}
so the leading Coulombic binding energy is \cite{LandauLifshitzQM,MessiahQM}
\begin{equation}
E_{B,n}
=
\frac{\mu(C_F\alpha_s)^2}{2n^2}
=
\frac{m_XC_F^2\alpha_s^2}{4n^2}.
\label{eq:B2binding}
\end{equation}
For
\begin{widetext}
\begin{equation}
g_3(M_\chi)=g_{G(2)}(M_\chi)=0.565,
\qquad
\alpha_s=0.0254031,
\qquad
m_X=5.7\times10^{13}\ {\rm GeV},
\end{equation}
\end{widetext}
the $1S$ estimate gives
\begin{widetext}
\begin{equation}
E_B=1.6348\times10^{10}\ {\rm GeV},\qquad
\frac{E_B}{m_X}=2.8681\times10^{-4},\qquad
M_{2X}=2m_X-E_B
=1.13984\times10^{14}\ {\rm GeV}.
\label{eq:B2mass}
\end{equation}
\end{widetext}
The corresponding Bohr radius is
\begin{equation}
a_0=
\frac{1}{\mu C_F\alpha_s}
=
1.036\times10^{-12}\ {\rm GeV}^{-1}
=
2.04\times10^{-28}\ {\rm m}.
\end{equation}

The numerical value is useful only inside its domain of validity.  The
Coulombic description requires the Bohr radius to be short compared with the
confinement length.  Introducing
\begin{equation}
x\equiv\frac{\sqrt\sigma}{m_X},
\end{equation}
one finds
\begin{equation}
{
a_0\sqrt\sigma=\frac{2x}{C_F\alpha_s},
\qquad
x_{a_0\sqrt\sigma=1}=\frac{C_F\alpha_s}{2}=0.01694.
}
\label{eq:CoulombValidityX}
\end{equation}
The weak-binding estimate is therefore controlled for $x\ll1$, including
the literal ordinary-QCD realization, but it is not controlled in the
$x=O(0.1)$ regime used below as a spectroscopic deformation.  Once confinement
becomes comparable with $m_X$, the physical two-vector pole mass must instead
be regarded as the nonperturbative function
\begin{equation}
{\mu_2(x)\equiv\frac{M_{2X}(x)}{m_X}.}
\label{eq:mu2Definition}
\end{equation}

\paragraph{Domain of validity of the reference interpolation.}

At intermediate $x$ two different approximations will be combined: lattice
ratios imported from decoupled pure $SU(3)$ Yang--Mills theory and constituent
potential models for the heavy-vector sector.  Each is controlled in a
different limit.  The pure-glue inputs require $k_Jx\ll1$, whereas the
constituent inequalities are rigorous only within the explicit fixed-$m_X$
Hamiltonians in which they are derived.  Their crossings at intermediate $x$
are therefore \emph{reference interpolations}: they identify where a change
of ordering could occur and define quantitative targets for a dedicated
$G(2)+\mathbf{7}$ lattice generalized eigenvalue problem (GEVP), but they are not first-principles predictions of
the full gauge--Higgs spectrum.
Here and below, \(G(2)+\mathbf{7}\) denotes the isolated
\(G(2)\) gauge--Higgs theory containing a scalar
\(\chi\) in the fundamental \(\mathbf{7}\) of \(G(2)\), with
\(\langle\chi\rangle=v_\chi e_7\) breaking
\(G(2)\to SU(3)_C\).  This shorthand refers to the broken
gauge--Higgs effective theory itself, prior to imposing its full
\(E_6/E_7\) ultraviolet parent completion and the Class-B chiral
projector.

Within the genuinely weak-confinement regime, however, the conclusion is
simple: an $X\bar X$ pole is dynamically plausible and its mass lies well
below the naive three-constituent scale
$3m_X\simeq1.71\times10^{14}\,$GeV.

The annihilation properties again depend on the symmetry sector.  For the
unprotected $C$-even scalar, the perturbative estimate
\begin{equation}
\Gamma_{2g}\sim\alpha_s^5m_X
\end{equation}
remains the relevant weak-coupling scale.  For the
$1^{+-}$ odd level, two on-shell identical gauge bosons are excluded by the
spin-one selection rule, while a $C$-odd three-gluon channel is possible \cite{Landau1948,Yang1950}.
As a purely perturbative scale estimate, adding one power of $\alpha_s$ to
Eq.~\eqref{eq:XXbar2glifetime} gives
\begin{equation}
\Gamma_{1^{+-},3g}\sim4.54\times10^3\ {\rm GeV},
\qquad
{\tau_{1^{+-},3g}\sim1.45\times10^{-28}\ {\rm s}.}
\label{eq:XXbar3glifetime}
\end{equation}
This last expression must not be interpreted as the physical width of the
lightest exactly odd hadron.  In the confining theory free gluons are not the
asymptotic final states, and exact dark $G$-parity stabilizes the lowest odd
\emph{gauge-invariant} eigenstate.  The real question is therefore spectral:
does the two-vector-dominated $1^{+-}$ pole lie below, above, or strongly mix
with a lighter odd glue-dominated state?

\subsection{Competition with the $1^{+-}$ glue sector and the physical stability criterion}
\label{sec:odd-correlator}

The two-vector channel does not live in isolation.  The same gauge-invariant
odd sector contains an operator whose leading FMS expansion can be built
entirely from the unbroken daughter gauge fields.  Ref.~\cite{ButtazzoG2Higgs}
identifies the corresponding baryonic structure as
\begin{equation}
\mathcal B_G
=
\epsilon_{i_1\cdots i_7}
S^{i_1}G^{i_2i_3}G^{i_4i_5}G^{i_6i_7}.
\label{eq:SGGGoperator}
\end{equation}
After inserting the Higgs vacuum $S=we_7$, this operator does not require
any broken-vector field at leading order: its internal tensor has nonzero
components involving only the eight unbroken $SU(3)$ generators.  In the same
exact basis used above, one explicit witness is
\begin{equation}
\epsilon_{7jklmns}
(T_1)_{jk}(T_1)_{lm}(T_{10})_{ns}
=
-\frac{4\sqrt3\,i}{3}.
\label{eq:AAAexactWitness}
\end{equation}
The full contraction is more informative than a single nonzero component.  Let
$t_a$ be the restriction of the eight unbroken generators to the
$H_\chi=+1$ three-dimensional subspace.  They obey
$\operatorname{tr}_3(t_at_b)=\delta_{ab}/2$.  Defining
\begin{equation}
C_{abc}=\epsilon_{7jklmns}(T_a)_{jk}(T_b)_{lm}(T_c)_{ns},
\end{equation}
the exact certificate finds a totally symmetric tensor with
\begin{equation}
{
C_{abc}=-4i\,d_{abc}
}
\label{eq:AAAfullDsymbol}
\end{equation}
in the present phase convention, and hence
\begin{equation}
\sum_{abc}|C_{abc}|^2=\frac{640}{3}
=16\sum_{abc}d_{abc}^2.
\end{equation}
Hence the same exact odd baryonic operator has an unsuppressed,
normalization-fixed three-gluon FMS component, together with mixed
$A X\bar X$ terms.  The $1^{+-}$ spectral problem must therefore allow
mixing between heavy-vector and glue-dominated interpolators from the outset.

The decoupled pure-$SU(3)$ glueball spectrum provides the cleanest reference
limit for this competition.  It is well established by lattice calculations
and has also been compared with constituent and continuum bound-state descriptions \cite{TeperGlueball1998,MorningstarPeardon1997,MorningstarPeardon1999,LuciniTeperWenger2004,MeyerGlueballs2005,ChenGlueballs2006,LuciniRagoRinaldi2010,MathieuBuisseret2009,BoulangerBuisseretMathieu2008,BuisseretMathieuSemay2009,EichmannGlueball2021,MeyersSwanson2013}.  Their use here requires an explicit qualification.  Pure-Yang--Mills masses
are controlled inputs for the broken gauge--Higgs theory only when
$m_X\gg\sqrt\sigma$, equivalently when every imported glue level obeys
$k_Jx\ll1$.  At intermediate $x$ the heavy vectors are no longer
parametrically decoupled from the nominal glue spectrum.  In that regime the
lattice ratios below are retained only as deformation diagnostics, not as
first-principles masses of the full $G(2)+\mathbf{7}$ theory.  In the decoupled daughter theory, the spectrum contains a light
$1^{+-}$ glueball with
\begin{equation}
\frac{m_{1^{+-}}}{\sqrt\sigma}=k_1=6.065(40)
\label{eq:SU3oneplusminus}
\end{equation}
\cite{AthenodorouTeper2020}.  The comparison with the heavy-vector branch should therefore be written in
a form that does not assume the Coulombic result outside its domain:
\begin{equation}
{\mu_2(x)<k_1x}
\label{eq:twoVectorStabilityThreshold}
\end{equation}
is the sufficient condition for the two-vector branch to lie below the
lightest pure-gauge $1^{+-}$ competitor in the decoupled-daughter limit.  Equation~\eqref{eq:twoVectorStabilityThreshold}
is controlled only together with the pure-Yang--Mills decoupling condition; a numerical crossing near
$x\simeq0.330$ resulted from setting $\mu_2=1.999713$ even though
Eq.~\eqref{eq:CoulombValidityX} gives $a_0\sqrt\sigma\simeq19.5$ there.

\begin{figure}[htbp]
\centering
\includegraphics[width=\linewidth]{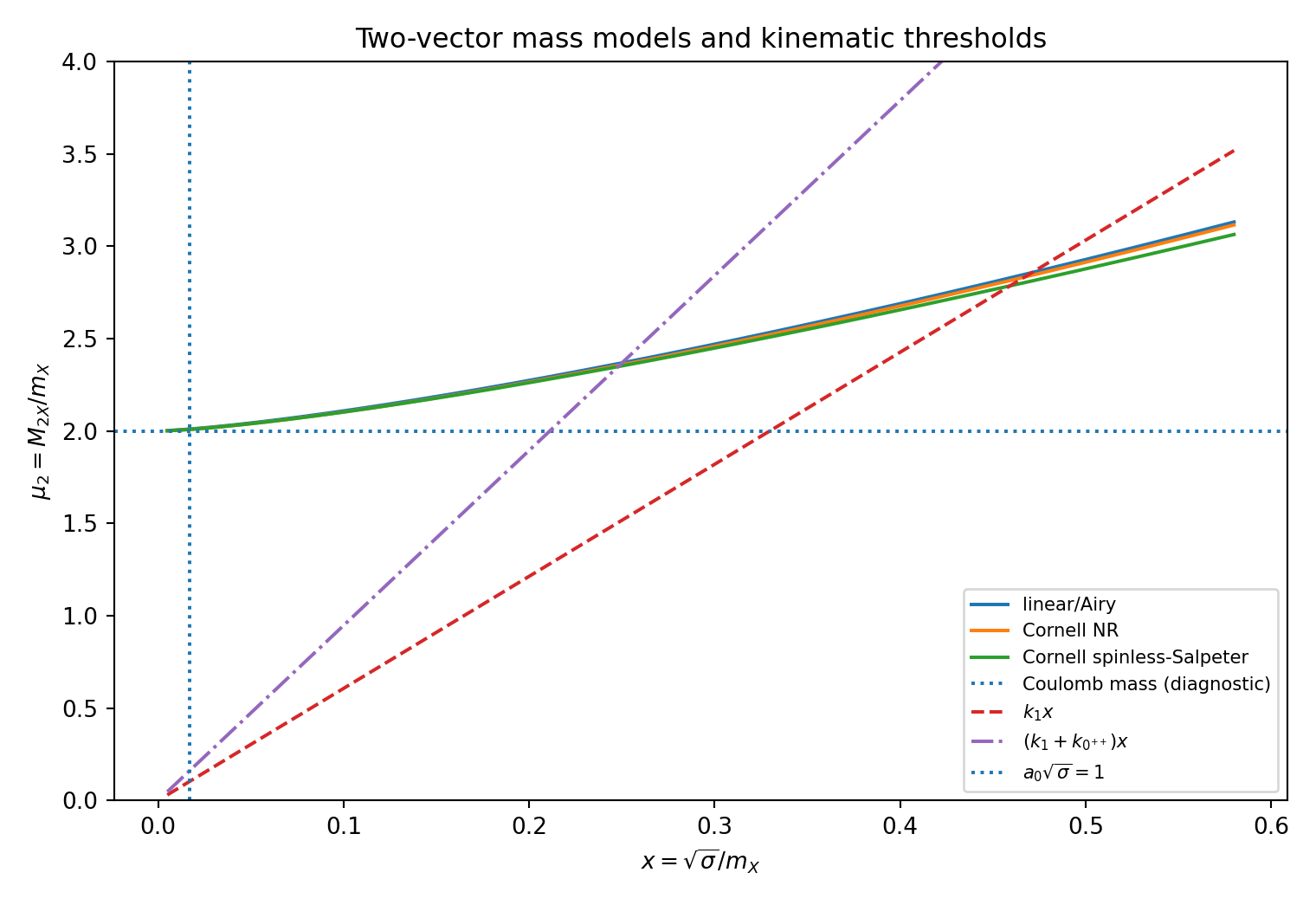}
\caption{Illustrative two-vector mass functions compared with the decoupled pure-$SU(3)$
odd threshold $k_1x$ and the pure-glue specialization of the lighter-odd-plus-even threshold
$(k_1+k_{0^{++}})x$.  A lighter radial scalar lowers this second threshold.  The vertical line marks $a_0\sqrt\sigma=1$; the
Coulombic stability interval lay far to its right and was therefore outside
the Coulomb domain.}
\label{fig:mu2PotentialModels}
\end{figure}

To visualize how the ordering might evolve once confinement is no longer a
small perturbation, we use three standard potential models as deliberately
limited diagnostics, with conventional variational and angular-momentum
methods \cite{LandauLifshitzQM,MessiahQM,Varshalovich1988}.  A pure
linear potential gives the Airy ground-state estimate
\begin{equation}
\mu_2^{\rm Airy}(x)=2+|a_1|x^{4/3},
\qquad |a_1|=2.338107\ldots .
\label{eq:AiryMu2}
\end{equation}
A nonrelativistic (NR) Gaussian variational treatment of the Cornell potential\cite{EichtenCornell1978}
$V(r)=-C_F\alpha_s/r+\sigma r$ gives, with $t=b/m_X$,
\begin{equation}
\mu_2^{\rm C,NR}(x)=2+\min_{t>0}
\left[
\frac32t^2-\frac{2C_F\alpha_s}{\sqrt\pi}t
+\frac{2x^2}{\sqrt\pi\,t}
\right].
\label{eq:CornellNRMu2}
\end{equation}
Replacing the nonrelativistic kinetic term by the spinless-Salpeter expectation
in the same Gaussian state defines a third diagnostic,
\begin{align}
\mu_2^{\rm C,Sal}(x)=2+\min_{t>0}\Bigg\{&2\left[
\int_0^\infty\frac{4q^2e^{-q^2}}{\sqrt\pi}
\sqrt{1+t^2q^2}\,dq-1\right]\nonumber\\
&-\frac{2C_F\alpha_s}{\sqrt\pi}t
+\frac{2x^2}{\sqrt\pi\,t}\Bigg\}.
\label{eq:CornellSalMu2}
\end{align}
None of these potentials is a substitute for the gauge-invariant
$G(2)$--Higgs spectrum.  Their role is only to estimate how the heavy
two-vector branch can depart from the Coulombic value once confinement becomes
comparable with $m_X$.  Conversely, the pure-$SU(3)$ glue masses lose their
parametric decoupling justification in exactly this same intermediate-$x$
region.  The crossings produced by combining the two descriptions must
therefore be read as reference interpolations and as targets for a future
$G(2)+\mathbf{7}$ lattice calculation, not as predictions of the full theory.
Table~\ref{tab:mu2ModelCrossings} lists the three reference crossings and
Fig.~\ref{fig:mu2PotentialModels} displays the corresponding mass functions.

\begin{table}[htbp]
	\centering
	\small
	\setlength{\tabcolsep}{4pt}
	\renewcommand{\arraystretch}{1.25}
	\caption{Zero-mixing reference crossings for the three potential models.
		The second line in each numerical entry gives the conservative $1\sigma$
		envelope propagated from $k_1=6.065(40)$ and $k_{0^{++}}=3.405(21)$.
		Potential-model systematics are not included, so the entries are diagnostic
		rather than first-principles boundaries of the full $G(2)+\mathbf{7}$ theory.}
	\label{tab:mu2ModelCrossings}
	\begin{tabular}{@{}lcc@{}}
		\toprule
		model & $x_{\rm closed}$ & $x_{\rm lightest}$\\
		\midrule
		linear/Airy
		& \shortstack{$0.25010$\\[-1pt]{\scriptsize $[0.24808,\,0.25215]$}}
		& \shortstack{$0.47106$\\[-1pt]{\scriptsize $[0.46594,\,0.47630]$}}\\
		Cornell (NR)
		& \shortstack{$0.24877$\\[-1pt]{\scriptsize $[0.24677,\,0.25080]$}}
		& \shortstack{$0.46714$\\[-1pt]{\scriptsize $[0.46210,\,0.47230]$}}\\
		Cornell (Salpeter)
		& \shortstack{$0.24796$\\[-1pt]{\scriptsize $[0.24598,\,0.24996]$}}
		& \shortstack{$0.45925$\\[-1pt]{\scriptsize $[0.45451,\,0.46409]$}}\\
		\bottomrule
	\end{tabular}
\end{table}
%\begin{table}[htbp]
%\centering
%\caption{Illustrative zero-mixing reference crossings obtained by combining
%three potential models with decoupled pure-$SU(3)$ glue inputs.  They are not
%first-principles crossings of the full $G(2)+\mathbf{7}$ theory at intermediate $x$.  The ranges in parentheses are conservative
%linear one-sigma envelopes from the quoted lattice errors
%$k_1=6.065(40)$ and $k_{0^{++}}=3.405(21)$; potential-model uncertainty is
%larger.  Here $x_{\rm closed}$ is the lower boundary at which the
%pure-glue specialization of the lower-odd-plus-even decay channel closes as $x$ increases, while
%$x_{\rm lightest}$ is the upper boundary at which the two-vector branch
%itself becomes the lightest odd level.}
%\label{tab:mu2ModelCrossings}
%\resizebox{\columnwidth}{!}{%
%\begin{tabular}{lcc}
%\toprule
%model & $x_{\rm closed}$ & $x_{\rm lightest}$\\
%\midrule
%linear/Airy & $0.25010\;(0.24808\!-
%0.25215)$ & $0.47106\;(0.46594\!-
%0.47630)$\\
%Cornell, nonrelativistic Gaussian & $0.24877\;(0.24677\!-
%0.25080)$ & $0.46714\;(0.46210\!-
%0.47230)$\\
%Cornell, spinless Salpeter & $0.24796\;(0.24598\!-
%0.24996)$ & $0.45925\;(0.45451\!-
%0.46409)$\\
%\bottomrule
%\end{tabular}%
%}
%\end{table}

The limitations are visible numerically: at the upper Salpeter crossing the
Gaussian momentum scale has already reached
$p_{\rm rms}/m_X\simeq0.57$.  Even the relativized-potential result is
therefore not a precision boundary.  What survives the model comparison is a
more modest statement: a finite high-confinement region in which the ordering
changes is not excluded for an appropriate full-theory mass function
$\mu_2(x)$, but the boundaries and the mixed glue/constituent levels are
intrinsically nonperturbative.

Figure~\ref{fig:mu2PotentialModels} should accordingly be read as a map of
the reference interpolation, not as a phase diagram of the complete theory.

A useful analytic stress test is
$\mu_2=2+c x^{4/3}$.  A crossing with $k_1x$ exists only for
\begin{equation}
{
c\le c_{\max}=\left(\frac{27k_1^4}{512}\right)^{1/3}=4.148
}
\label{eq:cmaxWindow}
\end{equation}
at the central lattice value.  Even the existence of an interval in which the heavy two-vector branch is
the lightest odd level is therefore a dynamical question, not a consequence
of dimensional analysis alone.

The model crossings only motivate the calculation that actually decides the
spectrum: the gauge-invariant odd-sector correlator matrix.  A minimal
$1^{+-}$ variational basis contains
\begin{equation}
\resizebox{0.98\columnwidth}{!}{$\displaystyle
\mathcal O_1^i
=
\frac12\epsilon^{ijk}
O_{abc}S^a(D_jS)^b(D_kS)^c,
\qquad
\mathcal O_2^i
=
\left[\epsilon SGGG\right]_{1^{+-}}^i,
$}
\end{equation}
supplemented on the lattice by optimized three-gluon/Wilson-loop operators
and by further derivative $OSSS$ operators.  Define
\begin{equation}
C_{mn}(t)
=
\langle
\mathcal O_m(t)\mathcal O_n^\dagger(0)
\rangle .
\end{equation}
The physical poles follow from the generalized eigenvalue problem
\begin{equation}
\resizebox{0.98\columnwidth}{!}{$\displaystyle
{
C(t)\,v_n
=
\lambda_n(t,t_0)\,
C(t_0)\,v_n,
\qquad
\lambda_n(t,t_0)
\stackrel{t\gg t_0}{\sim}
e^{-M_n(t-t_0)}.
}
$}
\label{eq:oddGEVP}
\end{equation}
The eigenvalues determine the finite-volume spectrum, while the eigenvectors
measure the operator overlaps and therefore quantify whether a given pole is
predominantly two-vector, glue, or an allowed same-$J^{PC}$ three-vector
projection.  At the analytic level a
two-channel truncation may be written
\begin{equation}
H_{\rm odd}^{(1^{+-})}
=
\begin{pmatrix}
M_{2X}&\Delta\\
\Delta&m_G
\end{pmatrix},
\qquad
m_G\equiv m_{1^{+-}}.
\label{eq:odd2x2Hamiltonian}
\end{equation}
Equation~\eqref{eq:oddGEVP} is the definitive nonperturbative formulation of
the $1^{+-}$ problem.  Before such correlators are available, the two-channel
Hamiltonian in Eq.~\eqref{eq:odd2x2Hamiltonian} isolates one exact piece of
kinematics that is often missed by constituent counting: the heavier odd state
need not decay simply because a lighter odd pole exists.

\subsection{Mixing, level repulsion and a second kinematic stability window}
\label{sec:odd-two-level-exact}

The two-channel model can be solved exactly and is useful because it
separates \emph{level ordering} from \emph{kinematic stability}.  Define the
unmixed level separation
\begin{equation}
\delta=m_G-M_{2X}.
\end{equation}
The exact eigenvalues are
\begin{equation}
{
E_\pm
=
\frac{M_{2X}+m_G}{2}
\pm
\frac12\sqrt{\delta^2+4\Delta^2}.
}
\label{eq:odd2x2eigenvalues}
\end{equation}
The two-vector probabilities in the lower and upper states are
\begin{align}
P_{2X}^{(-)}
&=
\frac12
\left(
1+\frac{\delta}{\sqrt{\delta^2+4\Delta^2}}
\right),\\
P_{2X}^{(+)}
&=
\frac12
\left(
1-\frac{\delta}{\sqrt{\delta^2+4\Delta^2}}
\right).
\label{eq:odd2x2probabilities}
\end{align}
These expressions make the avoided crossing explicit.  For $\delta>0$ the
lower eigenstate is predominantly two-vector, whereas for $\delta<0$ that
character migrates to the upper eigenstate.  At $\delta=0$ both states are
exactly $50$--$50$ mixtures and the level repulsion is $2|\Delta|$.

The crucial consequence is that being the \emph{lightest odd state} is a
sufficient condition for stability, but not a necessary one.
Suppose
\begin{equation}
m_G<M_{2X},
\end{equation}
where $m_G\equiv m_{1^{+-}}^{\rm(glue)}$
denotes the unmixed glue-dominated \(1^{+-}\) reference level while
\(M_{2X}\) denotes the unmixed two-vector \(X\bar X\)-dominated level, so that the latter is the upper eigenstate.  Exact dark $G$-parity forbids the upper odd level from decaying to an
all-even final state.  A transition to the lower odd state alone is also
kinematically impossible, because a massive one-particle state cannot decay
into a lighter one-particle state without carrying away additional
four-momentum.  The first allowed hidden final state must therefore contain
the lower odd eigenstate \emph{plus} at least one parity-even excitation of
the gauge--Higgs theory.
Define
\begin{equation}
 m_{\rm even}^{\min}
 \equiv\min\left(m_{0^{++},{\rm glue}},m_{\rm rad},\ldots\right),
 \qquad
 \epsilon_e(x)\equiv\frac{m_{\rm even}^{\min}}{m_X}.
 \label{eq:LightestEvenThreshold}
\end{equation}
Here $m_{\rm rad}$ is the radial Higgs excitation along the dark-breaking
direction, and the ellipsis denotes any still lighter even excitation present
in the full gauge--Higgs spectrum.  In the
decoupled pure-$SU(3)$ limit with all scalar excitations heavier than the glue
sector, $m_{\rm even}^{\min}=m_{0^{++},{\rm glue}}=3.405\sqrt\sigma$ and
$\epsilon_e=k_{0^{++}}x$.  The heavier odd level is kinematically stable whenever
\begin{equation}
E_+-E_-<m_{\rm even}^{\min}.
\end{equation}
Using Eq.~\eqref{eq:odd2x2eigenvalues},
\begin{equation}
\sqrt{(M_{2X}-m_G)^2+4\Delta^2}<m_{\rm even}^{\min}.
\label{eq:metastableActuallyStableCondition}
\end{equation}
Hence, when $m_G<M_{2X}$,
\begin{equation}
|\Delta|<\Delta_{\rm max}
=\frac12\sqrt{
(m_{\rm even}^{\min})^2-(M_{2X}-m_G)^2
}.
\label{eq:DeltaMaxOdd}
\end{equation}
Once the actual lightest even threshold is known, this inequality is exact
within the two-level odd-sector truncation.

At zero odd-sector mixing the physical content is especially transparent.
Using Eq.~\eqref{eq:mu2Definition}, the spectrum separates into three regimes:
\begin{enumerate}[label=(\roman*)]
\item $\mu_2(x)<k_1x$: the two-vector branch can be the lightest odd state;
\item $k_1x<\mu_2(x)<k_1x+\epsilon_e(x)$: it is heavier than the lightest odd
state but remains kinematically stable;
\item $\mu_2(x)>k_1x+\epsilon_e(x)$: the lower-odd plus even decay channel is open.
\end{enumerate}
The exact zero-mixing condition for the stable upper branch is therefore
\begin{equation}
k_1x<\mu_2(x)<k_1x+\epsilon_e(x).
\label{eq:secondStabilityBoundary}
\end{equation}
If the lightest even state is specialized to the decoupled pure-glue
$0^{++}$ level, $\epsilon_e=k_{0^{++}}x$, and the Coulombic heavy mass is
inserted, one obtains the illustrative interval
$0.211163<x<0.329714$.  That number is \emph{not} a controlled prediction:
both the Coulombic constituent mass and the pure-Yang--Mills threshold are
being extrapolated outside their common decoupling regime.  In the full
Gauge--Higgs theory a lighter radial scalar, or any other lighter even state,
can narrow or eliminate the window.

With mixing, the exact two-level condition is correspondingly
\begin{equation}
\sqrt{\left[\mu_2(x)-k_1x\right]^2
+4\left(\frac{\Delta}{m_X}\right)^2}<\epsilon_e(x).
\label{eq:DeltaMaxMu2}
\end{equation}
Equation~\eqref{eq:DeltaMaxMu2} is therefore the robust result.  Unlike the
pure-glue numerical interval, it remains valid when the microscopic identity
of the lightest even state changes.

For any future nonperturbative determination of $\mu_2(x)$ and
$\epsilon_e(x)$, Eq.~\eqref{eq:DeltaMaxMu2} immediately gives the maximum
mixing compatible with absolute stability of the upper odd level.  We do not
quote a numerical $\Delta_{\max}$ table because doing so would simply
repackage the same uncontrolled Coulomb extrapolation.  The Airy, Cornell and
spinless-Salpeter results above are retained only as diagnostics of where the
nonperturbative calculation becomes relevant.

The conceptual lesson is simple but important.  An exact $\mathbb Z_2$
guarantees the stability of the lightest odd state, yet it does not force every
heavier odd excitation to decay.  A heavier odd pole is also absolutely stable
whenever the first parity-allowed multiparticle threshold lies above it.  This
``second window'' will be kept distinct from the stronger condition that the
heavy two-vector branch itself be the lightest odd state.

\subsection{Three-vector states and the separation of $J^{PC}$ blocks}
\label{sec:odd-three-channel}

The odd three-vector sector must be treated separately from the
$1^{+-}$ two-vector/glue block because odd $G$-parity alone does not fix the
spin.  The protected real combination
\begin{equation}
B_+=\frac{XXX+\bar X\bar X\bar X}{\sqrt2}
\end{equation}
is $C$ odd, but its $J^P$ assignment depends on the orbital and spin structure
of the interpolator.  For the simplest local nonrelativistic $L=0$ configuration the color
wavefunction is antisymmetric.  Bose symmetry then requires the spin
wavefunction to be totally antisymmetric as well, selecting the unique scalar
combination of three spin-one polarizations \cite{Varshalovich1988}.  The natural lowest assignment is
therefore
\begin{equation}
{J^{PC}=0^{--}}
\label{eq:3XSWaveJPC}
\end{equation}
This is a constituent-level inference for the local $S$ wave, not yet a
statement about the exact confined spectrum.  Derivatives or orbital
excitation can move three-vector strength into other $C=-$ sectors, including
$1^{+-}$.  Consequently a three-vector component may enter the $1^{+-}$ GEVP,
but the natural local three-vector scalar belongs to the independent
$0^{--}$ block.

\begin{widetext}
\begin{center}
\begin{minipage}{0.98\textwidth}

\centering
\small
\captionof{table}{Spin/parity refinement of the protected odd sector.  Exact $P_B$
stability classifies states by the discrete odd/even eigenvalue, while mixing
occurs only within a common $J^{PC}$ block.}
\label{tab:oddJPCsummary}
\begin{tabular}{p{0.29\textwidth}p{0.16\textwidth}p{0.39\textwidth}}
\toprule
interpolating component & $J^{PC}$ & present status \\
\midrule
$X_\mu\bar X^\mu$ & $0^{++}$ & $C$ even; not protected and rapidly annihilating \\
$OSDSDS\to X\bar X$ & $1^{+-}$ & exact gauge-invariant odd FMS channel proved here \\
pure-$SU(3)$ odd glue operator & $1^{+-}$ & competing odd spectral channel used in the GEVP \\
local $B_+\sim XXX+\bar X^3$ $S$ wave & $0^{--}$ & natural constituent assignment; does not mix with $1^{+-}$ \\
derivative/excited $B_+$ projection & $1^{+-}$ (possible) & may mix with the two-vector and glue $1^{+-}$ operators; overlap is spectral \\
\bottomrule
\end{tabular}
\end{minipage}
\end{center}
\end{widetext}
Table~\ref{tab:oddJPCsummary} summarizes the resulting separation.  The
important point is that the physically relevant competition is not simply
``two vectors versus three vectors'': it is a comparison of two distinct
$J^{PC}$ spectra, with mixing allowed only among operators inside a common
block.

\subsection{The physical spectroscopy problem: separate $1^{+-}$ and $0^{--}$ GEVPs}
\label{sec:1pm-vs-0mm}

Once the spin assignments are kept explicit, the spectroscopy problem
factorizes into two independent variational sectors.  The $1^{+-}$ block
contains the two-vector operator constructed above together with glue and
excited same-spin interpolators; the natural local three-vector scalar belongs
to a separate $0^{--}$ block.  Before comparing their spectra, it is useful to
make the scalar channel as explicit as the vector one.

\paragraph{Exact gauge-invariant interpolator for the $0^{--}$ channel.}
Define
\begin{equation}
{
\mathcal O_{0^{--}}(x)
=
\epsilon_{ijk}\,O_{abc}
(D_iS)^a(D_jS)^b(D_kS)^c .
}
\label{eq:O0mmExact}
\end{equation}
This operator is a rotational scalar.  The triple product of three polar
vectors makes it odd under spatial parity, while the three scalar factors make
it odd under the accidental sign $Z_S$ and hence under $G_B$.  In the isolated
Higgs theory it therefore interpolates the protected $J^{PC}=0^{--}$ sector.  Expanding about $S=w e_7$, each covariant derivative supplies one broken
vector.  With the explicit generators of Ref.~\cite{ButtazzoG2Higgs} one finds
\begin{equation}
K_{ABC}^{(3)}
=
O_{abc}(T_Ae_7)^a(T_Be_7)^b(T_Ce_7)^c,
\end{equation}
with $24$ nonzero ordered entries in the broken basis.  Rewriting them with the
$X,\bar X$ embedding gives the compact identity
\begin{equation}
\resizebox{0.98\columnwidth}{!}{$\displaystyle
{
\mathcal O_{0^{--}}
=
\frac{4\sqrt3}{9}g^3w^3\,
\epsilon_{ijk}\epsilon_{\alpha\beta\gamma}
\left(
X_i^\alpha X_j^\beta X_k^\gamma
+
\bar X_{i\alpha}\bar X_{j\beta}\bar X_{k\gamma}
\right)
+\cdots .
}
$}
\label{eq:O0mmFMS}
\end{equation}
Equation~\eqref{eq:O0mmFMS} removes any ambiguity about the existence of the
scalar odd channel: the local gauge-invariant $0^{--}$ operator has an exact
leading FMS component proportional to the $C$-odd
$XXX+\bar X\bar X\bar X$ combination discussed in
Ref.~\cite{ButtazzoG2Higgs}.  The scalar alternative is therefore genuine;
what remains uncertain is its mass relative to the $1^{+-}$ sector.

\paragraph{Two independent generalized-eigenvalue problems.}
The lowest useful operator bases may therefore be chosen independently as
\begin{widetext}
\begin{align}
\mathcal V_{1^{+-}}
&=
\left\{
\mathcal O_{2X}^{i}=\frac12\epsilon^{ijk}O_{abc}S^a(D_jS)^b(D_kS)^c,
\quad
\mathcal O_{G1}^{i}=[\epsilon SGGG]_{1^{+-}}^{i},\ldots
\right\},\\
\mathcal V_{0^{--}}
&=
\left\{
\mathcal O_{3X}=\mathcal O_{0^{--}},
\quad
\mathcal O_{G0}^{(D^2)}=[\epsilon S\,G\,D D G\,G]_{0^{--}},\ldots
\right\}.
\end{align}
\end{widetext}
For each set one constructs its own correlation matrix.  The resulting
lattice generalized-eigenvalue problems are \cite{Michael1985,LuscherWolff1990,Blossier2008,Blossier2009}
\begin{equation}
\resizebox{0.98\columnwidth}{!}{$\displaystyle
C^{(J^{PC})}(t)v_n
=\lambda_n(t,t_0)C^{(J^{PC})}(t_0)v_n,
\qquad
J^{PC}=1^{+-},0^{--},
$}
\label{eq:SeparateGEVPs}
\end{equation}
with no matrix elements connecting the two $J^{PC}$ blocks.  The notation
$\mathcal O_{G0}^{(D^2)}$ is schematic: it denotes any nonvanishing derivative-dressed
(or extended-loop) pure-glue interpolator projected to $0^{--}$.  The naive local
scalar $[\epsilon SGGG]_{0^{--}}$ is identically zero; this changes the useful operator
basis, not the existence of the $0^{--}$ spectral channel or of the exact three-vector
FMS interpolator $\mathcal O_{3X}$.  Derivative-dressed $0^{--}$ gluonic currents are
standard in the oddball literature \cite{PimikovEtAl2017Oddballs}.
A first-principles determination of their ordering requires Monte-Carlo
correlators for the broken $G(2)$ gauge--Higgs theory, which are not presently
available.  The analytic information can nevertheless be organized in a
controlled way by studying the leading Ritz projection of each block \cite{LandauLifshitzQM,MessiahQM}.

\paragraph{Two-state Ritz representation.}
After orthonormalizing the two reference states in each sector, write
\begin{align}
\frac{H_{1^{+-}}}{m_X}
&=
\begin{pmatrix}
M_{2X}/m_X & \delta_1\\
\delta_1 & k_1x
\end{pmatrix},
&
 k_1&=6.065,
\\[2mm]
\frac{H_{0^{--}}}{m_X}
&=
\begin{pmatrix}
M_{3X}/m_X & \delta_0\\
\delta_0 & k_0x
\end{pmatrix},
&
 x&\equiv\frac{\sqrt\sigma}{m_X}.
\label{eq:TwoSeparateRitzBlocks}
\end{align}
The first diagonal entry of the vector block is the weak-confinement result
$M_{2X}/m_X=1.99971319$ derived above.  The corresponding three-vector
reference mass can be estimated consistently in the same constituent regime.
For a color-singlet $XXX$ configuration every pair satisfies
$\langle T_i\!\cdot T_j\rangle=-2/3$.  A symmetric Gaussian variational wave
function in the two Jacobi coordinates gives
\begin{equation}
E_3(a)
=\frac{3a}{2m_X}
-2\alpha_s\sqrt{\frac{2a}{\pi}},
\end{equation}
whose minimum yields
\begin{equation}
\resizebox{0.98\columnwidth}{!}{$\displaystyle
{
\frac{E_{B,3}^{\rm var}}{m_X}
=\frac{4}{3\pi}\alpha_s^2
=2.7388\times10^{-4},
\qquad
\frac{M_{3X}^{\rm var}}{m_X}=2.99972612 .
}
$}
\label{eq:ThreeXVariationalBinding}
\end{equation}
The three-body binding is therefore of the same small fractional order as
the two-body Coulombic binding.  Ordinary perturbative binding cannot remove
the nearly one-$m_X$ constituent gap between the two reference thresholds.

\paragraph{Hall--Post bound in the fixed-constituent Hamiltonian.}
Before using constituent inequalities, their scope should be fixed sharply.
Throughout this paragraph $m_X$ is an additive constituent mass in a
nonrelativistic Hamiltonian.  The resulting bounds are rigorous \emph{inside
that Hamiltonian} in the Coulombic or pairwise-linear domains specified below;
they are not theorems about exact pole masses once strong confinement removes
a free-$X$ threshold.  The Gaussian calculation above supplies a variational
upper bound on the three-body energy.  A complementary lower bound follows
from the Hall--Post decomposition of an intrinsic three-body Hamiltonian \cite{HallPost1956,HallPost1967,HallPostReview2019}.
For three equal masses interacting through a common pair potential one has
\begin{equation}
E_3(m;g)\ge 3E_2(3m/2;g).
\label{eq:HallPostGeneral}
\end{equation}
In the color-singlet $XXX$ state,
$(T_1+T_2+T_3)^2=0$ implies
$\langle T_i\!\cdot T_j\rangle=-2/3$ for every pair, so the Coulomb strength is
$g=(2/3)\alpha_s$.  Since a two-body Coulomb system of equal constituent mass
$m$ has $E_2(m;g)=-m g^2/4$, Eq.~\eqref{eq:HallPostGeneral} gives
\begin{equation}
E_{3X}^{\rm Coul}\ge -\frac12\alpha_s^2m_X.
\end{equation}
Together with the Gaussian variational upper bound this brackets the exact
pairwise-Coulomb binding as
\begin{equation}
{
\frac{4}{3\pi}\alpha_s^2m_X
\le B_{3X}^{\rm Coul}
\le \frac12\alpha_s^2m_X .
}
\label{eq:ThreeXHallPostBracket}
\end{equation}
The two-body Coulomb result is exact,
$B_{2X}=(4/9)\alpha_s^2m_X$, and therefore the heavy thresholds obey
\begin{equation}
{
M_{3X}-M_{2X}
\ge m_X\left(1-\frac{\alpha_s^2}{18}\right)
=0.99996415\,m_X
}
\label{eq:HallPostHeavyGap}
\end{equation}
for $\alpha_s=0.0254031$.  Equation~\eqref{eq:HallPostHeavyGap} is therefore stronger than the Gaussian
estimate but narrower in scope: it is a rigorous statement for the
spin-independent pairwise-Coulomb Hamiltonian, not a lattice result for the
full confining theory.  Within that controlled heavy-particle limit, no more
accurate treatment of ordinary Coulombic three-body binding can invert the
$3X$ and $2X$ thresholds.  If the scalar sector is ultimately lighter, the
mechanism must be genuinely infrared and non-Coulombic.

The same Hall--Post identity can be used in a second, complementary model of
confinement.  The inequality itself is potential independent, but applying it
to strong confinement still requires a definite pairwise Hamiltonian; the
following statement is therefore a theorem of the linear $\Delta$-law model,
not of an arbitrary confining gauge theory.  In the linear
$\Delta$-law model, the baryonic pair color factor is one half of the mesonic
one, so take $v_{3X}(r)=(\sigma/2)r$ while the two-body singlet has
$v_{2X}(r)=\sigma r$.  For a linear two-body potential
\begin{equation}
E_2(M;k)=|a_1|\left(\frac{k^2}{M}\right)^{1/3},
\qquad |a_1|=2.338107\ldots ,
\end{equation}
where $M$ is the equal constituent mass entering the relative kinetic term.
Hall--Post therefore gives
\begin{equation}
\frac{M_{3X}-M_{2X}}{m_X}
\ge
1+|a_1|\left(\frac{3}{6^{1/3}}-1\right)x^{4/3},
\qquad x=\frac{\sqrt\sigma}{m_X}.
\label{eq:HallPostLinearGap}
\end{equation}
For illustration,
\begin{equation}
\resizebox{0.98\columnwidth}{!}{$\displaystyle
x=(0.25,0.35,0.47)
\quad\Longrightarrow\quad
\frac{M_{3X}-M_{2X}}{m_X}\ge(1.240,1.375,1.556).
$}
\end{equation}
The linear pairwise model therefore separates the thresholds even more
strongly than the Coulombic one.  Equation~\eqref{eq:HallPostLinearGap} is
rigorous for that nonrelativistic $\Delta$-law Hamiltonian, but its value is
precisely as a controlled reference limit; it is not promoted here to a
model-independent statement about the full gauge theory.

\paragraph{$Y$-junction confinement does not lower the three-body reference floor.}
A possible concern is that a genuine baryonic flux tube should be modeled by
a $Y$ junction rather than by a pairwise $\Delta$ potential.  The relevant
ordering can be fixed geometrically, without solving a second three-body
Schr\"odinger problem.  Let $L_Y$ be the total
length of any connected tree joining the three constituent positions.  The
unique tree path between every pair traverses each edge twice when the three
pairwise paths are summed, while the Euclidean distance is no larger than the
corresponding path length.  Hence
\begin{equation}
 r_{12}+r_{23}+r_{31}\le 2L_Y,
 \qquad
 L_Y\ge\frac12\sum_{i<j}r_{ij}.
 \label{eq:YTreeGeometricBound}
\end{equation}
For the standard linear $Y$ potential,
$V_Y=\sigma L_Y$, this gives the operator inequality
\begin{equation}
 V_Y\ge \frac{\sigma}{2}\sum_{i<j}r_{ij}=V_\Delta,
 \qquad
 E_{3X}^{Y}\ge E_{3X}^{\Delta}
 \label{eq:YAboveDelta}
\end{equation}
when the same constituent-mass and additive self-energy convention is used.
Therefore a genuine $Y$ junction cannot lower the coordinate-dependent
three-body confinement energy below the $\Delta$-law reference used in the
Hall--Post bound.  Equation~\eqref{eq:HallPostLinearGap} remains a lower floor
for this class of linear baryonic confinement models.
This direction of the inequality is consistent with lattice studies of the
static three-quark potential, which find a $Y$-type linear term governed by
the minimal flux-tube length and a string tension close to the mesonic one
\cite{TakahashiSuganuma2002}.
  
The recent Solenoidal Tracker at the Relativistic Heavy Ion Collider (STAR) baryon-transport measurements substantially strengthen the physical
motivation for this junction picture. In relativistic nuclear collisions the observed transport of baryon number is incompatible with the naive
scenario in which the valence quarks alone carry the transported baryon charge, and instead favors the baryon-junction description, in which the
nonperturbative $Y$-shaped gluonic configuration acts as the carrier and transporter of the ordinary baryon number~\cite{STARJunction2026}.
Thus a gauge-flux junction need not be merely a convenient representation of a multi-constituent state: in QCD it can encode the nonperturbative
transport of a conserved quantum number.
This result does not, however, imply the existence of an additional conserved ``junction number'' independent of baryon number.  Rather, the
junction is the proposed carrier of the already existing baryon charge.
Accordingly, the STAR result gives direct phenomenological support to the
physical relevance of $Y$-junction flux organization, but it cannot by
itself furnish an exact dark charge in the broken $G(2)$ theory.  Such a
conclusion would require an independently established dark symmetry or
topological charge whose carrier can then be identified with the dark
junction.

\paragraph{Modern pure-glue thresholds in the $0^{--}$ sector.}
The scalar block has an additional subtlety: the lowest quoted lattice
single-particle-like level is not the bottom of the full spectral support.  We
therefore begin with the modern pure-$SU(3)$ information and separate the
resonance-like doorway from the multiparticle threshold.  The
$A_1^{--}$ cubic-irrep ground level obeys
$m_{A_1^{--}}/\sqrt\sigma=9.78(46)$ \cite{AthenodorouTeper2020}.  It must be
distinguished from the bottom of the full $0^{--}$ spectral support.  The
$0^{--}$ channel already contains the two-glueball state
$1^{+-}+0^{++}$ in relative $L=1$: its charge conjugation is
$(-)(+)=-$, its parity is $(+)(+)(-1)^L=-$, and coupling spin one to
$L=1$ contains $J=0$.  In the decoupled pure-Yang--Mills daughter the
corresponding threshold is
\begin{equation}
 T_{0^{--}}^{(2G)}
 =m_{1^{+-}}+m_{0^{++}}
 =(6.065+3.405)\sqrt\sigma
 =9.470\sqrt\sigma .
 \label{eq:ScalarTwoGlueThreshold}
\end{equation}
The central $A_1^{--}$ level lies only $0.310\sqrt\sigma$ above this branch
point, about $0.67$ of its quoted $0.46\sqrt\sigma$ uncertainty.  It is
therefore useful as a single-particle-like doorway in a variational basis, but
it cannot be used as a lower bound on the full $0^{--}$ spectrum.  A scalar spectroscopy GEVP must include the
$1^{+-}0^{++}$ scattering operators explicitly.  In the full gauge--Higgs
theory the even partner in the lowest scalar continuum can instead be the
radial $0^{++}$ mode if it is lighter than the glueball.

For comparison with older literature, we retain the legacy
Hamiltonian-lattice ratios only as deliberately scalar-favorable stress tests \cite{HuLuoChen1996},
\begin{equation}
\frac{m_{0^{--}}}{m_{0^{++}}}=2.44\pm0.05\pm0.20,
\qquad
\frac{m_{1^{+-}}}{m_{0^{++}}}=1.91\pm0.05\pm0.12.
\label{eq:HistoricalOddGlueRatios}
\end{equation}
Multiplying the first ratio by the modern
$m_{0^{++}}/\sqrt\sigma=3.405(21)$ value gives the mixed diagnostic
\begin{equation}
k_0^{\rm mix}=8.3082\pm0.704,
\label{eq:k0Historical}
\end{equation}
where the quoted legacy and modern $0^{++}$ uncertainties are propagated in
quadrature.  Because this combines systematics from different calculations it
is not used as a precision continuum determination.  The historical double
ratio anchored to the modern $1^{+-}$ scale gives
\begin{equation}
k_0^{\rm dbl}=k_1\frac{2.44}{1.91}\simeq7.75,
\label{eq:k0DoubleRatio}
\end{equation}
and the deliberately low value $k_0=7.606$ is retained only as a
scalar-favorable stress-test floor.

The multiparticle threshold gives a cleaner and more model-independent way
to state what a scalar victory would require.
Let $E_1$ denote the exact lowest $1^{+-}$ pole and let $m_+$ be the lightest
$C$-even $0^{++}$ mass in the isolated confining theory, including the radial
Higgs excitation when it lies below the glueball.  The allowed
$0^{--}$ continuum starts no higher than
\begin{equation}
 T_0=E_1+m_+ .
\end{equation}
If a stable scalar pole $E_0$ is to be the lighter odd state, $E_0<E_1$, then
its binding below this continuum must satisfy
\begin{equation}
 B_0\equiv T_0-E_0
 =m_+ +(E_1-E_0)>m_+.
 \label{eq:ScalarWinnerBindingCriterion}
\end{equation}
In the decoupled pure-$SU(3)$ limit with the radial scalar heavier than the
glueball this becomes $B_0>3.405(21)\sqrt\sigma$; if the radial scalar is
lighter, the corresponding lower $m_+$ must be used instead.  The modern $A_1^{--}$ level therefore does not by itself exclude the scalar
candidate.  Instead it localizes the surviving loophole: the $0^{--}$ sector
would have to develop a genuinely subthreshold pole with binding larger than
the lightest even mass.  This is why both $1^{+-}$ and $0^{--}$ assignments are
retained until a scattering-aware lattice GEVP is available.

\paragraph{Modern glue thresholds and heavy-$X$ mixing power counting.}
The $9.78(46)\sqrt\sigma$ $A_1^{--}$ level remains useful as a discrete
doorway diagnostic, while Eq.~\eqref{eq:ScalarTwoGlueThreshold} is the robust
reference for the onset of scalar spectral support.  Keeping those roles
separate avoids treating the cubic-irrep level as a nonexistent spectral
floor.

%A second piece of information comes from weak-Higgs power counting.  It does
%not determine the nonperturbative mixing matrix, but it indicates which
%hierarchy would be natural if the heavy-particle expansion remains meaningful.  For a tree made from Yang--Mills cubic and quartic vertices, with
%$E$ external gauge bosons, the graph identities imply
%\begin{equation}
% V_3+2V_4=E-2,
% \qquad {\cal A}_E\propto g^{E-2}.
% \label{eq:TreeGaugePowerCounting}
%\end{equation}
%Consequently the hard conversion of an $X\bar X$ configuration into a
%three-gluon operator begins one gauge coupling earlier than the corresponding
%$XXX$ conversion,
%\begin{equation}
% X\bar X\to3g:\ {\cal A}_1\sim g^3,
% \qquad
% XXX\to3g:\ {\cal A}_0\sim g^4 .
%\end{equation}
%For normalized nonrelativistic $S$ waves the contact factors scale as
%$\psi_2(0)\sim(m_Xv)^{3/2}$ and
%$\Psi_3(0,0)\sim(m_Xv)^3$.  Heavy-particle naive dimensional analysis (NDA) therefore gives
%\begin{equation}
% \frac{|\Delta_0|}{|\Delta_1|}
% \sim
% g\,v^{3/2}\,
% \left|\frac{c_0Z_{G0}}{c_1Z_{G1}}\right|,
% \label{eq:HeavyXMixingHierarchy}
%\end{equation}
%where $Z_{GJ}$ denotes the reduced nonperturbative glue overlap and the
%$c_J$ are order-one hard matching coefficients in the natural heavy regime.
A second piece of information comes from weak-Higgs power counting.  It does
not determine the nonperturbative mixing matrix, but it indicates which
hierarchy would be natural if the heavy-particle expansion remains meaningful.
For a connected tree constructed from Yang--Mills cubic and quartic vertices,
with $E$ external gauge bosons, the usual tree-graph identities give
\cite{ManganoParke1991}
\begin{equation}
	V_3+2V_4=E-2,
	\qquad
	{\cal A}_E\propto g^{E-2}.
	\label{eq:TreeGaugePowerCounting}
\end{equation}
Indeed, $3V_3+4V_4=2I+E$ together with
$I=V_3+V_4-1$ gives the first relation, while each cubic and quartic
Yang--Mills vertex contributes respectively one and two powers of $g$.

Consequently the hard conversion of an $X\bar X$ configuration into a
three-gluon operator begins one gauge coupling earlier than the corresponding
$XXX$ conversion,
\begin{equation}
	X\bar X\to3g:\ {\cal A}_1\sim g^3,
	\qquad
	XXX\to3g:\ {\cal A}_0\sim g^4 .
\end{equation}
For normalized nonrelativistic $S$ waves, standard velocity power counting
gives
\cite{BodwinBraatenLepage1995,LukeManohar1997}
\begin{equation}
	\psi_2(0)\sim(m_Xv)^{3/2},
	\qquad
	\Psi_3(0,0)\sim(m_Xv)^3 .
\end{equation}
The first scaling follows from normalization in one relative coordinate,
whereas the three-body wave function depends on two independent Jacobi
coordinates.  Combining the short-distance gauge-coupling counting with
the nonrelativistic contact factors, and using naive dimensional analysis
for the hard matching coefficients
\cite{ManoharGeorgi1984,LukeManohar1997}, gives
\begin{equation}
	\frac{|\Delta_0|}{|\Delta_1|}
	\sim
	g\,v^{3/2}\,
	\left|\frac{c_0Z_{G0}}{c_1Z_{G1}}\right|,
	\label{eq:HeavyXMixingHierarchy}
\end{equation}
where $Z_{GJ}$ denotes the reduced nonperturbative glue overlap and the
$c_J$ are order-one hard matching coefficients in the natural heavy regime.

With $g=\sqrt{4\pi\alpha_s}=0.565$ and the controlled Coulomb momentum
$v\simeq C_F\alpha_s/2=0.01694$, the explicit prefactor is
$1.25\times10^{-3}$.  Even the deliberately relativistic stress-test value
$p_{\rm rms}/m_X\simeq0.57$ gives only $0.24$.  Equation~\eqref{eq:HeavyXMixingHierarchy} is not a theorem at strong
confinement because the ratio $Z_{G0}/Z_{G1}$ is genuinely nonperturbative.
Nevertheless, its natural heavy-particle scaling points in the \emph{opposite}
direction from that required for a scalar inversion.  Together with the
basis-independent criterion derived below, this means that a $0^{--}$ ground
state would require an anomalously enhanced scalar self-energy or threshold
effect, not merely a modest refinement of perturbative heavy-vector dynamics.

\paragraph{Reference gap before scalar-specific binding.}
The two controlled pieces of information can now be combined without
assuming which particular doorway dominates either spin sector: the heavy
constituent hierarchy and the physical two-glue threshold.  In the
linear $\Delta$-law model, and therefore also for the coordinate-dependent
part of a genuine $Y$ junction, introduce the reference energies
\begin{equation}
\resizebox{0.98\columnwidth}{!}{$\displaystyle
 r_0(x)=\min\!\left(M_{3X}^{\Delta,Y},\,
 (k_1+k_+)x\,m_X\right),
 \qquad
 r_1(x)=\min\!\left(M_{2X}^{\Delta},\,k_1x\,m_X\right),
$}
\end{equation}
where $k_1=6.065$ and $k_+=m_{0^{++}}/\sqrt\sigma=3.405(21)$.  The two pairwise
comparisons are
\begin{equation}
\resizebox{0.98\columnwidth}{!}{$\displaystyle
 M_{3X}^{\Delta,Y}-M_{2X}^{\Delta}
 \ge m_X\left[1+1.5220226\,x^{4/3}\right],
 \qquad
 T_{0^{--}}^{(2G)}-k_1x\,m_X=k_+x\,m_X .
$}
\end{equation}
The elementary inequality
\begin{equation}
\resizebox{0.98\columnwidth}{!}{$\displaystyle
 \min(a,b)-\min(c,d)\ge\min(A,B)
 \quad\hbox{if}\quad a-c\ge A,\ b-d\ge B
$}
\end{equation}
therefore yields the exact comparison of these \emph{reference configurations},
\begin{equation}
\resizebox{0.98\columnwidth}{!}{$\displaystyle
 r_0-r_1\ge m_X\,\Delta_{\rm ref}(x),\qquad
 \Delta_{\rm ref}(x)=\min\!\left[3.405x,\,
 1+1.5220226\,x^{4/3}\right].
$}
\label{eq:DoorwayGapTheorem}
\end{equation}
The two branches of this lower reference gap cross at $x\simeq0.4460$.
Representative values are collected in Table~\ref{tab:referenceGapValues}.
\begin{table}[htbp]
\centering
\caption{Reference-configuration gap values used only as diagnostics of the combined constituent/pure-glue interpolation.}
\label{tab:referenceGapValues}
\resizebox{0.92\columnwidth}{!}{%
\begin{tabular}{cccc}
\toprule
$x$ & $\Delta_{\rm ref}(x)$ & conservative glue-threshold value &
$\Delta_{\rm ref}/\alpha_s$\\
\midrule
0.10 & 0.341 & 0.338 & 13.4\\
0.20 & 0.681 & 0.677 & 26.8\\
0.25 & 0.851 & 0.846 & 33.5\\
0.30 & 1.022 & 1.015 & 40.2\\
0.35 & 1.192 & 1.184 & 46.9\\
0.40 & 1.362 & 1.354 & 53.6\\
0.47 & 1.556 & 1.556 & 61.3\\
\bottomrule
\end{tabular}%
}
\end{table}
The conservative column lowers only $k_+$ to $3.405-0.021$; the same measured
$k_1$ occurs on both sides of the two-glue threshold and therefore cancels in
the glue-branch gap.  If a variational realization is used for the vector
constituent entry, its upper-bound character is in the direction needed for
this comparison; the Hall--Post/$Y$ three-body term is the corresponding lower
floor on the scalar constituent reference.

The interpretation of Eq.~\eqref{eq:DoorwayGapTheorem} is deliberately
limited.  It is a lower gap between the stated \emph{reference
configurations}, not a lower bound on the exact $0^{--}$ spectrum.  Residual attraction in the
$1^{+-}0^{++}$ channel can bind a scalar below its continuum threshold, and
such a subthreshold state is precisely the loophole quantified by
Eq.~\eqref{eq:ScalarWinnerBindingCriterion}.  Its value is to show that all ordinary constituent and unbound two-glue
reference configurations begin parametrically above the corresponding vector
reference throughout the controlled comparison.  Any scalar inversion must
therefore be produced by additional scalar-specific attraction.

\paragraph{Exact Feshbach criterion for a scalar inversion.}
The remaining question can be posed without restricting the Hilbert space
to a finite two-state Ritz ansatz.  The Feshbach projection formalism
\cite{Feshbach1958,Feshbach1962} partitions the full state space into a
chosen ``doorway'' sector \(P_J\), containing the explicitly retained
operators in a fixed \(J^{PC}\) channel, and its orthogonal complement
\(Q_J=1-P_J\), which contains all omitted states.  Eliminating the
\(Q_J\) sector exactly produces an energy-dependent effective Hamiltonian
acting only in the doorway space.  Its second term therefore measures the
net self-energy, and hence the additional attraction or repulsion,
generated by all states that were omitted from the finite variational
basis.  
Let $P_J$ project onto a chosen finite-volume doorway subspace in the
$J$ sector, including the relevant scattering operators, and let $Q_J=1-P_J$.
After diagonalizing the doorway block, denote its lowest variational energy by
$d_J$.  The exact Feshbach operator is
\begin{equation}
 H^{\rm eff}_J(E)
 =P_JH_JP_J+P_JH_JQ_J
 \frac{1}{E-Q_JH_JQ_J}
 Q_JH_JP_J .
 \label{eq:ExactFeshbachOperator}
\end{equation}
If the exact lowest pole lies below the chosen doorway spectrum, define the
net lowering relative to the lowest doorway eigenvalue by
\begin{equation}
 {\cal A}^{\rm net}_J\equiv d_J-E_J.
\end{equation}
Then, identically,
\begin{equation}
 E_{0^{--}}-E_{1^{+-}}
 =(d_0-d_1)-({\cal A}^{\rm net}_0-{\cal A}^{\rm net}_1),
\end{equation}
so a scalar inversion requires
\begin{equation}
 {\cal A}^{\rm net}_{0^{--}}-{\cal A}^{\rm net}_{1^{+-}}>d_0-d_1.
 \label{eq:FeshbachDifferentialAttraction}
\end{equation}
For a multidimensional doorway subspace, this net lowering must be
distinguished from the projected self-energy because the exact state need not
align with the lowest doorway eigenvector.  Let $A_J=P_JH_JP_J$ and let
$|\psi_J(E_J)\rangle$ be the normalized $P_J$-space component of the exact
Feshbach eigenstate.  Define
\begin{align}
 \Delta_{P,J}&\equiv\langle\psi_J|A_J|\psi_J\rangle-d_J\ge0,\\
 \widetilde{\cal A}_J(E_J)&\equiv
 -\langle\psi_J|\Sigma_J(E_J)|\psi_J\rangle\ge0,
\end{align}
where
\begin{equation}
 \Sigma_J(E)=P_JH_JQ_J(E-Q_JH_JQ_J)^{-1}Q_JH_JP_J.
\end{equation}
The exact relation is
\begin{equation}
 {\cal A}^{\rm net}_J=\widetilde{\cal A}_J-\Delta_{P,J}.
 \label{eq:FeshbachNetLowering}
\end{equation}
The exact inversion criterion therefore compares \emph{differential}
spectral attraction between the two spin sectors and also accounts for
misalignment inside each doorway space.  The reference gap of
Eq.~\eqref{eq:DoorwayGapTheorem} remains a useful diagnostic scale, but the
quantities entering the exact criterion can only be extracted from
scattering-aware GEVPs.

If the omitted $Q_J$ spectrum begins at $T_J>E_J$, the resolvent gives
\begin{equation}
 \widetilde{\cal A}_J\le
 \frac{\|P_JH_JQ_J\|^2}{T_J-E_J}.
 \label{eq:FeshbachResolventBound}
\end{equation}
This resolvent bound is least restrictive precisely in the scalar channel,
its decoupled pure-$SU(3)$ reference branch point lies at
$T_0\simeq E_1+m_+$, below the central single-particle-like $A_1^{--}$
doorway.  Hence $T_0-E_0$ can be small for a near-threshold scalar pole, and
the resolvent enhancement can be large.  A near-threshold enhancement of the $0^{--}$ spectral density is therefore
the sharpest surviving mechanism for reversing the reference ordering in the
decoupled-glue limit.
After projection, the spectral theorem gives a positive state-dependent
measure $\rho_{J,\psi_J}$,
\begin{equation}
 \widetilde{\cal A}_J(E_J)=
 \int_{T_J}^{\infty}d\omega\,
 \frac{\rho_{J,\psi_J}(\omega)}{\omega-E_J}.
\end{equation}
The inversion condition can then be written directly as a comparison of the
two projected spectral measures:
\begin{align}
 &\left[\int d\omega\frac{\rho_{0,\psi_0}(\omega)}{\omega-E_0}-\Delta_{P,0}\right]
 -\left[\int d\omega\frac{\rho_{1,\psi_1}(\omega)}{\omega-E_1}-\Delta_{P,1}\right]\nonumber\\
 &\hspace{45mm}>d_0-d_1.
 \label{eq:SpectralDensityInversionTest}
\end{align}
For a one-dimensional doorway $\Delta_{P,J}=0$, recovering the familiar
self-energy formula.  The practical consequence is unambiguous: a lattice
calculation that aims to decide the scalar ordering must include the
$1^{+-}0^{++}$ scattering operators explicitly in the $0^{--}$ variational
basis.  In
finite volume the continuum is a tower of discrete levels; the quoted
$9.470\sqrt\sigma$ value is the corresponding infinite-volume branch point,
so the finite-volume level pattern must be analyzed rather than interpreting
the $A_1^{--}$ energy as an isolated scalar mass.  This finite-volume scattering problem is the genuine nonperturbative wall.
It is the precise reason the present paper favors, but does not mathematically
prove, the $1^{+-}$ assignment.

\paragraph{Spin-dependent corrections do not overturn the reference hierarchy.}
One might ask whether conventional spin-dependent forces could undo the
reference ordering before genuinely nonperturbative threshold effects become
important.  They cannot do so parametrically.  Chromomagnetic, tensor and
hyperfine interactions are suppressed by $1/m_X^2$.  Their characteristic heavy-particle splittings scale as
$\alpha_s^4m_X$ in the Coulombic regime and as
$\Lambda^2/m_X=x^2m_X$ in a heavy confining regime.  At the controlled
Coulomb boundary the first scale is only
$\alpha_s^4m_X\simeq4.2\times10^{-7}m_X$, whereas the glue-threshold branch of
Eq.~\eqref{eq:DoorwayGapTheorem} is approximately
$3.405x\,m_X\simeq5.77\times10^{-2}m_X$.  In the confining regime the ratio of
a natural $O(x^2m_X)$ spin splitting to this reference branch is at most
$x/3.405$, about $0.073$ at $x=0.25$ and $0.103$ at $x=0.35$.  These scales are far smaller than the reference separation.  Ordinary
spin-dependent corrections therefore cannot by themselves generate the deep
subthreshold binding required by
Eq.~\eqref{eq:ScalarWinnerBindingCriterion}.  The surviving scalar route is
intrinsically a strong, near-threshold effect.

\paragraph{Status of the numerical Ritz stress tests.}
The numerical Ritz values quoted later are useful only if their status is
kept explicit.  The Coulombic constituent entries
$M_2\simeq1.999713$ and $M_3\simeq2.999726$ are controlled only for
$a_0\sqrt\sigma\lesssim1$, i.e. $x\lesssim0.01694$ in the present benchmark.
The tables below extend these numbers to $x=0.05$--$0.40$ only as a
\emph{stress-test interpolation}; their individual decimal entries are not
first-principles mass predictions in that region.  Importantly, the exact equal-mixing ordering theorem below does
\emph{not} depend on extending these Coulombic numbers into the strong
confinement regime: it needs only
$M_3>M_2$ and $k_0>k_1$.  The first inequality holds rigorously in the
pairwise-Coulomb domain by Eq.~\eqref{eq:HallPostHeavyGap} and throughout the
pairwise linear $\Delta$-law model by Eq.~\eqref{eq:HallPostLinearGap}.  The
unequal-mixing critical values quoted below should therefore be read as
diagnostics of the size of mixing required in that interpolation, not as
controlled strong-coupling numbers.

Within this explicitly defined Ritz model, both projected GEVPs can now be
diagonalized analytically.  Their lower eigenvalues are
\begin{align}
\mu_1^-(x)
&=\frac12\left[
M_2+k_1x-
\sqrt{(M_2-k_1x)^2+4\delta_1^2}
\right],\\
\mu_0^-(x)
&=\frac12\left[
M_3+k_0x-
\sqrt{(M_3-k_0x)^2+4\delta_0^2}
\right],
\label{eq:SeparateLowerEigenvalues}
\end{align}
where $M_2$ and $M_3$ now denote the corresponding masses in units of $m_X$.
The equal-mixing case admits an exact ordering theorem.  If
$|\delta_0|=|\delta_1|\equiv\delta$, a phase redefinition makes the two
off-diagonal entries equal and
\begin{equation}
H_{0^{--}}-H_{1^{+-}}
=m_X
\begin{pmatrix}
M_3-M_2&0\\
0&(k_0-k_1)x
\end{pmatrix}.
\end{equation}
Both diagonal entries are strictly positive for $x>0$, even if the deliberately scalar-favorable $k_0=7.606$ floor is used.  By the min--max theorem,
\begin{equation}
\resizebox{0.98\columnwidth}{!}{$\displaystyle
{
\mu_1^-(x)<\mu_0^-(x)
\qquad
\text{for every }x>0\text{ and every common mixing }\delta .
}
$}
\label{eq:EqualMixingOrderingTheorem}
\end{equation}
Thus equal level repulsion, however large, cannot reverse the ordering as
long as the scalar reference entries exceed the vector ones.  A scalar
inversion necessarily requires a \emph{differential} enhancement of the
$0^{--}$ sector.

\paragraph{Basis-independent necessary condition for a scalar inversion.}
The requirement of anomalously strong scalar mixing is not an artifact of
using only two operators.  A basis-independent necessary condition can be
obtained for an arbitrary finite Ritz space.  Let an
arbitrary finite Ritz basis in each $J^{PC}$ sector be written as
\begin{equation}
H_J=D_J+V_J,
\end{equation}
where all diagonal threshold/self-energy renormalizations are absorbed into
$D_J$ and $V_J$ is, by definition, purely off diagonal in the chosen threshold
basis.  Denote $d_J=\lambda_{\min}(D_J)$.  With this convention the trial vector
that realizes $d_1$ has vanishing diagonal expectation value of $V_1$, while
Weyl's inequality controls the scalar block; hence the Rayleigh--Ritz principle
implies \cite{KatoPerturbation,ReedSimonIV}
\begin{equation}
\lambda_{\min}(H_{0^{--}})\ge d_0-\|V_0\|,
\qquad
\lambda_{\min}(H_{1^{+-}})\le d_1.
\end{equation}
Consequently a scalar ground state is possible only if
\begin{equation}
{
\|V_0\|>d_0-d_1 .
}
\label{eq:OperatorNormScalarNecessary}
\end{equation}
This condition is deliberately conservative in favor of the scalar: it
allows the full off-diagonal norm of the $0^{--}$ block to lower its state but
ignores any additional lowering in the competing $1^{+-}$ block.  Taking the most scalar-favorable
inputs used here,
\begin{equation}
\resizebox{0.98\columnwidth}{!}{$\displaystyle
d_0=\min\!\left(3-\frac12\alpha_s^2,\;7.606x\right),
\qquad
d_1=\min\!\left(2-\frac49\alpha_s^2,\;6.065x\right),
$}
\end{equation}
gives the finite-basis diagnostic summarized in Table~\ref{tab:operatorNormGap}.
\begin{table}[h]
\centering
\caption{Scalar-favorable finite-basis gap diagnostic entering the operator-norm inversion criterion.}
\label{tab:operatorNormGap}
\resizebox{0.72\columnwidth}{!}{%
\begin{tabular}{ccc}
\toprule
$x$ & $(d_0-d_1)/m_X$ & $(d_0-d_1)/(\alpha_sm_X)$\\
\midrule
0.20 & 0.3082 & 12.1\\
0.25 & 0.3853 & 15.2\\
0.30 & 0.4623 & 18.2\\
0.3297 & 0.5081 & 20.0\\
0.40 & 0.99996 & 39.4\\
\bottomrule
\end{tabular}%
}
\end{table}
Across the confinement range relevant to the heavy two-vector deformation,
the scalar block therefore needs a collective off-diagonal scale of several
tenths of $m_X$ merely to become competitive in this conservative bound.  The
statement is independent of the number of interpolators retained in the finite
Ritz basis.

\subsection{Where the analytic ordering argument ends.}
The preceding bounds narrow the problem but do not close it.  Equations~\eqref{eq:FeshbachDifferentialAttraction}--\eqref{eq:SpectralDensityInversionTest}
identify the unresolved quantity exactly: a scalar ground state requires a
sufficiently larger integrated $0^{--}$ spectral attraction than the
corresponding $1^{+-}$ attraction.  No known positivity, reflection-positivity, or group-theoretic identity
orders the required spectral attraction between two different rotational
sectors.  The generalized-eigenvalue method itself only guarantees
variational convergence within each sector
\cite{Michael1985,LuscherWolff1990,Blossier2008,Blossier2009}.  Therefore the
remaining loophole is sharply localized: a near-threshold or deeply bound
$0^{--}$ pole generated by an anomalously large scalar spectral enhancement.
The analytic evidence therefore favors $1^{+-}$, but it does not constitute
a theorem that the vector channel is the ground state.  The sign of the exact
difference is encoded in the two unknown Euclidean correlator matrices of
Eq.~\eqref{eq:SeparateGEVPs}.  This is a genuine nonperturbative limitation,
not missing algebra.

Within the two-state stress-test model, the unequal-mixing crossing can
nevertheless be solved exactly.  Once
$\mu_1^-$ is fixed, the scalar state crosses below it only if
\begin{equation}
{
|\delta_0|^2
>
\left(M_3-\mu_1^-\right)
\left(k_0x-\mu_1^-\right) .
}
\label{eq:ScalarWinCriticalMixing}
\end{equation}
At the benchmark weak-Higgs reference
$|\delta_1|=\alpha_s=0.0254031$, the illustrative Coulomb-extrapolated Ritz
numbers are collected in Appendix~\ref{app:RitzStressTable}, rather than in
the main text because they lie outside the controlled Coulomb domain once
$x\gtrsim0.017$.  Their only intended use is to quantify the size of mixing
required in that stress-test interpolation.  The conclusion is unchanged if
every choice is biased in favor of the scalar:
set $\delta_1=0$ and take the deliberately scalar-favorable $k_0=7.606$ floor.  Over
$x=0.05$--$0.40$ the required scalar mixing is still
\begin{equation}
|\delta_0|_{\rm crit}\simeq0.46\text{--}1.02,
\end{equation}
i.e. roughly $18$--$40$ times $\alpha_s$.  A literal Euclidean two-operator
GEVP constructed from
$C(t)=Z\exp(-Ht)Z^T$ reproduces these Ritz eigenvalues independently of the
invertible operator-overlap matrix $Z$.  This separate illustrative check uses
$k_0=k_0^{\rm mix}=8.3082$ and $\delta_0=\delta_1=\alpha_s$; at $x=0.25$ it gives
\begin{equation}
\frac{E_{1^{+-},0}}{m_X}\simeq1.515,
\qquad
\frac{E_{0^{--},0}}{m_X}\simeq2.076.
\label{eq:LiteralGEVPBenchmark}
\end{equation}

The spectroscopy result can therefore be summarized without overclaiming:
\begin{enumerate}[label=(\roman*)]
\item the scalar $0^{--}$ channel exists as an exact gauge-invariant spectral sector;
\item the controlled heavy-particle and junction limits favor the $1^{+-}$ channel;
\item the decoupled pure-glue limit independently favors the $1^{+-}$ channel.
\end{enumerate}
A scalar ground state remains possible only through physics not captured by
the controlled reference limits: an anomalously light $0^{--}$ pole below the
known $1^{+-}$ glue level, or a strongly nonperturbative $0^{--}$ mixing of
order $m_X$ that is parametrically much larger than the corresponding
$1^{+-}$ mixing.  Neither possibility is forbidden mathematically, but neither
is supported by the present weak-Higgs or lattice information.  A first-principles decision therefore requires Monte-Carlo evaluation of the
two scattering-aware correlator matrices in Eq.~\eqref{eq:SeparateGEVPs}.
No further rearrangement of the existing analytic inputs can substitute for
those correlators.

\subsection{Microscopic form of the protecting $G(2)$ transformation}

The spectroscopy discussion can now be tied back to the exact symmetry
operator.  The two three-vector baryonic structures combine into
\begin{equation}
B_+=\operatorname{Re}(XXX)
=\frac{XXX+\bar X\bar X\bar X}{\sqrt2},
\label{eq:Bplus_v3}
\end{equation}
which is $C$ odd and corresponds to the protected Higgs-phase combination
emphasized in Ref.~\cite{ButtazzoG2Higgs}.  The orthogonal real combination is
$C$ even and can decay through the characteristic cubic interactions of the
broken $G(2)$ gauge sector.  After confinement, however, the protected sector is described by a
correlation matrix involving derivative $SSS$, $SGGG$, and other operators
with the same quantum numbers.  Equation~\eqref{eq:B2FMSleading} is the
explicit reason why the exact operator classification does not uniquely fix
the number of heavy vectors that dominate a particular FMS wavefunction.

To prepare the ultraviolet lifting problem, the two ingredients of the
isolated protecting symmetry should now be separated explicitly: the gauge
transformation inside $G(2)$ and the accidental sign of the fundamental
scalar.  In the conventions of Ref.~\cite{ButtazzoG2Higgs}, the unbroken scalar transformation is
\begin{equation}
\eta=\operatorname{diag}(-1,-1,-1,+1,+1,+1,+1).
\label{eq:etaButtazzo}
\end{equation}
The nonzero octonionic triples may be chosen as
\begin{equation}
O_{123}=O_{516}=O_{624}=O_{435}=O_{471}=O_{673}=O_{572}=1
\label{eq:octonionTriples}
\end{equation}
up to antisymmetry.  Define
\begin{equation}
U_B=-\eta
=\operatorname{diag}(+1,+1,+1,-1,-1,-1,-1).
\label{eq:UBexplicit}
\end{equation}
For each nonzero octonionic triple in Eq.~\eqref{eq:octonionTriples},
\begin{equation}
(U_B)_{ii}(U_B)_{jj}(U_B)_{kk}=+1,
\end{equation}
so $U_B$ preserves $O_{ijk}$ and therefore belongs to the compact real $G(2)$.
Conjugation of the gauge generators by $U_B$ is identical to conjugation by
$\eta$, because the central factor $-\mathbf{1}_7$ cancels in
$U_BTU_B^{-1}$.  Thus $U_B$ is an explicit inner element of the compact $G(2)$ that realizes
the conjugation of Eq.~\eqref{eq:ButtazzoCfull} on the gauge fields.

The vacuum symmetry is nevertheless not supplied by this gauge element
alone, because its action on the fundamental Higgs is different.  The isolated
fundamental-$\mathbf{7}$ Lagrangian has an accidental global sign $Z_S:S\to-S$.  Since
$U_B e_7=-e_7$, the inner element $U_B$ normalizes the residual $SU(3)$ but
does \emph{not} itself stabilize the chosen Higgs vacuum
$\langle S\rangle=w e_7$.  The full unbroken Higgs-phase parity is instead
\begin{equation}
{P_B=Z_S\circ U_B},
\qquad P_B(S)=\eta S,
\end{equation}
for which $P_B\langle S\rangle=\langle S\rangle$.  This makes the conceptual distinction precise.  The nontrivial normalizer
class of the residual $SU(3)$ must not be confused with an unbroken discrete
gauge subgroup of the Higgs vacuum: the physical $\mathbb Z_2$ uses the accidental
scalar sign together with the inner $G(2)$ transformation.  The distinction becomes decisive in the exceptional uplift: $U_B$ is guaranteed to lie
in the original gauge group $G(2)$ and to normalize the residual $SU(3)$,
whereas the accidental sign of an isolated fundamental scalar need not
automatically extend to every component and coupling of a full $E_6$ $\mathbf{650}$
parent.  
The distinction can also be seen directly in the renormalizable
multi-Higgs effective potential of the underlying Class-B construction
\cite{MasiE6E7}.  In the notation of that work,
\(\chi\sim(\mathbf{7},\mathbf{1}_A)_0\subset\mathbf{650}_{0,H}\)
is the real \(G(2)\)-fundamental scalar whose vacuum expectation value
breaks \(G(2)\to SU(3)_C\), while
\(\Omega\sim(\mathbf{1},\mathbf{8}_A)_0
\subset\mathbf{650}_{0,H}\) is the adjoint \(SU(3)_A\)-breaking scalar.
The field \(\Phi\) denotes here the high-scale \(E_6\)-breaking scalar
coordinate associated with the singlet direction
\(S\sim(\mathbf{1},\mathbf{1}_A)_0\subset\mathbf{650}_{0,H}\),
and \(\Theta\) denotes the hypercharge-neutral scalar direction responsible
for the subsequent abelian breaking
\(U(1)_A\times U(1)_X\to U(1)_Y\).
Restricting the Class-B potential to the terms containing the dark-breaking
field \(\chi\) gives
\begin{widetext}
	\begin{equation}
		V_{\rm trunc}\supset
		\frac12m_\chi^2\,\chi^T\chi
		+\frac{\lambda_\chi}{4}(\chi^T\chi)^2
		+\kappa_{\chi\Omega}(\chi^T\chi)
		\operatorname{Tr}(\Omega^\dagger\Omega)
		+\kappa_{\Phi\chi}|\Phi|^2(\chi^T\chi)
		+\kappa_{\Theta\chi}(\Theta^\dagger\Theta)(\chi^T\chi).
		\label{eq:chiPotentialTruncation}
	\end{equation}
\end{widetext}
Here \(m_\chi^2\) is the quadratic scalar-potential parameter
(not the breaking scale \(M_\chi\)), \(\lambda_\chi\) is the
\(\chi\) self-quartic coupling, and
\(\kappa_{\chi\Omega}\), \(\kappa_{\Phi\chi}\), and
\(\kappa_{\Theta\chi}\) are the corresponding renormalizable portal
couplings.  The contraction \(\chi^T\chi\) is the invariant norm of the
real \(\mathbf7\) of \(G(2)\), while the trace is over the
\(SU(3)_A\) adjoint indices of \(\Omega\).

Every term retained in Eq.~\eqref{eq:chiPotentialTruncation} contains
\(\chi\) only through the even combination \(\chi^T\chi\).  The truncated
low-energy scalar potential is therefore invariant under the accidental
coordinate reflection
\(\chi\to-\chi\).  This observation is deliberately weaker than a symmetry statement about the complete exceptional scalar sector:
%Indeed, the displayed low-energy multi-Higgs truncation
%can be chosen even in the isolated $\chi$ direction,
%\begin{widetext}
%\begin{equation}
%V_{\rm trunc}\supset
%\frac12m_\chi^2\,\chi^T\chi
%+\frac{\lambda_\chi}{4}(\chi^T\chi)^2
%+\kappa_{\chi\Omega}(\chi^T\chi)\operatorname{Tr}(\Omega^\dagger\Omega)
%+\kappa_{\Phi\chi}|\Phi|^2(\chi^T\chi)
%+\kappa_{\Theta\chi}(\Theta^\dagger\Theta)(\chi^T\chi),
%\end{equation}
%\end{widetext}
the parent \(\mathbf{650}_{0,H}\) contains many additional \(G(2)\times SU(3)_A\) components, including
$2(\mathbf{7},\mathbf{8})$, $(\mathbf{7},\mathbf{10})$, $(\mathbf{7},\overline{\mathbf{10}})$, $(\mathbf{27},\mathbf{1})$ and $(\mathbf{27},\mathbf{8})$, and the full \(E_6/E_7\)-invariant potential need not preserve a sign acting only on the selected
\(\chi\) direction. Therefore the isolated sign $S\to-S$ cannot simply be assumed to survive
inside the parent representation. It must be tested against the complete
$E_6/E_7$ invariant potential, the full set of scalar components, and the
chosen vacuum.  This is the precise ultraviolet question addressed in the
next part of the paper, and it is independent of the fact that the gauge
conjugation itself is realized by the inner element $U_B$.

\section{Exceptional parent structure and the Class-B chiral assignment}
\label{sec:group-theory}

%\paragraph{Physical role of the construction.}
The ultraviolet stability problem is tied directly to the visible-sector
embedding.  The same exceptional representations that reproduce the
Standard Model charges and chiral zero modes also determine whether the dark
sector can communicate with light matter.  We therefore introduce the Class-B
branching structure first and use it, in the following sections, both to fix
the visible assignment and to identify the dark--visible couplings that are
allowed before compactification effects are included.

\subsection{$E_7$ uplift, representation conventions, and hypercharge}

We begin by fixing the representation conventions used throughout the
ultraviolet analysis.  Under $E_6\times U(1)_X$, the $E_7$ fundamental
branches as
\begin{equation}
\mathbf{56}\longrightarrow \mathbf{27}_{-1}\oplus\overline{\mathbf{27}}_{+1}\oplus \mathbf{1}_{+3}\oplus \mathbf{1}_{-3}.
\end{equation}
The adjoint branches as
\begin{equation}
\mathbf{133}\longrightarrow \mathbf{78}_0\oplus \mathbf{1}_0\oplus \mathbf{27}_{+2}\oplus\overline{\mathbf{27}}_{-2}.
\end{equation}
The Class-B construction of Ref.~\cite{MasiE6E7} then uses the special,
non-regular embedding
\begin{equation}
E_6\longrightarrow G(2)\times SU(3)_A.
\end{equation}
The detailed branching conventions follow the standard exceptional-group
literature and computational tables \cite{Slansky1981,Dynkin1952,McKayPatera1981,AdamsExceptional,GeorgiLieAlgebras,CahnSemiSimple,SpringerVeldkamp,BaezOctonions,GurseyRamondSikivie1976,HewettRizzo1989,LieART}.
Because conjugate-label conventions differ across the literature, we fix the
$E_6\to G(2)\times SU(3)_A$ convention explicitly:
\begin{align}
\mathbf{27}&\longrightarrow (\mathbf{7},\mathbf{3}_A)\oplus(\mathbf{1},\overline{\mathbf{6}}_A),\\
\overline{\mathbf{27}}&\longrightarrow (\mathbf{7},\overline{\mathbf{3}}_A)\oplus(\mathbf{1},\mathbf{6}_A),
\label{eq:27SpecialConvention}
\end{align}
while
\begin{equation}
\mathbf{78}\longrightarrow (\mathbf{14},\mathbf{1}_A)\oplus(\mathbf{1},\mathbf{8}_A)\oplus(\mathbf{7},\mathbf{8}_A).
\end{equation}
The placement of the bars is not arbitrary.  It is fixed by the $E_6$ cubic
invariant: $\mathbf{3}_A\otimes\mathbf{3}_A=\mathbf{6}_A\oplus\overline{\mathbf{3}}_A$, so the mixed cubic singlet
uses the $\overline{\mathbf{6}}_A$ component.  We likewise fix the $\mathbf{351}$-dimensional notation once and use it consistently
throughout the paper:
\begin{equation}
{\Lambda^2 \mathbf{27}\equiv\overline{\mathbf{351}},\qquad
\Lambda^2\overline{\mathbf{27}}\equiv\mathbf{351},}
\label{eq:351Convention}
\end{equation}
The inequivalent symmetric-traceless representation is denoted $\mathbf{351}'$, so that
\begin{equation}
\mathrm{Sym}^2 \mathbf{27}=\overline{\mathbf{27}}\oplus\overline{\mathbf{351}}'.
\end{equation}
This convention removes an otherwise persistent ambiguity in the
$\mathbf{351}/\overline{\mathbf{351}}$ labels.  With Eq.~\eqref{eq:351Convention}, the parent
assignments in Table~\ref{tab:ClassB} agree with the published Class-B
construction.  Physically, the two subgroup factors have distinct roles:
$G(2)$ contains the strong/dark sector, while $SU(3)_A$ is the electroweak
ancestor.

The abelian embedding is then fixed as follows.  Let $t_A^8$ be the diagonal
$SU(3)_A$ generator in the standard normalization and define
\begin{equation}
A=\sqrt3\,t_A^8,
\qquad
q_A=2A.
\end{equation}
The hypercharge embedding is
\begin{equation}
{
Y=\frac13(A+X)=\frac16(q_A+2X).
}
\label{eq:hypercharge}
\end{equation}
The $U(1)_X$ supplied by $E_7$ therefore performs two jobs at once: it
participates directly in hypercharge and it distinguishes the charged parent
branches on which the chirality selector acts.

For the Class-B selector it is convenient to express the same normalization in
the Cartan-divisor basis.  The divisor measuring the physical
$X$ charge is
\begin{equation}
D_X^{\rm ch}=\frac12D_X^{\rm int},
\end{equation}
where
\begin{equation}
{
D_X^{\rm int}=2D_1+3D_2+4D_3+6D_4+5D_5+4D_6+3D_7.
}
\label{eq:DXint}
\end{equation}
The coefficient vector follows directly from the $E_7$ Cartan matrix in the
chosen Bourbaki-type ordering \cite{BourbakiLie46,Slansky1981}. In Bourbaki-type ordering with edges
\begin{equation}
1-3,
\quad
3-4,
\quad
4-5,
\quad
5-6,
\quad
6-7,
\quad
2-4,
\end{equation}
the Cartan matrix is
\begin{equation}
A_{E_7}=\begin{pmatrix}
2&0&-1&0&0&0&0\\
0&2&0&-1&0&0&0\\
-1&0&2&-1&0&0&0\\
0&-1&-1&2&-1&0&0\\
0&0&0&-1&2&-1&0\\
0&0&0&0&-1&2&-1\\
0&0&0&0&0&-1&2
\end{pmatrix}.
\end{equation}
Solving
\begin{equation}
A_{E_7}x=e_7
\end{equation}
gives
\begin{equation}
x=\left(1,\frac32,2,3,\frac52,2,\frac32\right).
\end{equation}
Multiplying by two gives the integral vector
\begin{equation}
c=(2,3,4,6,5,4,3),
\end{equation}
which is precisely Eq.~\eqref{eq:DXint}.  It satisfies
\begin{equation}
A_{E_7}c=(0,0,0,0,0,0,2),
\end{equation}
and
\begin{equation}
c^TA_{E_7}c=6.
\end{equation}
Thus the integral coroot combination is $D_X^{\rm int}$, while the
charge-normalized divisor used in the physical flux formulas is
\begin{equation}
D_X^{\rm ch}=\frac12D_X^{\rm int}.
\end{equation}
This factor of two fixes the charge normalization used in the selector flux below.

\subsection{The Class-B Standard Model branch table}

With these conventions fixed, the selected Class-B branches can be stated
without further ambiguity.  Table~\ref{tab:ClassB} summarizes the assignment
of Ref.~\cite{MasiE6E7}; every hypercharge follows from
Eq.~\eqref{eq:hypercharge}.  We retain both $A$ and $X$ explicitly because the
same charges enter the later operator and flux selection rules.

\begin{table*}[t]
\centering
\renewcommand{\arraystretch}{1.22}
\caption{Parity-adapted Class-B branch assignment. The first block lists
	the selected light chiral branches.  The final two rows display the heavy
	vectorlike quark-doublet ancestors already contained in the same
	\(\overline{\mathbf{351}}_{+1}\oplus\mathbf{351}_{-1}\) parent sector.
	They are not additional chiral selector branches: the
	\(Q_{\mathbf6}\) component is required because the exact parent Clebsch
	for the nominal down-type branch vanishes, and it contributes to the
	physical light doublet only through vectorlike mixing discussed below.
	  The bars on the quark
$SU(3)_A$ ancestors are essential.  The alternative
$u^c\subset(\mathbf{7},\mathbf{3}_A)$, present in the broader representation scan, is removed here
because $\overline{\mathbf{3}}_A\otimes\mathbf{3}_A$ contains a singlet and therefore admits a direct
$Q u\chi$ component.  The $\Theta_-$ conjugate direction required by an exact
high-scale parity completion is discussed separately and need not remain light.}
\label{tab:ClassB}
\begin{tabular}{c c c c c c}
\toprule
field & parent branch & $G(2)\times SU(3)_A$ & $A$ & $X$ & $Y=(A+X)/3$\\
\midrule
$Q_L$ & $\overline{\mathbf{351}}_{+1}$ & $(\mathbf{7},\overline{\mathbf{3}}_A)$ & $-1/2$ & $+1$ & $+1/6$\\
$u^c$ & $\mathbf{351}_{-1}$ & $(\mathbf{7},\mathbf{15}_A)$ & $-1$ & $-1$ & $-2/3$\\
$d^c$ & $\mathbf{351}_{-1}$ & $(\mathbf{7},\overline{\mathbf{6}}_A)$ & $+2$ & $-1$ & $+1/3$\\
$L$ & $\mathbf{2925}_{-3}$ & $(\mathbf{1},\mathbf{8}_A)$ & $+3/2$ & $-3$ & $-1/2$\\
$e^c$ & $\mathbf{2925}_{+3}$ & $(\mathbf{1},\mathbf{8}_A)$ & $0$ & $+3$ & $+1$\\
$N^c$ & $\mathbf{2925}_{+3}$ & $(\mathbf{1},\mathbf{10}_A)$ & $-3$ & $+3$ & $0$\\
$\Theta_+$ & $\mathbf{2925}_{+3}$ & $(\mathbf{1},\mathbf{10}_A)$ & $-3$ & $+3$ & $0$\\
$H$ & $\mathbf{650}_{0,H}$ & $(\mathbf{1},\mathbf{8}_A)$ & $+3/2$ & $0$ & $+1/2$\\
\bottomrule
\midrule
\multicolumn{6}{c}{\itshape vectorlike completion (not additional chiral branches)}\\
\midrule
$Q_{\mathbf6}$ &
$\overline{\mathbf{351}}_{+1}$ &
$(\mathbf7,\mathbf6_A)$ &
$-1/2$ & $+1$ & $+1/6$\\
$\widetilde Q_{\mathbf6}$ &
$\mathbf{351}_{-1}$ &
$(\mathbf7,\overline{\mathbf6}_A)$ &
$+1/2$ & $-1$ & $-1/6$\\
\end{tabular}
\end{table*}

The table distinguishes the selected chiral spectrum from one heavy
vectorlike component that will be needed by the exact parent-Clebsch
analysis.  In particular,
\begin{equation}
Q_{\mathbf6}\subset
(\mathbf7,\mathbf6_A)_{+1}
\subset\overline{\mathbf{351}}_{+1},
\qquad
\widetilde Q_{\mathbf6}\subset
(\mathbf7,\overline{\mathbf6}_A)_{-1}
\subset\mathbf{351}_{-1}.
\end{equation}
Thus \(Q_{\mathbf6}\) is not a new \(E_6\) matter representation or a
seventh chiral Class-B branch: the label \(\mathbf6\) refers to its
\(SU(3)_A\) ancestor.  Its weak-doublet component has
\(A=-1/2\), \(X=+1\), and hence \(Y=+1/6\), exactly the quantum numbers
of \(Q_L\).  Together with its conjugate it belongs to the vectorlike
massive sector already contained in the real
\(\mathbf{27664}_{E_7}\).  The reason this otherwise heavy component is
required will emerge from the exact \(E_6\) Clebsches below: the nominal
\(Q_Ld^cH^\dagger\) parent projection vanishes, whereas
\(Q_{\mathbf6}d^cH^\dagger\) is nonzero.

The charged matter branches are not independent additions: they sit inside a
single real $E_7$ macro-parent \cite{MasiE6E7},
\begin{equation}
\overline{\mathbf{351}}_{+1}\oplus\mathbf{351}_{-1}\oplus\mathbf{2925}_{-3}\oplus\mathbf{2925}_{+3}
\subset\mathbf{27664}_{E_7}.
\end{equation}
Using $\Lambda^3 \mathbf{56}=\mathbf{27664}\oplus\mathbf{56}$ and
$\mathbf{56}\to\mathbf{27}_{-1}\oplus\overline{\mathbf{27}}_{+1}\oplus\mathbf{1}_{+3}\oplus\mathbf{1}_{-3}$,
the complete charged parent can be organized as \cite{Slansky1981,LieART,ExceptionalReductions}
\begin{align}
\mathbf{27664}\to{}&
\overline{\mathbf{351}}_{A,-5}\oplus\mathbf{351}_{A,+5}\nonumber\\
&\oplus(\mathbf{78}\oplus\mathbf{650}\oplus\mathbf{2925})_{-3}\nonumber\\
&\oplus(\mathbf{78}\oplus\mathbf{650}\oplus\mathbf{2925})_{+3}\nonumber\\
&\oplus\mathcal R_{-1}\oplus\overline{\mathcal R}_{+1},
\end{align}
with
\begin{equation}
\mathcal R=\mathbf{27}\oplus2(\mathbf{351})\oplus\mathbf{1728}\oplus\mathbf{7371}.
\end{equation}
The repeated $\mathbf{351}$ in $\mathcal R$ is intentional in the convention
Eq.~\eqref{eq:351Convention}: in the $X=-1$ sector,
$\Lambda^2 \mathbf{27}\otimes\overline{\mathbf{27}}=\overline{\mathbf{351}}\otimes\overline{\mathbf{27}}$
contains $\mathbf{27}\oplus\mathbf{351}\oplus\mathbf{1728}\oplus\mathbf{7371}$, while
$\Lambda^2\overline{\mathbf{27}}\otimes\mathbf{1}_{-3}$ supplies the second $\mathbf{351}$; subtracting
the $\mathbf{27}_{-1}$ contained in the $\mathbf{56}$ leaves precisely the displayed
$\mathcal R$.  The $X=+1$ sector is its conjugate.
The dimension check is
\begin{equation}
\resizebox{0.98\columnwidth}{!}{$\displaystyle
2(\mathbf{351})+2(78+650+2925)+2(27+2\times351+1728+7371)=27664.
$}
\end{equation}
The finite-dimensional $E_7$ parent is therefore exactly charge-conjugation
paired under $(R,q)\leftrightarrow(\overline R,-q)$.  Chirality is not a
property of this representation content alone; it must be generated by the
compactification data and the resulting sheaf/projector structure.

The $X$ charges relevant for chirality are
\begin{equation}
X(Q_L)=+1,
\qquad
X(u^c)=X(d^c)=-1,
\end{equation}
\begin{equation}
X(L)=-3,
\qquad
X(e^c)=X(N^c)=+3.
\end{equation}
These charges explain the logic of the Class-B selector.  The matter
supports are chosen so that the integral of the $X$ flux over each curve
produces the required signed index, yielding three candidate zero modes on
each selected branch.  The next sections test whether this branch-level
picture survives the more demanding component and parity constraints.

\subsection{Compact base and selector supports relevant to stability}
\label{sec:compact-base}
The full compactification of Ref.~\cite{MasiE6E7} contains more geometric
data than are needed for the present stability problem.  Here we isolate the
load-bearing ingredients: the compact gauge surface, the selector flux, the
matter-support classes, and the divisor arithmetic that constrains any repair
of the visible-sector projector.  The gauge surface is $S=dP_2$, and the
physical selector divisor is
\begin{equation}
F_X=E_1-E_2,
\qquad
\sigma:E_1\leftrightarrow E_2,
\qquad
\sigma^*F_X=-F_X .
\end{equation}
Here \(E_1\) and \(E_2\) are the two exceptional divisors of
\(dP_2=\mathrm{Bl}_{p_1,p_2}\mathbb P^2\).  The nontrivial base
involution used by the Class-B selector exchanges the two blow-up
divisors,
\begin{equation}
\sigma:E_1\leftrightarrow E_2,
\end{equation}
and therefore sends
\begin{equation}
F_X=E_1-E_2
\quad\longmapsto\quad
\sigma^*F_X=-F_X .
\end{equation}
The six reduced Class-B supports are
\begin{widetext}
\begin{align}
\Sigma_Q&=3(H-E_1),&
\Sigma_u&=3(H-E_2),&
\Sigma_d&=3(H-E_2),\\
\Sigma_L&=H-E_2,&
\Sigma_e&=H-E_1,&
\Sigma_N&=H-E_1,
\end{align}
\end{widetext}
with selector intersections $(+3,-3,-3,-1,+1,+1)$.

The total $\mathbf{56}$ matter-support class on the gauge surface is
%\begin{equation}
% [P_{\mathbf{56}}]=14H-6E_1-6E_2,
%\label{eq:P56classCompact}
%\end{equation}
\begin{equation}
	[P_{\mathbf{56}}]=14H-6E_1-6E_2,
	\label{eq:P56classCompact}
\end{equation}
which is manifestly invariant under \(E_1\leftrightarrow E_2\), whereas the six supports displayed above sum to
\begin{equation}
 \sum_{a=Q,u,d,L,e,N}\Sigma_a=12H-5E_1-7E_2.
\label{eq:SixSupportSumCompact}
\end{equation}
A later parent-Clebsch calculation will force an additional quark-doublet
ancestor $Q_{\mathbf{6}}\subset(\mathbf{7},\mathbf{6}_A)_{+1}$.  It is useful to test already at the level
of divisor classes whether three chiral families of this branch can fit into
the fixed factorization.  Three chiral $Q_6$ families would require another
$3(H-E_1)$ support, leaving
\begin{equation}
\Sigma'_{\rm ex}
=[P_{\mathbf{56}}]-\sum_a\Sigma_a-3(H-E_1)
=-H+2E_1+E_2,
\end{equation}
whose intersection with the nef class $H$ is negative.  Hence the residual
divisor class is not effective, and three chiral
$Q_{(\mathbf{7},\mathbf{6})}$ families do not fit the present
$P_{\mathbf{56}}$ factorization.  In
the fixed construction $Q_6$ can therefore appear only as a charge-paired,
vectorlike massive state rather than as a second independent chiral family
support.  The divisor-budget conclusion is retained because it fixes the role
of the $Q_6$ ancestor used by the parent-Clebsch analysis below.

\subsection{Exact parent Clebsches and the component-resolved matter assignment}
\label{sec:component-exotic-analysis}
Each selected $G(2)\times SU(3)_A$ representation contains several descendants
after the final gauge breaking.  After $G(2)\to SU(3)_C$ and
$SU(3)_A\to SU(2)_L\times U(1)$, color and electroweak projectors resolve the
individual descendants inside each source branch.  The finite-dimensional
question is whether the \emph{actual parent Clebsch maps} reach the unwanted
components while leaving precisely the desired Standard Model kernel.  In
physical terms, this asks whether the unwanted descendants can be made heavy
without simultaneously removing the Standard Model modes.  The
full descendant tables and projector eigenvalues are retained in the technical
archive; the main text keeps the parent Clebsches and ranks that carry the
physical argument.

Subgroup tensor products alone are insufficient because an allowed
$SU(3)_A$ contraction can still be annihilated by the full $E_6$ Clebsch \cite{DeppischE6Tensors,Slansky1981}.
We therefore restrict the actual parent maps explicitly.  With
$\overline{\mathbf{351}}\simeq\Lambda^2 \mathbf{27}$ and $\mathbf{351}\simeq\Lambda^2\overline{\mathbf{27}}$, take the two independent $E_6$-covariant maps to the neutral $\mathbf{650}$ as $K_1,K_2$.  The explicit exterior tensors give for the stability-selected up branch
\begin{equation}
\resizebox{0.98\columnwidth}{!}{$\displaystyle
{
K_1[Q_{(\mathbf{7},\bar{\mathbf{3}})}u_{(\mathbf{7},\mathbf{15})}H_{(\mathbf{1},\mathbf{8})}]
=-\frac{\sqrt{105}}{70},\qquad
K_2[Q_{(\mathbf{7},\bar{\mathbf{3}})}u_{(\mathbf{7},\mathbf{15})}H_{(\mathbf{1},\mathbf{8})}]
=+\frac{\sqrt{105}}{168}.
}
$}
\label{eq:parentUpClebsch}
\end{equation}
The up-type coupling is therefore nonzero already at the parent level; no
subgroup-only assumption enters this conclusion.

The nominal down-type branch behaves differently.  The same parent
calculation gives the exact zero
\begin{equation}
{
K_1[Q_{(\mathbf{7},\bar{\mathbf{3}})}d_{(\mathbf{7},\bar{\mathbf{6}})}H_{(\mathbf{1},\mathbf{8})}]
=K_2[Q_{(\mathbf{7},\bar{\mathbf{3}})}d_{(\mathbf{7},\bar{\mathbf{6}})}H_{(\mathbf{1},\mathbf{8})}]=0.
}
\label{eq:parentDownZero}
\end{equation}
Although the relevant $SU(3)_A$ tensor product contains an octet, the full
$E_6$ intertwiner projects that subgroup component out.  A nonzero down-type Clebsch is recovered only for the second quark-doublet
ancestor
\begin{equation}
Q_{\mathbf{6}}\subset(\mathbf{7},\mathbf{6}_A)_{+1},
\end{equation}
for which, in the same normalization,
\begin{equation}
K_1[Q_6dH]=\sqrt{\frac5{42}},\qquad
K_2[Q_6dH]=\sqrt{\frac7{30}}.
\label{eq:parentDownQ6}
\end{equation}
The same divisor-budget result shows that $Q_6$ cannot be introduced as a
seventh chiral zero-mode branch on the present base.  The
consistent possibility is mixing with a charge-paired vectorlike state.  Let
\begin{equation}
Q_6\sim(\mathbf{7},\mathbf{6}_A)_{+1},\qquad
\widetilde Q_{\mathbf{6}}\sim(\mathbf{7},\bar{\mathbf{6}}_A)_{-1}
\end{equation}
denote a massive pair in the real $\mathbf{27664}$.  After $SU(3)_A$ breaking the
subgroup contraction
\begin{equation}
(\mathbf{7},\bar{\mathbf{3}})_{+1}\,(\mathbf{7},\bar{\mathbf{6}})_{-1}\,(\mathbf{1},\mathbf{8})_0
\end{equation}
contains a singlet because $\bar{\mathbf{3}}\otimes\bar{\mathbf{6}}\supset\mathbf{8}$ and
$\mathbf{7}\otimes\mathbf{7}\supset\mathbf{1}$ \cite{GeorgiLieAlgebras,Slansky1981}.  Writing its parent-restricted coefficient as
$c_\Omega$, the relevant mass terms are
\begin{equation}
\mathcal L_{\rm mix}\supset
M_6 Q_6\widetilde Q_{\mathbf{6}}+
\Delta_Q Q_{(\mathbf{7},\bar{\mathbf{3}})}\widetilde Q_{\mathbf{6}}+\mathrm{h.c.},
\qquad
\Delta_Q=c_\Omega\langle\Omega\rangle .
\label{eq:Q6VectorlikeMixing}
\end{equation}
For one family the heavy mass is
$M_H=(|M_6|^2+|\Delta_Q|^2)^{1/2}$ and the surviving light doublet may be
written
\begin{equation}
\resizebox{0.98\columnwidth}{!}{$\displaystyle
{
Q_{\rm phys}=\cos\theta_Q\,Q_{(\mathbf{7},\bar{\mathbf{3}})}
-e^{i\phi_Q}\sin\theta_Q\,Q_{(\mathbf{7},\mathbf{6})},\qquad
\tan\theta_Q=\frac{|\Delta_Q|}{|M_6|},
}
$}
\label{eq:QphysicalMixture}
\end{equation}
with $\phi_Q=\arg(\Delta_Q/M_6)$.  In the minimal two-channel limit, if $Y_u$ denotes the nonzero
$Q_{(\mathbf{7},\bar{\mathbf{3}})}u^cH$ parent coupling and $Y_6$ the nonzero
$Q_6d^cH^\dagger$ coupling of Eq.~\eqref{eq:parentDownQ6}, then
\begin{equation}
Y_u^{\rm eff}=Y_u\cos\theta_Q,\qquad
Y_d^{\rm eff}=-e^{i\phi_Q}Y_6\sin\theta_Q .
\label{eq:Q6EffectiveYukawas}
\end{equation}
For three generations the same mechanism becomes matrix valued: a rank-three down
matrix requires the vectorlike mixing spurion $\Delta_QM_6^{-1}$ not to be
rank deficient.  The present finite-dimensional calculation proves the required
$Q_6d^cH^\dagger$ Clebsch, but it does \emph{not} evaluate the separate
parent coefficient $c_\Omega$ or its global wavefunction overlap.
The down-type Yukawa repair is therefore a \emph{conditional} result: the
required parent Clebsch exists, but a viable three-family mass matrix still
requires a nonzero, sufficiently ranked vectorlike mixing spurion.  It is not
an additional chiral selector branch.

The lepton sector provides a useful cross-check that the parent map has the
required tensor structure rather than merely the correct dimensions.  Restricting the parent $\mathbf{2925}_{-3}\otimes\mathbf{2925}_{+3}\to\mathbf{650}$ contraction to the unique $(\mathbf{1},\mathbf{8})$ states gives
\begin{equation}
{
\alpha_A=-i\sqrt{\frac5{84}},\qquad
\alpha_S=\sqrt{\frac3{35}},
}
\label{eq:leptonOctetClebsches}
\end{equation}
for the raw antisymmetric and symmetric $\mathbf{8}\otimes\mathbf{8}\to\mathbf{8}$ tensors \cite{GeorgiLieAlgebras,Slansky1981}.  Our
normalization is
\begin{equation}
\sum_{abc}f_{abc}^2=24,\qquad
\sum_{abc}d_{abc}^2=\frac{40}{3}.
\end{equation}
Consequently
\begin{equation}
\frac{\alpha_A}{\alpha_S}
\sqrt{\frac{\sum f^2}{\sum d^2}}
=-\,i\,\frac{\sqrt5}{2},
\end{equation}
and in normalized coisometric maps the selected direction is
\begin{equation}
\frac13\left(2C_S-i\sqrt5\,C_A\right).
\end{equation}
The symmetric octet component is therefore definitely present, closing the
specific Clebsch ambiguity that otherwise remained in the charged-lepton
splitting argument.  The mixed neutrino $\mathbf{10}$ is also selected nontrivially, with raw coefficient $-\sqrt{2/35}$; the second $\mathbf{10}$ contained purely in the cubic singlet block is annihilated by this parent map.

%\subsubsection{Restricted Clebsch mass matrix: algebraic result}
%\label{sec:A2-clebsch-rank}
Collecting the exact parent maps over all six selected source branches gives
the restricted descendant mass operator
\begin{equation}
\operatorname{rank}\mathbb M_{\rm res}=178,
\qquad
\dim\ker\mathbb M_{\rm res}=16,
\label{eq:A2RestrictedFullRank}
\end{equation}
exactly the dimensions of $Q_L,u^c,d^c,L,e^c,N^c$ for one family.  This is a positive finite-dimensional result: the allowed parent maps reach
all unwanted descendants while leaving a kernel of exactly the dimension of
one SM family plus $N^c$.  It answers the mass-lifting question, but not the
chirality question.  Whether the compactification assigns the required net
chiral index to those surviving modes is independent and is tested below.

\section{The stability-relevant Yukawa selection rules}
\label{sec:dark-component-selection}
The Class-B light spectrum, including $N^c$, is anomaly complete, while the
additional $Q_{\mathbf{6}}\oplus\widetilde Q_{\mathbf{6}}$ pair is vectorlike and contributes no
net anomaly.  The ordinary anomaly and flavor bookkeeping was carried out in
Ref.~\cite{MasiE6E7} and is not repeated here.  What matters for the present
paper is a narrower question: after the exact parent Clebsches and the
conditional down-type mixing are included, does the
selected light Standard Model subspace couple at renormalizable level to a
single insertion of the dark-breaking scalar $\chi$?  The answer provides the first low-energy stability test of the Class-B assignment.
%\subsection{Why the selected light Yukawa sector has no renormalizable one-$\chi$ coupling}
Such a term would bypass the
seclusion that motivates the odd confined sector.  We therefore fix the dark
component
\begin{equation}
\chi\sim(\mathbf{7},\mathbf{1}_A)_0\subset\mathbf{650}_0.
\end{equation}
The $G(2)$ factor by itself does not forbid the coupling, since
\begin{equation}
\mathbf{7}\otimes\mathbf{7}=\mathbf{1}\oplus\mathbf{7}\oplus\mathbf{14}\oplus\mathbf{27},
\end{equation}
and the octonionic invariant contracts three fundamentals \cite{Slansky1981,AdamsExceptional}.  The decisive
selection rule must therefore come from the $SU(3)_A$ quantum numbers and,
where necessary, from the restriction to the actual Standard Model weights.

For the preferred up-type ancestor \cite{GeorgiLieAlgebras,Slansky1981},
\begin{equation}
\overline{\mathbf{3}}_A\otimes\mathbf{15}_A
=\mathbf{27}_A\oplus\mathbf{10}_A\oplus\mathbf{8}_A.
\end{equation}
Hence
\begin{equation}
\operatorname{Hom}_{G(2)\times SU(3)_A}
\!\left[(\mathbf{7},\overline{\mathbf{3}}_A)\otimes(\mathbf{7},\mathbf{15}_A),(\mathbf{7},\mathbf{1}_A)\right]=0.
\end{equation}
Similarly
\begin{equation}
\overline{\mathbf{3}}_A\otimes\overline{\mathbf{6}}_A
=\overline{\mathbf{10}}_A\oplus\mathbf{8}_A
\end{equation}
implies
\begin{equation}
\operatorname{Hom}_{G(2)\times SU(3)_A}
\!\left[(\mathbf{7},\overline{\mathbf{3}}_A)\otimes(\mathbf{7},\overline{\mathbf{6}}_A),(\mathbf{7},\mathbf{1}_A)\right]=0.
\end{equation}
Thus the preferred up and down source branches have no subgroup singlet that
can feed the dark $(\mathbf{7},\mathbf{1}_A)$ direction.  Both independent parent maps
$\overline{\mathbf{351}}\otimes\mathbf{351}\to\mathbf{650}$ consequently obey
\begin{equation}
{
P_{(\mathbf{7},\mathbf{1})_\chi}Y_{q,r}\big|_{Qu}=0,\qquad
P_{(\mathbf{7},\mathbf{1})_\chi}Y_{q,r}\big|_{Qd}=0,\qquad r=1,2.
}
\label{eq:lightchiClebschzero}
\end{equation}
This zero is not a statement that the ordinary Yukawa map itself vanishes:
the up-type projection onto the electroweak Higgs is explicitly nonzero in
Eq.~\eqref{eq:parentUpClebsch}.  For the down channel the subgroup octet is projected out by both parent maps, Eq.~\eqref{eq:parentDownZero}; the nonzero down Yukawa instead requires the $Q_{(\mathbf{7},\mathbf{6})}$ ancestor of Eq.~\eqref{eq:QphysicalMixture}.

The vectorlike $Q_6$ ancestor requires a more refined check, because its
full $SU(3)_A$ branch \emph{does} admit a singlet with $\chi$:
\begin{equation}
\mathbf{6}_A\otimes\bar{\mathbf{6}}_A=\mathbf{1}_A\oplus\mathbf{8}_A\oplus\mathbf{27}_A
\end{equation}
does contain a singlet.  The potentially dangerous branch-level invariant, however, does not survive
projection onto the retained Standard Model weights.  With our conventions
\begin{equation}
\mathbf{6}_A\to\mathbf{3}_{+2}\oplus\mathbf{2}_{-1}\oplus\mathbf{1}_{-4},
\end{equation}
so the $Q_6$ doublet is the $\mathbf{2}_{-1}$ weight \cite{GeorgiLieAlgebras,Slansky1981}.  The selected singlets are
$u^c\subset\mathbf{1}_{-2}\subset\mathbf{15}_A$ and
$d^c\subset\mathbf{1}_{+4}\subset\bar{\mathbf{6}}_A$.  Hence
\begin{equation}
\resizebox{0.98\columnwidth}{!}{$\displaystyle
q_A(Q_6u^c\chi)=-1-2=-3,\qquad
q_A(Q_6d^c\chi)=-1+4=+3,
$}
\end{equation}
and in both cases the weak product is $\mathbf{2}\otimes\mathbf{1}$, not a singlet.  Since an
$SU(3)_A$ invariant is weight conserving, the exact parent restriction onto
these selected weights vanishes:
\begin{equation}
P_{Q_{6,L}}Y_\chi P_{u^c}=0,\qquad
P_{Q_{6,L}}Y_\chi P_{d^c}=0.
\label{eq:Q6SelectedChiZero}
\end{equation}
Hence the full branch Hom space is nonzero, but its nonzero components
connect unwanted descendants rather than the retained weak doublet and singlet
quarks.  Combining
Eq.~\eqref{eq:Q6SelectedChiZero} with the two
$Q_{(\mathbf{7},\bar{\mathbf{3}})}$ zeros above shows that the vectorlike admixture in
Eq.~\eqref{eq:QphysicalMixture} does not spoil the selected-light
one-$\chi$ zero.

This protection is also what selects the preferred up-type ancestor.  The
alternative assignment $u^c\subset(\mathbf{7},\mathbf{3}_A)$ behaves differently:
\begin{equation}
\overline{\mathbf{3}}_A\otimes\mathbf{3}_A=\mathbf{1}_A\oplus\mathbf{8}_A.
\end{equation}
Because the $G(2)$ invariant $O_{ijk}$ contracts three fundamentals, the
component
\begin{equation}
O_{ijk}\,\chi^iQ^j u^k
\end{equation}
is allowed in this alternative branch.  A direct $Q u\chi$ coupling is therefore allowed for that alternative.
Dark sector stability itself selects
\begin{equation}
{u^c\subset(\mathbf{7},\mathbf{15}_A)}
\end{equation}
and removes the $(\mathbf{7},\mathbf{3}_A)$ alternative.

The lepton sector is even simpler: $L,e^c,N^c$ are $G(2)$ singlets, so
$\mathbf{1}\otimes\mathbf{1}\otimes\mathbf{7}$ contains no singlet.  The entire selected light subspace
therefore satisfies
\begin{equation}
{
P_{\rm light}Y_\chi P_{\rm light}=0,
\qquad
P_{\rm light}Y_H P_{\rm light}\neq0.
}
\label{eq:lightmaster}
\end{equation}

The component zero is stable against insertions of the neutral high-scale
VEVs used elsewhere in Class B.  This can be seen directly from the abelian
spurion charges.  For the selected Standard
Model components
\begin{widetext}
\begin{equation}
(A_Q,X_Q)=(-1/2,+1),\qquad
(A_u,X_u)=(-1,-1),\qquad
(A_d,X_d)=(+2,-1).
\end{equation}
\end{widetext}
Hence
\begin{equation}
(A_{Qu},X_{Qu})=(-3/2,0),\qquad
(A_{Qd},X_{Qd})=(+3/2,0).
\end{equation}
The $SU(3)_A$-breaking VEV lies in the $A=0$ singlet of
\begin{equation}
\mathbf{8}_A\to\mathbf{3}_0\oplus\mathbf{2}_{+3/2}\oplus\mathbf{2}_{-3/2}\oplus\mathbf{1}_0,
\end{equation}
so any number of $\langle\Omega\rangle$ insertions cannot close the $\pm3/2$
charge mismatch of the selected $Qu$ or $Qd$ pair.  A parity-symmetric
$\Theta_+\Theta_-$ background is also neutral in both $A$ and $X$.
Neutral high-scale insertions therefore cannot repair the missing charge.
Any coupling of a selected chiral bilinear to a single $\chi$ must also
involve the electroweak doublet direction $H$ or $H^\dagger$.  This is the
origin of the additional suppression in the restricted loop mechanism studied
later.

%\subsection{Role of the parity-completed $\Theta_+\leftrightarrow\Theta_-$ background}
The high-scale repair uses the hypercharge-neutral pair
$\Theta_+\subset\mathbf{2925}_{+3}$ and $\Theta_-\subset\mathbf{2925}_{-3}$ with equal
magnitude.  Indeed, on these two directions the abelian D-terms are proportional to
$D_X\propto3(|\Theta_+|^2-|\Theta_-|^2)$ and
$D_A\propto-3(|\Theta_+|^2-|\Theta_-|^2)$, so equal magnitudes solve both;
because the two VEV directions have $Y=0$, hypercharge remains unbroken \cite{MasiE6E7,WijnholtHiggsBundles}.  For
the present stability argument its important property is neutrality with respect
to the charge mismatch just derived: insertions of the $\Theta_+\Theta_-$
background cannot replace the electroweak $H/H^\dagger$ spurion needed to
connect a selected light bilinear to one $\chi$.

\section{Selector flux and the chiral projector}
\label{sec:selector-flux}

%\paragraph{Physical role of the construction.}
%The parent-Clebsch and renormalizable stability questions have been treated
%above.  What remains from the published Class-B selector \cite{MasiE6E7} is the chiral-index
%problem: the selected branches must produce three mirror-free SM families
%without introducing extra light matter that would interfere with the dark
%sector.  We therefore retain the selector ingredients needed to define
%the light chiral projector used in the ultraviolet parity test \cite{VafaFTheory,MorrisonVafaI,MorrisonVafaII,BHV,DonagiWijnholt}.

The parent-Clebsch and renormalizable stability questions have been treated
above.  What remains from the published Class-B selector of
Ref.~\cite{MasiE6E7} is the chiral-index problem: the selected branches must
produce three mirror-free SM families without introducing extra light matter
that would interfere with the dark sector.  We therefore retain only the
selector ingredients needed to define the light chiral projector used in the
ultraviolet parity test.  The use of gauge fluxes, localized matter curves and
Higgs-bundle zero modes follows the standard F-theory framework
\cite{VafaFTheory,MorrisonVafaI,MorrisonVafaII,BHV,DonagiWijnholt}.

\subsection{Selector vector and uniqueness}

At the branch level the selector is exceptionally simple.  For a matter
branch $R_a$ of $U(1)_X$ charge $X_a$, the flux contribution to the chiral
index is 
\begin{equation}
\chi(R_a)=X_a n_a,
\qquad
n_a=\int_{\Sigma_a}F_X.
\end{equation}
This is the standard matter-curve flux contribution to the four-dimensional
chiral index in the local/semi-local F-theory description
\cite{BHV,DonagiWijnholt,MarsanoSaulinaSchaferNameki}.
Requiring three net generations gives
\begin{equation}
X_a n_a=3.
\end{equation}
Using
\begin{equation}
\resizebox{0.98\columnwidth}{!}{$\displaystyle
X_Q=+1,
\quad
X_u=X_d=-1,
\quad
X_L=-3,
\quad
X_e=X_N=+3,
$}
\end{equation}
we get
\begin{equation}
{(n_Q,n_u,n_d,n_L,n_e,n_N)=(+3,-3,-3,-1,+1,+1).}
\label{eq:selector-vector}
\end{equation}
Under the degree-zero internal line-bundle assumption used in the Class-B
construction \cite{MasiE6E7}, Eq.~\eqref{eq:selector-vector} is the unique six-branch index
vector that gives three generations.  The vectorlike
$Q_{(\mathbf{7},\mathbf{6})}$ ancestor required later for the down Yukawa is not an additional
entry of this chiral index vector; the divisor-budget argument above shows that promoting it to a seventh chiral
branch is incompatible with the present
$P_{\mathbf{56}}$ budget.  Indeed, if $n_a'$ gives the same three-generation spectrum, then
\begin{equation}
X_a(n'_a-n_a)=0.
\end{equation}
Since none of the Standard Model branches has $X_a=0$, this implies
\begin{equation}
n'_a=n_a
\end{equation}
for all selected branches.  The qualification is important: this proves uniqueness of the \emph{index
vector}, not uniqueness of the geometric representatives of the matter
supports or of the complete Higgs/sheaf complex.

\subsection{Cartan flux and Sheaf description}

Combining the surface flux with the charge-normalized Cartan divisor of
Sec.~\ref{sec:group-theory} gives the four-form selector flux 
\begin{equation}
{
G_X=F_X\wedge D_X^{\rm ch}=\frac12(E_1-E_2)\wedge D_X^{\rm int},
}
\label{eq:GX}
\end{equation}
with $D_X^{\rm int}$ given in Eq.~\eqref{eq:DXint}. The normalization in Eq.~\eqref{eq:GX} is the one used in the Class-B
construction \cite{MasiE6E7}; the realization of gauge flux through
Cartan-resolved four-form classes is standard in F-theory GUT constructions
\cite{BHV,DonagiWijnholt,MarsanoSaulinaSchaferNameki}.  In the chosen Cartan
basis,
\begin{equation}
\resizebox{0.98\columnwidth}{!}{$\displaystyle
G_X=\frac12(E_1-E_2)\wedge
(2D_1+3D_2+4D_3+6D_4+5D_5+4D_6+3D_7).
$}
\end{equation}

The branch indices must ultimately arise from zero-mode cohomology, not from
flux integers in isolation.  In the Higgs-bundle/sheaf description \cite{BHV,DonagiWijnholtHiggs2009,WijnholtHiggsBundles,HayashiTatarWatari}, the
zero-mode sheaf on a branch $R_a$ is 
\begin{equation}
\mathcal F_a=K_{\Sigma_a}^{1/2}\otimes\mathcal L_X^{X_a}\otimes\mathcal M_a,
\qquad
c_1(\mathcal L_X)=F_X.
\end{equation}
The factor $\mathcal M_a$ encodes the internal branch choice inside the
corresponding $E_6$ parent representation.  We take
\begin{equation}
\mathcal M_a\in\mathrm{Pic}^0(\Sigma_a)
\end{equation}
for the selected branches.  Riemann--Roch gives \cite{Hartshorne1977}
\begin{equation}
\chi(R_a)=\deg \mathcal F_a+1-g(\Sigma_a).
\end{equation}
With the spin factor included, this reduces to
\begin{equation}
\chi(R_a)=X_a\int_{\Sigma_a}F_X+\deg\mathcal M_a.
\end{equation}
For $\deg\mathcal M_a=0$,
\begin{equation}
\chi(R_a)=X_aF_X\cdot\Sigma_a.
\end{equation}
Thus the branch-level index calculation reproduces the desired value three
on every selected support.

One notational point is essential for the subsequent zero-mode count.  
The symbol \(3C\) denotes three reduced sheaf components over the same
reduced curve \(C\), equivalently a rank-three sheaf with associated graded
\(\mathcal O_C^{\oplus3}\) \cite{HuybrechtsLehn2010,AndersonHeckmanKatz}; it does not denote the nonreduced thickening
\(\mathcal O_{3C}\).
%The
%symbol $3C$ denotes three reduced spectral/sheaf components over a reduced
%curve $C$---equivalently a rank-three T-brane sheaf whose associated graded is
%\begin{equation}
%\mathrm{gr}\,\mathcal S\simeq\mathcal O_C^{\oplus 3}.
%\end{equation}
Confusing the reduced rank-three object with the nonreduced thickening
$\mathcal O_{3C}$ changes the cohomology.  For a full thickening,
\begin{equation}
\mathrm{gr}\,\mathcal O_{mE}=\bigoplus_{k=0}^{m-1}\mathcal O_E(-kE).
\end{equation}
Because $E^2=-1$, such a thickening would introduce additional positive-degree
pieces and would not reproduce the intended three-family zero-mode census.

\subsection{Explicit zero-mode cohomology on the ruling curves}

The branch-level index can now be checked directly at the level of
cohomology.  The relevant supports are the ruling curves
\begin{equation}
C_1=H-E_1,\qquad C_2=H-E_2,
\end{equation}
rather than the exceptional curves $E_i$.  Both satisfy $C_i\simeq\mathbb P^1$, and
\begin{equation}
K_{C_i}^{1/2}=\mathcal O_{\mathbb P^1}(-1).
\end{equation}
Moreover
\begin{equation}
F_X\cdot C_1=+1,\qquad F_X\cdot C_2=-1.
\end{equation}
For $\mathcal O(k)$ on $\mathbb P^1$ \cite{Hartshorne1977},
\begin{equation}
h^0(\mathbb P^1,\mathcal O(k))=k+1\quad(k\ge0),
\qquad h^0=0\quad(k<0),
\end{equation}
and
\begin{equation}
\resizebox{0.98\columnwidth}{!}{$\displaystyle
h^1(\mathbb P^1,\mathcal O(k))=0\quad(k\ge-1),
\qquad h^1=-k-1\quad(k\le-2).
$}
\end{equation}

For $Q_L$ on three reduced copies of $C_1$ with $X=+1$,
\begin{equation}
\mathcal F_Q|_{C_1}
=\mathcal O(-1)\otimes\mathcal O(+1)=\mathcal O,
\end{equation}
so
\begin{equation}
h^0(Q_L)=3,\qquad h^1(Q_L)=0.
\end{equation}
For $u^c,d^c$ on three reduced copies of $C_2$ with $X=-1$,
\begin{equation}
\mathcal F_{u,d}|_{C_2}
=\mathcal O(-1)\otimes\mathcal O(+1)=\mathcal O,
\end{equation}
hence
\begin{equation}
h^0(u^c)=h^0(d^c)=3,\qquad
h^1(u^c)=h^1(d^c)=0.
\end{equation}
For $L$ on $C_2$ with $X=-3$,
\begin{equation}
\mathcal F_L|_{C_2}
=\mathcal O(-1)\otimes\mathcal O(+3)=\mathcal O(2),
\end{equation}
so
\begin{equation}
h^0(L)=3,\qquad h^1(L)=0.
\end{equation}
For $e^c,N^c$ on $C_1$ with $X=+3$,
\begin{equation}
\mathcal F_{e,N}|_{C_1}
=\mathcal O(-1)\otimes\mathcal O(+3)=\mathcal O(2),
\end{equation}
and hence
\begin{equation}
h^0(e^c)=h^0(N^c)=3,\qquad
h^1(e^c)=h^1(N^c)=0.
\end{equation}
Hence the explicit cohomology calculation, rather than the index alone,
gives
\begin{equation}
{
h^0(Q,u,d,L,e,N)=3,\qquad
h^1(Q,u,d,L,e,N)=0.
}
\label{eq:correctedcohomology}
\end{equation}

\subsection{Index obstruction and the component-resolved projector}
\subsubsection{Why vectorlike mixing is not enough}

The positive Clebsch rank does not solve the more basic index problem.  A vectorlike massive tower has zero net chiral index in every representation
of the unbroken Standard Model.  Mixing can rotate which linear combination is
massless, as in Eq.~\eqref{eq:QphysicalMixture}, but it cannot create or erase a topological index \cite{FriedmanMorganWitten,DonagiWijnholtHiggs2009,WijnholtHiggsBundles}.

To make the obstruction maximally conservative, temporarily grant the
Clebsch-forced $Q_{(\mathbf{7},\mathbf{6})}$ an independent chiral index, even though the divisor
budget already forbids it in the present factorization---one may include it
together with the
current Class-B set
\begin{equation}
\resizebox{0.98\columnwidth}{!}{$\displaystyle
Q_{(\mathbf{7},\bar{\mathbf{3}})},\quad Q_{(\mathbf{7},\mathbf{6})},\quad u_{(\mathbf{7},\mathbf{15})},\quad d_{(\mathbf{7},\bar{\mathbf{6}})},\quad
L_{(\mathbf{1},\mathbf{8})},\quad e_{(\mathbf{1},\mathbf{8})},\quad N_{(\mathbf{1},\mathbf{10})}.
$}
\end{equation}
Branch every representation to $SU(3)_C\times SU(2)_L\times U(1)_Y$, form the net complex-representation index vector, and collect the seven ancestor indices in $n$.  The exact integer matrix equation
\begin{equation}
M n=y_{\rm SM+N}
\end{equation}
has
\begin{equation}
{
\operatorname{rank}M=5,\qquad
\operatorname{rank}(M|y_{\rm SM+N})=6,
}
\label{eq:ClassBIndexRankNoGo}
\end{equation}
and hence no solution even over $\mathbb Q$.  The rank mismatch can also be understood analytically from the reality of
$G(2)$ representations.  We use the
signed-index convention
\begin{equation}
 \chi[(r,Y)]\equiv N_L(r,Y)-N_L(\bar r,-Y),
 \label{eq:SignedIndexConvention}
\end{equation}
so that a physical $(\mathbf{3},Y)$ contribution is represented as minus the
conjugate-labelled $(\bar{\mathbf{3}},-Y)$ row.  Every complete $G(2)$ ancestor is
color-conjugation balanced and therefore obeys the row-space covector
\begin{equation}
\resizebox{0.98\columnwidth}{!}{$\displaystyle
 \ell^T M=0,\qquad
 \ell_{(\bar{\mathbf{3}},\mathbf{1})_{-1/3}}=\ell_{(\bar{\mathbf{3}},\mathbf{1})_{+1/3}}=1,
 \quad \ell_\rho=0\ \hbox{otherwise},
$}
\label{eq:ClassBLeftNull}
\end{equation}
equivalently
$\chi[(\bar{\mathbf{3}},\mathbf{1})_{-1/3}]+\chi[(\bar{\mathbf{3}},\mathbf{1})_{+1/3}]=0$.
One Standard Model+$N^c$ family violates this relation because of $d^c$.  The displayed
covector proves the no-solution statement directly: the rank $5\to6$ computation is an independent finite branching-matrix check.

Thus algebraic descendant lifting plus vectorlike mixing is not sufficient. The resulting topological statement is that the current split Class-B
whole-branch selector cannot by itself yield exactly three SM
families (including $N^c$) after the final breaking.
This is not a no-go theorem for every possible chiral projector inside the
$E_7$ macro-parent.  It is a no-go for the present split whole-branch
factorization.  A repair must add chiral ancestor/sheaf terms with nonzero Standard Model-component index, or replace the split construction by a genuinely different non-split Higgs complex whose graded terms contain those cancellations.  Purely vectorlike completion and post hoc Clebsch tuning are insufficient.

Importantly, this visible-sector index obstruction is logically separate
from the local dark-stability selection rule: the preferred $u^c\subset(\mathbf{7},\mathbf{15}_A)$ still eliminates the direct selected-light $Qu\chi$ coupling.  The six branch cohomologies can therefore retain their useful local
stability property while still failing to define a complete global chiral
compactification.

\subsubsection{Whole-branch index obstruction in the $\mathbf{27664}$}
\label{sec:full27664-index-nogo}
The seven-branch calculation might still be suspected of missing a repair
hidden elsewhere in the large $E_7$ parent.  To remove that ambiguity, we
repeat the index test on the complete charged decomposition of $\mathbf{27664}$:
\begin{widetext}
\begin{equation}
\mathbf{27664}\to\overline{\mathbf{351}}_{-5}\oplus\mathbf{351}_{+5}
\oplus(\mathbf{78}\oplus\mathbf{650}\oplus\mathbf{2925})_{-3}
\oplus(\mathbf{78}\oplus\mathbf{650}\oplus\mathbf{2925})_{+3}
\oplus \mathcal R_{-1}\oplus\overline{\mathcal R}_{+1},
\end{equation}
\end{widetext}
with
\begin{equation}
\mathcal R=\mathbf{27}\oplus2(\mathbf{351})\oplus\mathbf{1728}\oplus\mathbf{7371}.
\end{equation}
For the $X=\pm1$ sector it is convenient to reconstruct the full subgroup content directly from the representation-ring identity
\begin{equation}
\mathcal R_{-1}\simeq(\overline{\mathbf{351}}\otimes\overline{\mathbf{27}})\oplus\mathbf{351},
\qquad \dim \mathcal R_{-1}=9828,
\end{equation}
up to the global conjugation convention; $\overline{\mathcal R}_{+1}$ is the conjugate sector.  Branching every available $G(2)\times SU(3)_A$ term to
$SU(3)_C\times SU(2)_L\times U(1)_Y$ produces an exact complex-index matrix
with $109$ distinct whole-branch columns and $188$ complex SM
channels.  Its rational ranks are
\begin{equation}
{
\operatorname{rank}M_{27664}=51,
\qquad
\operatorname{rank}(M_{27664}|y_{\rm SM+N})=52.
}
\label{eq:Full27664RankNoGo}
\end{equation}
The one-unit rank mismatch is a useful exact computational check, but the
no-go does not rely on it.  The analytic origin is the same color-balance
functional implied by complete $G(2)$ representations.
Hence
\begin{equation}
{y_{\rm SM+N}\notin\operatorname{Im}M_{27664}}
\label{eq:Full27664NoSolution}
\end{equation}
even over $\mathbb Q$, and a fortiori there is no integral whole-branch solution.

To display that analytic obstruction explicitly, use the signed-index
convention of Eq.~\eqref{eq:SignedIndexConvention} and define the following
row-space covector on the $188$ complex channels:
\begin{widetext}
\begin{equation}
 \ell^{(C)}_\rho=
 \begin{cases}
 1,&\rho=(\bar{\mathbf{3}},\mathbf{1})_{-1/3},\ (\bar{\mathbf{3}},\mathbf{1})_{+1/3},\\
 0,&\text{otherwise}.
 \end{cases}
 \qquad
 (\ell^{(C)})^TM_{27664}=0,\quad
 (\ell^{(C)})^Ty_{\rm SM+N}\ne0.
 \label{eq:Full27664ColorWitness}
\end{equation}
\end{widetext}
Equivalently,
$\chi[(\bar{\mathbf{3}},\mathbf{1})_{-1/3}]+\chi[(\bar{\mathbf{3}},\mathbf{1})_{+1/3}]=0$
for every whole $G(2)\times SU(3)_A$ column, whereas one Standard Model family has a
nonzero value because of $d^c$.  The plus sign in this functional is not a hypercharge sum; it follows from
the conjugate relabelling in Eq.~\eqref{eq:SignedIndexConvention}: a physical
$(\mathbf{3},Y)$ state contributes with the opposite sign to the
$(\bar{\mathbf{3}},-Y)$ row.  Thus the two terms are the fixed-hypercharge color-balance
functional, not a sum of unrelated Standard Model hypercharges.  The $51\to52$ rank mismatch is therefore an independent computational
audit of the same analytic obstruction.  Its origin is purely representation theoretic.  Every complete $G(2)$ irrep restricts to $SU(3)_C$ with equal multiplicities of each complex color representation and its conjugate \cite{Slansky1981,AdamsExceptional}.  For example
\begin{equation}
\mathbf{7}\to\mathbf{1}\oplus\mathbf{3}\oplus\bar{\mathbf{3}},
\qquad
\mathbf{14}\to\mathbf{8}\oplus\mathbf{3}\oplus\bar{\mathbf{3}},
\end{equation}
and the same conjugation balance holds in the higher $G(2)$ irreps appearing in $\mathbf{27664}$.  Any color-blind sheaf index assigned to a complete $G(2)$ multiplet must
therefore satisfy Eq.~\eqref{eq:Full27664ColorWitness}.  Changing only the
Higgs differential or an extension class can reshuffle representatives but
cannot change this graded $K$-theory identity \cite{HuybrechtsLehn2010,CecottiCordovaHeckmanVafa,AndersonHeckmanKatz}.

%\paragraph{Consequence for the chiral projector.}
%The obstruction is therefore structural rather than a numerical accident of
%the original six-branch choice: as long as color triplets and antitriplets inherit one common index from a complete $G(2)$ multiplet, the visible spectrum cannot become chiral in the Standard Model sense.  A viable selector must resolve color components only after the
%$G(2)\to SU(3)_C$ decomposition has occurred at the level of the relevant
%graded data.  
\paragraph{Consequence for the chiral projector.}
The obstruction is therefore structural rather than a numerical accident
of the original six-branch choice.  Every complete \(G(2)\) representation
contains complex \(SU(3)_C\) representations together with their conjugates.
Consequently, if a single sheaf index is assigned to the whole
\(G(2)\times SU(3)_A\) multiplet, its \(\mathbf3_C\) and
\(\overline{\mathbf3}_C\) descendants remain constrained by the
color-balance relation
\begin{equation}
(\ell^{(C)})^T M_{27664}=0 .
\end{equation}
The desired Standard Model+\(N^c\) spectrum violates this relation,
\begin{equation}
(\ell^{(C)})^T y_{\rm SM+N}\neq0 ,
\end{equation}
and therefore cannot be obtained from graded terms consisting only of
complete \(G(2)\times SU(3)_A\) bundles drawn from the existing
\(\mathbf{27664}_{E_7}\).
Changing only the mass/Higgs map or a non-split extension can alter the
representatives of the zero-mode complex, but not the underlying graded
\(K\)-class.  Such deformations alone therefore
cannot remove the color-balance obstruction.  To obtain the mirror-free
chiral spectrum, the compactification data must distinguish the
\(\mathbf3_C\) and \(\overline{\mathbf3}_C\) descendants at the level of
the graded sheaf data itself.  This requires a component-resolved
modification of the \(K\)-class, rather than merely a different mass or
Higgs map within an unchanged whole-\(G(2)\) complex.  
%A component-resolved
%realization of this type was explored in the companion construction; 
A full geometric completion lies outside the present dark matter analysis: for the stability argument below, only the mirror-free chiral projector
specified by the Class-B assignment and its transformation under $U_B$ are used.

The result should therefore be read in two layers.  The Class-B
assignment defines the mirror-free light projector relevant for the
particle-physics stability analysis and retains the exact
renormalizable light--light--\(\chi\) zero derived above.  What fails is
the attempt to obtain that projector from a single chiral index assigned
to complete \(G(2)\times SU(3)_A\) multiplets: a genuinely
component-resolved ultraviolet realization is required.

In the remainder of the stability analysis we take the mirror-free
Class-B projector as the intended low-energy EFT spectrum and do not
assume that the whole-branch selector has completed its ultraviolet
geometric realization.  The next question is independent of that
index obstruction: even when the direct light--light--\(\chi\) coupling
vanishes, can integrating out the massive vectorlike sector regenerate
dark-\(G\)-odd operators?

%The required repair must therefore change the graded $K$-class seen by the
%color descendants, rather than merely tuning masses inside a whole-$G(2)$
%complex.

\section{Massive spectral pairing and cancellation of dark-$G$-odd heavy effects}
\label{sec:massive-pairing}

%\paragraph{From the light-kernel zero to the massive spectrum.}
The vanishing of the renormalizable light kernel is encouraging, but it does
not by itself control the heavy vectorlike spectrum.  Once the massive states are
integrated out they can still generate operators that are odd under the dark
conjugation.  We therefore study the positive-mass charge-$q$ and charge-$-q$
sectors and ask whether they pair so that the one-$\chi$ contributions cancel.
The pairing can be shown explicitly in the reduced sheaf/Higgs complex used by
the selector; the subsequent question is whether the exceptional vacuum itself
enforces the same relation.

\subsection{Serre--Hodge isospectrality of the positive-mass sector}

The cancellation of dark-\(G\)-odd heavy contributions requires more than
the absence of a light one-\(\chi\) coupling: the massive charge-\(q\) and
charge-\(-q\) sectors must also have identical positive-mass spectra.
For a matter curve \(C\), let \(\mathcal E_q\) denote the holomorphic
bundle or locally free sheaf describing the charge-\(q\) branch.  We choose
the opposite-charge branch to be its Serre dual \cite{Hartshorne1977,HuybrechtsLehn2010},
\begin{equation}
	\mathcal E_{-q}^{S}=K_C\otimes\mathcal E_q^\vee .
	\label{eq:SerrePartner}
\end{equation}
For the rank-one selector sheaves
\[
\mathcal F_q
=
K_C^{1/2}\otimes\mathcal L_X^q\otimes\mathcal M_q ,
\]
this requires
\(\mathcal M_{-q}\simeq\mathcal M_q^{-1}\).
This is a condition on the deliberately parity-completed massive sector,
not a consequence of the chiral selector by itself.

With dual Hermitian metrics, Hodge duality intertwines the Dolbeault
operators on the two branches \cite{Simpson1988,HuybrechtsLehn2010}.  Equivalently, there exists an antiunitary
map \(\mathcal J\) such that the corresponding positive operators are
isospectral.  Hence every nonzero mode has an opposite-charge partner with
the same mass,
\begin{equation}
	m_{n,-q}=m_{n,q},
	\qquad m_n>0 .
	\label{eq:massPairing}
\end{equation}
This is the positive-mass pairing used in the determinant cancellation
below.

Importantly, Eq.~\eqref{eq:massPairing} does not imply a vectorlike
four-dimensional zero-mode spectrum.  Serre duality exchanges the two
cohomological degrees,
\begin{equation}
	H^0(C,\mathcal E_{-q}^{S})
	\simeq H^1(C,\mathcal E_q)^*,
	\qquad
	H^1(C,\mathcal E_{-q}^{S})
	\simeq H^0(C,\mathcal E_q)^* .
	\label{eq:SerreKernels}
\end{equation}
A chiral kernel can therefore coexist with an exactly paired
positive-mass tower.  The explicit singular-mode construction is standard
Hodge theory and is not needed separately here; the determinant cancellation
below depends only on this duality relation.
The pairing condition is a condition on the complete massive complexes,
not on a particular split presentation of them.  More generally, a
non-split extension can be included provided the opposite-charge complex
is chosen as its Serre-dual extension \cite{HuybrechtsLehn2010,AndersonHeckmanKatz}.  We shall therefore formulate the
heavy-sector cancellation directly in terms of the duality relation
between the full positive-mass operators, without assuming a specific
non-split realization.
\subsection{Mode-by-mode cancellation of the parity-odd heavy determinant}

We can now formulate the cancellation directly at operator level.  Let
$\mathbb M_q(\chi,\mathcal V)$ be the reduced massive operator for charge
$q$, where $\mathcal V$ collectively denotes backgrounds chosen
even under the lifted conjugation (the singlet $S$, the parity-even
$\Omega$ direction, and a paired $\Theta_+/\Theta_-$ background).
An exactly dual parity-completed heavy sector obeys
\begin{equation}
{
\mathbb M_{-q}(-\chi,\mathcal V)
=
\mathcal J\,
\mathbb M_q(\chi,\mathcal V)^\dagger
\mathcal J^{-1}
}
\label{eq:massOperatorParity}
\end{equation}
on the positive-mass subspace.  Differentiating with respect to the
$P$-odd $\chi$ direction gives
\begin{equation}
Y_{\chi,-q}
=
-\mathcal J\,Y_{\chi,q}^\dagger\mathcal J^{-1}.
\label{eq:YchiPair}
\end{equation}

A representative dangerous Wilson coefficient is the one-$\chi$,
three-gauge-field term.  On the positive-mass subspace its heavy contribution
has the matrix form
\begin{equation}
\resizebox{0.98\columnwidth}{!}{$\displaystyle
\Delta_\Pi^{\rm heavy}
=
M_*^3
\sum_{q>0}
\operatorname{ReTr}'\!\left[
Y_{\chi,q}
(\mathbb M_q^\dagger\mathbb M_q)^{-3/2}
+
Y_{\chi,-q}
(\mathbb M_{-q}^\dagger\mathbb M_{-q})^{-3/2}
\right],
$}
\label{eq:DeltaMatrix}
\end{equation}
where the prime omits kernels.  Equations
\eqref{eq:massPairing} and \eqref{eq:YchiPair} give
\begin{equation}
{\Delta_\Pi^{\rm heavy}=0.}
\label{eq:DeltaHeavyZero}
\end{equation}
The vanishing is mode by mode.  Parity-even backgrounds may split different
heavy multiplets, but they split each Serre-dual pair identically and therefore
do not spoil the cancellation.

The same reasoning applies to the complete primed Euclidean determinant, not
only to the first operator in its expansion \cite{HenningLuMurayama2018}:
\begin{equation}
\Gamma_{\rm heavy}[\chi,\mathcal V]
=
-\log\det{}'\mathbb M_q(\chi,\mathcal V)
-\log\det{}'\mathbb M_{-q}(\chi,\mathcal V),
\end{equation}
Eq.~\eqref{eq:massOperatorParity} implies
\begin{equation}
{
\Gamma_{\rm heavy}[\chi,\mathcal V]
=
\Gamma_{\rm heavy}[-\chi,\mathcal V].
}
\label{eq:evenHeavyDet}
\end{equation}
The reduced determinant is therefore an even functional of $\chi$.  Every
heavy-sector Wilson coefficient containing an odd number of $\chi$
insertions vanishes in this deliberately parity-completed reduced complex.
The operator-completeness analysis below shows that the physical pure-dark odd
basis is empty through dimension seven and begins at dimension nine.  The
paired determinant therefore removes the entire first nonzero basis, including
the mixed-helicity $[3,1]$ $\chi F^4$ channel of
Eq.~\eqref{eq:O9F4}, rather than merely one selected representative.  The
simpler scalar $\chi F^3$ structure vanishes already by Lorentz and
representation symmetry, independently of this determinant theorem.

The logical scope of the theorem is important.  It pairs the \emph{massive}
sector; it does not turn the chiral four-dimensional kernel into a parity
multiplet.  The map $\mathcal J$ is the anti-linear
Serre-Hodge duality exchanging degree zero and degree one; it is precisely why
Eq.~\eqref{eq:SerreKernels} can coexist with three chiral families.  The theorem
obtained here is therefore a positive-mass determinant theorem, not a proof
that the entire chiral compactification possesses the isolated $G(2)$ internal
parity.

\subsection{Why the canonical $E_7$ charge flip does not enforce the pairing}
\label{sec:E7lift-nogo}

The physical question is now simple: can the exceptional theory itself
\emph{force} the $q\leftrightarrow -q$ pairing assumed above?  To do so, the
candidate transformation must both exchange the conjugate massive sectors and
leave the symmetry-breaking vacuum invariant.  The calculation below shows
that the second requirement fails in the minimal one-copy scalar
architecture.

The natural candidate for enforcing the duality is the canonical charge-flip
normalizer in $E_7$.  A first branching-only argument might suggest that the
nontrivial $E_6$ outer automorphism already fails because $\mathbf{27}$ and
$\overline{\mathbf{27}}$ restrict differently to the special subgroup
\begin{equation}
H_s=G(2)\times SU(3)_A\subset E_6
\end{equation}
to a second inner-inequivalent embedding.  That conclusion is not correct at subgroup level: an
outer automorphism can preserve $H_s$ as a set while acting by the outer
conjugation $\mathbf{3}_A\leftrightarrow\overline{\mathbf{3}}_A$ on its $A_2$ factor.  The correct test must therefore be performed on the actual scalar carrier
and vacuum, not inferred from the abstract subgroup embedding.

\paragraph{The special subgroup preservation.}
Choose the standard outer involution $\theta$ of $E_6$ whose connected fixed
subgroup is $F_4$.  The $G(2)$ in the special $A_2\times G(2)$ embedding is in
the same conjugacy class as the standard $G(2)\subset F_4$.  A finite branching
certificate is obtained by restricting the $A_2$ factor to its principal
$A_1$, for which $\mathbf{3}\to\mathbf{3}$ and $\overline{\mathbf{6}}\to\mathbf{5}\oplus\mathbf{1}$:
\begin{widetext}
\begin{align}
\mathbf{27}&=(\mathbf{7},\mathbf{3})\oplus(\mathbf{1},\overline{\mathbf{6}})
\longrightarrow (\mathbf{7},\mathbf{3})\oplus(\mathbf{1},\mathbf{5})\oplus(\mathbf{1},\mathbf{1})=\mathbf{26}\oplus\mathbf{1},\\
\mathbf{78}&=(\mathbf{14},\mathbf{1})\oplus(\mathbf{1},\mathbf{8})\oplus(\mathbf{7},\mathbf{8})\nonumber\\
&\longrightarrow
\big[(\mathbf{14},\mathbf{1})\oplus(\mathbf{1},\mathbf{3})\oplus(\mathbf{7},\mathbf{5})\big]
\oplus\big[(\mathbf{1},\mathbf{5})\oplus(\mathbf{7},\mathbf{3})\big]=\mathbf{52}\oplus\mathbf{26}.
\label{eq:F4SpecialCertificate}
\end{align}
\end{widetext}
These are precisely the standard $E_6\downarrow F_4$ restrictions \cite{CacciatoriE6Geometry}.  Moreover,
the $A_2$ factor is the centralizer of this $G(2)$ in $E_6$
\cite{Rubenthaler2008}.  Since $\theta$ fixes the chosen $G(2)\subset F_4$
pointwise, it preserves its centralizer and therefore
\begin{equation}
{\theta(H_s)=H_s\quad\text{setwise}.}
\label{eq:specialSubgroupOuterStable}
\end{equation}
The special $G(2)\times SU(3)_A$ subgroup is therefore compatible with this
outer involution.  The obstruction, if any, must lie in the transformation of
the actual Higgs representation and its vacuum direction.

\paragraph{The key test: the neutral $\mathbf{650}$ carrier and its vacuum direction.}
Subgroup invariance is only a necessary condition: the candidate transformation
must also act with the correct signs on the actual Higgs components that acquire
VEVs.  Let $P_7^{(0)}$ denote the canonical normalizer element of
$E_6\times U(1)_X\subset E_7$ that reverses $X$ charge.  Its action is fixed most
cleanly in the symplectic $\mathbf{56}$, whose decomposition is
\begin{equation}
\mathbf{56}=\mathbf{27}_{-1}\oplus\overline{\mathbf{27}}_{+1}\oplus\mathbf{1}_{+3}\oplus\mathbf{1}_{-3}
\end{equation}
and which carries the invariant $E_7$ symplectic form.  In a Darboux basis the
charge flip exchanges every pair of opposite $U(1)_X$ charge with the relative
minus sign required by symplectic invariance.  Consequently its action on the
neutral cross block
\begin{equation}
\mathbf{27}\otimes\overline{\mathbf{27}}\simeq\mathrm{End}(\mathbf{27})
=\mathbf{1}\oplus\mathbf{78}\oplus\mathbf{650}
\end{equation}
this induces, after an $F_4$-adapted identification of $\mathbf{27}$ with its dual,
\begin{equation}
{P_7^{(0)}:A\longmapsto-A^T.}
\label{eq:E7MinusTranspose}
\end{equation}
The minus-transpose sign in Eq.~\eqref{eq:E7MinusTranspose} is not a convention
that can be changed independently on the $\mathbf{650}$: it follows from the embedding
of the neutral carrier in the symplectic $\mathbf{56}$.  Multiplying the transformation
of an isolated real $\mathbf{650}$ by an additional overall sign would amount to
postulating a new discrete action, rather than using the canonical one inherited
from $E_7$.  The $\mathbf{56}$ realization and the $E_6\times U(1)$ decomposition are
standard \cite{CacciatoriE7MagicSquare,ExceptionalReductions,DeppischE6Tensors}.

Now decompose the $\mathbf{27}$ as
\begin{equation}
\resizebox{0.98\columnwidth}{!}{$\displaystyle
\mathbf{27}=A\oplus B,
\qquad A=(\mathbf{7},\mathbf{3}_A),\quad \dim A=21,
\qquad B=(\mathbf{1},\overline{\mathbf{6}}_A),\quad \dim B=6.
$}
\end{equation}
This decomposition makes the special-breaking singlet transparent.  The only
$H_s$-invariant endomorphisms are the two block projectors $\Pi_A$ and $\Pi_B$.
Their sum is the identity and therefore belongs to the $E_6$ singlet, whereas
the orthogonal traceless combination is the unique $H_s$ singlet carried by
the $\mathbf{650}$.  A convenient representative is
\begin{equation}
S=2\Pi_A-7\Pi_B,
\qquad \operatorname{Tr} S=0,
\qquad S^T=S.
\label{eq:Special650SingletMatrix}
\end{equation}
Equation~\eqref{eq:E7MinusTranspose} therefore gives
\begin{equation}
{P_7^{(0)}S=-S.}
\label{eq:P7Ssign}
\end{equation}

The dark-breaking direction transforms differently.  The
$G(2)$ fundamental $\chi\sim(\mathbf{7},\mathbf{1}_A)\subset\mathbf{650}$ is represented inside the
$A=(\mathbf{7},\mathbf{3}_A)$ block by
\begin{equation}
\chi_n=J_n\otimes\mathbf{1}_3,
\qquad (J_n)_{ij}=O_{nij},
\qquad J_n^T=-J_n.
\label{eq:Chi650Matrix}
\end{equation}
Because $\mathbf{78}|_{H_s}$ contains $(\mathbf{14},\mathbf{1}_A)$ but no $(\mathbf{7},\mathbf{1}_A)$, this antisymmetric
cross-block component belongs to the $\mathbf{650}$ rather than to the adjoint.  The
minus-transpose action therefore yields the opposite eigenvalue from the
symmetric singlet:
\begin{equation}
{P_7^{(0)}\chi=+\chi.}
\label{eq:P7Chisign}
\end{equation}
The explicit inner element $U_B\in G(2)$ fixes the singlet $S$ and sends the
chosen $n=e_7$ direction to $-n$.  Consequently
\begin{equation}
\resizebox{0.98\columnwidth}{!}{$\displaystyle
{
P_7^{(0)}:(S,\chi)\mapsto(-S,+\chi),
\qquad
U_BP_7^{(0)}:(S,\chi)\mapsto(-S,-\chi).
}
$}
\label{eq:E7specialNoGo}
\end{equation}
Equation~\eqref{eq:E7specialNoGo} is the central finite-dimensional obstruction:
the canonical charge flip is odd on the special-breaking singlet $S$, and
multiplication by the isolated-theory element $U_B$ changes the sign of
$\chi$ but leaves the unwanted sign of $S$ untouched.  Thus neither canonical
one-copy candidate preserves the vacuum $\langle\Phi\rangle=v_SS$.

\paragraph{No one-copy normalizer correction can restore the required signs.}
It remains to ask whether the charge flip can be multiplied by some other
$h\in E_6$ so that the special VEV is restored.  If $hP_7^{(0)}$ preserved the
VEV, then
$P_7^{(0)}S=-S$, this would require
\begin{equation}
hS=-S.
\label{eq:hSminusS}
\end{equation}
The stabilizers of $S$ and $-S$ coincide and are both
$H_s=G(2)\times SU(3)_A$.  Therefore any such $h$ would have to map this
special subgroup to itself, and Eq.~\eqref{eq:hSminusS} implies
\begin{equation}
hH_sh^{-1}=H_s;
\end{equation}
that is, $h$ must lie in the normalizer $N_{E_6}(H_s)$.  A disconnected
normalizer component capable of compensating the outer action would have to
implement $\mathbf{3}_A\leftrightarrow\bar{\mathbf{3}}_A$.  But on the same $\mathbf{27}$ one has
\begin{equation}
27\downarrow H_s=(\mathbf{7},\mathbf{3}_A)\oplus(\mathbf{1},\bar{\mathbf{6}}_A),
\end{equation}
whereas the putative conjugated action would require
\begin{equation}
(\mathbf{7},\bar{\mathbf{3}}_A)\oplus(\mathbf{1},\mathbf{6}_A).
\end{equation}
The two restrictions are not isomorphic as $H_s$ representations, so no inner
normalizer element can implement the required compensating conjugation.  Two
remaining elementary possibilities also fail: the $E_6$ center is trivial on
$\mathbf{650}\subset\mathbf{27}\otimes\overline{\mathbf{27}}$, while an independent global sign
$\Phi\mapsto-\Phi$ is not a symmetry of the generic one-$\mathbf{650}$ potential because
cubic invariants are allowed unless a new discrete symmetry is imposed.
Therefore the obstruction is not an artifact of choosing the particular
representative $U_B$; it is a one-copy statement:
the conclusion is independent of the representative chosen for the normalizer: no
exact ultraviolet $\mathbb Z_2$ of the minimal one-copy $E_7/E_6$ vacuum
can make $S$ even while making $\chi$ odd.
This theorem closes the finite-dimensional parity search in the minimal one-copy
scalar architecture.

The finite one-copy result can therefore be summarized as:
\begin{enumerate}[label=(\roman*)]
\item the isolated $G(2)$ parity of Ref.~\cite{ButtazzoG2Higgs} is exact at the
renormalizable level;
\item the subgroup $H_s=G(2)\times SU(3)_A$ may be chosen setwise stable under
the relevant outer action;
\item the canonical $E_7$ charge flip does not preserve the special
$\mathbf{650}$ vacuum;
\item no one-copy normalizer/stabilizer repair removes this failure;
\item equivalently, the one-copy carrier cannot make $S$ even and $\chi$ odd
simultaneously.
\end{enumerate}

%\paragraph{Implication for the heavy determinant.}
The practical consequence is that the exact determinant cancellations derived
for the paired massive complex must always be labelled by their hypothesis.
Relations such as $\Delta_\Pi^{\rm heavy}=0$ are exact once the conjugate
complexes and their extension data are chosen as duals, but the present
one-copy exceptional vacuum does not furnish an unbroken canonical $E_7$
symmetry that forces Nature to make that choice.  This distinction between an
exact conditional theorem and an exact symmetry of the ultraviolet vacuum is
kept explicit in the remainder of the paper.

\section{Ultraviolet decay operators and state-resolved lifetime bounds}
\label{sec:dark-decay}

The spectroscopy problem and the ultraviolet lifetime problem are related but
not identical.  In the isolated bosonic theory, $C$ and $P$ are useful exact
labels and the relevant odd channels are $1^{+-}$ and $0^{--}$.  Once the full
mirror-free chiral embedding is restored, ordinary $C$ and $P$ are no longer
exact symmetries of the complete theory.  We therefore retain $1^{+-}$ and
$0^{--}$ below as \,\emph{channel-ancestry labels}: they identify the isolated-theory
spectral blocks to which the interacting poles continuously connect, not exact
quantum numbers of the full chiral Lagrangian.

The one-copy parity tests above make an operator-by-operator analysis unavoidable.
Because no unbroken canonical $E_7$ parity forbids all odd-dark couplings,
every interaction allowed by the exact gauge symmetries must be checked at the
parent level and then matched to the physical confined pole.  The distinction
between these two steps---existence of a ultraviolet tensor and its nonperturbative pole
overlap---is central throughout this section.

\subsection{Dimension-five visible-sector connectors: parent nonzero versus pole matching}
\label{sec:d5-generic-connector}

The exact renormalizable zero
$P_{\rm light}Y_\chi P_{\rm light}=0$ is an important selection rule, but it
does not postpone the generic effective operator basis to dimension nine.
Once the electroweak Higgs direction is available, the broken-phase gauge
quantum numbers already allow the dimension-five structures
\begin{align}
\mathcal O_{5,u}^{ij}&=
\frac{c_{5,u}^{ij}}{M_*}\,Q_{(\mathbf{7},\bar{\mathbf{3}})}^i u_{(\mathbf{7},\mathbf{15})}^{c\,j}\,\chi H,
\label{eq:O5uQG}\\
\mathcal O_{5,d}^{ij}&=
\frac{c_{5,d}^{ij}}{M_*}\,Q_{(\mathbf{7},\mathbf{6})}^i d_{(\mathbf{7},\bar{\mathbf{6}})}^{c\,j}\,\chi H^\dagger,
\label{eq:O5dQG}
\end{align}
with the subgroup contractions understood.  It is important that these are
not artifacts of working only at the $G(2)\times SU(3)_A$ level.  The explicit
parent-tensor analysis finds nonzero $E_6$ contractions in both channels, and a
direct $E_7$ completion in
$\mathbf{27664}\otimes\mathbf{27664}\otimes\mathbf{1463}\otimes\mathbf{1463}$ gives the normalized component
matrix elements
\begin{equation}
\mathcal I_u^{(E_7)}=-\frac{\sqrt{210}}{126},
\qquad
\mathcal I_d^{(E_7)}=+\frac{\sqrt{105}}{126}.
\label{eq:E7d5Witnesses}
\end{equation}
The two witnesses satisfy the independent normalization check
\begin{equation}
 \frac{\mathcal I_u^{(E_7)}}{\mathcal I_d^{(E_7)}}=-\sqrt2,
 \label{eq:E7d5WitnessRatio}
\end{equation}
which follows directly from their common parent normalization and provides an
additional checksum on the relative quark-channel conventions.
The meaning of Eq.~\eqref{eq:E7d5Witnesses} is precise: the relevant parent
Clebsches are nonzero, so the operators are not forbidden by representation
theory.  It does \emph{not} determine the Wilson coefficients $c_{5,u,d}$,
which depend on ultraviolet masses, couplings, localization data and global
wavefunction overlaps.  In particular, the down-type witness uses the same
$Q_{(\mathbf{7},\mathbf{6})}$ ancestor already required by the ordinary down Yukawa; projecting
to the physical mixed quark doublet therefore does not recover a structural
zero.

\paragraph{The selected light projection is radial, not an odd-pole interpolator.}
The existence of the parent operator is only the first half of the lifetime
problem.  After electroweak breaking, Eqs.~\eqref{eq:O5uQG}--\eqref{eq:O5dQG}
violate the putative odd-$\chi$ rule, but the selected light--light component
does not directly interpolate either odd confined state.  The reason follows
from the unique $G(2)$ singlet in $\mathbf{7}\otimes\mathbf{7}\otimes\mathbf{7}$, namely the octonionic
tensor.  Under
$SU(3)_C\subset G(2)$ it contains schematically
\begin{equation}
 (\mathbf{7}\otimes\mathbf{7}\otimes\mathbf{7})_{\bf1}
 \supset \epsilon(\mathbf{3},\mathbf{3},\mathbf{3})+\epsilon(\bar{\mathbf{3}},\bar{\mathbf{3}},\bar{\mathbf{3}})
 +\mathbf{1}\cdot\mathbf{3}\cdot\bar{\mathbf{3}} .
 \label{eq:G2CubicSU3Decomp}
\end{equation}
The selected light quark colors are $Q_L\subset\mathbf{3}_C$ and
$u^c,d^c\subset\bar{\mathbf{3}}_C$.  Their light--light projection therefore uses only
the color-singlet scalar component $\chi_1\parallel e_7$,
\begin{equation}
 {\cal O}_5\big|_{\rm light-light}
 \propto Q_{\mathbf{3}_C}u^c_{\bar{\mathbf{3}}_C}\chi_1H
 \quad(\text{or }Q_{\mathbf{3}_C}d^c_{\bar{\mathbf{3}}_C}\chi_1H^\dagger).
 \label{eq:D5LightColorProjection}
\end{equation}
The VEV/radial $\chi_1$ fluctuation is $0^{++}$ and even under the residual
pure-dark grading.  Consequently, in the unperturbed confining theory,
\begin{equation}
 \langle0|\chi_1|1^{+-}\rangle=
 \langle0|\chi_1|0^{--}\rangle=0.
 \label{eq:D5DirectOddPoleZero}
\end{equation}
Equation~\eqref{eq:D5DirectOddPoleZero} eliminates the simplest factorized
vacuum-to-pole matrix element.  
%It is not a theorem that the complete decay
%amplitude vanishes.  
The connected matrix element
$\langle f|Q u^c\chi_1H|B_J\rangle$ can receive nonfactorizing contributions
from the interacting bound state, including heavy-descendant propagation,
Goldstone or longitudinal conversion, additional gauge insertions, loop
matching, and genuinely connected confined matrix elements.  A lifetime
estimate must therefore keep the parent Wilson coefficient and the
state-dependent pole matching conceptually separate.

We encode that second step in a single state-resolved factor
$\kappa_{5,J}^{\rm eff}$, defined to include the \emph{entire} matching from
the parent operator to the physical pole.  No universal dynamical suppression pattern or fixed power is assumed.  With this deliberately conservative definition,
naive dimensional analysis gives \cite{ManoharGeorgi1984,LukeManohar1997}
\begin{equation}
\Gamma_{B_J}^{(5)}\sim
\frac{\kappa_{5,J}^{\rm eff}}{8\pi}
\left|\frac{c_5v}{M_*}\right|^2m_{B_J}.
\label{eq:d5NDAwidth}
\end{equation}
For $m_B=1.71\times10^{14}\,$GeV and $M_*=M_{E_6}=6.9\times10^{15}\,$GeV,
cosmological survival to the age of the Universe requires approximately
\begin{equation}
|c_5|\sqrt{\kappa_{5,J}^{\rm eff}}\lesssim1.3\times10^{-14},
\end{equation}
while the illustrative longevity benchmark $\tau_B\gtrsim10^{26}\,$s requires
\begin{equation}
|c_5|\sqrt{\kappa_{5,J}^{\rm eff}}\lesssim8.7\times10^{-19}.
\label{eq:d5LifetimeBound}
\end{equation}
Even if the cutoff is raised to the reduced Planck scale, the illustrative
$10^{26}\,$s requirement still demands a product of order $3\times10^{-16}$.
For comparison, the simple ratios
$m_X/M_{E_6}\simeq8.3\times10^{-3}$ and
$g_{G(2)}^2/(16\pi^2)\simeq2.0\times10^{-3}$ show that one or two ordinary
kinematic or loop suppressions are not automatically enough.  They can weaken
a unit-overlap estimate by several orders of magnitude, but cosmological
longevity still requires either a much stronger state-overlap suppression or
a structural selection rule.

\paragraph{No model-independent lifetime ordering between the $J=1$ and $J=0$ poles.}
The previous discussion also prevents us from inferring a spin hierarchy from
the parent Clebsches alone.  The two channels require independent matching
factors,
\begin{equation}
\kappa_{5,1^{+-}}^{\rm eff},\qquad
\kappa_{5,0^{--}}^{\rm eff}.
\end{equation}
Neither is fixed by the finite parent Clebsch alone.  If a local
two-fermion operator is generated after the heavy matching, its on-shell
spin-zero and spin-one amplitudes may contain the usual scalar/pseudoscalar
and vector/axial/tensor structures; there is no universal $P$-wave theorem
for the vector.  But Eq.~\eqref{eq:D5DirectOddPoleZero} shows that these
local structures arise only \emph{after} the indirect matching just described,
not from the bare light--light projection of $\mathcal O_5$.

The exact parent witnesses establish three separate facts:
\begin{enumerate}[label=(\roman*)]
\item the dimension-five parent operator breaks the would-be dark grading;
\item its direct selected light--light projection has zero overlap with either
odd pole;
\item the parent data imply no model-independent ordering of
$\kappa_{5,1}^{\rm eff}$ and $\kappa_{5,0}^{\rm eff}$.
\end{enumerate}
The spectroscopy can therefore favor a $1^{+-}$ ground state without implying
that the $J=1$ pole is parametrically longer lived than the $0^{--}$ pole.
The ultraviolet survival criterion is state specific and depends on the full
combination $|c_5|\sqrt{\kappa_{5,J}^{\rm eff}}$.

This dimension-five problem motivates the separated-$\mathbf{1539}_H$ construction
studied later.  If $\mathbb Z_{2,D}$ is genuinely gauged, the entire odd-$\Xi$
dimension-five family is forbidden before dark Higgsing.  Operators even in
$\Xi$, however, remain allowed.  Section~\ref{sec:1539EvenDressing} performs the
corresponding dimension-six analysis.  Its key result is an exact residual-color
zero for the selected-light term with one dark VEV and one transverse
$\mathbf{3}_C\oplus\bar{\mathbf{3}}_C$ fluctuation; radial and two-transverse channels survive and
are instead encoded in state-dependent $\kappa_{6,J}^{\rm eff}$ factors.

\subsection{Lowest purely dark symmetry-violating operators: on-shell basis through dimension nine}
\label{sec:pureDarkOperatorBasis}
The physical question is whether the dark grading can be violated by a local
operator even when the visible-sector connectors are absent.  To answer it
without overcounting redundant structures, the completeness question is
formulated in an on-shell operator basis, where integration-by-parts and
equation-of-motion redundancies are removed at the amplitude level.  We follow the modern Young-tableau and
Hilbert-series organization of such bases
\cite{HenningLuMeliaMurayama2016,HenningLuMeliaMurayama2017,
LiRenXiaoYuZheng2022,DongMaShu2023,GrafHenningLuMeliaMurayama2023}.

The visible-sector connectors are not the only possible source of instability.
One must also classify local Lorentz scalars built only from the dark Higgs
$\chi$, its covariant derivatives and the $G(2)$ field strength.  A simple
tensor-product singlet count is not sufficient for this purpose: Lorentz
permutation symmetry, Bose symmetry, integration by parts (IBP), Bianchi
identities and equations of motion (EOM) can remove operators that are
algebraically nonzero off shell.  We therefore organize the pure-dark sector as
an on-shell local contact-term basis, where equations of motion and integration
by parts are removed through the standard operator--amplitude correspondence
\cite{HenningLuMeliaMurayama2016,HenningLuMeliaMurayama2017}, and then match
the permutation symmetry of the kinematic polynomial to the corresponding
$G(2)$ Schur functor.  This gives a completeness statement through canonical
dimension nine rather than a list of selected candidate operators.

\paragraph{Dimensions below seven.}
Let $n_\chi,n_D,n_F$ denote the number of dark-Higgs fields, covariant
derivatives and field strengths.  The canonical dimension is
\begin{equation}
 d=n_\chi+n_D+2n_F .
\label{eq:OddOperatorDimensionCount}
\end{equation}
Lorentz invariance requires the total number of Lorentz indices to be even.
Since each field strength contributes two indices, this implies that $n_D$ is
even.  Equation~\eqref{eq:OddOperatorDimensionCount} therefore gives
$d\equiv n_\chi\pmod 2$: a dark-odd operator, with odd $n_\chi$, can occur
only at odd canonical dimension.  The low-dimensional
internal channels already eliminate the obvious possibilities.  In particular,
\begin{equation}
 \mathbf{7}\not\subset\mathbf{14}\otimes\mathbf{14}
\end{equation}
forbids $\chi F^2$, while
\begin{equation}
 \mathrm{Sym}^2\mathbf{7}=\mathbf{1}\oplus\mathbf{27},
 \qquad
 \mathrm{Sym}^3\mathbf{7}=\mathbf{7}\oplus\mathbf{77}'
\label{eq:G2LowSymPowers}
\end{equation}
removes the corresponding Bose-symmetrized odd scalar possibilities \cite{Slansky1981,AdamsExceptional}.  No
independent pure-dark odd contact term occurs at $d=3$ or $d=5$.

\paragraph{Exact zero of the apparent $\chi F^3$ operator.}
Three adjoints are the first tensor power for which a $\mathbf{7}$ occurs, but
the permutation symmetry is decisive.  The exact plethysms are
\begin{widetext}
\begin{equation}
 \Lambda^3\mathbf{14}
 =\mathbf{1}\oplus\mathbf{27}\oplus\mathbf{77}\oplus\mathbf{77}'\oplus\mathbf{182},
 \qquad
 \mathrm{Sym}^3\mathbf{14}
 =\mathbf{7}\oplus\mathbf{14}\oplus\mathbf{77}'\oplus\mathbf{189}\oplus\mathbf{273}.
\label{eq:G2AdjointCubePlethysms}
\end{equation}
\end{widetext}
The $\mathbf{7}$ occurs in the symmetric cube but not in the exterior cube \cite{FultonHarris1991,LieART}.
Writing $F_a$ as an antisymmetric Lorentz matrix,
\begin{equation}
 L_{abc}\equiv
 F_{\mu}{}^{\nu a}F_{\nu}{}^{\rho b}F_{\rho}{}^{\mu c}
 =\operatorname{tr}(F_aF_bF_c),
\end{equation}
transposition and cyclicity give $L_{abc}=-L_{bac}$, and therefore total
antisymmetry in the three adjoint labels.  The octonionic contraction of one
$\chi$ with three adjoints is instead symmetric in those slots; in the daughter
basis it is proportional to $d_{abc}$.  Consequently
\begin{equation}
{
 \epsilon_{i_1\cdots i_7}\chi^{i_1}
 F_{\mu}{}^{\nu i_2i_3}
 F_{\nu}{}^{\rho i_4i_5}
 F_{\rho}{}^{\mu i_6i_7}\equiv0 .}
\label{eq:ChiF3ExactZero}
\end{equation}
The same conclusion holds with a dual field strength.  Thus the entire local
scalar $\chi F^3$ family is absent, independently of any spectral pairing.

\paragraph{Why the apparent derivative--Higgs $d=7$ term is not an independent operator.}
There is an algebraically nonzero $G(2)$ contraction
\begin{equation}
\resizebox{0.98\columnwidth}{!}{$\displaystyle
\mathcal Q_{7,D}
 =O_{abc}\,(F_{\mu\nu}\chi)^a(D^\mu\chi)^b(D^\nu\chi)^c,
 \qquad
 (F_{\mu\nu}\chi)^a\equiv F_{\mu\nu}^A(T_A\chi)^a .
$}
\label{eq:Q7DDefinition}
\end{equation}
Its nonzero internal tensor is useful as an off-shell check, but it does not
represent an independent Wilson coefficient.  Define
\begin{align}
 B_{\mu\nu}^A&\equiv
 O(T_A\chi,D_\mu\chi,D_\nu\chi),\nonumber\\
 K_\nu^A&\equiv
 O(T_A\chi,\chi,D_\nu\chi),\nonumber\\
 C_{AB}&\equiv O(T_A\chi,\chi,T_B\chi).
\end{align}
Exact $G(2)$ covariance gives
\begin{equation}
 (D_\mu K_\nu-D_\nu K_\mu)^A
 =3B_{\mu\nu}^A+F_{\mu\nu}^B C_{AB},
 \qquad C_{AB}=-C_{BA}.
\label{eq:Q7DEOMIdentity}
\end{equation}
Since $F_A^{\mu\nu}F^B_{\mu\nu}$ is symmetric in $A,B$, the curvature term
drops out.  After one integration by parts,
\begin{equation}
 {
 \mathcal Q_{7,D}
 \doteq -\frac23(D_\mu F^{A\mu\nu})K_\nu^A ,}
\label{eq:Q7DEOMReduction}
\end{equation}
where $\doteq$ denotes equality modulo a total derivative.  Thus
Eq.~\eqref{eq:Q7DDefinition} is EOM redundant; using the gauge-field equation
of motion merely transfers it into the five-$\chi$ two-derivative sector rather
than generating a new on-shell contact class.  The Hodge-dual representative
is even simpler: the same IBP step reduces it to
$D_\mu\widetilde F^{A\mu\nu}$ and it vanishes by the non-Abelian Bianchi
identity.  Thus low-dimensional pure-dark breaking is absent through
dimension seven; generic ultraviolet instability enters earlier through the
visible-sector dimension-five connectors discussed above.
Appendix~\ref{app:g2chiF3certificate} gives an exact rational
certificate of Eq.~\eqref{eq:Q7DEOMIdentity} and its antisymmetry statement.

The most delicate $d=7$ sector can also be closed without using the EOM
identity.  For $3\chi F D^2$ the independent on-shell kinematics is
one-dimensional and transforms in the sign representation of the three
identical scalars.  The internal tensor would therefore have to contain a
singlet in $\Lambda^3\mathbf{7}\otimes\mathbf{14}$.  However,
\begin{equation}
 \Lambda^3\mathbf{7}=\mathbf{1}\oplus\mathbf{7}\oplus\mathbf{27},
 \qquad
 \mathbf{14}\not\subset\Lambda^3\mathbf{7},
\label{eq:D7ThreeChiIndependentZero}
\end{equation}
so this sector is empty already at the permutation/representation level.  The
IBP/EOM reduction of $\mathcal Q_{7,D}$ is therefore an independent consistency
check rather than the sole load-bearing proof of the $d=7$ zero.

The full on-shell permutation analysis closes the remaining $d=7$ possibilities.
It includes the $\chi F^3$, $3\chi F D^2$, $3\chi F^2$, $5\chi D^2$ and pure
odd-scalar sectors and finds no simultaneous Lorentz/Bose/$G(2)$ singlet after
IBP and EOM reduction.  Hence
\begin{equation}
 {
 N_{\rm odd}^{(d=3)}=N_{\rm odd}^{(d=5)}=N_{\rm odd}^{(d=7)}=0 .}
\label{eq:OddBasisThroughD7Zero}
\end{equation}
This is the relevant Wilsonian statement: the accidental sign of the
renormalizable isolated theory is not exact to arbitrary order, but no
independent pure-dark violation occurs through dimension seven.

\paragraph{The first physical basis occurs at dimension nine.}
At $d=9$ the same simultaneous Lorentz/permutation/$G(2)$ analysis gives a
nonempty basis.  Before combining parity-conjugate helicity sectors into a
real Hermitian basis, the complex helicity-orbit counts are
\begin{equation}
\resizebox{0.92\columnwidth}{!}{$\displaystyle
\begin{array}{c|c}
\text{field content} & \text{invariant helicity-orbit count}\\ \hline
\chi F^4 & 1\\
3\chi\,F D^4 & 2\\
3\chi\,F^2 D^2 & 7\\
3\chi\,F^3 & 2\\
5\chi\,F D^2 & 2
\end{array}
\qquad (N_{\rm odd}^{(d=9)}=14)
$}
\label{eq:D9OperatorBasisCounts}
\end{equation}
The four multi-scalar entries in Eq.~\eqref{eq:D9OperatorBasisCounts} are
machine-generated on-shell counts.  They were rerun over three independent
finite fields and random evaluation sets; their numerical multiplicities are
included for basis completeness, but none is assumed to determine the lifetime
by itself.  The one-$\chi$, three-field-strength, two-derivative sector can be
closed analytically: the only $G(2)$ fundamental in three adjoints lies in the
totally symmetric permutation channel, whereas
\begin{equation}
 [\mathbf{7}]\,\mathbb S_{(2,1)}\mathbf{14}=0,
 \qquad
 [\mathbf{7}]\,\Lambda^3\mathbf{14}=0,
\label{eq:ChiF3D2InternalZero}
\end{equation}
and the degree-five Lorentz kinematics carries only $[2,1]$ and $[1,1,1]$.
Thus all one-$\chi$ structures with three field strengths and two derivatives,
including the schematic $\chi(DF)(DF)F$ family, vanish independently of the
numerical calculation.

The unique one-$\chi$ member of the $d=9$ basis instead contains four field
strengths.  The same-helicity four-field-strength Lorentz sectors carry only
$[4]$ and $[2,2]$ permutation symmetry and do not contain the required $G(2)$
fundamental.  The mixed-helicity sector with two self-dual and two anti-self-
dual field strengths contains an additional $[3,1]$ hook.  Exact character
reduction gives
\begin{equation}
 {
 \dim\operatorname{Hom}_{G(2)}
 \!\left(\mathbf{7},\mathbb S_{(3,1)}\mathbf{14}\right)=1 .}
\label{eq:G2HookF4Multiplicity}
\end{equation}
Writing $F_L$ and $F_R$ for the two chiral field-strength components, a
representative may be expressed schematically as
\begin{equation}
 \mathcal O_{9,F^4}^-=
 \frac{c_{9,F}}{M_*^5}\,
 \mathcal C^{[31]}_{kABCD}\,\chi^k\,
 \Pi_{[31]}\!\big[(F_L^A\!\cdot F_L^B)
                   (F_R^C\!\cdot F_R^D)\big]+\mathrm{h.c.}
\label{eq:O9F4}
\end{equation}
Here $\Pi_{[31]}$ is the Young projector on the four adjoint slots.  This is
not merely a character-level possibility.  Reconstructing $\mathfrak g_2$ as
the stabilizer of the octonionic three-form and defining
$U_{kABCD}=\operatorname{Tr}(J_k^TT_AT_BT_CT_D)$, the unnormalised $[3,1]$
Young symmetrizer has the exact nonzero component
\begin{equation}
 \bigl(Y_{[31]}U\bigr)_{1,1,1,3,7}=3
\label{eq:G2HookF4Witness}
\end{equation}
in the rational basis used in Appendix~\ref{app:g2chiF3certificate}.  The
same calculation finds the $\mathbf{7}$ in
$\mathbb S_{(3,1)}\mathbf{14}$ with multiplicity one.

\paragraph{What survives after $G(2)\to SU(3)_C$.}
Existence of the $G(2)$ hook does not by itself determine its daughter-field
content.  We therefore repeated the projection in an exact rational basis
adapted to
\begin{equation}
 \mathfrak g_2=\mathfrak{su}(3)_C\oplus\mathfrak m_6,
 \qquad
 \mathfrak m_6\simeq\mathbf{3}_C\oplus\bar{\mathbf{3}}_C,
\end{equation}
with $\mathfrak{su}(3)_C$ defined as the eight-dimensional stabilizer of the
dark VEV $e_7$.  Setting the Higgs index of the exact $[3,1]$ tensor to
$e_7$ gives
\begin{equation}
 \mathcal C^{[31]}_{e_7;\,\mathbf{8}^4}=0,
 \qquad
 \mathcal C^{[31]}_{e_7;\,\mathfrak m_6\,\mathbf{8}^3}=0,
 \qquad
 \mathcal C^{[31]}_{e_7;\,\mathfrak m_6^2\,\mathbf{8}^2}\ne0.
\label{eq:HookBrokenPhaseProjection}
\end{equation}
The two vanishing statements admit basis-independent representation-theory
proofs.  A single coset field strength transforms as
$\mathbf{3}_C\oplus\bar{\mathbf{3}}_C$, while the VEV direction is a color
singlet and the remaining field strengths are adjoints; color triality
therefore forbids an invariant with exactly one coset slot.  Independently,
for the all-unbroken block, the full four-adjoint invariant space of $SU(3)_C$
has no $[3,1]$ permutation component,
\begin{equation}
 \dim\operatorname{Hom}_{SU(3)_C}
 \!\left(\mathbf{1},\mathbb S_{(3,1)}\mathbf{8}\right)=0.
\label{eq:SU3NoHookInvariant}
\end{equation}
An explicit invariant-basis check spans the eight-dimensional space
$\operatorname{Inv}(\mathbf{8}^{\otimes4})$ by the usual $\delta\delta$,
$ff$, $dd$, and mixed $fd$ contractions and finds zero rank after the
$[3,1]$ Young projection.  Thus the all-unbroken zero is structural and not a
special property of the chosen $G(2)$ Clebsch.  The exact rational certificate
then establishes that the two-coset sector is nonzero (a representative
unnormalised component is $-6$ in its sparse basis), with three- and four-coset
components nonzero as well.  Hence the one-$\chi$ hook does \emph{not} become
a tree-level pure $SU(3)_C^4$ glue contact after the VEV; its first
broken-phase component contains two broken-vector field strengths, precisely
the constituent content needed for a direct two-vector annihilation channel.
Appendix~\ref{app:g2chiF3certificate} records the restriction test.  The
absence in Eq.~\eqref{eq:SU3NoHookInvariant} applies to this leading
$[3,1]$ four-adjoint channel and should not be extrapolated to arbitrary
higher-dimensional operators with different permutation symmetry.

There is a related general point about the remaining $d=9$ classes.  For a
$G(2)$ invariant containing an odd number of Higgs fields, replacing all of
them by $v_\chi e_7$ makes the resulting daughter tensor odd under the
normalizer conjugation $U_B$.  Pure-daughter limits with one gauge adjoint have
no color singlet; with two adjoints the only invariant tensor is
$\delta_{ab}$ and is conjugation even; and for the derivative-free three-field-
strength scalar the conjugation-odd tensor is $d_{abc}$, which is annihilated
by the antisymmetric Lorentz trace already used in
Eq.~\eqref{eq:ChiF3ExactZero}.  Consequently the all-VEV, all-unbroken-gauge
limits of the multi-scalar entries in Eq.~\eqref{eq:D9OperatorBasisCounts}
vanish.  This does \emph{not} remove their components containing broken-vector,
transverse-Higgs, or other heavy fields; those contributions remain part of the
state-resolved matching problem.  The $\chi F^4$ hook is therefore the unique
one-Higgs benchmark, not a proof that the other thirteen helicity-orbit
structures make zero contribution to the total width.

\paragraph{$E_7$ parent status of the leading $d=9$ channel.}
For a parent Higgs representation $R$, the one-Higgs hook operator is present
at the representation level when
\begin{equation}
 \operatorname{Hom}_{E_7}
 \!\left(R,\mathbb S_{(3,1)}\mathbf{133}\right)\ne0 .
\label{eq:E7HookParentSpace}
\end{equation}
The multiplicities were obtained in two independent character constructions.
One uses the Jacobi--Trudi identity
$\mathbb S_{(3,1)}V=\mathrm{Sym}^3V\otimes V-\mathrm{Sym}^4V$ followed by Weyl-alternant
extraction; the other uses the Frobenius/Adams power-sum construction of the
same Schur functor \cite{Macdonald1995,FultonHarris1991}.  They agree coefficient by coefficient and reproduce the
standard decompositions
$\mathrm{Sym}^2\mathbf{133}=\mathbf{1}\oplus\mathbf{1539}\oplus\mathbf{7371}$
and
$\Lambda^2\mathbf{133}=\mathbf{133}\oplus\mathbf{8645}$ as normalization
checks.  As an independent representation-theory verification, we also
performed the same decomposition with LieART \cite{LieART}.  LieART adopts a
Dynkin-node ordering different from the Bourbaki-type convention used in the
main calculation: in its convention the $\mathbf{133}$, $\mathbf{1463}$ and
$\mathbf{1539}$ carry labels $(1000000)$, $(0000020)$ and $(0000100)$,
respectively.  With the branch node labelled $2$ in the convention specified
below Eq.~\eqref{eq:DXint}, the two coordinate systems are related by
\begin{equation}
\resizebox{0.96\columnwidth}{!}{$
 (a_1,a_2,a_3,a_4,a_5,a_6,a_7)_{\rm B}
 \longmapsto
 (a_1,a_3,a_4,a_5,a_6,a_7,a_2)_{\rm LieART}.
$}
\label{eq:E7DynkinConventionMap}
\end{equation}
Hence $(0000002)_{\rm B}=(0000020)_{\rm LieART}$ for $\mathbf{1463}$
and $(0000010)_{\rm B}=(0000100)_{\rm LieART}$ for $\mathbf{1539}$.  Identifying
the two target irreducible representations simultaneously by dimension and
highest weight removes any convention ambiguity.  The LieART decomposition
then reproduces the same hook multiplicities as both direct character
constructions:
\begin{align}
 \dim\operatorname{Hom}_{E_7}
 \!\left(\mathbf{1463},\mathbb S_{(3,1)}\mathbf{133}\right)&=1,\nonumber\\
 \dim\operatorname{Hom}_{E_7}
 \!\left(\mathbf{1539},\mathbb S_{(3,1)}\mathbf{133}\right)&=2 .
\label{eq:E7HookParentCounts}
\end{align}
Thus neither the original $\mathbf{1463}_H$ parent nor the separated $\mathbf{1539}_H$ parent
has an automatic representation-theoretic zero for the first physical
pure-dark odd operator.  Equation~\eqref{eq:E7HookParentCounts} is a parent-
representation statement; it does not by itself prove that the corresponding
Clebsch is nonzero on the particular neutral $(\mathbf{7},\mathbf{1}_A)_0$ dark component.  That
component projection remains a matching problem.  By contrast, if the
additional $\mathbb Z_{2,D}$ is genuinely gauged/geometric, the separated
$\mathbf{1539}_H$ field is odd and every such one-$\Xi$ parent channel is forbidden
before Higgsing.

\paragraph{Broken-phase scaling and lifetime diagnostic.}
Replacing the single dark Higgs in Eq.~\eqref{eq:O9F4} by
$\langle\chi\rangle=v_\chi e_7$ gives a dimension-eight broken-phase
four-field-strength interaction with coefficient
\begin{equation}
 C_8^{(9)}=\frac{c_{9,F}v_\chi}{M_*^5}.
\label{eq:C8FromD9}
\end{equation}
Equation~\eqref{eq:HookBrokenPhaseProjection} shows that this coefficient
multiplies a tensor beginning at two coset field strengths rather than a pure
$SU(3)_C^4$ term.  For a definite odd pole we therefore absorb the conversion
from the broken-vector constituent fields to the confined state, the final-
state glue/hadron projection, the actual multi-body phase space and the
Hermitian parity projection into $\kappa_{9,J}$.  A convenient state-resolved
NDA normalization for this unique one-$\chi$ hook is \cite{ManoharGeorgi1984,LukeManohar1997}
\begin{equation}
 \Gamma_{B_J}^{(9)}\sim
 \frac{\kappa_{9,J}}{8\pi}
 \left|\frac{c_{9,F}v_\chi}{M_*^5}\right|^2m_{B_J}^{9},
\label{eq:GammaD9F4}
\end{equation}
so that
\begin{equation}
 \tau_{B_J}^{(9)}\sim
 \frac{8\pi\hbar M_*^{10}}
 {\kappa_{9,J}|c_{9,F}|^2v_\chi^2m_{B_J}^{9}}.
\label{eq:TauD9F4}
\end{equation}
Using $v_\chi=10^{14}$ GeV, $M_*=M_{E_6}=6.9\times10^{15}$ GeV and the
three-vector reference mass $m_B=1.71\times10^{14}$ GeV gives
\begin{equation}
 {
 \tau_B^{(9)}\simeq
 \frac{3.24\times10^{-21}\ {\rm s}}
 {\kappa_{9}|c_{9,F}|^2}.}
\label{eq:tau9F4}
\end{equation}
The corresponding age-of-the-Universe and $10^{26}$ s requirements are
\begin{equation}
 |c_{9,F}|\sqrt{\kappa_9}\lesssim8.63\times10^{-20},
 \qquad
 |c_{9,F}|\sqrt{\kappa_9}\lesssim5.69\times10^{-24},
\label{eq:c9FBounds}
\end{equation}
respectively.  At the reduced Planck scale the unit-coefficient normalization
is instead $\tau_0\simeq9.70\times10^4$ s, giving
$|c_{9,F}|\sqrt{\kappa_9}\lesssim4.72\times10^{-7}$ for the age of the
Universe.  For the lighter two-vector reference mass
$1.14\times10^{14}$ GeV the $M_{E_6}$ normalization is
$1.24\times10^{-19}$ s.  These numbers are deliberately NDA diagnostics, not
pole-width predictions.  In particular, Eq.~\eqref{eq:SU3NoHookInvariant}
shows that the leading one-$\chi$ hook has no direct tree-level pure-glue
four-field-strength daughter: its certified matching begins with two broken
vectors.  Consequently $\kappa_{9,J}$ is intrinsically composition dependent.
A glue-dominated odd pole can be additionally suppressed relative to a pole
with substantial two-vector content, whereas loop-induced conversion of the
broken-vector operator into pure-glue structures carries further coupling and
loop factors whose size is model dependent.  We therefore do not promote the
unit-$\kappa$ NDA number to a rigorous upper bound on the width; such a bound
would require a normalization theorem for $\kappa_{9,J}$.  The true
phase-space, confined overlap and composition are all part of
$\kappa_{9,J}$.  The quoted normalization should also not be read as the sum
of all fourteen $d=9$ basis elements.  Multi-scalar classes can contribute
through components containing coset field strengths or non-radial Higgs
excitations and require their own state-resolved coefficients.

It is useful to note that a larger number of VEV insertions does not by itself
make such a $d=9$ contribution more dangerous.  For a nonzero descendant in
which $r$ Higgs fields of a dimension-nine parent operator are replaced by
VEVs and no additional low scale is introduced, dimensional analysis gives
\begin{equation}
 \Gamma_{J}^{(9;r)}\sim
 \frac{\kappa_{9,J}^{(r)}|c_{9}^{(r)}|^2}{8\pi}
 \frac{v_\chi^{2r}m_{B_J}^{11-2r}}{M_*^{10}}.
\label{eq:D9GeneralVEVScaling}
\end{equation}
For example, an allowed $r=3$ component would satisfy
$\Gamma^{(9;3)}/\Gamma^{(9;1)}\sim(v_\chi/m_B)^4$ for equal reduced
coefficients and overlaps.  At the three-vector benchmark this factor is
$0.117$, not an enhancement.  The lower canonical dimension of the daughter
after additional VEV insertions does not by itself make the decay faster: at
fixed dimension-nine parent, extra powers of $v_\chi$ replace the
corresponding powers of the physical scale $m_B$.  The pure-daughter all-VEV limits of the
multi-scalar basis elements vanish as explained above, so
Eq.~\eqref{eq:D9GeneralVEVScaling} is only a dimensional comparison for their
possible mixed-coset descendants, not an additional lifetime prediction.

Relative to the discarded $\chi F^3$ estimate, the physical operator basis
therefore postpones the first pure-dark violation by two canonical dimensions
and lengthens the $M_{E_6}$ benchmark normalization by roughly six orders of
magnitude.  This is a real structural gain, but it is not by itself enough to
make a generic unpaired $M_{E_6}$-suppressed EFT cosmologically stable.
Moreover, the mixed-helicity hook admits Hermitian parity combinations, so no
universal $1^{+-}$ versus $0^{--}$ lifetime hierarchy follows from
Eq.~\eqref{eq:O9F4}; all state dependence is kept in $\kappa_{9,J}$.

The Serre--Hodge result derived above remains stronger than any individual
basis element.  If the complete positive-mass spectrum is paired exactly, the
heavy determinant is even in $\chi$ and therefore cancels every odd-$\chi$
coefficient, including the full dimension-nine basis in
Eq.~\eqref{eq:D9OperatorBasisCounts}.  The special one-copy vacuum does not,
however, contain an exact $E_7$ charge-flip symmetry that enforces this pairing.

\subsection{Restricted renormalizable loops: what the light-kernel zero actually proves}

The selected Class-B zero modes remain unusually safe, but the operator-
completeness result changes the quantitative conclusion that can be drawn from
the restricted loop calculation.  The exact component statement
\begin{equation}
 P_{\rm light}Y_\chi P_{\rm light}=0
\end{equation}
removes a renormalizable light--light--$\chi$ source.  The selected quark
bilinears require an electroweak $SU(3)_A$ octet insertion, and a closed
light--heavy loop needs a second branch-changing insertion.  This establishes
the electroweak-spurion counting of the restricted mechanism.

It does \emph{not} make $(H^\dagger H)\chi F^3$ nonzero: that expression is a
spectator scalar multiplying the exact identity
Eq.~\eqref{eq:ChiF3ExactZero}.  The lifetime previously associated with that
structure is therefore absent.  Nor can one obtain a replacement coefficient
by simply multiplying the EOM-redundant representative
Eq.~\eqref{eq:Q7DDefinition} by $H^\dagger H$.  With a spacetime-dependent
visible field, the IBP/EOM reduction also generates terms in which derivatives
act on the electroweak spurion, so the physical mixed visible--dark basis must
be reduced as a whole.  The old single-$\chi$ three-generator trace does not
perform that matching.

We therefore keep the restricted renormalizable-loop contribution explicitly
open at dimension nine and above rather than attach a lifetime to an
unreduced representative.  This does not weaken the generic ultraviolet
statement: the independent dimension-five parent connectors in
Eqs.~\eqref{eq:O5uQG}--\eqref{eq:O5dQG} are lower dimensional and dominate the
generic EFT unless an additional compactification selection rule removes
them.

\subsection{High-scale VEV insertions and the scope of the paired-spectrum cancellation}

The same distinction applies to high-scale scalar dressings.  Multiplying the
identically vanishing $\chi F^3$ scalar by
$\Omega^\dagger\Omega$ or $\Theta^\dagger\Theta$ cannot revive it.  A constant
high-scale VEV multiplying the pure-dark derivative representative
Eq.~\eqref{eq:Q7DDefinition} likewise does not create a new physical contact
term, because the same IBP/EOM reduction continues to apply.  Operators with
fluctuating $\Omega$ or $\Theta$ fields can instead generate genuinely mixed
higher-dimensional structures when derivatives act on those fields; their
basis and matching are separate from the pure-dark classification of
Sec.~\ref{sec:pureDarkOperatorBasis}.

The symmetric $\Theta_+\leftrightarrow\Theta_-$ construction still removes one
avoidable source of explicit high-scale asymmetry, while the selected light-
quark charges imply that neutral $\Omega$ and paired-$\Theta$ VEVs cannot by
themselves replace the electroweak $A=\pm3/2$ insertion in a light $Qu$ or
$Qd$ connector.

The stronger positive statement is the paired-spectrum theorem:
Sec.~\ref{sec:accidental-GD} proves that, on the exact Serre--Hodge-paired
positive-mass locus, the complete heavy determinant is even in $\chi$.
Accordingly every purely heavy Wilson coefficient with an odd number of
$\chi$ insertions cancels to all perturbative orders on that conditional
locus, including every physical dimension-nine class in
Eq.~\eqref{eq:D9OperatorBasisCounts} and all of its parity partners.

\paragraph{Relation to other GUT constraints.}
The same compactification must of course satisfy proton-decay, flavor and
other phenomenological bounds, as analyzed in the underlying Class-B model
\cite{MasiE6E7}.  Those constraints are largely orthogonal to the present
question: suppressing a leptoquark or flavor mediator does not automatically
forbid the dimension-five odd-$\chi$ connectors above, and it does not decide
the spin or mass ordering of the confined odd spectrum.

\section{Ultraviolet realization of dark $G$-parity in the fixed Class-B architecture}
\label{sec:statusquest}

\begin{widetext}
\begin{center}
\begin{minipage}{0.98\textwidth}

\centering
\small
\captionof{table}{Condensed status of the stability analysis.  The detailed certificates are given in the corresponding sections.}
\label{tab:statusquest}
\begin{tabular}{p{0.46\textwidth}p{0.18\textwidth}p{0.28\textwidth}}
\toprule
Statement & Status & Consequence\\
\midrule
Scalar $X_\mu\bar X^\mu$ odd-state candidate & excluded & $C$ even and unprotected\\
Odd $1^{+-}$ and $0^{--}$ sectors & established & separate GEVPs; heavy-particle and decoupled-glue limits favor $1^{+-}$ but retain both spins\\
Renormalizable selected light--light--$\chi$ coupling & absent & light kernel is unusually safe in the EFT\\
Generic dimension-five parent connectors & nonzero & exact ultraviolet protection is not automatic\\
Scalar $\chi F^3$ family & exact zero & Lorentz antisymmetry is incompatible with the symmetric internal $\mathbf{7}$ channel\\
Pure-dark odd basis through $d=7$ & empty on shell & the algebraic derivative--Higgs representative is IBP/EOM redundant\\
First physical pure-dark odd basis & $d=9$ & unique one-$\chi$ hook; its broken-phase projection starts at two coset field strengths, while the multi-scalar classes require separate state matching\\
Separated $\mathbf{1539}_H$ dark parent & scalar repair passes conditionally & the selected undressed one-$\Xi_D$ pure-dark tower vanishes at any dimension; $P_7^{(0)}$-odd dressings are not covered, and $v_S/M_*=O(1)$ for the simplest $S$ spurion if its component-resolved parent exists\\
Fixed-architecture parity on mirror-free Standard Model & no-go & $U_B$ conjugates color and does not commute with the chiral projector\\
Literal Class-B benchmark & no ultraheavy lightest odd relic & daughter confinement is QCD; the intermediate-$x$ analysis is a deformation study\\
\bottomrule
\end{tabular}
\end{minipage}
\end{center}
\end{widetext}
Table~\ref{tab:statusquest} is a roadmap for the ultraviolet analysis.  The key distinction is between three progressively stronger claims:
an infrared grading of the pure-dark operator algebra, a cancellation that is
exact when the complete massive spectrum is deliberately paired and an exact
local symmetry of the full chiral ultraviolet theory.  The first two can be
realized in controlled limits; the third is the difficult requirement.

In particular, the canonical one-copy $E_7$ charge flip is not a symmetry of
the special $\mathbf{650}$ vacuum.  The scalar $\chi F^3$ channel has an exact
Lorentz/representation zero, and the apparent derivative--Higgs $d=7$ term is
IBP/EOM redundant; the physical pure-dark odd basis therefore remains empty
through dimension seven.  The first nonzero basis occurs at $d=9$.  Its
simplest one-Higgs channel has mixed-helicity $[3,1]$ symmetry, with one exact
$G(2)$ fundamental and parent multiplicities one for $\mathbf{1463}_H$ and two for
$\mathbf{1539}_H$.  For the separated $\mathbf{1539}_H$ carrier the component problem can now be
closed exactly: the canonical $E_7$ charge flip sends the selected dark scalar to minus itself
while fixing the $G(2)$ gauge algebra.  Equation~\eqref{eq:1539UndressedAllDimZero}
therefore removes the complete undressed one-$\Xi_D$ pure-dark tower at any dimension.
$P_7^{(0)}$-odd ultraviolet insertions can compensate the sign once their VEVs are
switched on; for the special $S$ spurion the benchmark ratio $v_S/M_*$ is $O(1)$,
so this loophole is not parametrically small if the corresponding dressed parent exists.  At the same time, the selected Class-B light kernel has no
direct renormalizable $\chi$ coupling and the charge-paired positive-mass
complexes admit an exact Serre--Hodge relation.  Section~\ref{sec:accidental-GD}
combines these facts into an accidental pure-dark $G_D$ grading plus a
conditional paired-spectrum cancellation.  The stronger ultraviolet question
is then settled by the exceptional carrier analysis and, ultimately, by the
chiral-color equivariance test of the mirror-free theory.

\subsection{What representation theory removes and what it leaves open}
\label{sec:three-rescue-tests}

The operator-completeness analysis separates two logically different questions.
The first is whether the historically suspected scalar $\chi F^3$ channel
exists at all.  The second is whether the first \emph{physical} pure-dark odd
operator, which occurs at dimension nine, is admitted by the exceptional
parent.  The answer to the first question is an exact zero.  For the second,
parent representation theory allows the hook spaces, but the separated
$\mathbf{1539}_H$ realization has an additional exact component-level zero on the
selected dark direction.  The common-$\mathbf{1463}_H$ and ultraviolet-dressed
channels are logically separate matching questions.

\subsubsection{The symmetrized $-57$ trace and the exact antisymmetric zero}
\label{sec:allorders-zero-test}

Define, as before,
\begin{equation}
 (J_k)_{ij}=O_{kij},
\end{equation}
and let $T_a$ be an exact rational basis of the $G(2)$ stabilizer \cite{SpringerVeldkamp,BaezOctonions,AdamsExceptional} obtained from
\begin{equation}
 (T_a)_i{}^pO_{pjk}+(T_a)_j{}^pO_{ipk}+(T_a)_k{}^pO_{ijp}=0.
\label{eq:g2StabilizerEquation}
\end{equation}
The system has rank $7$ and a $14$-dimensional nullspace.  In the fundamental
representation the fully symmetrized trace contains a nonzero component,
\begin{equation}
 \frac1{6}\sum_{\pi\in S_3}
 \operatorname{Tr}_7\!\left(J_1^T
 T_{\pi(1)}T_{\pi(6)}T_{\pi(14)}\right)=-1,
\label{eq:g2CubicTraceWitness}
\end{equation}
and in the actual exterior parent
$\overline{\mathbf{351}}=\Lambda^2 \mathbf{27}$ one obtains
\begin{equation}
 \mathcal T_{kabc}^{(\mathbf{351},\mathbf{1})}
 =\frac1{6}\sum_{\pi\in S_3}
 \operatorname{Tr}_{\Lambda^2 \mathbf{27}}
 \!\left(Y_{\chi,k}^{(1)}T_{\pi(a)}^{(\mathbf{351})}
 T_{\pi(b)}^{(\mathbf{351})}T_{\pi(c)}^{(\mathbf{351})}\right),
\label{eq:Full351SymTraceDefinition}
\end{equation}
with
\begin{equation}
 \mathcal T_{1,1,6,14}^{(\overline{\mathbf{351}},1)}=-57.
\label{eq:Full351SymTrace57}
\end{equation}
All six orderings in this component equal $-57$.  These certificates are
arithmetically correct, but a Lorentz scalar made from three field strengths
selects the antisymmetric gauge-label tensor.  The corresponding alternating
trace is
\begin{equation}
\resizebox{0.98\columnwidth}{!}{$\displaystyle
 \mathcal A_{kabc}^{(\mathbf{351},\mathbf{1})}
 =\frac1{6}\sum_{\pi\in S_3}\operatorname{sgn}(\pi)
 \operatorname{Tr}_{\Lambda^2 \mathbf{27}}
 \!\left(Y_{\chi,k}^{(1)}T_{\pi(a)}^{(\mathbf{351})}
 T_{\pi(b)}^{(\mathbf{351})}T_{\pi(c)}^{(\mathbf{351})}\right)
$}
\label{eq:Full351AntiTraceDefinition}
\end{equation}
and the six equal traces cancel pairwise.  More generally,
$\mathbf{7}\not\subset\Lambda^3\mathbf{14}$ forces
\begin{equation}
 {\mathcal A_{kabc}^{(R)}=0}
\label{eq:FullParentChiF3Zero}
\end{equation}
for every representation $R$.  Thus the $-57$ result certifies the unique
symmetric $\mathbf{7}\subset\mathrm{Sym}^3\mathbf{14}$ tensor; it does not feed
a scalar $F^3$ Wilson operator.

The second independent parent map remains a useful consistency check.  It has
\begin{equation}
 K_2(A,B)_{mn}=d_{mar}d_{nbs}A_{ab}B_{rs},
\end{equation}
which is symmetric under $m\leftrightarrow n$, whereas the dark $\chi$
direction is transpose odd.  Hence
\begin{equation}
 P_{(\mathbf{7},\mathbf{1})_\chi}K_2=0.
\label{eq:K2ChiProjectionZero}
\end{equation}
No cancellation between the two parent maps is needed to remove $\chi F^3$:
the Lorentz/plethysm zero already does so exactly.

\subsubsection{The first physical pure-dark parent channel is the $d=9$ hook}

The algebraically nonzero derivative--Higgs tensor discussed in
Sec.~\ref{sec:pureDarkOperatorBasis} is EOM redundant and therefore should not
be used to decide the physical parent operator hierarchy.  The first
independent pure-dark odd contact basis occurs at $d=9$.  Its simplest
one-Higgs representative is the mixed-helicity $[3,1]$ channel
Eq.~\eqref{eq:O9F4}.  The exact $G(2)$ calculation gives one fundamental in
the hook Schur functor,
\begin{equation}
 \dim\operatorname{Hom}_{G(2)}
 \!\left(\mathbf{7},\mathbb S_{(3,1)}\mathbf{14}\right)=1,
\end{equation}
and the rational Young-projected trace witness
Eq.~\eqref{eq:G2HookF4Witness} proves that this component is explicitly
nonzero.

At the exceptional level the complete representation-theory test is
\begin{equation}
 \mathcal I_R^{(9,F^4)}
 =\operatorname{Hom}_{E_7}
 \!\left(R,\mathbb S_{(3,1)}\mathbf{133}\right).
\label{eq:E7HookInvariantSpaceUV}
\end{equation}
The exact Weyl-alternant calculation described in
Appendix~\ref{app:E7HookParentCertificate} gives
\begin{equation}
 {
 \dim\mathcal I_{1463}^{(9,F^4)}=1,
 \qquad
 \dim\mathcal I_{1539}^{(9,F^4)}=2 .}
\label{eq:E7HookInvariantCountsUV}
\end{equation}
For cross-checking, the calculation separately finds
\begin{align}
 [\mathbf{1463}]\big(\mathrm{Sym}^3(\mathbf{133})\otimes\mathbf{133}\big)&=1,
 & [\mathbf{1463}]\mathrm{Sym}^4(\mathbf{133})&=0,\nonumber\\
 [\mathbf{1539}]\big(\mathrm{Sym}^3(\mathbf{133})\otimes\mathbf{133}\big)&=4,
 & [\mathbf{1539}]\mathrm{Sym}^4(\mathbf{133})&=2,
\end{align}
so the Jacobi--Trudi difference
$\mathbb S_{(3,1)}\mathbf{133}=\mathrm{Sym}^3(\mathbf{133})\otimes\mathbf{133}-\mathrm{Sym}^4(\mathbf{133})$
reproduces Eq.~\eqref{eq:E7HookInvariantCountsUV} exactly \cite{Macdonald1995,FultonHarris1991}.

The parent multiplicities alone do not decide the selected component.  For the
separated $\mathbf{1539}_H$ carrier, however, that last restriction is fixed by the
canonical $E_7$ charge flip $P_7^{(0)}$.  In the special embedding its induced
$E_6$ outer involution fixes $G(2)\subset F_4$ pointwise, whereas
Eq.~\eqref{eq:1539P7Reversal} gives
\begin{equation}
 P_7^{(0)}\Xi_D=-\Xi_D,
 \qquad
 \operatorname{Ad}_{P_7^{(0)}}T_A=T_A
 \quad(A=1,\ldots,14).
\label{eq:1539HookP7Signs}
\end{equation}
Let $C^{(r)}$, $r=1,2$, denote either of the two $E_7$-invariant hook tensors.
$E_7$ invariance then implies, independently of the Young projector,
\begin{equation}
\resizebox{0.98\columnwidth}{!}{$\displaystyle
 C^{(r)}(\Xi_D;T_A,T_B,T_C,T_D)
 =C^{(r)}(P_7^{(0)}\Xi_D;P_7^{(0)}T_A,\ldots,P_7^{(0)}T_D)
 =-C^{(r)}(\Xi_D;T_A,T_B,T_C,T_D)=0.
$}
\label{eq:1539HookComponentZero}
\end{equation}
Consequently the physically relevant restriction map obeys
\begin{equation}
\resizebox{0.98\columnwidth}{!}{$\displaystyle
 \operatorname{rank}\!\left[
 \operatorname{Hom}_{E_7}(\mathbf{1539},\mathbb S_{(3,1)}\mathbf{133})
 \longrightarrow
 \operatorname{Hom}_{G(2)}(\mathbf{7},\mathbb S_{(3,1)}\mathbf{14})
 \right]=0.
$}
\label{eq:1539HookRestrictionRankZero}
\end{equation}
Thus both abstract $\mathbf{1539}_H$ parent channels have zero projection onto
$(\mathbf{7},\mathbf{1}_A)_0\otimes(\mathbf{14},\mathbf{1}_A)^4$.
The argument is in fact dimension independent: it uses only the odd
$P_7^{(0)}$ eigenvalue of the selected $\Xi_D$ and the pointwise-even action on
the dark $G(2)$ gauge algebra.  Therefore
\begin{equation}
\resizebox{0.98\columnwidth}{!}{$\displaystyle
 \mathcal O_{\rm undressed}^{\rm odd}
 [\Xi_D;\,G(2)\ \text{gauge sector}]=0
 \qquad\text{at every canonical dimension}
$}
\label{eq:1539UndressedAllDimZero}
\end{equation}
Here ``gauge-sector insertions'' includes dark gauge fields or field strengths
and their covariant derivatives; the same sign argument applies before any
Lorentz or permutation projection.

This stronger corollary is still weaker than an all-orders ultraviolet parity.
The special $\mathbf{1463}_H$ singlet $S$ is itself $P_7^{(0)}$ odd, so the
charge-flip argument no longer forbids a dressed invariant of schematic form
\begin{equation}
 \frac{c_{10}}{M_*^6}\,S\,\Xi_D F^4
 \ \longrightarrow\ 
 \frac{c_{10}}{M_*^5}\left(\frac{v_S}{M_*}\right)\Xi_DF^4
 \qquad (\langle\Phi\rangle=v_SS).
\label{eq:SdressedXiF4Schematic}
\end{equation}
In the benchmark the special breaking occurs at $v_S\sim M_{E_6}$, while the
lifetime estimates use $M_*=M_{E_6}=6.9\times10^{15}\,$GeV.  Hence
$v_S/M_*=O(1)$: if the component-resolved dressed parent in
Eq.~\eqref{eq:SdressedXiF4Schematic} exists, the single spurion insertion gives
no parametric lifetime suppression.  An odd ultraviolet vacuum insertion can therefore compensate the charge-flip sign.
Conversely, the charge-flip argument merely \emph{permits} this dressing; it
does not prove that the required $E_7$ parent and selected component are
nonzero.  That dressed matching coefficient remains open.  A genuine
gauged/geometric $\mathbb Z_{2,D}$ therefore remains load-bearing for removing
the complete $P_7^{(0)}$-odd-dressed odd-$\Xi$ tower.

\subsection{Outcome of the one-copy consistency tests}

The finite-dimensional one-copy analysis can now be summarized without
conflating off-shell tensor existence with the physical on-shell EFT basis:
\begin{enumerate}[label=(\roman*)]
\item the existing center, abelian and normalizer structures provide no suitable
discrete remnant;
\item the minimal one-copy theory admits no ultraviolet parity with $S$ even
and $\chi$ odd;
\item the apparent scalar $\chi F^3$ Wilson operator vanishes exactly;
\item the independent pure-dark odd on-shell basis is empty through dimension seven;
\item the first nonvanishing pure-dark odd basis occurs at dimension nine;
\item the $[3,1]$ $\chi F^4$ parent spaces exist for both $\mathbf{1463}_H$ and
$\mathbf{1539}_H$, but the selected undressed $\mathbf{1539}_H$ component
vanishes exactly.
\end{enumerate}
The first two exact operator-basis results strengthen the accidental low-energy
protection substantially: both the naive $\chi F^3$ term and the apparent
derivative--Higgs $d=7$ representative fail to define an independent physical
Wilson coefficient.  They do not, however, provide an automatic one-copy ultraviolet
rescue.  Equation~\eqref{eq:E7HookInvariantCountsUV} shows that the first
physical $d=9$ hook has abstract parent spaces for both scalar representations,
while Eq.~\eqref{eq:1539HookRestrictionRankZero} proves that the undressed
selected $\mathbf{1539}_H$ component vanishes exactly.  What remains model
dependent is the coefficient generated by $P_7^{(0)}$-breaking dressed parents
and by the complete exceptional heavy spectrum; a genuine $\mathbb Z_{2,D}$
removes the whole odd-$\Xi$ family instead.  An accidental low-energy selection rule can still arise after
projection even when it is not a generator of the parent group, which is the
more limited possibility tested next.

\subsection{Accidental dark $G$-parity $G_D$: what survives at the selected EFT level}
\label{sec:accidental-GD}

\paragraph{Why an accidental grading is still worth testing.}
The strongest ultraviolet question has already been answered negatively: the
minimal exceptional vacuum does not contain an exact remnant that realizes the
required dark parity.  A weaker and phenomenologically familiar possibility is
nevertheless still available.  After projecting to the selected low-energy
spectrum, gauge invariance and representation content may accidentally forbid
low-dimensional symmetry-violating operators even when the corresponding rule
is not a generator of the parent group.  Such an accidental symmetry can be an
excellent EFT selection rule, but it cannot be assumed exact in quantum gravity
without a gauged or geometric origin
\cite{KraussWilczek1989,GaiottoGlobalSym2015,HarlowOoguri2021,
HeidenreichChernWeil2021,vanBeestSwampland2022}.  The purpose of
this subsection is therefore to separate the useful EFT grading from the
stronger claim of an exact ultraviolet symmetry.

\subsubsection{Definition of $G_D$ and its domain of validity}

Let $U_B\in G(2)$ be the inner gauge element that implements the conjugation
identified in the isolated gauge--Higgs theory of
Ref.~\cite{ButtazzoG2Higgs}, and let $Z_\chi$ denote the accidental sign of the
\emph{retained} $G(2)$-fundamental breaking coordinate,
\begin{equation}
Z_\chi:\qquad \chi\longmapsto-\chi .
\end{equation}
We define the low-energy accidental realization of dark $G$-parity by
\begin{equation}
G_D\equiv Z_\chi U_B .
\label{eq:GDdefinition}
\end{equation}
This is the EFT realization of the dark $G$-parity already introduced in the
notation dictionary.  Its exact domain must be stated carefully.  On
\emph{purely dark gauge-invariant operators}, every $G(2)$ invariant is
$U_B$ invariant, so the grading reduces to the sign of the retained dark
coordinate and is exact within that operator algebra.  By contrast, the action
on chiral matter cannot be chosen independently by declaring all visible fields even.  The $U_B$ factor conjugates color
representations and the selected mirror-free light projector is not closed
under that action; Sec.~\ref{sec:chiralColorParityNoGo} proves the resulting
failure of equivariance.  Accordingly, three statements are established in this
subsection: the pure-dark grading, the absence of a selected renormalizable
one-$\chi$ vertex, and the determinant cancellation when the massive complexes
are exactly paired.  None of them should be read as an exact symmetry of the
full mirror-free Class-B Lagrangian.  The pure-dark channel assignments are
summarized in Table~\ref{tab:pureDarkGrading}.
\begin{table}[t]
\centering
\caption{Pure-dark grading and channel ancestry in the isolated bosonic sector.}
\label{tab:pureDarkGrading}
\resizebox{\columnwidth}{!}{%
\begin{tabular}{c c c c}
\toprule
channel & dominant overlap & $J^{PC}$ & $G_D$\\
\midrule
$X_\mu\bar X^\mu$ & two-vector scalar & $0^{++}$ & $+1$ (unprotected)\\
$OSDSDS$ block & two-vector/glue & $1^{+-}$ & $-1$\\
$SGGG$ block & glue dominated & $1^{+-}$ & $-1$\\
$OSSS$ $S$-wave block & three-vector & $0^{--}$ & $-1$\\
\bottomrule
\end{tabular}%
}
\end{table}
The table makes explicit why $G_D$ is a symmetry classification, not a
constituent-number assignment.  Both the two-vector/glue $1^{+-}$ block and
the three-vector $0^{--}$ block are odd, whereas the naive scalar
$X\bar X$ channel is even.  Their relative ordering remains the
nonperturbative GEVP problem of Eq.~\eqref{eq:oddGEVP}.

\subsubsection{Why the selected renormalizable light kernel respects the accidental grading}

The first EFT ingredient is the exact component zero already established by
the parent-Clebsch calculation:
\begin{equation}
P_{\rm light}Y_\chi P_{\rm light}=0.
\label{eq:GDlightzero}
\end{equation}
Thus no renormalizable one-$\chi$ vertex connects two selected chiral zero
modes.  This is a genuine consequence of the Class-B component assignment: the
alternative $u^c\subset(\mathbf{7},\mathbf{3}_A)$ would permit such a coupling and is therefore
rejected independently by the stability analysis, leaving
$u^c\subset(\mathbf{7},\mathbf{15}_A)$.

There is also a purely $G(2)$ reason why the retained scalar coordinate has an
accidental sign on the actual vacuum.  All other background directions used in
the Class-B chain are $G(2)$ singlets.  Consequently
\begin{equation}
\operatorname{Hom}_{G(2)}(\mathbf{7}\otimes\mathbf{1}\otimes\mathbf{1},\mathbf{1})=0,
\qquad
\mathbf{1}\not\subset\operatorname{Sym}^3 \mathbf{7} .
\label{eq:G2oddchiTreeZero}
\end{equation}
Because the other active background directions are $G(2)$ singlets, these
representation-theory zeros forbid one or three powers of the retained $\mathbf{7}$ in
a renormalizable scalar interaction evaluated on the Class-B vacuum \cite{Slansky1981,AdamsExceptional}.  The
selected scalar potential is therefore even in $\chi$ at the renormalizable
level.  This statement concerns the retained dark coordinate; it does not yet
imply a sign symmetry of the full exceptional Higgs representation.

\subsubsection{Why the accidental sign does not extend to the entire $\mathbf{1463}_H$ parent}

It is tempting to promote the retained-coordinate sign to
$\mathbf{1463}_H\to-\mathbf{1463}_H$, but the full parent potential does not allow this in the
minimal construction.  The $\mathbf{650}_0$ scalar sector is embedded in
$\mathbf{1463}_H\subset\mathrm{Sym}^2 \mathbf{56}$, while the first breaking
$E_7\to E_6\times U(1)_X$ may be implemented by an adjoint $\mathbf{133}_H$ VEV along
its $\mathbf{1}_0$ direction, as in the construction of Ref.~\cite{MasiE6E7}.  The exact
$E_7$ product table gives \cite{Slansky1981}
\begin{widetext}
\begin{align}
\mathbf{1463}\otimes\mathbf{1463}={}&1_s\oplus\mathbf{133}_a\oplus\mathbf{1463}_a\oplus\mathbf{1539}_s
\oplus\mathbf{7371}_s\oplus\mathbf{150822}_s\nonumber\\
&\oplus\mathbf{152152}_a\oplus\mathbf{293930}_s\oplus\mathbf{617253}_s\oplus\mathbf{915705}_a .
\label{eq:1463square}
\end{align}
\end{widetext}
The absence of $\mathbf{1463}$ from $\mathrm{Sym}^2 \mathbf{1463}$ means that one commuting
$\mathbf{1463}_H$ has no cubic \emph{self}-invariant.  That observation is useful but
insufficient: the Class-B scalar sector also contains $\mathbf{133}_H$, and mixed
invariants can violate the common sign even when the isolated self-cubic is
absent.
Using the explicit $\mathbf{56}$ realization in which
$\mathrm{Sym}^2 \mathbf{56}=\mathbf{133}\oplus\mathbf{1463}$, an $E_7$-invariant quartic is
\begin{equation}
I_{3,1}=\operatorname{Tr}_{\mathbf{56}}\!\left(\Phi_{1463}^3\Sigma_{133}\right),
\label{eq:1463cub133quartic}
\end{equation}
and the explicit parent projector gives
$\operatorname{Tr}_{\mathbf{56}}(R_{\mathbf{1463}}^3T_A)\neq0$ for generic $R_{\mathbf{1463}}$.  A second
allowed structure with one $\mathbf{1463}$ and three adjoints is likewise generically
nonzero.  Thus the common sign $\Phi_{1463}\to-\Phi_{1463}$ is not a symmetry of the
generic $\mathbf{133}_H+\mathbf{1463}_H$ potential.
After $\langle133_H\rangle\subset\mathbf{1}_0$, Eq.~\eqref{eq:1463cub133quartic}
reproduces the generic $E_6$-level cubic structures of the neutral $\mathbf{650}$.

There is no contradiction with the retained-coordinate zero in
Eq.~\eqref{eq:G2oddchiTreeZero}.  After the adjoint acquires its $E_6$-singlet
VEV, the remaining Class-B background directions are still $G(2)$ singlets,
so an induced invariant cannot contain an odd number of the particular
$\chi\sim(\mathbf{7},\mathbf{1}_A)_0$ coordinate without another $G(2)$-charged insertion.  The
correct statement is therefore narrower: the selected dark $\mathbf{7}$ has an
accidental renormalizable sign, whereas the complete $\mathbf{1463}_H$ does not.

\subsubsection{Exact determinant cancellation when the complete massive spectrum is paired}

The scalar $\chi F^3$ coefficient vanishes before any spectral sum is taken,
and the apparent derivative--Higgs $d=7$ representative is EOM/IBP redundant.
The first physical pure-dark odd Wilson coefficients therefore occur at
$d=9$, beginning with the mixed-helicity hook operator of
Eq.~\eqref{eq:O9F4}.  The relevant EFT question is whether the \emph{sum over
the complete positive-mass spectrum} cancels this entire odd basis when
opposite-charge complexes are exact Serre--Hodge partners.  For the
Serre--Hodge-related $q$ and $-q$ complexes derived above,
\begin{equation}
\mathbb M_{-q}(-\chi,\mathcal V)
=\mathcal J\,\mathbb M_q(\chi,\mathcal V)^\dagger\mathcal J^{-1},
\end{equation}
where $\mathcal V$ denotes the $G_D$-even background.  Hence
\begin{equation}
\Gamma_{\rm heavy}[\chi,\mathcal V]
=\Gamma_{\rm heavy}[-\chi,\mathcal V] .
\label{eq:GDheavyEven}
\end{equation}
Equation~\eqref{eq:GDheavyEven} is stronger than a one-loop trace
cancellation: it states that the entire controlled reduced positive-mass
determinant is even in $\chi$ \cite{HenningLuMurayama2018}.  Consequently every perturbative Wilson
coefficient generated purely by this paired heavy tower and containing an odd
number of $\chi$ insertions vanishes.  In particular, the complete $d=9$
basis summarized in Eq.~\eqref{eq:D9OperatorBasisCounts}, including
$\mathcal O_{9,F^4}^-$ and its Hermitian parity partner, is absent on the
paired locus.  The determinant theorem is therefore strictly stronger than the
individual $\chi F^3$ zero: once the complete positive-mass spectrum is
arranged in the required dual pairs, \emph{every} perturbative Wilson
coefficient odd in $\chi$ vanishes, independently of its Lorentz structure.

\subsubsection{Restricted renormalizable-loop violation after the operator-basis correction}

Once the paired pure-heavy determinant is removed, explicit violations in the
restricted renormalizable construction must involve the selected chiral
kernel.  The charge analysis still requires the electroweak doublet direction
for a single-$\chi$ connector and two branch-changing insertions in a closed
light--heavy loop.  What changes is the local dark operator onto which this
spurion can match.  The previously written
$(H^\dagger H)\chi F^3$ structure is identically zero by
Eq.~\eqref{eq:ChiF3ExactZero}.

The first physical pure-dark target available to such a matching is now the
$d=9$ basis of Eq.~\eqref{eq:D9OperatorBasisCounts}.  However, the old
single-$\chi$ three-generator trace does not determine the coefficient of the
$[3,1]$ $\chi F^4$ channel, and multiplying the EOM-redundant
Eq.~\eqref{eq:Q7DDefinition} by $H^\dagger H$ is not a substitute for a
mixed-basis reduction: integration by parts also generates terms in which
derivatives act on the electroweak spurion.  A new covariant-derivative or
amplitude-basis matching calculation is therefore required \cite{HenningLuMurayama2018,HenningLuMeliaMurayama2016,HenningLuMeliaMurayama2017}.  We do not quote
a replacement lifetime for this restricted mechanism.  This is a loss of one
numerical estimate, not of the exact light-kernel result: the independent
dimension-five parent operators in Eqs.~\eqref{eq:O5uQG}--\eqref{eq:O5dQG}
remain lower dimensional and dominate the generic ultraviolet EFT unless an
additional compactification selection rule removes them.

\subsubsection{Exact scope of the accidental-$G_D$ result}

The content of the accidental-$G_D$ statement can now be given without
ambiguity.  The pure-dark operator algebra carries a well-defined $G_D$ grading
and the selected one-$\chi$ kernel has an exact renormalizable zero; when the conjugate massive complexes are chosen as Serre--Hodge partners,
the controlled positive-mass determinant is likewise even in $\chi$.  It is
not an exact symmetry of the full chiral Class-B Lagrangian because the matter
projector is not $U_B$ equivariant.  It does \emph{not} define an exact
discrete symmetry of the generic higher-dimensional ultraviolet action.  The
explicit $E_7$ witnesses in Eq.~\eqref{eq:E7d5Witnesses} prove that the latter
contains symmetry-allowed dimension-five breaking channels.  Thus the unresolved problem is not an ordinary low-order loop violation of the
paired pure-dark EFT.  It is more fundamental: the ultraviolet theory would need an additional exact
selection rule that removes the parent-allowed dimension-five operators themselves.

The selected EFT grading and its limitations can therefore be stated as follows:
\begin{enumerate}[label=(\roman*)]
\item in the pure-dark algebra, $G_D$ is a well-defined grading and its first
physical odd operator basis occurs at dimension nine;
\item in the complete chiral EFT, $G_D$ is not an exact symmetry because
$[P_{\rm light},U_B]\neq0$;
\item the direct selected light--light--$\chi$ coupling is absent and the paired
heavy determinant is even in $\chi$;
\item the restricted $F^3$ loop connector vanishes, whereas dimension-nine
rematching remains state dependent;
\item the generic $E_7$ chiral connector already appears at dimension five, so
the one-copy ultraviolet theory has no exact protection.
\end{enumerate}

\section{Separated $\mathbf{1539}_H$ and the bosonic $G$-parity completion}
\label{sec:1539escape}

The previous analysis shows where the original one-copy scalar assignment fails.  The problem is not the presence of a $G(2)$ fundamental by itself, but the fact that the special-breaking field $S$, the electroweak Higgs $H$ and the dark-breaking field $\chi$ all descend from the same neutral $\mathbf{650}_0\subset\mathbf{1463}_H$.  With a common parent, one discrete sign cannot distinguish the dark direction from the vacuum-even directions.

This observation suggests a minimal modification which keeps the published Class-B gauge chain and visible matter assignment unchanged: the dark-breaking field can be moved to an inequivalent $E_7$ parent.  At the \emph{bosonic scalar} level the construction works.  The separated $\mathbf{1539}_H$ contains the required neutral $G(2)$ fundamental, reverses the finite charge-flip sign, removes the common-$\mathbf{1463}_H$ Bose-symmetric dressing channel and admits an all-orders stationary dark direction with a nonempty open region of locally stable broken-phase vacua.  If an additional gauged or geometric $\mathbb Z_{2,D}$ is present, the odd-$\Xi$ operator tower is removed before dark Higgsing and a diagonal bosonic remnant survives afterwards.  Even-$\Xi$ operators are still allowed and are studied separately below.  In this way the scalar repair can be followed step by step before the independent chiral-equivariance condition is imposed.

\subsection{Why the $\mathbf{1539}_H$ contains the required dark direction}

The starting point is the standard exterior-square decomposition
\begin{equation}
\Lambda^2 \mathbf{56}=\mathbf{1}\oplus\mathbf{1539}.
\label{eq:56wedge1539}
\end{equation}
With
\begin{equation}
\mathbf{56}=\mathbf{27}_{-1}\oplus\overline{\mathbf{27}}_{+1}\oplus\mathbf{1}_{+3}\oplus\mathbf{1}_{-3},
\end{equation}
the exterior-square bookkeeping gives \cite{Slansky1981,LieART,ExceptionalReductions}
\begin{align}
\mathbf{1539}\to{}&(\mathbf{1}\oplus\mathbf{78}\oplus\mathbf{650})_0\nonumber\\
&\oplus(\overline{\mathbf{351}}\oplus\overline{\mathbf{27}})_{-2}
\oplus(\mathbf{351}\oplus\mathbf{27})_{+2}\nonumber\\
&\oplus\mathbf{27}_{-4}\oplus\overline{\mathbf{27}}_{+4}.
\label{eq:1539E6branch}
\end{align}
The branching is internally consistent: the dimensions close exactly,
\begin{equation}
729+378+378+27+27=1539,
\end{equation}
and the neutral sector obeys
\begin{equation}
27^2+1-1_{E_7}=729=1+78+650.
\end{equation}
To identify the actual dark-breaking component, we next restrict the neutral $\mathbf{650}_0$ through the special $E_6\to G(2)\times SU(3)_A$ embedding:
\begin{align}
\mathbf{650}\to{}&(\mathbf{1},\mathbf{1})\oplus(\mathbf{7},\mathbf{1})\oplus(\mathbf{1},\mathbf{8})\nonumber\\
&\oplus2(\mathbf{7},\mathbf{8})\oplus(\mathbf{7},\mathbf{10})\oplus(\mathbf{7},\overline{\mathbf{10}})\nonumber\\
&\oplus(\mathbf{14},\mathbf{8})\oplus(\mathbf{27},\mathbf{1})\oplus(\mathbf{1},\mathbf{27})\oplus(\mathbf{27},\mathbf{8}),
\label{eq:650specialFor1539}
\end{align}
This branching contains exactly one neutral dark fundamental,
\begin{equation}
\Xi_D\sim(\mathbf{7},\mathbf{1}_A)_0\subset\mathbf{650}_0\subset\mathbf{1539}_H.
\label{eq:chi1539}
\end{equation}
The special-breaking and electroweak directions can remain
\begin{equation}
S,H\subset\mathbf{650}_0\subset\mathbf{1463}_H.
\end{equation}
The crucial point is therefore not merely that $\mathbf{1539}_H$ contains another $\mathbf{650}_0$, but that the dark direction now lives in an irreducible $E_7$ representation inequivalent to the one containing $S$ and $H$.  The single-$\mathbf{1463}_H$ Schur obstruction is removed without introducing a second copy of the $\mathbf{1463}_H$ or altering the visible-sector parent assignment.

\subsection{Bose symmetry and the disappearance of the common-parent channel}
\label{sec:1539BoseAudit}

Moving $\chi$ to a different parent is useful only if subsequent vacuum insertions do not reconstruct the same forbidden channel indirectly.  Because the scalar fields commute, the relevant test is the \emph{symmetric} square of the $\mathbf{1539}$.  We independently recomputed the full $\mathbf{1539}\otimes\mathbf{1539}$ character with the $E_7$ Weyl formula and Freudenthal recursion and then separated the exchange-even and exchange-odd characters; the result agrees with the LieART conventions \cite{LieART,FultonHarris1991,McKayPatera1981}:
\begin{widetext}
\begin{align}
\mathrm{Sym}^2(\mathbf{1539})={}&\mathbf{1}\oplus2(\mathbf{1539})\oplus\mathbf{7371}\oplus\mathbf{40755}
\oplus\mathbf{150822}\oplus\mathbf{365750}\oplus\mathbf{617253},
\label{eq:Sym21539}\\
\Lambda^2(\mathbf{1539})={}&\mathbf{133}\oplus\mathbf{1463}\oplus\mathbf{8645}\oplus\mathbf{40755}
\oplus\mathbf{152152}\oplus\mathbf{980343}.
\label{eq:Wedge21539}
\end{align}
\end{widetext}
The dimension checks are
\begin{widetext}
\begin{equation}
\dim\mathrm{Sym}^2(\mathbf{1539})=\frac{1539\,1540}{2}=1185030,
\qquad
\dim\Lambda^2(\mathbf{1539})=\frac{1539\,1538}{2}=1183491.
\end{equation}
\end{widetext}

As an independent checksum of the exchange split, use the Dynkin-index
normalization $T(\mathbf{56})=6$, for which $T(\mathbf{133})=h^\vee(E_7)=18$ and
$T(\mathbf{1539})=324$ \cite{Slansky1981,McKayPatera1981}.  This is half of the convention in which the adjoint index is
written as $2h^\vee=36$, a distinction worth stating explicitly when
comparing with tabulations that use the doubled normalization.  The proposed constituents
obey
\begin{widetext}
\begin{align}
\sum_{\mathrm{Sym}^2}T
&=2(324)+2106+12870+56700+148500+278460\nonumber\\
&=499284=(1539+2)\,324,\\
\sum_{\Lambda^2}T
&=18+330+2340+12870+54912+427518\nonumber\\
&=497988=(1539-2)\,324.
\label{eq:1539IndexChecks}
\end{align}
\end{widetext}
The dimension equalities verify completeness of the two exchange sectors.  The Dynkin-index sums provide an independent checksum of the same split, while the direct Weyl-character subtraction supplies an independent decomposition certificate rather than only a dimension test.  A further structural check is automatic because the real orthogonal $\mathbf{1539}$ must contain the infinitesimal $E_7$ action, and hence the adjoint $\mathbf{133}$, in $\Lambda^2(\mathbf{1539})$.  The same calculation also gives the narrower self-map statement
\begin{equation}
\mathbf{1539}\not\subset\Lambda^2(\mathbf{1539}),
\label{eq:1539NoAntisymSelfMap}
\end{equation}
which is useful as a representation diagnostic but is not an on-shell operator-basis theorem.  These checks establish the fact that matters for the scalar obstruction:
\begin{equation}
\mathbf{1463}\not\subset\mathrm{Sym}^2(\mathbf{1539}).
\label{eq:no1463Sym21539}
\end{equation}
Because two identical scalar insertions probe only $\mathrm{Sym}^2(\mathbf{1539})$, Eq.~\eqref{eq:no1463Sym21539} means that a pair of $\mathbf{1539}_H$ fields cannot feed the $\mathbf{1463}$ channel responsible for the common-parent $S$-dressing failure.  This is an exact Bose-symmetry statement at the renormalizable tensor level, not a cancellation between coefficients.  It removes the \emph{common-parent} obstruction, but it does not yet forbid every operator containing an odd number of $\Xi$ fields; that stronger statement requires the discrete gauge factor introduced below.

\subsubsection{Odd-$\Xi$ operators before dark Higgsing}
The exchange decomposition also determines which mixed renormalizable cubics can exist.  
The symmetric square contains two copies of the $\mathbf{1539}$ itself but contains
neither the $\mathbf{133}$ nor the $\mathbf{1463}$.  Hence, for a commuting scalar
$\Xi\sim\mathbf{1539}_H$, the $E_7$ tensor algebra permits cubic $\Xi^3$
structures but gives
\begin{equation}
\resizebox{0.98\columnwidth}{!}{$\displaystyle
\operatorname{Hom}_{E_7}(\mathrm{Sym}^2 \mathbf{1539},\mathbf{133})=0,
\qquad
\operatorname{Hom}_{E_7}(\mathrm{Sym}^2 \mathbf{1539},\mathbf{1463})=0.
$}
\label{eq:1539NoMixedCubics}
\end{equation}
Thus neither $\mathbf{133}_H\,\Xi^2$ nor $\mathbf{1463}_H\,\Xi^2$ is a Bose-symmetric renormalizable cubic.  The remaining $\Xi^3$ structures are allowed by $E_7$ itself, so Bose symmetry alone is not an all-orders parity.  If the proposed $\mathbb Z_{2,D}$ is genuinely gauged, however, those cubic terms are odd and disappear as well.  More generally,
with the current matter and $\mathbf{133}_H,\mathbf{1463}_H,\Theta$ sectors even, every local
operator containing an odd number of $\mathbf{1539}_H$ fields is $\mathbb Z_{2,D}$
odd.  Consequently, in the un-Higgsed Lagrangian every genuine $E_7$ invariant
with an odd number of $\Xi$ fields is forbidden, including
\begin{equation}
\resizebox{0.98\columnwidth}{!}{$\displaystyle
\Psi\Psi\Xi,
\qquad \Psi\Psi\Phi^n\Xi^{2k+1},
\qquad \bigl[\mathcal O_{9,F^4}[\Xi]\bigr]_{E_7},
\qquad (\Phi^\dagger\Phi)^n
\bigl[\mathcal O_{9,F^4}[\Xi]\bigr]_{E_7}
$}
\label{eq:1539OddTower}
\end{equation}
whenever the corresponding parent invariant is present.  The on-shell
operator analysis shows that the pure-dark odd basis is empty through $d=7$,
so the first physical one-$\Xi$ gauge operator to test is the $d=9$ hook.
Equation~\eqref{eq:E7HookInvariantCountsUV} finds two $\mathbf{1539}_H$ parent channels
in $\mathbb S_{(3,1)}\mathbf{133}$, but
Eq.~\eqref{eq:1539HookRestrictionRankZero} proves that both have zero projection
onto the selected dark $(\mathbf{7},\mathbf{1}_A)_0$ component when all four gauge
slots lie in $G(2)$.  This undressed component zero follows from the canonical
$P_7^{(0)}$ carrier signs and does not require the additional $\mathbb Z_{2,D}$.
The genuine $\mathbb Z_{2,D}$ remains load-bearing at the stronger all-orders
level because $P_7^{(0)}$ is broken by the special $S$ background and odd-$\Xi$
operators dressed by $P_7^{(0)}$-odd insertions are not removed by the rank-zero
theorem.  The ordinary
$\Psi\Psi\Phi$ Yukawa channel remains even.  Even-$\Xi$ operators are
therefore a separate question: after dark Higgsing they can generate vertices
with a single fluctuation when one or more $\Xi$ fields are replaced by
$\langle\Xi_D\rangle$.  Their visible-sector fate cannot be inferred from
$P_D$, because Sec.~\ref{sec:chiralColorParityNoGo} proves that the residual
normalizer action does not close on the mirror-free chiral light kernel.  We
instead test these terms directly under the exact unbroken $SU(3)_C$ below.
The conclusion of this first operator analysis is therefore sharply delimited: the separated parent removes the complete odd-$\Xi$ tower \emph{before} dark Higgsing, conditional on a genuine discrete gauge factor, whereas even-$\Xi$ operators remain and require an independent analysis after the VEV is inserted.

\subsubsection{Reversed canonical charge-flip sign of the dark direction}
A second, independent advantage of the exterior-square realization is that it reverses the canonical charge-flip eigenvalue
of the dark direction.  The sign is fixed by exchange symmetry: on the
neutral $\mathbf{27}\otimes\overline{\mathbf{27}}\simeq\mathrm{End}(\mathbf{27})$ cross carrier,
$P_7^{(0)}$ exchanges the two factors, while the $\mathbf{1463}$ and $\mathbf{1539}$ descend
from the symmetric and antisymmetric two-$\mathbf{56}$ carriers.  Thus their neutral
cross blocks acquire opposite transpose signs.  Appendix~\ref{app:1539-sign-certificate}
spells out this cross-block calculation in the same convention used for the
$\mathbf{1463}$ certificate.  In the conventions used here the canonical $E_7$ charge
flip acts as
\begin{equation}
M_{1463}\longmapsto-M_{1463}^{T},
\qquad
M_{1539}\longmapsto+M_{1539}^{T}.
\label{eq:1539TransposeSigns}
\end{equation}
The $(\mathbf{7},\mathbf{1}_A)$ representative is transpose odd.  Hence
\begin{equation}
P_7^{(0)}\chi_{1463}=+\chi_{1463},
\qquad
P_7^{(0)}\chi_{1539}=-\chi_{1539}.
\label{eq:1539P7Reversal}
\end{equation}
The separated parent therefore passes the finite carrier-sign test: the dark direction has precisely the relative sign that failed in the symmetric-square $\mathbf{1463}_H$.  This is still only a representation-theory statement.  It does not by itself construct a symmetry of the four-dimensional chiral vacuum, but it shows that the common-$\mathbf{1463}_H$ sign obstruction was specific to the common $\mathbf{1463}_H$ carrier rather than to the $G(2)$ breaking itself.

\subsection{Conditional gauged realization of bosonic dark $G$-parity}

Representation theory has now removed the common-parent scalar obstruction,
but it has not by itself produced an exact protecting symmetry.  The next
step is therefore conditional: we ask what follows if the separated
$\mathbf{1539}_H$ carries an independent discrete gauge charge,
\begin{equation}
\resizebox{0.98\columnwidth}{!}{$\displaystyle
\mathbb Z_{2,D}:\qquad
\mathbf{1539}_H\mapsto-\mathbf{1539}_H,
\qquad
\mathbf{133}_H,\ \mathbf{1463}_H,\ \mathbf{27664}_F,\Theta\mapsto+1.
$}
\label{eq:Z2D1539charges}
\end{equation}
For this sign to define an ultraviolet-exact selection rule rather than an
accidental global symmetry, we assume that it is a genuine discrete gauge
factor \cite{KraussWilczek1989}.  Its microscopic origin is not constructed
here; the purpose of the present test is to determine the particle-physics
consequences if such a factor exists.

Let $U_B\in G(2)$ be Eq.~\eqref{eq:UBexplicit}.  Since $U_Be_7=-e_7$, the dark vacuum
\begin{equation}
\langle\Xi_D\rangle=v_\chi e_7
\end{equation}
does not preserve the factor $z_D$ by itself:
$z_D\langle\Xi_D\rangle=-\langle\Xi_D\rangle$.  Rather, the dark
Higgsing leaves the diagonal bosonic transformation
\begin{equation}
P_D=z_DU_B,
\qquad
P_D\langle\Xi_D\rangle=\langle\Xi_D\rangle,
\qquad P_D^2=1,
\label{eq:PD1539}
\end{equation}
while the $G(2)$-singlet VEVs $S,H,\Theta$ remain even.  The symmetry logic is therefore two-stage.  Before dark Higgsing, $z_D$ forbids every operator with an odd number of $\Xi$ fields.  The VEV $\langle\Xi_D\rangle$ then Higgses the bare $z_D$ factor, but the combination with $U_B$ remains unbroken and defines the bosonic residual parity $P_D$.  The radial fluctuation along $e_7$ is $P_D$ even.  Importantly, we do \emph{not} use this bosonic statement as a selection rule on the mirror-free visible fermions; Sec.~\ref{sec:chiralColorParityNoGo} shows that the normalizer action does not close on that chiral subspace.  Visible even-$\Xi$ dressings must therefore be tested directly under the exact unbroken color group.

The corresponding symmetry-breaking manifold is particularly simple,
\begin{equation}
\frac{G(2)\times\mathbb Z_{2,D}}
     {SU(3)_C\times\mathbb Z_{2,D}^{\rm diag}}
\simeq
\frac{G(2)}{SU(3)_C}
\simeq S^6,
\label{eq:1539VacuumManifold}
\end{equation}
so $\pi_0=\pi_1=0$: no domain walls or strings are generated by this stage \cite{AdamsExceptional,SpringerVeldkamp,BaezOctonions}.
All fermions are $z_D$ even in Eq.~\eqref{eq:Z2D1539charges}, so the
four-dimensional matter spectrum has no fermionic mixed/discrete or
gravitational anomaly for this $\mathbb Z_2$ assignment \cite{HsiehDiscreteAnomalies}.  This does not
replace the still-missing geometric construction of the discrete gauge
factor.

Conditional on a genuine gauged $\mathbb Z_{2,D}$, the bosonic effective action about this vacuum is exactly $P_D$ invariant.  Purely dark operators with odd $P_D$ charge are then absent, while the FMS classification of the isolated gauge--Higgs spectrum is unchanged:
\begin{equation}
\resizebox{0.98\columnwidth}{!}{$\displaystyle
P_D(0^{--})=-1,
\qquad
P_D(1^{+-})=-1,
\qquad
P_D(0^{++})=+1,
$}
\label{eq:1539JPCparities}
\end{equation}
so the three-vector-dominated scalar and two-vector/glue axial-vector blocks
remain the protected candidates at the bosonic level.

\subsection{Even-$\Xi$ portals: exact residual-color zero and surviving channels}
\label{sec:1539EvenDressing}

The discrete gauge charge solves only the odd-$\Xi$ problem.  The remaining
physical question is whether allowed even-$\Xi$ operators can regenerate a
dangerous one-particle portal after the dark VEV is inserted.  Operators
containing an even number of $\Xi$ fields are allowed, and after one or more
fields are replaced by the dark VEV they can in principle induce vertices
with a single fluctuation.  The leading parent example already appears at dimension six.  Because
$\mathbf{1}\subset\mathrm{Sym}^2(\mathbf{1539})$, any nonzero ordinary $E_7$ Yukawa tensor
$\mathcal Y_{E_7}(\Psi,\Psi,H)$ generates the parent invariant
\begin{equation}
\mathcal O_{6}^{(1)}=
\frac{c_6}{M_*^2}\,
\mathcal Y_{E_7}(\Psi,\Psi,H)\,(\Xi\otimes\Xi)_{1},
\label{eq:EvenXiParentSinglet}
\end{equation}
up to the normalization absorbed into $c_6$.  Thus even-$\Xi$ dressing is a genuine operator-level effect and cannot be dismissed by the discrete charge.  The potentially dangerous \emph{selected-light} component is nevertheless much more constrained.  Along the
dark $G(2)$ fundamental,
\begin{equation}
\mathbf{7}\longrightarrow\mathbf{1}_C\oplus\mathbf{3}_C\oplus\bar{\mathbf{3}}_C,
\qquad
\Xi_D=(v_\chi+\sigma)e_7+\pi_{\mathbf{3}}+\pi_{\bar{\mathbf{3}}},
\label{eq:XiDarkColorDecomp}
\end{equation}
where $\sigma$ is the radial color singlet and the six
$\pi_{\mathbf{3}}\oplus\pi_{\bar{\mathbf{3}}}$ modes are the Goldstone directions eaten by the
massive broken vectors.  The selected light quark colors obey
$Q_L\subset\mathbf{3}_C$ and $u^c,d^c\subset\bar{\mathbf{3}}_C$, while $H$ and every dark VEV
insertion are color singlets.  A term containing one transverse fluctuation
and any odd number of dark VEV insertions would therefore require
\begin{equation}
\operatorname{Hom}_{SU(3)_C}
\!\left[\mathbf{3}\otimes\bar{\mathbf{3}}\otimes(\mathbf{3}\oplus\bar{\mathbf{3}}),\mathbf{1}\right]=0,
\label{eq:EvenXiSingleTransverseColorZero}
\end{equation}
because
\begin{align}
(\mathbf{1}\oplus\mathbf{8})\otimes\mathbf{3}&=\mathbf{3}\oplus\mathbf{3}\oplus\bar{\mathbf{6}}\oplus\mathbf{15},\\
(\mathbf{1}\oplus\mathbf{8})\otimes\bar{\mathbf{3}}&=\bar{\mathbf{3}}\oplus\bar{\mathbf{3}}\oplus\mathbf{6}\oplus\overline{\mathbf{15}}.
\label{eq:EvenXiTrialityProducts}
\end{align}
Equivalently, the total color triality is $\pm1$ rather than zero \cite{GeorgiLieAlgebras,Slansky1981}.  Since every
$E_7$ invariant restricts to an invariant of the unbroken $SU(3)_C$, this is
not merely a subgroup example or a particular Clebsch cancellation.  For
\emph{every} parent contraction and every $k\ge1$,
\begin{equation}
\left.
\mathcal O_{2k}^{\rm vis}
\right|_{Q_{\mathbf{3}_C}\,u^c_{\bar{\mathbf{3}}_C}\ {\text{or}}\ d^c_{\bar{\mathbf{3}}_C};\,
 v_\chi^{\,2k-1}\pi_{\mathbf{3}\oplus\bar{\mathbf{3}}}}=0.
\label{eq:EvenXiAllOrderOneTransverseZero}
\end{equation}
Because every $E_7$ invariant restricts to an invariant of the unbroken color group, Eq.~\eqref{eq:EvenXiAllOrderOneTransverseZero} is an exact residual-symmetry theorem rather than a special Clebsch cancellation.  It is also logically independent of the later failure to extend $P_D$ to the mirror-free chiral theory.  For visible fields, this color theorem is therefore the correct statement to use in place of an assumed parity selection rule.

The theorem does \emph{not} remove every even-$\Xi$ channel.  The singlet contraction in
Eq.~\eqref{eq:EvenXiParentSinglet} contains schematically
\begin{equation}
(\Xi\Xi)_{\mathbf{1}}\supset
(v_\chi+\sigma)^2+2\,\pi_{\mathbf{3}}\!\cdot\!\pi_{\bar{\mathbf{3}}}+\cdots,
\label{eq:EvenXiSurvivingChannels}
\end{equation}
so there is no $v_\chi\pi_\perp$ term, whereas a radial
$v_\chi\sigma$ vertex and two-transverse vertices are allowed.  The radial
field is a $0^{++}$ excitation and has zero direct vacuum-to-$1^{+-}$ or
vacuum-to-$0^{--}$ overlap in the isolated bosonic theory, by the same
factorized argument as Eq.~\eqref{eq:D5DirectOddPoleZero}.  The surviving
visible portals can nevertheless contribute through heavy-descendant,
gauge, derivative, loop, or nonfactorizing confined matching.  We therefore
parameterize their odd-pole matching by independent factors
\begin{equation}
\kappa_{6,1^{+-}}^{\rm eff},\qquad
\kappa_{6,0^{--}}^{\rm eff},
\label{eq:kappa6EvenXi}
\end{equation}
rather than assigning unit overlap to a constituent-level conversion.
Absorbing order-one singlet normalization into $c_6$, the radial matching
carries the parametric suppression
\begin{equation}
\frac{v_\chi}{M_*}=1.449\times10^{-2},
\qquad
\ln\!\frac{M_*}{v_\chi}=4.234
\label{eq:EvenXiNumericalSuppression}
\end{equation}
for $v_\chi=10^{14}\,$GeV and $M_*=6.9\times10^{15}\,$GeV.  The numerical suppression is therefore potentially relevant for metastability, but it is not a direct one-$X$ decay amplitude and cannot be converted into a universal lifetime without the state-dependent matching factors $\kappa_{6,J}^{\rm eff}$.

\paragraph{Local interpolators versus a gauged discrete charge.}
If $\mathbb Z_{2,D}$ is genuinely gauged, a state carrying the nontrivial residual
$P_D$ charge belongs to a discrete-gauge superselection sector.  The local FMS
operators used above are local invariants of the connected continuous gauge group
and diagnose the spectral channel, but a fully gauge-invariant charged state in
the gauged-$\mathbb Z_2$ theory requires the usual nonlocal dressing (equivalently,
a line to infinity or an operator creating a charge-anticharge pair) \cite{KraussWilczek1989,GaiottoGlobalSym2015}.  This
standard distinction does not alter the local mass analysis or the chiral-color
equivariance theorem; it only prevents us from treating a $P_D$-odd local field
as a standalone gauge-invariant observable once the conditional discrete factor
is promoted from a global grading to a gauge symmetry.

Having controlled the visible even-$\Xi$ portal, we next ask whether the
separated parent can actually support the required dark vacuum.  This is a
different question from operator selection: the dark direction must be
stationary and the physical scalar fluctuations around it must have
nonnegative masses.
Writing the $\mathbf{1463}$ tensor as a symmetric two-form $\Phi$ and the $\mathbf{1539}$ tensor
as an antisymmetric two-form $\Xi$, define
\begin{equation}
R_\Phi=\Omega^{-1}\Phi,
\qquad K_\Xi=\Omega^{-1}\Xi.
\end{equation}
Then
\begin{equation}
I_{\Phi\Xi}=\operatorname{Tr}_{\mathbf{56}}
(R_\Phi K_\Xi R_\Phi K_\Xi)
\label{eq:PhiXiQuartic}
\end{equation}
is $E_7$ invariant and even in $\Xi$.  On the neutral block, with
\begin{widetext}
\begin{equation}
\Sigma_{650}^{(0)}=2\Pi_A-7\Pi_B,
\qquad A=(\mathbf{7},\mathbf{3}_A),\ \dim A=21,
\qquad B=(\mathbf{1},\overline{\mathbf{6}}_A),\ \dim B=6,
\end{equation}
\end{widetext}
its normalized support eigenvalues are
\begin{equation}
A\to A:4,
\qquad B\to B:49,
\qquad A\leftrightarrow B:-14.
\label{eq:1539splittingEigenvalues}
\end{equation}
The unequal eigenvalues show explicitly that $\langle S\rangle$ can split different $\mathbf{1539}_H$ support blocks, so the separated parent is not forced to remain degenerate after the special breaking.  However,
$(\mathbf{7},\mathbf{1}_A)$ and $(\mathbf{27},\mathbf{1}_A)$ both lie in the $A\to A$ block.  The remaining
degeneracy is removed by an independent self-quartic.  For
$K_\Xi=\Omega^{-1}\Xi$,
\begin{equation}
I_{\Xi^4}=\operatorname{Tr}_{\mathbf{56}}K_\Xi^4
\label{eq:Xi4Invariant}
\end{equation}
is again $E_7$ invariant.  On one neutral $\mathbf{27}$ block, normalizing by the fourth power of the Frobenius
norm gives
\begin{widetext}
\begin{equation}
\rho_{27}(\mathbf{7},\mathbf{1})=
\frac{\operatorname{Tr}_{27}M_{(\mathbf{7},\mathbf{1})}^4}
     {[\operatorname{Tr}_{27}(M_{(\mathbf{7},\mathbf{1})}^TM_{(\mathbf{7},\mathbf{1})})]^2}=\frac1{18},
\qquad
\left.\rho_{27}(\mathbf{27},\mathbf{1})\right|_{T=\mathrm{diag}(a,-a,0,\ldots,0)}
=\frac16.
\label{eq:1539SevenTwentySevenSplit}
\end{equation}
\end{widetext}
The neutral element occurs with the conjugate block in the full $\mathbf{56}$, so the
corresponding full-$\mathbf{56}$ normalized ratios for
$I_{\Xi^4}=\operatorname{Tr}_{\mathbf{56}}K_\Xi^4$ are one half of these values,
$1/36$ and $1/12$, respectively.  The common factor does not affect the
non-degeneracy conclusion.  The first value is orbit independent for the dark
$\mathbf{7}$; the second is a representative value on the $\mathbf{27}$ and is not constant
over that whole multiplet.
Thus the $(\mathbf{7},\mathbf{1}_A)$ and $(\mathbf{27},\mathbf{1}_A)$ directions are not symmetry-forced to be degenerate.  These invariant tests provide the mass-splitting handles needed for a local vacuum analysis, but they do not pretend to specify the most general renormalizable unbroken-$E_7$ potential.  The appropriate question is instead whether the broken-phase Wilsonian theory admits an aligned, locally stable vacuum.

\subsection{Dark-ray stationarity and the broken-phase scalar spectrum}
\label{sec:1539FullHessian}

After $\langle S\rangle$, the symmetry relevant for the dark-stage scalar alignment problem is
\begin{equation}
H_s\times U(1)_X=G(2)\times SU(3)_A\times U(1)_X.
\end{equation}
The $\mathbf{1539}$ branches completely as \cite{Slansky1981,LieART}
\begin{widetext}
\begin{align}
\mathbf{1539}_0={}&2(\mathbf{1},\mathbf{1})\oplus(\mathbf{14},\mathbf{1})\oplus2(\mathbf{1},\mathbf{8})\oplus3(\mathbf{7},\mathbf{8})
\oplus(\mathbf{7},\mathbf{1})\oplus(\mathbf{7},\mathbf{10})\oplus(\mathbf{7},\overline{\mathbf{10}})\nonumber\\
&\oplus(\mathbf{14},\mathbf{8})\oplus(\mathbf{27},\mathbf{1})\oplus(\mathbf{1},\mathbf{27})\oplus(\mathbf{27},\mathbf{8}),\label{eq:1539Hs0}\\
\mathbf{1539}_{+2}={}&(\mathbf{7},\overline{\mathbf{6}})\oplus(\mathbf{14},\overline{\mathbf{6}})\oplus(\mathbf{1},\mathbf{3})
\oplus(\mathbf{27},\mathbf{3})\oplus(\mathbf{7},\mathbf{15})\oplus2(\mathbf{7},\mathbf{3})\oplus(\mathbf{1},\mathbf{15})\oplus(\mathbf{1},\overline{\mathbf{6}}),
\label{eq:1539HsP2}\\
\mathbf{1539}_{-2}={}&(\mathbf{7},\mathbf{6})\oplus(\mathbf{14},\mathbf{6})\oplus(\mathbf{1},\overline{\mathbf{3}})
\oplus(\mathbf{27},\overline{\mathbf{3}})\oplus(\mathbf{7},\overline{\mathbf{15}})
\oplus2(\mathbf{7},\overline{\mathbf{3}})\oplus(1,\overline{\mathbf{15}})\oplus(\mathbf{1},\mathbf{6}),
\label{eq:1539HsM2}\\
\mathbf{1539}_{-4}={}&(\mathbf{7},\mathbf{3})\oplus(\mathbf{1},\overline{\mathbf{6}}),\qquad
\mathbf{1539}_{+4}=(\mathbf{7},\overline{\mathbf{3}})\oplus(\mathbf{1},\mathbf{6}).
\label{eq:1539HsPM4}
\end{align}
\end{widetext}
The dimensions of the five charge sectors are respectively $729,378,378,27,27$ and sum to $\mathbf{1539}$, providing a direct branching check.  Most importantly, the desired
\begin{equation}
D\equiv(\mathbf{7},\mathbf{1}_A)_0
\end{equation}
occurs with multiplicity one.

The branching also identifies the two neutral $H_s$ singlets that are potentially dangerous for the residual parity,
\begin{equation}
\xi_{\mathbf{1}}\subset\mathbf{1}_0\subset\mathbf{1539},
\qquad
\xi_{650}\subset(\mathbf{1},\mathbf{1})_0\subset\mathbf{650}_0\subset\mathbf{1539}.
\label{eq:1539TwoOddSinglets}
\end{equation}
These are the most consequential orthogonal alignment directions because $U_B$ acts trivially on them whereas $z_D$ reverses their sign.  Any
$\langle\xi_1\rangle$ or $\langle\xi_{\mathbf{650}}\rangle$ would therefore break
the residual $P_D$.  Their neutral-singlet $2\times2$ Hessian block must be
positive definite; the projector point below satisfies this automatically.

Because the dark representation occurs with multiplicity one, it defines a canonical $H_s\times U(1)_X$-equivariant projector
$\Pi_D$ onto the dark subspace.  It may be written entirely in terms of
commuting representation operators.  With $\widehat X$ denoting the $X$
charge operator,
\begin{equation}
P_{X=0}=
\left(1-\frac{\widehat X^2}{4}\right)
\left(1-\frac{\widehat X^2}{16}\right),
\label{eq:PXzero1539}
\end{equation}
and, in the normalization \cite{Slansky1981,GeorgiLieAlgebras}
$C_2^{G(2)}(\mathbf{7})=2$, $C_2^{G(2)}(\mathbf{14})=4$,
$C_2^{G(2)}(\mathbf{27})=14/3$,
\begin{equation}
P^{G(2)}_{7}
=
\frac{3}{32}\,
C_2^{G(2)}
\left(C_2^{G(2)}-4\right)
\left(C_2^{G(2)}-\frac{14}{3}\right).
\label{eq:PG27}
\end{equation}
The $SU(3)_A$ singlet projector is similarly the finite Lagrange polynomial
\begin{equation}
P^A_1=
\prod_{c\in{\cal C}_A^\times}
\left(1-\frac{C_2^A}{c}\right),
\label{eq:PA1}
\end{equation}
where ${\cal C}_A^\times$ is the finite set of nonzero quadratic-Casimir
eigenvalues of the $SU(3)_A$ irreps occurring in
Eqs.~\eqref{eq:1539Hs0}--\eqref{eq:1539HsPM4}.  Hence
\begin{equation}
\Pi_D=P_{X=0}P^{G(2)}_7P^A_1,\qquad
\Pi_D^2=\Pi_D,
\label{eq:PiDprojector1539}
\end{equation}
and $\Pi_\perp\equiv1-\Pi_D$ projects onto the remaining $1532$ real scalar
directions.

The multiplicity-one statement does more than simplify the quadratic analysis: it also controls all higher-order tadpoles at the dark-breaking stage.  Let
$\eta\in \Pi_\perp(\mathbf{1539})$.  A term linear in $\eta$ in the derivative of any
$z_D$-even scalar invariant evaluated on the dark ray must transform as
\begin{equation}
\mathrm{Sym}^{2k+1}D\otimes\eta,
\qquad k\ge0,
\label{eq:1539TadpoleTensor}
\end{equation}
because varying one field in an even-$\Xi$ invariant leaves an odd number of
background dark fundamentals.  For the $G(2)$ fundamental,
\begin{equation}
\mathrm{Sym}^{m}7
=
\bigoplus_{j=0}^{\lfloor m/2\rfloor}V_{(m-2j,0)}.
\label{eq:G2SymmetricPowers7}
\end{equation}
For completeness, the decomposition in Eq.~\eqref{eq:G2SymmetricPowers7} follows from the standard realization of $G(2)$ harmonics on the fundamental orbit $G(2)/SU(3)\simeq S^6$ \cite{AdamsExceptional,SpringerVeldkamp,BaezOctonions}.  The highest-weight vector in the traceless
rank-$n$ symmetric tensors generates $V_{(n,0)}$, and the Weyl dimension
formula gives
\begin{equation}
\resizebox{0.98\columnwidth}{!}{$\displaystyle
\dim V_{(n,0)}
=
\frac{(n+1)(n+2)(n+3)(n+4)(2n+5)}{120}
=
\binom{n+6}{6}-\binom{n+4}{6},
$}
\end{equation}
which is exactly the dimension of the traceless symmetric rank-$n$ tensors
of a seven-dimensional vector.  Removing traces recursively proves
Eq.~\eqref{eq:G2SymmetricPowers7}.  The multiplicities are therefore sharper
than the exclusion alone: for every odd $m$ the fundamental $7=V_{(1,0)}$
appears exactly once, while $1$, $\mathbf{14}$ and $\mathbf{27}$ are absent; for every even
$m\ge2$ the singlet and $\mathbf{27}=V_{(2,0)}$ each appear exactly once.  The adjoint
$\mathbf{14}=V_{(0,1)}$ never occurs in any symmetric power.  The singlet statement
also follows geometrically from the transitive action
$G(2)/SU(3)\simeq S^6$: polynomial invariants of one fundamental vector are
functions of its norm and hence occur only in even degree.  Thus odd
symmetric powers begin with dimensions $7,77,378,\ldots$ and have a unique
$G(2)$-fundamental channel at every odd order.
Eqs.~\eqref{eq:1539Hs0}--\eqref{eq:1539HsPM4} show that the complete
$\mathbf{1539}$ contains only $G(2)$ irreps $\mathbf{1},\mathbf{7},\mathbf{14},\mathbf{27}$, and the only neutral
$(\mathbf{7},\mathbf{1}_A)_0$ is $D$ itself.  Therefore
\begin{equation}
\operatorname{Hom}_{H_s\times U(1)_X}
\!\left(\mathrm{Sym}^{2k+1}D,\Pi_\perp(\mathbf{1539})\right)=0
\quad\forall k\ge0.
\label{eq:1539AllOrderTadpoleZero}
\end{equation}
Hence $\Pi_\perp\langle\Xi\rangle=0$ is exactly stationary for every $z_D$-even potential at the $\chi$-breaking stage, including arbitrary insertions of already-condensed $H_s$-singlet backgrounds.  This is an all-orders stationarity theorem, not a statement about the signs of the quadratic eigenvalues.  Local stability must still be demonstrated separately, and the theorem is intentionally scoped to the stage before the later $SU(3)_A$-breaking VEVs are switched on.

\subsubsection{Later visible VEVs and parity-preserving bending of the dark ray}
\label{sec:1539LaterVEVAlignment}
Once $SU(3)_A$ is broken at later stages, the stronger statement $\Pi_\perp\langle\Xi\rangle=0$ need no longer hold exactly.  What matters physically is whether the induced displacement can force color or the residual bosonic parity to break.  In the present chain all later scalar backgrounds are $G(2)$ singlets and are even under $z_D$.  Varying an
even-$\Xi$ invariant with respect to an orthogonal $\mathbf{1539}_H$ component still
leaves an odd symmetric power of the dark fundamental under $G(2)$.
Equation~\eqref{eq:G2SymmetricPowers7} therefore continues to exclude
$G(2)$ irreps $1$, $\mathbf{14}$ and $\mathbf{27}$ from a tadpole.  More strongly, the
fundamental occurs with multiplicity one in every odd symmetric power, so the
allowed $G(2)$ tensor type is unique at every order.  The only possible induced
$\mathbf{1539}_H$ directions transform as a $G(2)$ fundamental.

The $X$-charge bookkeeping makes this more restrictive.  The later octet
backgrounds have $X=0$, while the paired $\Theta_\pm$ VEVs carry charges
$\pm3$.  Any monomial built from these backgrounds has total $X$ charge a
multiple of three.  Since the $\mathbf{1539}_H$ sectors have
$X=0,\pm2,\pm4$, a term linear in an induced $\mathbf{1539}_H$ VEV can therefore
involve only the neutral sector.  Combining this with
Eq.~\eqref{eq:1539Hs0}, the only orthogonal representations which can be
sourced are
\begin{equation}
 3(\mathbf{7},\mathbf{8})_0\oplus(\mathbf{7},\mathbf{10})_0\oplus(\mathbf{7},\overline{\mathbf{10}})_0.
 \label{eq:LaterVEVAllowedXiDirections}
\end{equation}
This restriction is independent of the order at which the visible VEVs enter.

After $\langle\chi\rangle$ the visible backgrounds preserve
$SU(3)_C$.  Since
\begin{equation}
 \mathbf{7}\downarrow SU(3)_C=\mathbf{1}\oplus\mathbf{3}\oplus\bar{\mathbf{3}},
\end{equation}
any induced VEV in Eq.~\eqref{eq:LaterVEVAllowedXiDirections} must lie along
the unique color-singlet direction $e_7$; a colored component cannot acquire a
tadpole from color-singlet backgrounds.  On this component
$U_Be_7=-e_7$, while $z_D$ acts as $-1$ on the entire $\mathbf{1539}_H$.  Hence
\begin{equation}
 P_D(e_7\hbox{ component of every }7)=(-1)(-1)=+1.
\end{equation}
Thus later visible vacuum expectation values may bend the single
$(\mathbf{7},\mathbf{1})_0$ ray into other neutral $G(2)$-fundamental
directions, but they do not force either color breaking or $P_D$ breaking in
the bosonic scalar sector.
This is an all-orders symmetry/representation statement conditional on the
bosonic $z_D$ assignment; it does not cure the independent chiral-color
no-go for extending $P_D$ to the complete mirror-free theory.

The theorem allows bending, but it is useful to exhibit the first concrete source.  It appears already at quartic order through the invariant
$I_{\Phi\Xi}$ of Eq.~\eqref{eq:PhiXiQuartic} contains two visible $\mathbf{1463}_H$
insertions and two $\mathbf{1539}_H$ insertions.  For two insertions of the same commuting
$SU(3)_A$ octet background only the symmetric product contributes,
\begin{equation}
 \operatorname{Sym}^2 \mathbf{8}=\mathbf{1}\oplus\mathbf{8}_s\oplus\mathbf{27}.
\end{equation}
After one $\Xi$ is replaced by $v_\chi e_7$, this quartic therefore gives an
explicit source for a $(\mathbf{7},\mathbf{8})_0$ admixture.  The $(\mathbf{7},\mathbf{10})_0$ and
$(\mathbf{7},\overline{\mathbf{10}})_0$ directions in
Eq.~\eqref{eq:LaterVEVAllowedXiDirections} remain allowed by the all-orders
selection theorem, but they require higher-order and/or distinct visible
insertions rather than this identical-octet quadratic VEV.  Writing the
orthogonal mass of the explicitly sourced octet channel as $m_8^2$,
\begin{equation}
 t_8\sim\kappa_8 v_\chi v_8^2,
 \qquad
 \frac{\delta v_8^{(\Xi)}}{v_\chi}
 \sim-\kappa_8\frac{v_8^2}{m_8^2}.
 \label{eq:LaterVEVInducedAlignment}
\end{equation}
For $m_8\sim v_\chi$ and the benchmark high-scale octet/paired-$\Theta$
scale $v_8\sim1.03\times10^{13}\,$GeV, one finds
\begin{equation}
 \left|\frac{\delta v_8^{(\Xi)}}{v_\chi}\right|
 \sim1.06\times10^{-2}|\kappa_8|,
\end{equation}
whereas an electroweak $H$ insertion gives only
$6.05\times10^{-24}|\kappa_8|$.  Because the induced directions are
$SU(3)_A$-orthogonal to the original singlet, their contribution to the
broken-$G(2)$ vector mass begins quadratically; an order-percent induced VEV
therefore changes $m_X^2$ only at the $10^{-4}$ level for order-one reduced
coefficients.  The exact Clebsches and orthogonal masses are model dependent, so these numbers are scaling estimates rather than a precision spectrum.  Their role is only to show that the first allowed parity-preserving bending can naturally be small for the published hierarchy.

\subsubsection{Explicit Wilsonian potential and a nonempty stability cone}
The all-orders tadpole result must now be complemented by a genuine Hessian test.  A transparent point in the allowed Wilsonian coupling space is provided by
the following $\mathbb Z_{2,D}$-even potential below the special-breaking
scale:
\begin{align}
V_D={}&-\frac12\mu_D^2\,\Xi^\dagger\Pi_D\Xi
+\frac{\lambda_D}{4}\left(\Xi^\dagger\Pi_D\Xi\right)^2
+\frac12M_\perp^2\,\Xi^\dagger\Pi_\perp\Xi\nonumber\\
&+\frac{\kappa_D}{2}
\left(\Xi^\dagger\Pi_D\Xi\right)
\left(\Xi^\dagger\Pi_\perp\Xi\right)
+\frac{\lambda_\perp}{4}
\left(\Xi^\dagger\Pi_\perp\Xi\right)^2 .
\label{eq:1539ProjectorPotential}
\end{align}
The potential is manifestly invariant under $G(2)\times SU(3)_A\times U(1)_X$ and under $\Xi\mapsto-\Xi$.  On the single dark $\mathbf{7}$, the invariant polynomial ring is generated by the norm; the resulting accidental $O(7)$ of this restricted block does not introduce extra physical Goldstones, because its six angular directions coincide exactly with $\dim[G(2)/SU(3)_C]=6$ and are the six gauge Goldstones eaten by the broken vectors.
The parent invariants
Eqs.~\eqref{eq:PhiXiQuartic} and \eqref{eq:Xi4Invariant} show explicitly that
the ultraviolet tensor algebra does not lock all of the coefficients entering
this broken-phase potential.  Higher $S$-dressed even invariants simply
renormalize the same Wilsonian coefficients; the gauged
$\mathbb Z_{2,D}$ forbids the odd-$\Xi$ terms.

For
\begin{equation}
\mu_D^2>0,\qquad \lambda_D>0,
\end{equation}
there is a stationary orbit
\begin{equation}
\resizebox{0.98\columnwidth}{!}{$\displaystyle
\Pi_D\langle\Xi\rangle=v_\chi n,\qquad
\Pi_\perp\langle\Xi\rangle=0,\qquad
n^Tn=1,\qquad
v_\chi^2=\frac{\mu_D^2}{\lambda_D}.
$}
\label{eq:1539ProjectorVacuum}
\end{equation}
Choosing $n=e_7$ gives the desired $G(2)\to SU(3)_C$ vacuum.  The exact Hessian
at this point is block diagonal:
\begin{equation}
{\cal H}_{1539}
=
2\mu_D^2\,P_{\rm rad}
+0\cdot P_{\rm G}
+\left(M_\perp^2+\kappa_Dv_\chi^2\right)\Pi_\perp ,
\label{eq:1539FullHessian}
\end{equation}
where
\begin{equation}
\operatorname{rank}P_{\rm rad}=1,\qquad
\operatorname{rank}P_{\rm G}=6,\qquad
\operatorname{rank}\Pi_\perp=1532.
\end{equation}
The six zero eigenvalues are precisely the $G(2)/SU(3)_C$ gauge-orbit directions and are eaten by the six broken vectors.  No additional scalar flat direction is forced by this projector point.  The separated parent therefore need not leave light $\mathbf{1539}_H$ exotics: all $1532$ orthogonal modes can be lifted.  This statement concerns only the new $\mathbf{1539}_H$ sector; the $\mathbf{133}_H$ and $\mathbf{1463}_H$ spectra of the original Class-B construction remain the inputs of Ref.~\cite{MasiE6E7}.  It also does not imply that a complete $\mathbf{1539}_H$ multiplet may be kept light over a large renormalization-group interval.
In the convention $\beta(g)=-b_0g^3/(16\pi^2)$, one real $\mathbf{1539}$ would shift
the one-loop $E_7$ coefficient by \cite{MachacekVaughn1983}
\begin{equation}
 \Delta b_0=-\frac16T(\mathbf{1539})=-54,
\label{eq:1539BetaShift}
\end{equation}
using $T(\mathbf{1539})=324$.  Any four-dimensional realization in which the whole
multiplet propagates appreciably below the parent threshold must therefore
redo the high-scale running; the present scalar-stability theorem requires
only that the non-dark components be positive-mass modes, not that they remain
light.

\paragraph{Running constraint on the non-dark $\mathbf{1539}_H$ modes.}
Vacuum stability alone is not enough: the extra $\mathbf{1539}_H$ states must
also be heavy enough not to spoil the published gauge-coupling evolution.
The full-$E_7$ beta-function shift is only a warning; the sharper question
for the Class-B benchmark is how much of the neutral $\mathbf{1539}_0$ slice
may propagate between $M_\chi$ and $M_{E_6}$.  From
Eq.~\eqref{eq:1539Hs0}, using $T_{G(2)}(\mathbf{7})=1$, $T_{G(2)}(\mathbf{14})=4$,
$T_{G(2)}(\mathbf{27})=9$ and $T_A(\mathbf{8})=3$, $T_A(\mathbf{10})=15/2$, $T_A(\mathbf{27})=27$, one finds \cite{McKayPatera1981}
\begin{align}
T_{G(2)}(\mathbf{1539}_0)&=162,\\
T_A(\mathbf{1539}_0)&=324.
\label{eq:1539NeutralIndices}
\end{align}
The desired real $(\mathbf{7},\mathbf{1}_A)_0$ accounts for one unit of the first index and none
of the second.  The projected Class-B running already contains precisely one
real $G(2)$ fundamental $\chi$ and uses
\begin{equation}
b_{G(2)}=\frac{29}{2},\qquad b_A=\frac{13}{2},
\qquad b_{G(2)}-b_A=8.
\label{eq:1539BaselineClassBSlopes}
\end{equation}
Therefore the separated construction preserves the published one-loop slopes exactly if the only $\mathbf{1539}_H$ descendant active near $M_\chi$ is the desired real $(\mathbf{7},\mathbf{1}_A)_0$.

If all other neutral descendants enter at a common illustrative threshold
$M_\perp$ below the original meeting point, their additional real-scalar
contributions are
\begin{equation}
\Delta b_{G(2)}=-\frac{161}{6},\qquad
\Delta b_A=-\frac{324}{6}=-54,
\end{equation}
so the relative slope above $M_\perp$ becomes
\begin{equation}
(b_{G(2)}-b_A)_{\rm above}=\frac{211}{6}.
\end{equation}
Writing the unperturbed meeting as
$M_0=6.9\times10^{15}\,{\rm GeV}$, one-loop piecewise matching gives \cite{MachacekVaughn1983,MasiE6E7}
\begin{equation}
M_{\rm meet}=M_\perp
\left(\frac{M_0}{M_\perp}\right)^{48/211},
\qquad M_\chi<M_\perp<M_0.
\label{eq:1539ThresholdMeeting}
\end{equation}
Representative values are displayed in Table \ref{tab:1539ThresholdRunning}.
\begin{table}[htbp]
\centering
\small
\caption{One-loop sensitivity of the Class-B exceptional meeting to a common
threshold for the non-dark neutral $\mathbf{1539}_H$ descendants.  The calculation is
a threshold diagnostic, not a precision two-loop fit.}
\label{tab:1539ThresholdRunning}
\begin{tabular}{cc}
\toprule
$M_\perp\ [{\rm GeV}]$ & $M_{\rm meet}\ [{\rm GeV}]$\\
\midrule
$1.0\times10^{15}$ & $1.55\times10^{15}$\\
$3.0\times10^{15}$ & $3.63\times10^{15}$\\
$6.0\times10^{15}$ & $6.19\times10^{15}$\\
$6.5\times10^{15}$ & $6.59\times10^{15}$\\
\bottomrule
\end{tabular}
\end{table}
In this common-threshold diagnostic, retaining the original $6.9\times10^{15}\,$GeV meeting to within about ten percent requires $M_\perp\gtrsim6.1\times10^{15}\,$GeV, while $M_\perp\simeq6.5\times10^{15}\,$GeV leaves only a few-percent shift.  The charged $X=\pm2,\pm4$ branches must likewise be localized near the $E_7$ threshold if the short $U(1)_X$ running interval of Ref.~\cite{MasiE6E7} is to be retained.  This is a constraint on the split ultraviolet spectrum, not a proposal to add a new light multiplet.

The Wilsonian Hessian permits such a hierarchy.  For example, in units
$v_\chi=1$, the same open stable domain contains the point
\begin{equation}
\lambda_D=\lambda_\perp=1,\qquad
\kappa_D=\frac12,\qquad
\mu_D^2=1,\qquad
M_\perp^2=65^2,
\label{eq:1539RGECompatibleStablePoint}
\end{equation}
for which $m_{\rm rad}=\sqrt2$ while
$m_\perp=\sqrt{65^2+1/2}\simeq65.004$.  Identifying $v_\chi$ with the
benchmark dark-breaking scale for orientation places the orthogonal modes near
$6.5\times10^{15}\,$GeV while retaining the dark radial mode near
$M_\chi$.  More general irrep-resolved thresholds need not be degenerate.  The explicit point proves only what is needed here: the local scalar-stability domain and the split-threshold domain required by the published running have a nonempty intersection.  Returning now to the general Hessian, the physical scalar spectrum contains one radial mode with
\begin{equation}
m_{\rm rad}^2=2\mu_D^2
\end{equation}
and $1532$ orthogonal modes with the common sufficient-slice eigenvalue
\begin{equation}
m_\perp^2=M_\perp^2+\kappa_Dv_\chi^2 .
\label{eq:1539PerpMass}
\end{equation}
For the projector slice the local-minimum conditions are
\begin{equation}
\mu_D^2>0,\qquad
\lambda_D>0,\qquad
M_\perp^2+\kappa_Dv_\chi^2>0.
\label{eq:1539HessianConeLocal}
\end{equation}
The common orthogonal eigenvalue characterizes this symmetric projector slice; it is not the most general condition on all $1532$ orthogonal directions.  To state the general result, decompose
\begin{equation}
\Pi_\perp(\mathbf{1539})=\bigoplus_\rho V_\rho\otimes\mathbb F^{m_\rho},
\end{equation}
under $H_s\times U(1)_X$.  Equivariance gives the Schur form \cite{FultonHarris1991}
\begin{equation}
{\cal H}_\perp=
\bigoplus_\rho
\mathbf{1}_{V_\rho}\otimes{\cal M}_\rho^2,
\label{eq:1539GeneralHessianBlocks}
\end{equation}
with real-symmetric or Hermitian multiplicity matrices as appropriate.
Therefore the exact general local-stability condition is
\begin{equation}
m_{\rm rad}^2>0,
\qquad
{\cal M}_\rho^2\succ0
\label{eq:1539GeneralHessianConditions}
\end{equation}
for every orthogonal irrep/multiplicity block $\rho$.
In particular, the two $P_D$-odd neutral singlets of
Eq.~\eqref{eq:1539TwoOddSinglets} form a real $2\times2$ block requiring
$\operatorname{tr}{\cal M}_{(\mathbf{1},\mathbf{1})_0}^2>0$ and
$\det{\cal M}_{(\mathbf{1},\mathbf{1})_0}^2>0$.  At the projector point all orthogonal blocks
equal $m_\perp^2\mathbf{1}$, so the inequalities hold strictly and persist in
a full-dimensional open neighborhood of the irrep-resolved coupling space.

A sufficient boundedness condition for the quartic form is
\begin{equation}
\lambda_D>0,\qquad
\lambda_\perp>0,\qquad
\kappa_D>-\sqrt{\lambda_D\lambda_\perp}.
\label{eq:1539QuarticBoundedness}
\end{equation}

The local-stability region is demonstrably nonempty.  For example, in units where
$v_\chi=1$,
\begin{equation}
\lambda_D=\lambda_\perp=1,\qquad
\kappa_D=\frac12,\qquad
\mu_D^2=M_\perp^2=1
\label{eq:1539ExplicitStablePoint}
\end{equation}
gives
\begin{equation}
m_{\rm rad}^2=2,\qquad
m_\perp^2=\frac32,
\end{equation}
with only the six required gauge Goldstones at zero.  Because every physical
eigenvalue is separated from zero by a finite gap, the irrep-wise inequalities
in Eq.~\eqref{eq:1539GeneralHessianConditions} hold throughout an open
neighborhood.  Equation~\eqref{eq:1539AllOrderTadpoleZero} adds the stronger
statement that these deformations cannot generate an orthogonal tadpole at
the $\chi$ stage: the dark ray remains exactly stationary.

Equations~\eqref{eq:1539AllOrderTadpoleZero} and
\eqref{eq:1539FullHessian} therefore close the scalar-vacuum issue left open by
the orbit diagnostics: conditional on a gauged $\mathbb Z_{2,D}$, the
separated $\mathbf{1539}_H$ admits a locally stable
$G(2)\to SU(3)_C$ dark vacuum.
The scope of the result should be kept explicit.  It is a Wilsonian theorem after the special $E_6\to G(2)\times SU(3)_A$ breaking, not a classification of every coefficient in the most general renormalizable unbroken-$E_7$ potential.
Stationarity is structural and all-orders at the $\chi$ stage; stability
then requires the finite set of irrep-wise positive-matrix conditions in
Eq.~\eqref{eq:1539GeneralHessianConditions}.  The explicit projector point
proves that this domain is nonempty and open.  The independent chiral-color
theorem below, not the scalar Hessian, remains the obstruction to a full
mirror-free ultraviolet parity.

% Alternative visible assignments are outside the fixed architecture and are omitted.
\subsection{Chiral obstruction in the fixed Class-B architecture}
\label{sec:chiralColorParityNoGo}

At this point the scalar-sector question has been answered positively:
conditional on the additional discrete gauge factor, the separated
$\mathbf{1539}_H$ supports an exact \emph{bosonic} remnant after dark
Higgsing.  The final issue is no longer the scalar potential.  It is whether
the same transformation maps every selected light Standard Model field back
into the mirror-free light spectrum.  The symmetry that survives the dark Higgsing is the diagonal
\begin{equation}
P_D=z_DU_B .
\end{equation}
A phase such as $z_D$ cannot change the continuous gauge representation of a matter field.  The only nontrivial action of $P_D$ on color is therefore the action inherited from the $G(2)$ element $U_B$.  This fact turns the remaining problem into a finite equivariance question and allows it to be closed without assuming the canonical $E_7$ charge flip.

\paragraph{Uniqueness of the nontrivial $G(2)$ normalizer action.}
The compact simple algebra $\mathfrak g_2$ has no Dynkin-diagram automorphism \cite{AdamsExceptional,SpringerVeldkamp}, so
\begin{equation}
 \mathrm{Out}(\mathfrak g_2)=1,
 \qquad
 \mathrm{Aut}(\mathfrak g_2)=\mathrm{Inn}(\mathfrak g_2).
 \label{eq:G2AutomorphismLemma}
\end{equation}
Thus any internal invertible unitary zero-form symmetry acting linearly on the $G(2)$ gauge fields and preserving the Yang--Mills kinetic term is represented by an inner $G(2)$ action.  Requiring the residual color algebra to be preserved setwise restricts that representative to the normalizer of $SU(3)_C$.

Let $H_C=SU(3)_C\subset G(2)$ be the stabilizer of the unit vector
$n=e_7\in 7$.  The $H_C$-fixed subspace of the real fundamental is one
dimensional,
\begin{equation}
7^{H_C}=\mathbb R n.
\end{equation}
If $x\in N_{G(2)}(H_C)$, then $x$ preserves this fixed line and therefore
$x n=\pm n$.  For $xn=+n$ one has $x\in H_C$.  For $xn=-n$,
$xU_B^{-1}$ fixes $n$ and hence lies in $H_C$.  Consequently
\begin{equation}
\resizebox{0.98\columnwidth}{!}{$\displaystyle
N_{G(2)}(SU(3)_C)/SU(3)_C\simeq\mathbb Z_2,
\qquad
N_{G(2)}(SU(3)_C)=SU(3)_C\rtimes\langle U_B\rangle .
$}
\label{eq:G2NormalizerUnique}
\end{equation}
Equation~\eqref{eq:G2NormalizerUnique} exhausts the internal options within the fixed $G(2)$ factor: every local zero-form remnant that nontrivially grades the broken $G(2)/SU(3)_C$ vectors belongs to the same color-conjugating class represented by $U_B$.

\paragraph{Visible gauge automorphisms cannot compensate color conjugation.}
An internal invertible unitary zero-form symmetry that preserves the gauge
kinetic algebra induces an automorphism of
$\mathfrak{su}(3)_C\oplus\mathfrak{su}(2)_L\oplus\mathfrak u(1)_Y$.
The color action inherited from the broken $G(2)$ sector is already fixed by the
normalizer class above.  The $SU(2)_L$ doublet is self-conjugate, while an
automorphism of the hypercharge lattice can only preserve or reverse the
charge, $Y\mapsto\pm Y$ \cite{BourbakiLie46,GeorgiLieAlgebras}.  Hence even after allowing a simultaneous visible-gauge
automorphism the quark doublet must transform as
\begin{equation}
 (\mathbf{3},\mathbf{2},+1/6)\longmapsto(\bar{\mathbf{3}},\mathbf{2},\pm1/6).
 \label{eq:FullGaugeAutomorphismQ}
\end{equation}
Neither image occurs as a left-handed light multiplet in the mirror-free Standard Model.  Therefore no simultaneous automorphism of the visible gauge factors can compensate the unique color-conjugating $G(2)$ normalizer action.  Antilinear spacetime charge-parity (CP) is logically different and is tested separately below.

\paragraph{Equivariance of the chiral projector.}
Let $R$ be any complete $G(2)$ representation and let $\Pi$ be a local
spectral/cohomological projector selecting light four-dimensional states.  On
restriction to $SU(3)_C$, $U_B$ exchanges every complex color representation
with its conjugate.  If an exact unitary dark parity exists, equivariance
requires
\begin{equation}
[\Pi,P_D]=0.
\label{eq:ProjectorEquivarianceCondition}
\end{equation}
Because the extra factor $z_D$ acts only by phases on matter, this implies an
isomorphism between the projected paired color sectors,
\begin{equation}
 U_B:\operatorname{Im}\Pi|_{r_C}
 \xrightarrow{\ \simeq\ }
 \operatorname{Im}\Pi|_{\bar r_C},
\qquad
\operatorname{rank}\Pi|_{r_C}
=
\operatorname{rank}\Pi|_{\bar r_C}.
\label{eq:EquivariantProjectorColorBalance}
\end{equation}
A mirror-free Standard Model projector violates this condition.  For the
actual residual parity $P_D=z_DU_B$ one has, independently of any
$E_6$ charge flip,
\begin{align}
Q:(\mathbf{3},\mathbf{2},+1/6)&\longmapsto(\bar{\mathbf{3}},\mathbf{2},+1/6),\nonumber\\
u^c:(\bar{\mathbf{3}},\mathbf{1},-2/3)&\longmapsto(\mathbf{3},\mathbf{1},-2/3),\nonumber\\
d^c:(\bar{\mathbf{3}},\mathbf{1},+1/3)&\longmapsto(\mathbf{3},\mathbf{1},+1/3),
\label{eq:PDOnColoredSM}
\end{align}
while the displayed mirror representations are absent from the selected light
spectrum.  Therefore
\begin{equation}
[\Pi_{\rm ch},P_D]\neq0.
\label{eq:PDChiralProjectorNoGo}
\end{equation}
Assigning additional $z_D$ signs to the fermions cannot change this result,
because a sign cannot turn $\bar{\mathbf{3}}_C$ back into $\mathbf{3}_C$.
This conclusion is independent of the particular Class-B assignment of
Standard Model families among the available \(E_6/E_7\) parent branches.
Reassigning \(Q\), \(u^c\), \(d^c\), or the leptonic ancestors can modify
individual Clebsch coefficients and portal selection rules, but it cannot
alter the equivariance condition above.  Any assignment that retains
\(SU(3)_{G(2)}=SU(3)_C\) and projects onto the same mirror-free
SM spectrum necessarily removes one member of at least one
\(U_B\)-conjugate color pair and therefore breaks the exact parity.  Evading
the obstruction requires changing one of the assumptions of the theorem,
rather than merely reshuffling the Class-B family representations.

\paragraph{Fixed-architecture parity no-go.}
The final equivariance result rests on four assumptions:
\begin{enumerate}[label=(\roman*)]
\item the sequential breaking pattern is retained;
\item the $SU(3)$ daughter of $G(2)$ is identified with QCD color;
\item the low-energy SM spectrum is mirror free;
\item no additional gauge factor changes the action on colored matter.
\end{enumerate}
Under these assumptions, no exact local internal invertible unitary zero-form
symmetry---including a geometric or discrete realization---can both grade the
dark broken-vector sector and commute with the chiral SM projector.  In the
discussion below we refer to this more simply as the absence of an exact
internal dark parity.
Equivalently, under these assumptions the condition $[\widehat P,\Pi_{\rm ch}]=0$ has no solution.  The earlier whole-$G(2)$ color-balance index theorem and the present parity-equivariance theorem are two formulations of the same obstruction: one in index language, the other in symmetry language.

The theorem also closes the apparent heavy-mirror loophole unless one of its assumptions is relaxed.  A mirror-completed spectrum can carry the symmetry before projection, but an exact equivariant projector must keep or remove both members of every $U_B$ pair.  Removing only the mirror breaks the parity; retaining it leaves a mirror state in the light spectrum.  Giving only one member of the pair a large mass while the parity remains unbroken is likewise impossible, because an exact unbroken symmetry relates the corresponding mass eigenstates.

\subsection{Why generalized CP does not rescue dark stability}
\label{sec:1539CPNoRescue}

Generalized CP is the most obvious remaining way to close the conjugated color representation labels, because it combines the internal gauge conjugation with spacetime complex conjugation.  Representation closure, however, is not the same as a stabilizing dark-number selection rule.  Let $\mathcal O_5$ denote either of the dangerous connectors and suppose the combined transformation gives
\begin{equation}
\mathcal O_5\longmapsto-\mathcal O_5^\dagger.
\end{equation}
For
\begin{equation}
\mathcal L_5=C_5\mathcal O_5+C_5^*\mathcal O_5^\dagger,
\end{equation}
CP invariance requires
\begin{equation}
C_5=-C_5^*,
\end{equation}
which permits an arbitrary nonzero purely imaginary coefficient
$C_5=i c$ with $c\in\mathbb R$.  Thus CP invariance relates the connector to its Hermitian conjugate but does not force its Wilson coefficient to vanish; a nonzero purely imaginary coefficient remains allowed.  Generalized CP therefore cannot provide the required dark-number protection.  Independently, the observed quark sector has a nonzero Jarlskog invariant, $J=(3.12^{+0.13}_{-0.12})\times10^{-5}$ \cite{PDGCKM2024}, so exact unbroken CP is not a symmetry of the physical low-energy theory in any case.

The consequence is therefore unambiguous within the fixed Class-B architecture.  The $\mathbf{1539}_H$ removes the scalar-parent obstruction and supplies the required tensor signs and vacuum-splitting freedom, but generalized CP does not repair the remaining chiral-color mismatch.  We therefore do not introduce a different gauge architecture in this paper.  Once the exact-symmetry search is closed, the unresolved questions inside the specified chain are quantitative: the ordering and composition of the $1^{+-}$ and $0^{--}$ poles, their state-resolved ultraviolet form factors, and the still-open compactification coefficients of the allowed higher-dimensional operators.

\subsection{The role of the separated $\mathbf{1539}_H$}

\begin{table*}[t]
\centering
\small
\caption{Principal consequences of the separated-$\mathbf{1539}_H$ construction under the fixed Class-B assumptions.  The first four rows concern the bosonic scalar sector; the last row is the independent chiral closure result.}
\label{tab:1539Status}
\begin{tabular}{p{0.24\textwidth}p{0.70\textwidth}}
\toprule
sector & established result\\
\midrule
representation content & $\mathbf{1539}_H$ contains a neutral $\mathbf{650}_0$ and hence the required $(\mathbf7,\mathbf1_A)_0$ dark direction, allowing $S,H\in\mathbf{1463}_H$ and $\chi\in\mathbf{1539}_H$ to be separated.\\
operator selection & $\mathbf{1463}\not\subset\mathrm{Sym}^2(\mathbf{1539})$; the two $E_7$ parents of the dimension-nine $[3,1]$ $\Xi F^4$ structure restrict to zero on the selected undressed pure-dark component.  With a genuinely gauged or geometric $\mathbb Z_{2,D}$, the odd-$\Xi$ tower is removed before dark Higgsing.\\
broken phase & Dark Higgsing preserves the diagonal bosonic parity $P_D=z_DU_B$.  Unbroken $SU(3)_C$ triality removes every selected-light one-transverse even-$\Xi$ dressing, whereas radial and two-transverse portals remain allowed and require state-dependent matching.\\
vacuum and thresholds & The all-orders orthogonal-tadpole condition vanishes exactly, and the broken-phase Hessian is positive on a nonempty open domain.  Compatibility with the published one-loop Class-B running requires split $\mathbf{1539}_H$ thresholds, with the dark fundamental near $M_\chi$ and the remaining descendants lifted toward the exceptional scales.\\
full chiral theory & The residual $P_D$ cannot be extended to the mirror-free Standard Model projector when $SU(3)_{G(2)}=SU(3)_C$, because $U_B$ necessarily conjugates $\mathbf3_C\leftrightarrow\bar{\mathbf3}_C$.  Generalized charge-parity does not restore a stabilizing number symmetry.\\
\bottomrule
\end{tabular}
\end{table*}
Table~\ref{tab:1539Status} collects the logical consequences of the separated-parent construction.  The $\mathbf{1539}_H$ alternative is a genuine scalar-sector completion, but the chiral-color theorem remains independent of that repair and prevents it from becoming a complete ultraviolet stabilizing symmetry under the fixed assumptions.

The conclusion is therefore sharper than the original single-$\mathbf{1463}$ no-go.
The common $\mathbf{1463}_H$ is indeed the final obstruction of the \emph{scalar-parent}
architecture, and $\mathbf{1539}_H$ removes that obstruction cleanly.  The remaining
chiral-index problem is no longer an open equivariance problem under the fixed
assumptions: the fixed-architecture parity no-go above proves that a
component-resolved projector preserving the intermediate $G(2)$ and the
identification $SU(3)_{G(2)}=SU(3)_C$ cannot also preserve the residual dark
parity on a mirror-free spectrum.  Exact protection therefore requires relaxing at least one of those architectural assumptions.

\subsection{Baryon and lepton number as ultraviolet benchmarks}
\label{sec:BLbenchmark}

The status of dark $G$-parity is usefully compared with the familiar baryon
and lepton numbers.  In the renormalizable Standard Model, baryon number $B$
and total lepton number $L$ arise as accidental global symmetries rather than
as gauge principles.  Quantum electroweak effects violate $B+L$, whereas
$B-L$ is non-anomalous for the SM gauge interactions and becomes
fully gaugeable when one right-handed neutrino is included per family.
Higher-dimensional operators sharpen the distinction: the dimension-five
Weinberg operator violates lepton number, while dimension-six operators can
violate baryon and lepton number
\cite{PDGCKM2024,DevBNV2024,WangWanYou2022}.  Proton longevity therefore
does not imply an exact fundamental $U(1)_B$.

Grand unification generally weakens separate $B$ and $L$ conservation
because quarks and leptons are assembled into common unified multiplets, so
baryon- and lepton-number violating gauge or scalar transitions are usually
allowed \cite{DevBNV2024}.  This is not a theorem about every possible
unification: enlarged gauge structures can gauge $B$ and $L$ separately,
and spontaneous gauge breaking can leave exact discrete remnants
\cite{FileviezWise2010,KraussWilczek1989}.  Likewise, suitable $B-L$
breaking patterns can leave matter parity or related discrete gauge
symmetries \cite{AlvesMatterParity2017}.  The anomalous continuous $B+L$
symmetry of the Standard Model can also retain nontrivial discrete
subgroups, while several familiar simple GUT embeddings reduce the
compatible remnant essentially to fermion parity
\cite{WangWanYou2022}.  Exact baryon- or lepton-related discrete symmetries
are therefore model dependent, not a universal consequence of grand
unification.

The Class-B assignment contains a particularly useful $B-L$ structure.
For the selected SM family including $N^c$,
Table~\ref{tab:ClassB} gives
\begin{equation}
	\begin{aligned}
		X(Q,u^c,d^c,L,e^c,N^c)
		&=(+1,-1,-1,-3,+3,+3)\\
		&=3(B-L)(Q,u^c,d^c,L,e^c,N^c).
	\end{aligned}
	\label{eq:Xequals3BLselected}
\end{equation}
Thus the $U(1)_X$ charge inherited from
$E_7\to E_6\times U(1)_X$ realizes the usual anomaly-free
$3(B-L)$ assignment on the retained chiral family.  In this restricted
sense the Class-B construction gauges $B-L$ on the light spectrum, rather
than merely reproducing it as an accidental global number.  Equation
\eqref{eq:Xequals3BLselected} is nevertheless a projected identity on the
selected branches, not a global identification of the $E_7$ generator $X$
with $3(B-L)$ on every component of an exceptional representation; the
possible $B-L$ embeddings in $E_6$ are themselves nontrivial
\cite{HaradaE6BL2003}.

Separate $B$ and $L$ are not intrinsic symmetries of the unbroken $E_7$
parent used here.  The selected quark and lepton branches belong to the same
irreducible matter macro-parent $\mathbf{27664}_{E_7}$.  By Schur's lemma,
an additional internal symmetry commuting with unbroken $E_7$ acts as a
common phase on that irreducible multiplet and therefore cannot reproduce
the distinct baryon or lepton charges of its quark and lepton descendants \cite{FultonHarris1991}.
The $E_7$ center gives an even sharper statement for the actual field
content.  Since the $\mathbf{56}$ is center odd,
$\mathbf{27664}\subset\Lambda^3\mathbf{56}$ is center odd, whereas
$\mathbf{133}_H$, $\mathbf{1463}_H\subset\mathrm{Sym}^2\mathbf{56}$ and
$\mathbf{1539}_H\subset\Lambda^2\mathbf{56}$ are center even.  Hence the
nontrivial $Z(E_7)\simeq\mathbb Z_2$ element acts on the present spectrum
exactly as fermion parity $(-1)^F$ and supplies no additional baryon,
lepton, or dark-number selection rule \cite{Slansky1981,CacciatoriE7MagicSquare}.

The Abelian breaking also leaves a useful residual check.  In the present
chain the $X$-charged vacuum directions are
$X_{\Theta_\pm}=\pm3$, while the other vacuum directions relevant here are
$X$ neutral.  At the level of the $U(1)_X$ charge lattice \cite{KraussWilczek1989},
\begin{equation}
	U(1)_X\ \xrightarrow{\ \langle\Theta_\pm\rangle\ }\ 
	\mathbb Z_3,
	\qquad
	z_{BL}=\exp\!\left(\frac{2\pi iX}{3}\right).
	\label{eq:UXtoZ3}
\end{equation}
Using Eq.~\eqref{eq:Xequals3BLselected}, the same generator acts on the
selected light family as
%\begin{equation}
\begin{align}
	z_{BL}
	&\equiv
	\exp\!\left(\frac{2\pi iX}{3}\right)
	=
	\exp\!\left[2\pi i(B-L)\right],
	\nonumber\\
	Q&\longmapsto \omega\,Q,
	\qquad
	(u^c,d^c)\longmapsto
	\omega^{-1}(u^c,d^c),
	\nonumber\\
	(L,e^c,N^c)&\longmapsto(L,e^c,N^c),
	\qquad
	\omega=e^{2\pi i/3}.
	\label{eq:BLZ3action}
\end{align}
%\end{equation}
This order-three stabilizer must, however, be interpreted with the global
form of the exceptional subgroup in mind:
\begin{equation}
	E_7\supset \frac{E_6\times U(1)_X}{\mathbb Z_3}.
	\label{eq:E7E6U1Global}
\end{equation}
The order-three $U(1)_X$ element is therefore locked to the $E_6$ center
rather than defining an additional independent zero-form factor
\cite{CacciatoriE7MagicSquare}.  This also makes its dark sector blindness
structural in the present special embedding.  Under
$E_6\to G(2)\times SU(3)_A$, a central element of the $E_6$ image must
commute with both subgroup factors.  Since
\begin{equation}
	Z(G(2))=1,
\end{equation}
the nontrivial order-three center action cannot reside in the dark $G(2)$
factor; it is carried by the $SU(3)_A$ side of the embedding.  Thus no
discrete remnant inherited in this way can grade the $G(2)$ gauge sector.
The explicit charge assignment is consistent with this structural result:
the dark Higgs direction
$\chi\sim(\mathbf7,\mathbf1_A)_0$ and the $G(2)$ gauge fields have $X=0$.

Independently of the global-group identification, the same remnant does not
protect the proton because
\begin{align}
	&X(QQQL)=1+1+1-3=0,\;\\
	&X(u^cu^cd^ce^c)=-1-1-1+3=0.\nonumber 
	\label{eq:ClassBBLProtonOperators}
\end{align}
The standard $\Delta B=\Delta L$ proton-decay structures are therefore
neutral under the $B-L$-like charge.  Their suppression in Class B must
come from the special embedding, mediator structure and high scales, not
from exact baryon number.

The comparison with dark $G$-parity is consequently structural but not
quantitatively symmetric.  At the renormalizable level, $B$ and $L$ are
accidental SM numbers, with $B+L$ broken nonperturbatively and
higher-dimensional operators violating the corresponding continuous
selection rules.  In the isolated $G(2)$ gauge--Higgs theory the dark
grading is likewise exact at the renormalizable level; the independent
pure-dark odd on-shell operator basis is empty through dimension seven and
first appears at dimension nine.  In the complete Class-B theory, however,
visible-sector dimension-five connectors already violate the would-be exact
dark grading, so the dimension-nine statement applies only to the purely
dark operator sector.

The numerical difference explains why the same ultraviolet logic is mild
for ordinary baryons but severe for an ultraheavy dark state.  For the
common reference scale $M_*=M_{E_6}=6.9\times10^{15}\,\mathrm{GeV}$, a
schematic dimension-six proton-decay normalization may be written as \cite{DevBNV2024}
\begin{equation}
	\tau_p^{(6)}
	\sim
	1.3\times10^{42}\ {\rm s}\,
	\left(\frac{0.04}{\alpha_{\rm GUT}}\right)^2
	\left(\frac{M_*}{6.9\times10^{15}\ {\rm GeV}}\right)^4,
	\label{eq:ProtonDarkUVComparisonProton}
\end{equation}
where $\alpha_{\rm GUT}=0.04$ is only an illustrative normalization and
channel-dependent hadronic and renormalization factors are not included.
By contrast, Eq.~\eqref{eq:d5NDAwidth}, evaluated at
$m_B=1.71\times10^{14}\,\mathrm{GeV}$ with unit Wilson coefficient and
unit state-matching factor, gives
\begin{equation}
	\tau_B^{(5)}\big|_{|c_5|^2\kappa_{5,J}^{\rm eff}=1}
	\simeq7.6\times10^{-11}\ {\rm s}.
	\label{eq:ProtonDarkUVComparisonDark}
\end{equation}
At the reference value $\alpha_{\rm GUT}=0.04$, the two normalizations differ
by approximately $1.7\times10^{52}$.  This does not predict the physical
dark lifetime: the full result remains controlled by
$|c_5|^2\kappa_{5,J}^{\rm eff}$, and Eq.~\eqref{eq:d5LifetimeBound}
quantifies the suppression required for cosmological longevity.  It instead
isolates why the analogy has a sharp limit.  The dangerous dark connector
occurs already at dimension five and acts on a state of order
$10^{14}\,\mathrm{GeV}$, whereas conventional proton decay is dimension six
and acts on a GeV-scale state.

The ultraviolet failure of exact dark $G$-parity should therefore be read
in the same symmetry hierarchy familiar from baryon and lepton number: a
useful low-energy selection rule need not be a fundamental symmetry of the
unified theory.  
%What is lost in the dark case is an absolute stability
%theorem, and the very large mass of the candidate makes that loss
%phenomenologically much more consequential.  
A future ultraviolet completion with a genuine discrete gauge remnant could restore exact
protection, just as residual gauge parities can protect selected sectors in
other unified constructions
\cite{KraussWilczek1989,AlvesMatterParity2017}.

\section{Conclusions and outlook}

The aim of this work was to follow the proposed DM state from the
broken $G(2)$ theory up to its Class-B $E_6/E_7$ embedding.  Three questions
were central throughout the discussion: the identity and spin of the physical
gauge-invariant state, the scales controlling its mass and lifetime and the
survival of the protecting dark $G$-parity in the ultraviolet theory.  The
answer is necessarily layered.  The spectroscopy, the possible metastability,
the scalar-parent completion and the final chiral obstruction are separate
parts of the same construction and should not be compressed into a single
stability statement.

\paragraph{Odd-state spectroscopy.}
The first conclusion is a change in the identity of the protected candidate:
the scalar $0^{++}\sim X\bar X$ is even and therefore is not the protected
odd state.  The two relevant sectors are a
$1^{+-}$ two-vector/glue block and a $0^{--}$ three-vector block, described by
separate generalized-eigenvalue problems.  The controlled heavy-particle and decoupled-glue limits separately favor the vector: $M_{3X}>M_{2X}$ in the Coulombic
Hall--Post problem and in the linear $\Delta$ model, a genuine $Y$ junction
cannot lower the three-body floor, and in the decoupled pure-$SU(3)$ daughter the known glue spectrum places the
$1^{+-}+0^{++}$ scalar continuum below the central single-particle-like
$A_1^{--}$ level.  The intermediate-$x$ combinations of pure-glue and heavy-vector
inputs are reference interpolations rather than controlled full-theory masses.  This is not a proof that $1^{+-}$ is the ground state.  A
sufficiently strong near-threshold or subthreshold $0^{--}$ spectral
attraction remains possible, and the exact Feshbach formulation shows that
this is the specific loophole a scattering-aware lattice GEVP must test.
Accordingly, the controlled evidence favors the $1^{+-}$ channel but does not eliminate the $0^{--}$ alternative; both spin assignments must be retained until the scattering problem is solved nonperturbatively.

\paragraph{Ultraviolet stability and lifetime.}
The second conclusion is that spectroscopy does not determine the lifetime:
mass ordering and ultraviolet stability are separate questions.  The selected light kernel obeys
$P_{\rm light}Y_\chi P_{\rm light}=0$, but the parent theory admits the
nonzero dimension-five operators $Qu^c\chi H$ and
$Qd^c\chi H^\dagger$.  The naive factorized selected light-field contribution involves the
color-singlet radial $\chi_1$ direction and has zero direct vacuum-to-odd-pole
overlap.  The full connected decay amplitude nevertheless remains
interaction dependent; all heavy, gauge and nonfactorizing confined matching
is summarized by the state-dependent $\kappa^{\rm eff}_{5,J}$.  Thus the parent operator proves that the grading is
not exact, but it does not by itself fix a universal lifetime or a robust
ordering between the $J=0$ and $J=1$ decay form factors.

The purely dark operator basis contains a stronger exact selection rule than a
tensor-product count suggests.  The scalar $\chi F^3$ family vanishes
identically because the unique internal $\mathbf{7}$ lies in
$\mathrm{Sym}^3\mathbf{14}$ whereas the Lorentz scalar of three field strengths
selects $\Lambda^3\mathbf{14}$.  The algebraically nonzero derivative--Higgs
$d=7$ representative is in turn IBP/EOM redundant, and the complete on-shell
contact-term classification gives no independent pure-dark odd operator at
$d=3,5,7$.  The first physical basis occurs at $d=9$.  Its unique one-$\chi$
member is a mixed-helicity $[3,1]$ $\chi F^4$ operator with a unique exact
$G(2)$ hook tensor.  An exact broken-phase projection adds an important
qualification: after $\chi\to v_\chi e_7$ the hook vanishes for four
unbroken $SU(3)_C$ adjoints and also for one coset plus three unbroken adjoints;
its first nonzero component contains two broken-vector field strengths.  It is
therefore a direct heavy-vector annihilation operator rather than a pure-glue
contact.  At $M_*=M_{E_6}$ and the three-vector reference mass, its
state-resolved NDA lifetime normalization is
$\tau_0\simeq3.24\times10^{-21}$ s before the state-resolved factor is
included.  This is roughly six orders of magnitude longer than the discarded
$\chi F^3$ estimate, but it still requires substantial suppression for a
cosmologically long-lived generic unpaired EFT.  The other $d=9$ multi-scalar
classes have vanishing all-VEV pure-daughter limits but can contribute through
mixed-coset or non-radial components; their form factors are not fixed by the
one-$\chi$ benchmark.  The previously quoted
restricted-loop lifetime based on $(H^\dagger H)\chi F^3$ is withdrawn; its
physical mixed-sector replacement requires a new basis and heavy-spectrum
matching.

\paragraph{The separated $\mathbf{1539}_H$ completion.}
The separated-parent construction has one clear purpose: remove the
scalar-parent obstruction without changing the visible Class-B assignment.
Separating the dark scalar into $\mathbf{1539}_H$ does this: it removes the
specific common-$\mathbf{1463}_H$ scalar obstruction.  The antisymmetric-square parent contains
a unique neutral $(\mathbf{7},\mathbf{1}_A)_0$ direction and reverses the relevant canonical
charge-flip sign.  The independently recomputed square still gives
$\mathbf{1539}\not\subset\Lambda^2 \mathbf{1539}$, but the physical operator question is now
controlled by the $d=9$ hook rather than by the EOM-redundant derivative tensor.
The exact parent plethysm gives two $\mathbf{1539}_H$ channels in
$\mathbb S_{(3,1)}\mathbf{133}$ (and one for $\mathbf{1463}_H$), but the final component
restriction closes the undressed separated-parent question: both $\mathbf{1539}_H$ tensors
vanish on $(\mathbf{7},\mathbf{1}_A)_0\otimes(\mathbf{14},\mathbf{1}_A)^4$ because the canonical
$E_7$ charge flip makes the scalar odd and fixes the dark gauge algebra.  The same
argument is dimension independent and removes every undressed pure-dark operator with one
selected $\Xi_D$ and otherwise only dark gauge-sector insertions.  This closes
the undressed pure-dark question, but not the full ultraviolet one:
$P_7^{(0)}$-odd dressings are still allowed in principle.  In particular the special $S$ VEV satisfies
$v_S/M_*=O(1)$ in the benchmark, so an allowed $S$-dressed parent would not be
parametrically suppressed; its actual component-resolved existence remains open.  A genuine
gauged/geometric $\mathbb Z_{2,D}$ is therefore still needed for exact protection of the
complete dressed odd-$\Xi$ tower.  The corresponding even-$\Xi$ operator analysis sharpens the
result rather than overturning it.  Dimension-six even-$\Xi$ portals exist,
but after one dark VEV the selected-light coefficient of a single transverse
$\mathbf{3}_C\oplus\bar{\mathbf{3}}_C$ fluctuation vanishes for every $E_7$ parent contraction by
unbroken $SU(3)_C$ triality.  Radial and two-transverse portals survive; their
decay contribution is indirect and is parameterized by
$\kappa_{6,J}^{\rm eff}$ rather than by a unit constituent overlap.  At the
$\chi$ stage,
$\operatorname{Sym}^{2k+1}7$ contains a unique fundamental and no $1$, $\mathbf{14}$,
or $\mathbf{27}$, which gives an all-orders orthogonal-tadpole zero.  A general
irrep-resolved Hessian can be positive on a nonempty open domain and later
visible VEVs can bend the dark ray only into neutral $G(2)$-fundamental
components whose color-singlet directions are even under the residual
$P_D=z_DU_B$.  These are bosonic scalar-sector results and are conditional on
a genuinely gauged or geometric $\mathbb Z_{2,D}$, whose ultraviolet origin
is not specified here.

The scalar repair is compatible with the published one-loop Class-B running only in a split-threshold realization.  The dark $(\mathbf{7},\mathbf{1}_A)_0$ can
replace the original real $G(2)$ fundamental at $M_\chi$ without changing the
published slopes, whereas the remaining neutral $\mathbf{1539}_H$ modes must be lifted
close to $M_{E_6}$ and the $X\neq0$ branches localized near $M_{E_7}$.  The
explicit diagnostic in Eq.~\eqref{eq:1539ThresholdMeeting} places the common
neutral threshold above approximately $6.1\times10^{15}\,$GeV if the original
meeting is to remain within ten percent.  The Wilsonian scalar potential has an
open stable region containing such a hierarchy, so the renormalization-group evolution (RGE) check constrains
rather than excludes the $\mathbf{1539}_H$ alternative.

\paragraph{The remaining obstruction is chiral.}
After the scalar repair, the remaining obstruction to an exact dark $G$-parity is chiral.  Since
$7^{SU(3)_C}=\mathbb R e_7$,
\begin{equation}
N_{G(2)}(SU(3)_C)/SU(3)_C\simeq\mathbb Z_2,
\end{equation}
and the unique nontrivial normalizer class is represented by $U_B$, which
conjugates $\mathbf{3}_C\leftrightarrow\bar{\mathbf{3}}_C$.  Any exact internal dark parity that nontrivially grades the broken vectors
therefore induces the same color conjugation on matter.  A mirror-free SM projector cannot commute
with this action,
\begin{equation}
[\Pi_{\rm ch},P_D]\neq0.
\end{equation}
Consequently, if the sequential breaking pattern of Ref.~\cite{MasiE6E7} is kept,
$SU(3)_{G(2)}=SU(3)_C$, no new gauge factor is introduced and the low-energy
spectrum is mirror free, no exact internal dark $G$-parity survives in the
ultraviolet theory.  Generalized CP closes representation labels but does not
supply an independent stabilizing number parity.  Under these fixed assumptions the ordinary exact internal-parity search is
therefore closed.  No claim is made here to classify
exotic non-invertible symmetries; none is currently identified in the specified
construction.

\paragraph{Literal Class-B benchmark and interpretation.}
The literal Class-B benchmark must likewise be kept distinct from the intermediate-$x$ spectroscopy deformation.  Because the daughter
$SU(3)_C$ is ordinary QCD, the lightest odd hadronic scales are hadronic, not
$O(10^{14}\,{\rm GeV})$; the ultraheavy $2X/3X$ levels are not the lightest
odd relics of that benchmark.  Their production abundance is not predicted in
this paper and ordinary thermal freeze-out at the quoted ultraheavy mass is
not a viable mechanism.   The intermediate-$x$ analysis
should instead be read as a controlled deformation that isolates the
spin/mass problem of the broken gauge--Higgs system.  Accordingly, within the
literal fixed Class-B benchmark analyzed here there is no identified
ultraheavy symmetry-protected DM relic: the microscopic ultraheavy
candidate is defined only in the deformed confining regime.  Whether a
hadronic-scale odd state of the QCD daughter could play a cosmological role is
a different QCD-coupled problem and is not claimed in this work.

\paragraph{Open calculations.}
The remaining calculations are well defined and directly tied to the dark matter
question.  On the nonperturbative side one needs separate scattering-aware
$1^{+-}$ and $0^{--}$ correlation matrices, including $1^{+-}0^{++}$
operators in the scalar block, to determine the lowest pole and the mass
function $\mu_2(x)$.  On the ultraviolet side one needs the state-resolved
matching factors $\kappa^{\rm eff}_{5,J}$ and $\kappa^{\rm eff}_{6,J}$,
the pole form factors and the $P_7^{(0)}$-breaking dressed-parent matching
that can regenerate the $d=9$ low-energy hook after ultraviolet VEV
insertions.  These calculations determine spectroscopy and metastability;
they do not reopen the fixed-architecture chiral-equivariance theorem.

\paragraph{Artificial-intelligence-assisted computational methodology.}
Artificial intelligence (AI) tools by OpenAI ChatGPT
(GPT-5.6 Sol) were used as research aids for code debugging and numerical consistency checks. No AI-generated claim was retained solely on that basis: the author selected the assumptions, specified and directed the calculations and rechecked every retained correction through analytic derivations, executable certificates, independent representation-theory constructions, or direct
comparison with cited sources. The author takes full responsibility for the scientific content and final wording.

\paragraph{Data and software availability.}
The numerical certificates supplied with the submission include a complete
on-shell dark-operator analysis through $d=9$, an independent verification of the
permutation ranks over three finite-field primes entering the multi-scalar counts, an exact IBP/EOM
certificate for the algebraically nonzero but redundant derivative--Higgs
representative, an exact rational $G(2)$ Young-projected witness for the
nonzero $[3,1]$ $\chi F^4$ tensor and an exact broken-phase restriction showing
that this hook first appears with two coset field strengths after
$\chi\to v_\chi e_7$.  Independent $E_7$ Weyl-character scripts recompute the
complete symmetric and exterior squares of the $\mathbf{1463}$ and $\mathbf{1539}$, while a
separate Weyl-alternant calculation gives the hook multiplicities in
Eq.~\eqref{eq:E7HookInvariantCountsUV}. Independent Wolfram/LieART scripts are also archived; that runs
reproduce the $\mathbf{133}$ tensor-square checks, all symmetric-power
dimensions, the $39698505$-dimensional hook character and the
$\mathbf{1463}/\mathbf{1539}$ multiplicities $1/2$, while explicitly checking
the LieART dimensions and Dynkin labels of both target irreps.  A separate
support-boundary certificate records the $3025/28911/162417$ adjoint-weight
sumsets and the $126$ extremal cancellations yielding the $162291$-weight
hook support.  The final component certificate contains an explicit Darboux-basis matrix realization of
$P_7^{(0)}$ on the neutral carriers, reconstructs $G(2)$ exactly, checks
$P_7^{(0)}\Xi_D=-\Xi_D$ and
$\operatorname{Ad}_{P_7^{(0)}}T_A=T_A$ on all fourteen dark generators, and
verifies the rank-zero theorem in Eq.~\eqref{eq:1539HookRestrictionRankZero};
a separate exact $E_6$ character check finds
$\dim\operatorname{Hom}_{E_6}(\mathbf{650},\mathbb S_{(3,1)}\mathbf{78})=3$,
showing that the zero is not an intermediate-$E_6$ multiplicity absence.
Their exact outputs, together with the figure-generation inputs, bibliography
and supporting source files, are included with the submission package and are
available from the author upon reasonable request.  The large restricted-mass and whole-branch rank statements are additionally
supported by the analytic block-eigenvalue and color-balance witnesses displayed
in the text; their machine outputs from the companion construction are retained
with the author reproducibility record.

\appendix

\section{Illustrative two-state Ritz stress-test values}
\label{app:RitzStressTable}
This appendix collects the numerical interpolation used only to stress-test the inversion criterion of Eq.~\eqref{eq:ScalarWinCriticalMixing}.  The values should not be read as first-principles masses outside the controlled Coulomb domain $x\lesssim0.01694$; their purpose is to quantify how large the scalar-channel mixing would have to become in the illustrative two-state Ritz model.  Table~\ref{tab:RitzStressValues} uses
$M_2=1.99971319$, $M_3=2.99972612$, $k_1=6.065$,
$k_0=k_0^{\rm mix}=8.3082$, and $\delta_1=\alpha_s$.
\begin{table}[htbp]
\centering
\caption{Illustrative two-state Ritz stress-test values.  The displayed precision is intentionally limited because the legacy input entering $k_0^{\rm mix}$ is not a precision continuum determination.}
\label{tab:RitzStressValues}
\resizebox{0.73\columnwidth}{!}{%
\begin{tabular}{ccccc}
\toprule
$x$ & $\mu_1^-$ & $\mu_0^-(\delta_0=\alpha_s)$ &
$|\delta_0|_{\rm crit}$ & $|\delta_0|_{\rm crit}/\alpha_s$\\
\midrule
0.05 & 0.3029 & 0.4152 & 0.551 & 21.7\\
0.10 & 0.6060 & 0.8305 & 0.734 & 28.9\\
0.20 & 1.2122 & 1.6612 & 0.896 & 35.3\\
0.25 & 1.5149 & 2.0764 & 0.914 & 36.0\\
0.30 & 1.8160 & 2.4912 & 0.895 & 35.2\\
0.3297 & 1.9743 & 2.7369 & 0.886 & 34.9\\
0.40 & 1.9982 & 2.9977 & 1.152 & 45.3\\
\bottomrule
\end{tabular}%
}
\end{table}

\section{Exterior-cube bookkeeping for the $\mathbf{27664}$ parent}
\label{app:27664-bookkeeping}

This appendix records the exterior-cube decomposition used in the complete-parent index test, so that the charge-pairing statements in the main text can be checked independently \cite{Slansky1981,LieART,ExceptionalReductions}.  Write
\begin{equation}
\resizebox{0.98\columnwidth}{!}{$\displaystyle
\mathbf{56}=A\oplus B\oplus c\oplus d,
\qquad
A=\mathbf{27}_{-1},\quad
B=\overline{\mathbf{27}}_{+1},\quad
c=\mathbf{1}_{+3},\quad d=\mathbf{1}_{-3}.
$}
\end{equation}
Then
\begin{equation}
\Lambda^3 \mathbf{56}=\mathbf{27664}\oplus\mathbf{56}.
\end{equation}
The nonzero charge sectors of
$\Lambda^3(A\oplus B\oplus c\oplus d)$ are listed in Table~\ref{tab:Lambda3ChargeSectors}.
\begin{table}[htbp]
\centering
\caption{Charge sectors in the exterior-cube bookkeeping of the $\mathbf{56}$.}
\label{tab:Lambda3ChargeSectors}
\resizebox{0.66\columnwidth}{!}{%
\begin{tabular}{c c c}
\toprule
term & $U(1)_X$ charge & dimension\\
\midrule
$\Lambda^3A$ & $-3$ & $\mathbf{2925}$\\
$\Lambda^3B$ & $+3$ & $\mathbf{2925}$\\
$\Lambda^2A\otimes B$ & $-1$ & $\mathbf{351}\times\mathbf{27}$\\
$A\otimes\Lambda^2B$ & $+1$ & $\mathbf{351}\times\mathbf{27}$\\
$\Lambda^2A\otimes c$ & $+1$ & $\mathbf{351}$\\
$\Lambda^2A\otimes d$ & $-5$ & $\mathbf{351}$\\
$\Lambda^2B\otimes c$ & $+5$ & $\mathbf{351}$\\
$\Lambda^2B\otimes d$ & $-1$ & $\mathbf{351}$\\
$A\otimes B\otimes c$ & $+3$ & $729$\\
$A\otimes B\otimes d$ & $-3$ & $729$\\
$A\otimes c\otimes d$ & $-1$ & $\mathbf{27}$\\
$B\otimes c\otimes d$ & $+1$ & $\mathbf{27}$\\
\bottomrule
\end{tabular}%
}
\end{table}
Using
\begin{equation}
\Lambda^3 \mathbf{27}=2925,
\qquad
\mathbf{27}\otimes\overline{\mathbf{27}}=\mathbf{1}\oplus\mathbf{78}\oplus\mathbf{650},
\end{equation}
the $\pm3$ sectors contain
\begin{equation}
\mathbf{1}\oplus\mathbf{78}\oplus\mathbf{650}\oplus\mathbf{2925}.
\end{equation}
Subtracting the singlets $1_{\pm3}$ belonging to the $\mathbf{56}$ leaves
\begin{equation}
(\mathbf{78}\oplus\mathbf{650}\oplus\mathbf{2925})_{-3}
\oplus
(\mathbf{78}\oplus\mathbf{650}\oplus\mathbf{2925})_{+3}.
\end{equation}
Similarly the $\pm1$ sectors lose one $\mathbf{27}$ or $\overline{\mathbf{27}}$ when the
$\mathbf{56}$ is subtracted.  The result is the charge-paired decomposition quoted in
the main text:
\begin{align}
\mathbf{27664}\to{}&
\overline{\mathbf{351}}_{A,-5}\oplus\mathbf{351}_{A,+5}\nonumber\\
&\oplus(\mathbf{78}\oplus\mathbf{650}\oplus\mathbf{2925})_{-3}\nonumber\\
&\oplus(\mathbf{78}\oplus\mathbf{650}\oplus\mathbf{2925})_{+3}\nonumber\\
&\oplus\mathcal R_{-1}\oplus\overline{\mathcal R}_{+1},
\end{align}
with
\begin{equation}
\mathcal R=\mathbf{27}\oplus2(\mathbf{351})\oplus\mathbf{1728}\oplus\mathbf{7371}.
\end{equation}
The exact dimension identity is
\begin{equation}
\resizebox{0.98\columnwidth}{!}{$\displaystyle
2(\mathbf{351})+2(78+650+2925)
+2(27+2\cdot351+1728+7371)=27664.
$}
\end{equation}
This establishes representation-level charge pairing, but
Sec.~\ref{sec:E7lift-nogo} shows why that statement must not be confused with
invariance of the selected special-$\mathbf{650}$ vacuum.

\section{Finite $E_7$ charge-flip certificate for the common-$\mathbf{1463}_H$ obstruction}
\label{app:special-no-go-certificate}

A comparison of the $\mathbf{27}$ and $\overline{\mathbf{27}}$ branchings alone does not determine the sign of the special $\mathbf{650}$ vacuum under the $E_6$ outer involution.  This appendix supplies the finite carrier-level certificate used in the main text.

First, restrict the special branching
\begin{equation}
\mathbf{27}=(\mathbf{7},\mathbf{3})\oplus(\mathbf{1},\overline{\mathbf{6}})
\end{equation}
to the principal $A_1\subset A_2$.  Since $\mathbf{3}\to\mathbf{3}$ and
$\overline{\mathbf{6}}\to\mathbf{5}\oplus\mathbf{1}$,
\begin{equation}
\mathbf{27}\to(\mathbf{7},\mathbf{3})\oplus(\mathbf{1},\mathbf{5})\oplus(\mathbf{1},\mathbf{1})=\mathbf{26}\oplus\mathbf{1}.
\end{equation}
Similarly $\mathbf{8}\to\mathbf{3}\oplus\mathbf{5}$ gives
\begin{widetext}
\begin{align}
\mathbf{78}&=(\mathbf{14},\mathbf{1})\oplus(\mathbf{1},\mathbf{8})\oplus(\mathbf{7},\mathbf{8})\\
&\to[(\mathbf{14},\mathbf{1})\oplus(\mathbf{1},\mathbf{3})\oplus(\mathbf{7},\mathbf{5})]
\oplus[(\mathbf{1},\mathbf{5})\oplus(\mathbf{7},\mathbf{3})]=\mathbf{52}\oplus\mathbf{26}.
\end{align}
\end{widetext}
These are the standard $E_6\downarrow F_4$ branchings \cite{CacciatoriE6Geometry,Slansky1981}.  Together with the
$A_2$--$G(2)$ centralizer duality \cite{Rubenthaler2008}, this identifies an
outer representative with $F_4$ fixed and $H_s$ setwise stable.

The decisive sign comes from the symplectic $\mathbf{56}$ of $E_7$ \cite{CacciatoriE7MagicSquare,ExceptionalReductions}.  Write a vector as
$(v,\phi,a,b)$ in
$\mathbf{27}_{-1}\oplus\overline{\mathbf{27}}_{+1}\oplus\mathbf{1}_{+3}\oplus\mathbf{1}_{-3}$.  In a Darboux
basis a charge-flip representative can be chosen as
\begin{equation}
P(v,\phi,a,b)=(J^{-1}\phi,-Jv,b,-a),
\end{equation}
which preserves the symplectic form.  Therefore on the neutral symmetric cross
term, identified with $A\in\mathrm{End}(\mathbf{27})$,
\begin{equation}
P(A)=-J^{-1}A^TJ.
\end{equation}
In an $F_4$-adapted basis $J=1$, yielding Eq.~\eqref{eq:E7MinusTranspose}.

With $\mathbf{27}=A\oplus B$, $\dim A=21$, $\dim B=6$, define the block projectors
$\Pi_A,\Pi_B$.  The unique traceless special singlet is
\begin{equation}
S=2\Pi_A-7\Pi_B,
\qquad S^T=S,
\end{equation}
whereas a representative of $\chi\sim(\mathbf{7},\mathbf{1}_A)$ is
\begin{equation}
\chi_n=J_n\otimes\mathbf{1}_3,
\qquad (J_n)_{ij}=O_{nij},
\qquad \chi_n^T=-\chi_n.
\end{equation}
Thus $P(S)=-S$ and $P(\chi)=+\chi$.  The explicit inner element $U_B$ fixes $S$
and flips the chosen $\chi$ direction, so $U_BP$ is odd on both.

As a dimension cross-check, $P=-T$ on $\mathrm{End}(\mathbf{27})$ has an even
antisymmetric subspace of dimension $27\cdot26/2=351$ and an odd symmetric
subspace of dimension $27\cdot28/2=378$.  The adjoint $\mathbf{78}$ decomposes under
the same involution as $\mathbf{52}_+\oplus\mathbf{26}_-$, and the identity singlet is odd.
Subtracting $\mathbf{1}\oplus\mathbf{78}$ therefore leaves
\begin{equation}
\mathbf{650}=\mathbf{299}_+\oplus\mathbf{351}_-
\end{equation}
under the canonical charge-flip action.  The special singlet lies in the odd
part and the $\chi$ direction in the even part, exactly as above.

The resulting no-go is therefore a $\mathbf{650}$ sign obstruction, not a statement that the special subgroup is moved to a second inner-conjugacy class.

\section{Cross-block certificate for the reversed $\mathbf{1539}_H$ charge-flip sign}
\label{app:1539-sign-certificate}
The sign reversal in the separated parent is most transparent on the neutral cross carrier.  Write the charged part of the $E_7$ fundamental as
\begin{equation}
 \mathbf{56}\supset U\oplus U^*,\qquad U=\mathbf{27}_{-1},\quad U^*=\overline{\mathbf{27}}_{+1}.
\end{equation}
The neutral cross carrier is $U\otimes U^*\simeq\mathrm{End}(\mathbf{27})$.  Let
$M=u\otimes\bar v$ denote its matrix representative and let the canonical
charge flip exchange $U\leftrightarrow U^*$.  The cross term in
$\mathrm{Sym}^2 \mathbf{56}$ is $u\odot\bar v$, whereas that in $\Lambda^2 \mathbf{56}$ is
$u\wedge\bar v$.  Under $U\leftrightarrow U^*$ the ordered factors are
interchanged; restoring the canonical order gives no extra sign for
$u\odot\bar v$ but gives one extra minus sign for
$u\wedge\bar v$.  Relative to the phase convention already fixed by the
$\mathbf{1463}_H$ certificate in Appendix~\ref{app:special-no-go-certificate}, this
single wedge-exchange minus is exactly what reverses the transpose sign.  The
exchange therefore acts on the two neutral carriers as
\begin{equation}
 M_{\rm sym}\mapsto-M_{\rm sym}^{T},\qquad
 M_{\rm anti}\mapsto+M_{\rm anti}^{T}.
\label{eq:CrossTransposeCertificate}
\end{equation}
The neutral adjoint/singlet pieces are removed by the standard
$\mathrm{Sym}^2 \mathbf{56}=\mathbf{133}\oplus\mathbf{1463}$ and
$\Lambda^2 \mathbf{56}=\mathbf{1}\oplus\mathbf{1539}$ projectors \cite{Slansky1981,CacciatoriE7MagicSquare}, which commute with the exchange, so the
relative transpose sign survives on the $\mathbf{1463}$ and $\mathbf{1539}$ components:
\begin{equation}
 M_{1463}\mapsto-M_{1463}^{T},\qquad
 M_{1539}\mapsto+M_{1539}^{T}.
\end{equation}
For the special neutral $\mathbf{650}$ representatives used in the main text,
$S^T=S$ while the dark fundamental satisfies $\chi^T=-\chi$.  Hence
\begin{equation}
\begin{array}{c|cc}
 &S&\chi\\ \hline
\mathbf{650}_0\subset\mathbf{1463}_H&-&+\\
\mathbf{650}_0\subset\mathbf{1539}_H&+&-
\end{array}
\end{equation}
which is Eq.~\eqref{eq:1539P7Reversal}.  The result is a finite carrier-sign
statement; by itself it does not establish an exact symmetry of the chiral
four-dimensional vacuum.

\section{Exact operator-basis certificates through dimension nine}
\label{app:g2chiF3certificate}

This appendix records the finite calculations behind
Eqs.~\eqref{eq:ChiF3ExactZero}, \eqref{eq:Q7DEOMReduction} and
\eqref{eq:G2HookF4Witness}.  They use the octonionic tensor convention of
Ref.~\cite{ButtazzoG2Higgs}.  Starting from the seven nonzero triples in
Eq.~\eqref{eq:octonionTriples}, write a general antisymmetric $7\times7$
matrix $T$.  The stabilizer equations
\begin{equation}
 T_i{}^pO_{pjk}+T_j{}^pO_{ipk}+T_k{}^pO_{ijp}=0
\end{equation}
have $21$ unknown antisymmetric entries, rank $7$ and nullity $14$, thereby
reconstructing $\mathfrak g_2$ without importing a generator basis.  For the
resulting rational basis $\{T_A\}$ define
\begin{equation}
 (J_i)_{mn}=O_{imn}.
\end{equation}
The matrices $J_i$ span the complementary $\mathbf{7}$ in
$\mathfrak{so}(7)=\mathfrak g_2\oplus\mathbf{7}$ \cite{AdamsExceptional,SpringerVeldkamp,BaezOctonions}.

\paragraph{Three-field-strength zero.}
The old symmetrized three-generator certificates remain arithmetically
correct.  In the first rational basis,
\begin{equation}
 \operatorname{Tr}(J_1^TT_1T_6T_{14})=-1,
\end{equation}
and in $\overline{\mathbf{351}}=\Lambda^2 \mathbf{27}$ the six ordered traces of the corresponding
component are
\begin{equation}
 (-57,-57,-57,-57,-57,-57).
\end{equation}
Their symmetric average is $-57$, whereas their alternating average is exactly
zero.  This is the finite parent-space manifestation of
$\mathbf{7}\not\subset\Lambda^3\mathbf{14}$ and is the channel selected by the
Lorentz scalar $\operatorname{tr}(F_aF_bF_c)$.

\paragraph{The algebraic $d=7$ tensor and its EOM reduction.}
Exact multiplication gives, for every $A,i,j,k$,
\begin{equation}
 \frac12\operatorname{Tr}_7\!\left(J_i[J_j,J_k]T_A\right)
 =-\frac32 O_{a jk}(T_A)^a{}_i .
\label{eq:Q7DTraceIdentity}
\end{equation}
The right-hand side is not identically zero; for example the first rational
stabilizer generator has a component
\begin{equation}
 O_{abc}(T_1e_1)^a(e_1)^b(e_6)^c=-1.
\end{equation}
Thus Eq.~\eqref{eq:Q7DDefinition} is a genuine off-shell $G(2)$ tensor.  Its
operator-basis status is fixed by a second exact identity.  With
$B_{\mu\nu}^A$, $K_\nu^A$ and $C_{AB}$ defined in the main text, direct
component algebra gives
\begin{equation}
 (D_\mu K_\nu-D_\nu K_\mu)^A
 =3B_{\mu\nu}^A+F_{\mu\nu}^BC_{AB},
 \qquad C_{AB}=-C_{BA}.
\end{equation}
The supplied exact script checks $4802$ basis components of the derivative
identity and $16464$ polynomial coefficients of the antisymmetry relation.
Contracting with $F_A^{\mu\nu}$ removes the curvature term and gives
Eq.~\eqref{eq:Q7DEOMReduction}.  The dual version reduces to the non-Abelian
Bianchi identity.  Hence the nonzero tensor is not an independent on-shell
$d=7$ Wilson operator.

\paragraph{Completeness through $d=9$.}
The second certificate constructs local massless contact polynomials for every
odd-$n_\chi$ field content through $d=9$, imposes momentum conservation and
little-group weights, decomposes the resulting permutation representation of
the identical fields, and matches it to the exact $G(2)$ Schur-functor
character.  This performs the EOM/IBP quotient at the amplitude level \cite{HenningLuMeliaMurayama2016,HenningLuMeliaMurayama2017,LiRenXiaoYuZheng2022,DongMaShu2023}.  It
finds
\begin{equation}
 N_{\rm odd}^{(3)}=N_{\rm odd}^{(5)}=N_{\rm odd}^{(7)}=0,
 \qquad N_{\rm odd}^{(9)}=14
\end{equation}
before Hermitian/parity pairing at $d=9$, with the field-content distribution
shown in Eq.~\eqref{eq:D9OperatorBasisCounts}.  In particular the
one-$\chi$, three-field-strength, two-derivative sector has zero matches, so
$\chi(DF)(DF)F$ is absent from the on-shell basis.

\paragraph{Exact $[3,1]$ witness for $\chi F^4$.}
The mixed-helicity four-field-strength kinematics contains a $[3,1]$ hook.
The exact $G(2)$ Schur decomposition has
\begin{equation}
 \mathbb S_{(3,1)}\mathbf{14}
 =\cdots\oplus\mathbf{7},
 \qquad
 [\mathbf{7}]\,\mathbb S_{(3,1)}\mathbf{14}=1.
\end{equation}
To exhibit the invariant directly, define
\begin{equation}
 U_{kABCD}=\operatorname{Tr}(J_k^TT_AT_BT_CT_D)
\end{equation}
and apply the unnormalised Young symmetrizer for the tableau of shape $[3,1]$
with row slots $(A,B,C)$ and column pair $(A,D)$ \cite{FultonHarris1991}.  Exact integer arithmetic
gives
\begin{equation}
 (Y_{[31]}U)_{1,1,1,3,7}=3\ne0,
\end{equation}
which is Eq.~\eqref{eq:G2HookF4Witness}.  This provides an explicit nonzero
component independent of the character multiplicity.

\paragraph{Broken-phase hook restriction.}
The same exact rational basis gives a direct check after the dark VEV is
chosen.  Solving $T e_7=0$ inside the fourteen-dimensional stabilizer produces
an eight-dimensional $\mathfrak{su}(3)_C$ subalgebra; a rational complement has
six generators whose images of $e_7$ span the broken directions.  Evaluating
the $[3,1]$ tensor with its Higgs index fixed to $e_7$ gives, in the
sparse rational basis used by the supplied certificate, the following numbers
of nonzero ordered components as the number $n_{\rm coset}$ of broken adjoint
slots is varied:
\begin{equation}
\begin{array}{c|ccccc}
 n_{\rm coset} & 0&1&2&3&4\\ \hline
 N_{\rm nz}^{\rm(cert\ basis)} &0&0&940&480&40 .
\end{array}
\label{eq:HookCosetComponentCount}
\end{equation}
Only the two zeros and the nonvanishing of the remaining blocks are
basis-independent; the raw nonzero-entry counts depend on the chosen adapted
basis and are reported only as a checksum for the certificate.  An exact
two-coset witness has unnormalised value $-6$ in that basis.  Thus the $G(2)$
hook is nonzero, but its restriction to four unbroken $SU(3)_C$ field
strengths vanishes identically; the first surviving broken-phase term contains
two coset field strengths.  This distinction is used in the lifetime
interpretation of Sec.~\ref{sec:pureDarkOperatorBasis}.

\section{Independent Weyl-character certificates for the $\mathbf{1463}$ and $\mathbf{1539}$ squares}
\label{app:E7SquareCertificates}

For completeness, this appendix records the independent character calculation
used in the corrected operator analysis and in the separated-parent check.
We use the paper's Bourbaki-type $E_7$ Cartan convention, in which the
fundamental highest weights have dimensions
\begin{equation}
(133,912,8645,365750,27664,1539,56),
\end{equation}
so that $\mathbf{1539}$ has Dynkin label $(0,0,0,0,0,1,0)$ and $\mathbf{1463}$ has
$(0,0,0,0,0,0,2)$.  These labels are related to the reordered LieART labels
quoted in Sec.~\ref{sec:pureDarkOperatorBasis} by
Eq.~\eqref{eq:E7DynkinConventionMap}; they denote the same irreducible
representations.  Starting from the Weyl orbit of the simple roots, weight
multiplicities are generated by Freudenthal recursion.  The symmetric and
exterior-square characters are then formed directly from the weight
multiplicities and reduced by successive highest-weight subtraction \cite{FultonHarris1991,McKayPatera1981}.

For the $\mathbf{1539}$ this gives exactly
\begin{equation}
\resizebox{0.98\columnwidth}{!}{$\displaystyle
\begin{aligned}
\mathrm{Sym}^2(\mathbf{1539})={}&\mathbf{1}\oplus2(\mathbf{1539})\oplus\mathbf{7371}\oplus\mathbf{40755}
\oplus\mathbf{150822}\oplus\mathbf{365750}\oplus\mathbf{617253},\\
\Lambda^2(\mathbf{1539})={}&\mathbf{133}\oplus\mathbf{1463}\oplus\mathbf{8645}\oplus\mathbf{40755}
\oplus\mathbf{152152}\oplus\mathbf{980343}.
\end{aligned}
$}
\end{equation}
For the $\mathbf{1463}$ the same independent calculation gives
\begin{equation}
\resizebox{0.98\columnwidth}{!}{$\displaystyle
\begin{aligned}
\mathrm{Sym}^2(\mathbf{1463})={}&\mathbf{1}\oplus\mathbf{1539}\oplus\mathbf{7371}\oplus\mathbf{150822}
\oplus\mathbf{293930}\oplus\mathbf{617253},\\
\Lambda^2(\mathbf{1463})={}&\mathbf{133}\oplus\mathbf{1463}\oplus\mathbf{152152}\oplus\mathbf{915705}.
\end{aligned}
$}
\end{equation}
The dimensions close respectively on $R(R+1)/2$ and $R(R-1)/2$ in all four
cases.  The exchange statements are therefore obtained without relying on dimension
sums alone:
\begin{equation}
\resizebox{0.98\columnwidth}{!}{$\displaystyle
\mathbf{1463}\not\subset\mathrm{Sym}^2(\mathbf{1539}),\qquad
\mathbf{1539}\not\subset\Lambda^2(\mathbf{1539}),\qquad
\mathbf{1463}\subset\Lambda^2(\mathbf{1463})\ \text{once}
$}
\end{equation}
The first statement is the load-bearing Bose-symmetry result for the scalar
repair.  The latter two diagnose particular antisymmetric self-maps and should
not be confused with an on-shell operator-basis theorem.  The first physical
pure-dark parent test is instead the $[3,1]$ hook calculation in the next
appendix.  The executable exact-integer character reductions and their outputs
are part of the reproducibility package.

\section{Exact $E_7$ hook-parent certificate for the dimension-nine operator}
\label{app:E7HookParentCertificate}

The first physical one-Higgs pure-dark odd operator transforms in the $[3,1]$
permutation channel of four adjoint field strengths.  Its parent test is
therefore
\begin{equation}
 \operatorname{Hom}_{E_7}\!\left(R,\mathbb S_{(3,1)}\mathbf{133}\right),
 \qquad R=\mathbf{1463},\mathbf{1539}.
\end{equation}
We use the same exact $E_7$ Cartan and Freudenthal weight recursion as in
Appendix~\ref{app:E7SquareCertificates}.  The Jacobi--Trudi identity gives \cite{Macdonald1995,FultonHarris1991}
\begin{equation}
 \mathbb S_{(3,1)}V
 =\mathrm{Sym}^3V\otimes V-\mathrm{Sym}^4V.
\label{eq:E7S31JacobiTrudi}
\end{equation}
For $V=\mathbf{133}$ the exact weight characters have
\begin{align}
 \dim\mathrm{Sym}^3(\mathbf{133})&=400995,\nonumber\\
 \dim\mathrm{Sym}^4(\mathbf{133})&=13633830.
\end{align}
and therefore
\begin{equation}
 \dim\mathbb S_{(3,1)}\mathbf{133}=39698505,
\end{equation}
in agreement with the hook-length formula
$133\cdot132\cdot134\cdot135/8$.

Rather than decomposing the full $3.97\times10^7$-dimensional character, the
target multiplicity is extracted exactly with the Weyl alternant \cite{FultonHarris1991,McKayPatera1981}.  For a
Weyl-invariant weight character $X=\sum_\mu m_X(\mu)e^\mu$,
\begin{equation}
 [V_\lambda]X=
 \sum_{\mu:\,\lambda+\rho-\mu=w\rho}
 \operatorname{sgn}(w)\,m_X(\mu).
\label{eq:WeylAlternantMultiplicity}
\end{equation}
The Weyl-orbit condition and the sign are obtained by exact simple-root
reflections to the dominant chamber.  The calculation gives
\begin{align}
[\mathbf{1463}]\bigl(\mathrm{Sym}^3(\mathbf{133})\otimes\mathbf{133}\bigr)&=1,
&[\mathbf{1463}]\mathrm{Sym}^4(\mathbf{133})&=0,\nonumber\\
[\mathbf{1539}]\bigl(\mathrm{Sym}^3(\mathbf{133})\otimes\mathbf{133}\bigr)&=4,
&[\mathbf{1539}]\mathrm{Sym}^4(\mathbf{133})&=2.
\end{align}
Consequently
\begin{equation}
 {
 [\mathbf{1463}]\\,\mathbb S_{(3,1)}\mathbf{133}=1,
 \qquad
 [\mathbf{1539}]\\,\mathbb S_{(3,1)}\mathbf{133}=2 .}
\end{equation}
As an independent character-level check, the full $[3,1]$ character was
reconstructed from the Frobenius/Adams power-sum formula rather than from
Jacobi--Trudi.  It agrees coefficient by coefficient on all $162291$ weights
and reproduces the same target multiplicities $1$ and $2$.  A further
support-only check independently generates the $126$ roots plus the zero weight
of the adjoint and finds $3025$, $28911$, and $162417$ distinct weights in the
two-, three-, and four-fold sumsets, respectively.  The hook support is smaller
by exactly $126$: the missing weights are the extremal $4w$ with $w$ a root,
for which $\mathrm{Sym}^3V\otimes V$ and $\mathrm{Sym}^4V$ each contribute
multiplicity one and cancel in the Jacobi--Trudi difference.  This explains
$162417-126=162291$ and independently checks the boundary of the hook
character support.  The alternant extractor is also self-tested on exact
irreducible characters before being applied to either hook construction.

A third implementation was then run independently in Wolfram Language with
LieART weight systems and Weyl-reflection routines \cite{LieART}.  It first
reproduces
\begin{equation}
 \mathbf{133}\otimes\mathbf{133}
 =\mathbf{1}\oplus\mathbf{133}\oplus\mathbf{1539}
  \oplus\mathbf{7371}\oplus\mathbf{8645},
\end{equation}
and the symmetric-power dimensions
\begin{align}
 \dim\mathrm{Sym}^2\mathbf{133}&=8911,\qquad
 \dim\mathrm{Sym}^3\mathbf{133}=400995,\nonumber\\
 \dim\mathrm{Sym}^4\mathbf{133}&=13633830.
\end{align}
As a known-answer test of the multiplicity extraction, the LieART/Wolfram run
returns multiplicities $(1,0,1,1,0)$ for
$R=(\mathbf{1},\mathbf{133},\mathbf{1539},\mathbf{7371},\mathbf{8645})$
in $\mathrm{Sym}^2\mathbf{133}$.  The target identities themselves are also
checked at input: $\mathrm{Irrep}[E7][1463]$ and
$\mathrm{Irrep}[E7][1539]$ return dimensions $1463$ and $1539$, with LieART
labels $(0000020)$ and $(0000100)$, respectively.  Thus the convention switch
cannot silently retarget the $\mathbf{1463}$ calculation.  The Jacobi--Trudi
calculation then gives
\begin{equation}
 \begin{array}{c|ccc}
 R & [R](\mathrm{Sym}^3\mathbf{133}\otimes\mathbf{133})
   & [R]\mathrm{Sym}^4\mathbf{133}
   & [R]\mathbb S_{(3,1)}\mathbf{133}\\ \hline
 \mathbf{1463} & 1 & 0 & 1\\
 \mathbf{1539} & 4 & 2 & 2
 \end{array}
\end{equation}
with $\dim\mathbb S_{(3,1)}\mathbf{133}=39698505$.  A direct
Frobenius/Adams construction in the same independent Wolfram/LieART run again
returns $1$ and $2$.  This is an external implementation reproduction of the
two parent multiplicities, although it uses the same underlying
Weyl-alternant mathematics rather than a logically different theorem.

The parent-space multiplicities do not by themselves fix a subgroup
component, but for the separated $\mathbf{1539}_H$ the restriction can be closed
without constructing either enormous parent tensor explicitly.  The canonical
$E_7$ normalizer element $P_7^{(0)}$ induces the $F_4$-fixed outer involution on
$E_6$.  Hence it fixes the chosen $G(2)\subset F_4$ pointwise, whereas the
neutral exterior-square carrier has the opposite transpose sign from the
$\mathbf{1463}$ and Eq.~\eqref{eq:1539P7Reversal} gives
$P_7^{(0)}\Xi_D=-\Xi_D$.  Every $E_7$ invariant with one selected $\Xi_D$ and
otherwise only $P_7^{(0)}$-even dark gauge-sector insertions is therefore equal
to its own negative.  This proves Eq.~\eqref{eq:1539HookRestrictionRankZero}
simultaneously for both parent channels and gives the dimension-independent
corollary Eq.~\eqref{eq:1539UndressedAllDimZero}.

Two executable checks isolate where the zero occurs.  First, an exact $E_6$
Jacobi--Trudi/Weyl-alternant calculation gives
\begin{equation}
 \dim\operatorname{Hom}_{E_6}
 \!\left(\mathbf{650},\mathbb S_{(3,1)}\mathbf{78}\right)=3,
\label{eq:E6Hook650Multiplicity}
\end{equation}
from multiplicities $5$ in
$\mathrm{Sym}^3(\mathbf{78})\otimes\mathbf{78}$ and $2$ in
$\mathrm{Sym}^4(\mathbf{78})$.  Thus no intermediate-$E_6$ absence explains the
result.  Second, in the Darboux $\mathbf{56}=\mathbf{28}\oplus\mathbf{28}^*$
carrier the selected dark scalar is represented by
$M=J_n\otimes\mathbf1_3$ on the $(\mathbf7,\mathbf3_A)$ block.  The exact
integer certificate checks the charge-flip signs on all fourteen reconstructed
$G(2)$ generators and, as a trace-level checksum, finds
\begin{equation}
 \operatorname{Tr}_{56}
 \!\left(K_{\Xi_D}T_AT_BT_CT_D\right)=0
 \qquad\forall\ A,B,C,D\in\mathfrak g_2.
\label{eq:1539FundamentalTraceChecksum}
\end{equation}
The trace identity is only a checksum for one realization; the charge-flip
argument is the theorem that covers the complete two-dimensional parent
space.  In the explicit Darboux carrier the zero arises by cancellation between
the $\mathbf{28}$ and $\mathbf{28}^*$ blocks (equivalently the forward and reversed
trace words), not because the individual block trace vanishes.  Because the
special $S$ direction is itself $P_7^{(0)}$ odd, the same argument does not
forbid tensors with an additional $S$ insertion.  The benchmark has
$v_S/M_*=O(1)$, but the existence and selected Clebsch of such an $S$-dressed
parent remain open.  Exact all-orders protection therefore still requires the
independently postulated gauged/geometric $\mathbb Z_{2,D}$.

\section{NDA normalization for the leading dimension-nine pure-dark operator}

For the one-$\chi$ hook representative in Eq.~\eqref{eq:O9F4}, replacing the
dark Higgs by its VEV gives the dimension-eight gauge coefficient
\begin{equation}
 C_8^{(9)}=\frac{c_{9,F}v_\chi}{M_*^5}.
\end{equation}
We normalize the state-dependent width as
\begin{equation}
 \Gamma_{B_J}^{(9)}=
 \frac{\kappa_{9,J}}{8\pi}
 \left|\frac{c_{9,F}v_\chi}{M_*^5}\right|^2m_{B_J}^{9},
\end{equation}
where $\kappa_{9,J}$ includes the confined overlap, the actual multiparticle
phase space and the projection onto the relevant Hermitian parity combination \cite{ManoharGeorgi1984,LukeManohar1997}.
Thus
\begin{equation}
 \tau_{B_J}^{(9)}=
 \frac{8\pi\hbar M_*^{10}}
 {\kappa_{9,J}|c_{9,F}|^2v_\chi^2m_{B_J}^{9}}.
\end{equation}
For $v_\chi=10^{14}$ GeV, $M_*=6.9\times10^{15}$ GeV and
$m_{B_J}=1.71\times10^{14}$ GeV this gives
\begin{equation}
 \tau_0=3.2369\times10^{-21}\ {\rm s},
\end{equation}
which is rounded to Eq.~\eqref{eq:tau9F4} in the text.  With
$M_*=2.435\times10^{18}$ GeV the same normalization is
$9.6968\times10^4$ s.  The corresponding age-of-the-Universe coefficient
bounds are $8.63\times10^{-20}$ and $4.72\times10^{-7}$, respectively.  For the separated $\mathbf{1539}_H$ the undressed $E_7$ parent coefficient of this
selected component is zero by Eq.~\eqref{eq:1539HookRestrictionRankZero}.
The normalization above therefore applies to a generic low-energy hook, to a
parent realization not subject to that component zero, or conditionally to a hook
regenerated by $P_7^{(0)}$-breaking ultraviolet dressings after their VEVs are
inserted.  For the simplest possible $S$ dressing, $v_S/M_*=O(1)$ in the benchmark,
so the component zero alone would not improve this normalization parametrically
\emph{if} that dressed parent and its selected Clebsch are nonzero.  Their existence
is not established here.  These are NDA diagnostics only; no assumption of unit
four-gauge-field overlap or literal two-body phase space is made.

\bibliographystyle{apsrev4-2}
\makeatletter
\immediate\write\@auxout{\string\citation{apsrev42Control}}
\makeatother
\bibliography{dark_stability_refs_FINAL}

@article{ButtazzoG2Higgs,
  author={Buttazzo, Dario and Di Luzio, Luca and Landini, Giacomo and Strumia, Alessandro and Teresi, Daniele},
  title={Dark Matter from self-dual gauge/Higgs dynamics},
  journal={JHEP}, volume={10}, pages={067}, year={2019},
  doi={10.1007/JHEP10(2019)067}, eprint={1907.11228}, archivePrefix={arXiv}
}

@article{ButtazzoScalarGauge2020,
  author={Buttazzo, Dario and Di Luzio, Luca and Ghorbani, Parsa and Gross, Christian and Landini, Giacomo and Strumia, Alessandro and Teresi, Daniele and Wang, Jin-Wei},
  title={Scalar gauge dynamics and Dark Matter},
  journal={JHEP}, volume={01}, pages={130}, year={2020}, doi={10.1007/JHEP01(2020)130}
}

@article{MasiG2,
  author={Masi, Nicolo}, title={The Resurgence of the {G(2)} Group for the Strong Sector and the Emergence of Dark Matter},
  journal={Nucl. Phys. B}, volume={1004}, pages={116562}, year={2024}, doi={10.1016/j.nuclphysb.2024.116562}, eprint={2406.15421}, archivePrefix={arXiv}
}

@article{MasiE6E7,
  author={Masi, Nicolo},
  title={Toward a special $E_6\to G(2)\times SU(3)_A$ embedding for Standard Model and dark matter and an $E_7$ completion proposal},
  journal={Eur. Phys. J. C}, volume={86}, pages={985}, year={2026},
  doi={10.1140/epjc/s10052-026-16104-1}, eprint={2603.15710}, archivePrefix={arXiv}, primaryClass={hep-ph}
}

@article{MasiG2SciRep2021,
  author={Masi, Nicolo}, title={An exceptional {G(2)} extension of the Standard Model from the correspondence with Cayley--Dickson algebras automorphism groups},
  journal={Sci. Rep.}, volume={11}, pages={22528}, year={2021}, doi={10.1038/s41598-021-01814-1}, eprint={2111.11849}, archivePrefix={arXiv}
}

@article{FrohlichMorchioStrocchi1980,
  author={Frohlich, J. and Morchio, G. and Strocchi, F.}, title={Higgs phenomenon without a symmetry breaking order parameter},
  journal={Phys. Lett. B}, volume={97}, pages={249--252}, year={1980}, doi={10.1016/0370-2693(80)90594-8}
}

@article{FrohlichMorchioStrocchi1981,
  author={Frohlich, J. and Morchio, G. and Strocchi, F.}, title={Higgs phenomenon without symmetry breaking order parameter},
  journal={Nucl. Phys. B}, volume={190}, pages={553--582}, year={1981}, doi={10.1016/0550-3213(81)90448-X}
}

@misc{MaasFMSReview,
  author={Maas, Axel}, title={The Frohlich--Morchio--Strocchi mechanism: an underestimated legacy}, year={2023}, eprint={2305.01960}, archivePrefix={arXiv}
}

@article{MaasSondenheimerTorek2017,
  author={Maas, Axel and Sondenheimer, Rene and Torek, Pascal}, title={On the observable spectrum of theories with a Brout--Englert--Higgs effect}, journal={Annals Phys.}, volume={402}, pages={18--44}, year={2019}, eprint={1709.07477}, archivePrefix={arXiv}
}

@misc{MaasSondenheimerTorekSUN2018,
  author={Maas, Axel and Sondenheimer, Rene and Torek, Pascal}, title={A study of how the particle spectra of SU(N) gauge theories with a fundamental Higgs emerge}, journal={Annals Phys.}, volume={402}, pages={18--44}, year={2019}, eprint={1710.01941}, archivePrefix={arXiv}
}

@article{MaasTorekSU3,
  author={Maas, Axel and Torek, Pascal}, title={The spectrum of an SU(3) gauge theory with a fundamental Higgs field}, journal={Annals Phys.}, volume={397}, pages={303--336}, year={2018}, eprint={1804.04453}, archivePrefix={arXiv}
}

@article{MaasPedro2016,
  author={Maas, Axel and Pedro, Leonardo}, title={Gauge invariance and the physical spectrum in the two-Higgs-doublet model}, journal={Phys. Rev. D}, volume={93}, pages={056005}, year={2016}, eprint={1601.02006}, archivePrefix={arXiv}
}

@article{AthenodorouTeper2020,
  author={Athenodorou, Andreas and Teper, Michael}, title={The glueball spectrum of SU(3) gauge theory in 3+1 dimensions}, journal={JHEP}, volume={11}, pages={172}, year={2020}, eprint={2007.06422}, archivePrefix={arXiv}
}

@article{MorningstarPeardon1999,
  author={Morningstar, Colin J. and Peardon, Mike}, title={The glueball spectrum from an anisotropic lattice study}, journal={Phys. Rev. D}, volume={60}, pages={034509}, year={1999}, doi={10.1103/PhysRevD.60.034509}, eprint={hep-lat/9901004}, archivePrefix={arXiv}
}

@article{MorningstarPeardon1997,
  author={Morningstar, Colin and Peardon, Mike}, title={Efficient glueball simulations on anisotropic lattices}, journal={Phys. Rev. D}, volume={56}, pages={4043--4061}, year={1997}, eprint={hep-lat/9704011}, archivePrefix={arXiv}
}

@article{ChenGlueballs2006,
  author={Chen, Y. and Alexandru, A. and Dong, S. J. and Draper, T. and Horvath, I. and Lee, F. X. and Liu, K. F. and Mathur, N. and Morningstar, C. and others}, title={Glueball spectrum and matrix elements on anisotropic lattices}, journal={Phys. Rev. D}, volume={73}, pages={014516}, year={2006}, doi={10.1103/PhysRevD.73.014516}, eprint={hep-lat/0510074}, archivePrefix={arXiv}
}

@article{LuciniTeperWenger2004,
  author={Lucini, Biagio and Teper, Michael and Wenger, Urs}, title={Glueballs and k-strings in SU(N) gauge theories: calculations with improved operators}, journal={JHEP}, volume={06}, pages={012}, year={2004}, eprint={hep-lat/0404008}, archivePrefix={arXiv}
}

@article{LuciniRagoRinaldi2010,
  author={Lucini, Biagio and Rago, Antonio and Rinaldi, Enrico}, title={Glueball masses in the large N limit}, journal={JHEP}, volume={08}, pages={119}, year={2010}, eprint={1007.3879}, archivePrefix={arXiv}
}

@article{MeyerGlueballs2005,
  author={Meyer, Harvey B.}, title={Glueball Regge trajectories}, journal={JHEP}, volume={01}, pages={071}, year={2005}, eprint={hep-lat/0412021}, archivePrefix={arXiv}
}

@misc{TeperGlueball1998,
  author={Teper, Michael J.}, title={Glueball masses and other physical properties of SU(N) gauge theories in D=3+1: a review of lattice results for theorists}, year={1998}, eprint={hep-th/9812187}, archivePrefix={arXiv}
}

@article{HuLuoChen1996,
  author={Hu, Lian and Luo, Xiang-Qian and Chen, Qi-Zhou and Fang, Xi-Yan and Guo, Shuo-Hong},
  title={Glueball Masses from Hamiltonian Lattice QCD}, journal={Commun. Theor. Phys.}, volume={28}, pages={327--332}, year={1997},
  eprint={hep-ph/9609435}, archivePrefix={arXiv}
}

@article{Michael1985,
  author={Michael, C.}, title={Adjoint sources in lattice gauge theory}, journal={Nucl. Phys. B}, volume={259}, pages={58--76}, year={1985}, doi={10.1016/0550-3213(85)90297-4}
}

@article{LuscherWolff1990,
  author={Luscher, Martin and Wolff, Ulli}, title={How to calculate the elastic scattering matrix in two-dimensional quantum field theories by numerical simulation}, journal={Nucl. Phys. B}, volume={339}, pages={222--252}, year={1990}, doi={10.1016/0550-3213(90)90540-T}
}

@article{Blossier2009,
  author={Blossier, Benoit and Della Morte, Michele and von Hippel, Georg and Mendes, Tereza and Sommer, Rainer}, title={On the generalized eigenvalue method for energies and matrix elements in lattice field theory}, journal={JHEP}, volume={04}, pages={094}, year={2009}, doi={10.1088/1126-6708/2009/04/094}, eprint={0902.1265}, archivePrefix={arXiv}
}

@misc{Blossier2008,
  author={Blossier, B. and von Hippel, G. and Mendes, T. and Sommer, R. and Della Morte, M.}, title={Efficient use of the Generalized Eigenvalue Problem}, year={2008}, eprint={0808.1017}, archivePrefix={arXiv}
}

@article{HallPost1956,
  author={Hall, R. L. and Post, H. R.}, title={Many-particle systems: II. The structure of the ground-state energy}, journal={Proc. Phys. Soc. A}, volume={69}, pages={439--444}, year={1956}
}

@article{HallPost1967,
  author={Hall, R. L. and Post, H. R.}, title={Many-particle systems: IV. Short-range interactions}, journal={Proc. Phys. Soc.}, volume={90}, pages={381--396}, year={1967}
}

@misc{HallPostReview2019,
  author={Richard, Jean-Marc and Valcarce, Alfredo and Vijande, Javier}, title={Hall--Post inequalities: review and application to molecules and tetraquarks}, year={2019}, eprint={1910.08295}, archivePrefix={arXiv}
}

@book{KatoPerturbation,
  author={Kato, Tosio}, title={Perturbation Theory for Linear Operators}, publisher={Springer}, address={Berlin}, year={1995}
}

@book{ReedSimonIV,
  author={Reed, Michael and Simon, Barry}, title={Methods of Modern Mathematical Physics IV: Analysis of Operators}, publisher={Academic Press}, year={1978}
}

@article{Slansky1981,
  author={Slansky, Richard}, title={Group Theory for Unified Model Building}, journal={Phys. Rept.}, volume={79}, pages={1--128}, year={1981}, doi={10.1016/0370-1573(81)90092-2}
}

@article{Dynkin1952,
  author={Dynkin, E. B.}, title={Semisimple subalgebras of semisimple Lie algebras}, journal={Mat. Sbornik}, volume={30}, pages={349--462}, year={1952}
}

@book{McKayPatera1981,
  author={McKay, W. G. and Patera, J.}, title={Tables of Dimensions, Indices, and Branching Rules for Representations of Simple Lie Algebras}, publisher={Marcel Dekker}, year={1981}
}

@article{LieART,
  author={Feger, Robert and Kephart, Thomas W.}, title={LieART---A Mathematica Application for Lie Algebras and Representation Theory}, journal={Comput. Phys. Commun.}, volume={192}, pages={166--195}, year={2015}, eprint={1206.6379}, archivePrefix={arXiv}
}

@book{SpringerVeldkamp,
  author={Springer, T. A. and Veldkamp, F. D.}, title={Octonions, Jordan Algebras and Exceptional Groups}, publisher={Springer}, year={2000}
}

@article{BaezOctonions,
  author={Baez, John C.}, title={The Octonions}, journal={Bull. Amer. Math. Soc.}, volume={39}, pages={145--205}, year={2002}, eprint={math/0105155}, archivePrefix={arXiv}
}

@book{AdamsExceptional,
  author={Adams, J. F.}, title={Lectures on Exceptional Lie Groups}, publisher={University of Chicago Press}, year={1996}
}

@book{GeorgiLieAlgebras,
  author={Georgi, Howard}, title={Lie Algebras in Particle Physics}, publisher={Westview Press}, edition={2}, year={1999}
}

@book{CahnSemiSimple,
  author={Cahn, Robert N.}, title={Semi-Simple Lie Algebras and Their Representations}, publisher={Benjamin/Cummings}, year={1984}
}

@article{Rubenthaler2008,
  author={Rubenthaler, Hubert}, title={The (A2,G2) duality in E6, octonions and the triality principle}, journal={Trans. Amer. Math. Soc.}, volume={360}, pages={347--367}, year={2008}
}

@article{GurseyRamondSikivie1976,
  author={Gursey, Feza and Ramond, Pierre and Sikivie, Pierre}, title={A Universal Gauge Theory Model Based on E6}, journal={Phys. Lett. B}, volume={60}, pages={177--180}, year={1976}
}

@article{HewettRizzo1989,
  author={Hewett, JoAnne L. and Rizzo, Thomas G.}, title={Low-Energy Phenomenology of Superstring Inspired E6 Models}, journal={Phys. Rept.}, volume={183}, pages={193--381}, year={1989}
}

@article{VafaFTheory,
  author={Vafa, Cumrun}, title={Evidence for F-Theory}, journal={Nucl. Phys. B}, volume={469}, pages={403--418}, year={1996}, eprint={hep-th/9602022}, archivePrefix={arXiv}
}

@article{MorrisonVafaI,
  author={Morrison, David R. and Vafa, Cumrun}, title={Compactifications of F-Theory on Calabi--Yau Threefolds I}, journal={Nucl. Phys. B}, volume={473}, pages={74--92}, year={1996}, eprint={hep-th/9602114}, archivePrefix={arXiv}
}

@article{MorrisonVafaII,
  author={Morrison, David R. and Vafa, Cumrun}, title={Compactifications of F-Theory on Calabi--Yau Threefolds II}, journal={Nucl. Phys. B}, volume={476}, pages={437--469}, year={1996}, eprint={hep-th/9603161}, archivePrefix={arXiv}
}

@article{DonagiWijnholt,
  author={Donagi, Ron and Wijnholt, Martijn}, title={Model Building with F-Theory}, journal={Adv. Theor. Math. Phys.}, volume={15}, pages={1237--1317}, year={2011}, eprint={0802.2969}, archivePrefix={arXiv}
}

@article{BHV,
  author={Beasley, Chris and Heckman, Jonathan J. and Vafa, Cumrun}, title={GUTs and Exceptional Branes in F-theory I}, journal={JHEP}, volume={01}, pages={058}, year={2009}, eprint={0802.3391}, archivePrefix={arXiv}
}

@article{DonagiWijnholtHiggs2009,
  author={Donagi, Ron and Wijnholt, Martijn}, title={Higgs Bundles and UV Completion in F-Theory}, journal={Commun. Math. Phys.}, volume={326}, pages={287--327}, year={2014}, eprint={0904.1218}, archivePrefix={arXiv}
}

@misc{WijnholtHiggsBundles,
  author={Wijnholt, Martijn}, title={Higgs Bundles and String Phenomenology}, year={2012}, eprint={1201.2520}, archivePrefix={arXiv}
}

@article{FriedmanMorganWitten,
  author={Friedman, Robert and Morgan, John and Witten, Edward}, title={Vector Bundles and F Theory}, journal={Commun. Math. Phys.}, volume={187}, pages={679--743}, year={1997}, eprint={hep-th/9701162}, archivePrefix={arXiv}
}

@article{MarsanoSaulinaSchaferNameki,
  author={Marsano, Joseph and Saulina, Natalia and Schafer-Nameki, Sakura}, title={F-theory Compactifications for Supersymmetric GUTs}, journal={JHEP}, volume={08}, pages={030}, year={2009}, eprint={0904.3932}, archivePrefix={arXiv}
}

@article{HayashiTatarWatari,
  author={Hayashi, Hirotaka and Kawano, Teruhiko and Tsuchiya, Yoichi and Watari, Taizan}, title={Flavor Structure in F-theory Compactifications}, journal={JHEP}, volume={08}, pages={036}, year={2010}, eprint={0910.2762}, archivePrefix={arXiv}
}

@article{CecottiCordovaHeckmanVafa,
  author={Cecotti, Sergio and Cordova, Clay and Heckman, Jonathan J. and Vafa, Cumrun}, title={T-Branes and Monodromy}, journal={JHEP}, volume={07}, pages={030}, year={2011}, eprint={1010.5780}, archivePrefix={arXiv}
}

@article{AndersonHeckmanKatz,
  author={Anderson, Lara B. and Heckman, Jonathan J. and Katz, Sheldon}, title={T-Branes and Geometry}, journal={JHEP}, volume={05}, pages={080}, year={2014}, eprint={1310.1931}, archivePrefix={arXiv}
}

@article{Simpson1988,
  author={Simpson, Carlos T.}, title={Constructing Variations of Hodge Structure Using Yang--Mills Theory and Applications to Uniformization}, journal={J. Amer. Math. Soc.}, volume={1}, pages={867--918}, year={1988}
}

@article{MathieuBuisseret2009,
  author={Mathieu, Vincent and Buisseret, Fabien and Semay, Claude and Silvestre-Brac, Bernard}, title={The Glueball Spectrum from Constituent Models}, journal={Phys. Rev. D}, volume={79}, pages={014005}, year={2009}, eprint={0811.2710}, archivePrefix={arXiv}
}

@article{BuisseretMathieuSemay2009,
  author={Buisseret, Fabien and Mathieu, Vincent and Semay, Claude}, title={Glueball phenomenology and the pomeron}, journal={Phys. Rev. D}, volume={80}, pages={074021}, year={2009}, eprint={0906.3098}, archivePrefix={arXiv}
}

@article{BoulangerBuisseretMathieu2008,
  author={Boulanger, Nicolas and Buisseret, Fabien and Mathieu, Vincent and Semay, Claude}, title={Constituent gluon interpretation of glueballs and gluelumps}, journal={Eur. Phys. J. A}, volume={38}, pages={317--330}, year={2008}, eprint={0806.3174}, archivePrefix={arXiv}
}

@article{EichmannGlueball2021,
  author={Eichmann, Gernot and Fischer, Christian S. and Heupel, Walter}, title={The light scalar glueball as a bound state of two gluons}, journal={Phys. Lett. B}, volume={753}, pages={282--287}, year={2016}, eprint={1508.07178}, archivePrefix={arXiv}
}

@article{MeyersSwanson2013,
  author={Meyers, Jason and Swanson, Eric S.}, title={Spin-zero glueballs in the Bethe--Salpeter formalism}, journal={Phys. Rev. D}, volume={87}, pages={036009}, year={2013}, eprint={1211.4648}, archivePrefix={arXiv}
}

@book{Varshalovich1988,
  author={Varshalovich, D. A. and Moskalev, A. N. and Khersonskii, V. K.}, title={Quantum Theory of Angular Momentum}, publisher={World Scientific}, year={1988}
}

@book{LandauLifshitzQM,
  author={Landau, L. D. and Lifshitz, E. M.}, title={Quantum Mechanics: Non-Relativistic Theory}, publisher={Pergamon Press}, edition={3}, year={1977}
}

@book{MessiahQM,
  author={Messiah, Albert}, title={Quantum Mechanics}, publisher={North-Holland}, year={1961}
}

@misc{CacciatoriE7MagicSquare,
  author={Cacciatori, Sergio L. and Dalla Piazza, Francesco and Scotti, Antonio},
  title={E7 groups from octonionic magic square},
  year={2010},
  eprint={1007.4758},
  archivePrefix={arXiv},
  primaryClass={math-ph}
}

@misc{ExceptionalReductions,
  author={Marrani, Alessio and Orazi, Emanuele and Riccioni, Fabio},
  title={Exceptional Reductions},
  year={2010},
  eprint={1012.5797},
  archivePrefix={arXiv},
  primaryClass={hep-th}
}

@misc{DeppischE6Tensors,
  author={Deppisch, Thomas},
  title={E6Tensors: A Mathematica Package for E6 Tensors},
  year={2016},
  eprint={1605.05920},
  archivePrefix={arXiv},
  primaryClass={hep-ph}
}

@misc{CacciatoriE6Geometry,
  author={Bernardoni, Fabio and Cacciatori, Sergio L. and Cerchiai, Bianca L. and Scotti, Antonio},
  title={Mapping the geometry of the E6 group},
  year={2007},
  eprint={0710.0356},
  archivePrefix={arXiv},
  primaryClass={math-ph}
}

@article{EichtenCornell1978,
  author={Eichten, E. and Gottfried, K. and Kinoshita, T. and Lane, K. D. and Yan, T.-M.},
  title={Charmonium: The Model}, journal={Phys. Rev. D}, volume={17}, pages={3090--3117}, year={1978},
  doi={10.1103/PhysRevD.17.3090}, note={Erratum: Phys. Rev. D 21, 313 (1980)}
}

@article{FradkinShenker1979,
  author={Fradkin, Eduardo and Shenker, Stephen H.},
  title={{Phase diagrams of lattice gauge theories with Higgs fields}},
  journal={Phys. Rev. D}, volume={19}, pages={3682--3697}, year={1979},
  doi={10.1103/PhysRevD.19.3682}
}

@article{OsterwalderSeiler1978,
  author={Osterwalder, Konrad and Seiler, Erhard},
  title={{Gauge field theories on a lattice}},
  journal={Ann. Phys.}, volume={110}, pages={440--471}, year={1978},
  doi={10.1016/0003-4916(78)90039-8}
}

@article{KraussWilczek1989,
  author={Krauss, Lawrence M. and Wilczek, Frank},
  title={{Discrete gauge symmetry in continuum theories}},
  journal={Phys. Rev. Lett.}, volume={62}, pages={1221--1223}, year={1989},
  doi={10.1103/PhysRevLett.62.1221}
}

@article{GaiottoGlobalSym2015,
  author={Gaiotto, Davide and Kapustin, Anton and Seiberg, Nathan and Willett, Brian},
  title={{Generalized global symmetries}},
  journal={JHEP}, volume={02}, pages={172}, year={2015},
  doi={10.1007/JHEP02(2015)172}, eprint={1412.5148}, archivePrefix={arXiv}, primaryClass={hep-th}
}

@article{HarlowOoguri2021,
  author={Harlow, Daniel and Ooguri, Hirosi},
  title={{Symmetries in quantum field theory and quantum gravity}},
  journal={Commun. Math. Phys.}, volume={383}, pages={1669--1804}, year={2021},
  doi={10.1007/s00220-021-04040-y}, eprint={1810.05338}, archivePrefix={arXiv}, primaryClass={hep-th}
}

@article{HayashiHiggsConfinement2024,
  author={Hayashi, Yui},
  title={{Higgs-confinement continuity and matching of Aharonov--Bohm phases}},
  journal={Phys. Rev. Lett.}, volume={132}, pages={221901}, year={2024},
  doi={10.1103/PhysRevLett.132.221901}, eprint={2303.02129}, archivePrefix={arXiv}, primaryClass={hep-th}
}

@article{HsiehDiscreteAnomalies,
  author={Hsieh, Chang-Tse},
  title={{Discrete gauge anomalies revisited}},
  journal={JHEP}, volume={09}, pages={022}, year={2018},
  doi={10.1007/JHEP09(2018)022}, eprint={1808.02881}, archivePrefix={arXiv}, primaryClass={hep-th}
}

@article{PDGCKM2024,
  author={{Particle Data Group} and Navas, S. and others},
  title={{Review of Particle Physics: CKM Quark-Mixing Matrix}},
  journal={Phys. Rev. D},
  volume={110},
  pages={030001},
  year={2024},
  doi={10.1103/PhysRevD.110.030001}
}

@article{TakahashiSuganuma2002,
  author={Takahashi, T. T. and Suganuma, H. and Nemoto, Y. and Matsufuru, H.},
  title={Detailed analysis of the three-quark potential in SU(3) lattice QCD},
  journal={Phys. Rev. D}, volume={65}, pages={114509}, year={2002},
  doi={10.1103/PhysRevD.65.114509}, eprint={hep-lat/0204011}, archivePrefix={arXiv}
}

@article{STARJunction2026,
  author={{STAR Collaboration}},
  title={Tracking the baryon number with nuclear collisions},
  journal={Science}, volume={393}, year={2026},
  doi={10.1126/science.ads5962}, eprint={2408.15441}, archivePrefix={arXiv}
}

@article{GriestKamionkowski1990,
  author={Griest, Kim and Kamionkowski, Marc},
  title={Unitarity Limits on the Mass and Radius of Dark-Matter Particles},
  journal={Phys. Rev. Lett.}, volume={64}, pages={615--618}, year={1990},
  doi={10.1103/PhysRevLett.64.615}
}

@article{ChungKolbRiotto1998,
  author={Chung, Daniel J. H. and Kolb, Edward W. and Riotto, Antonio},
  title={Superheavy dark matter},
  journal={Phys. Rev. D}, volume={59}, pages={023501}, year={1999},
  doi={10.1103/PhysRevD.59.023501}, eprint={hep-ph/9802238}, archivePrefix={arXiv}
}

@article{ChungKolbRiotto1999,
  author={Chung, Daniel J. H. and Kolb, Edward W. and Riotto, Antonio},
  title={Production of massive particles during reheating},
  journal={Phys. Rev. D}, volume={60}, pages={063504}, year={1999},
  doi={10.1103/PhysRevD.60.063504}, eprint={hep-ph/9809453}, archivePrefix={arXiv}
}

@article{ChungCrottyKolbRiotto2001,
  author={Chung, Daniel J. H. and Crotty, Patrick and Kolb, Edward W. and Riotto, Antonio},
  title={Gravitational production of superheavy dark matter},
  journal={Phys. Rev. D}, volume={64}, pages={043503}, year={2001},
  doi={10.1103/PhysRevD.64.043503}
}

@article{PimikovEtAl2017Oddballs,
  author={Pimikov, Alexandr and Lee, Hee-Jung and Kochelev, Nikolai and Zhang, Pengming},
  title={{Is the exotic $0^{--}$ glueball a pure gluon state?}},
  journal={Phys. Rev. D}, volume={95}, pages={071501}, year={2017},
  doi={10.1103/PhysRevD.95.071501}, eprint={1611.08698}, archivePrefix={arXiv}, primaryClass={hep-ph}
}

@article{HenningLuMeliaMurayama2016,
  author={Henning, Brian and Lu, Xiaochuan and Melia, Tom and Murayama, Hitoshi},
  title={{Hilbert series and operator bases with derivatives in effective field theories}},
  journal={Commun. Math. Phys.}, volume={347}, pages={363--388}, year={2016},
  doi={10.1007/s00220-015-2518-2}, eprint={1507.07240}, archivePrefix={arXiv}, primaryClass={hep-th}
}

@article{HenningLuMeliaMurayama2017,
  author={Henning, Brian and Lu, Xiaochuan and Melia, Tom and Murayama, Hitoshi},
  title={{Operator bases, $S$-matrices, and their partition functions}},
  journal={JHEP}, volume={10}, pages={199}, year={2017},
  doi={10.1007/JHEP10(2017)199}, eprint={1706.08520}, archivePrefix={arXiv}, primaryClass={hep-th}
}

@article{CarenzaGlueball2022,
  author={Carenza, Pierluca and Pasechnik, Roman and Salinas, Gustavo and Wang, Zhi-Wei},
  title={Glueball Dark Matter Revisited},
  journal={Phys. Rev. Lett.}, volume={129}, pages={261302}, year={2022},
  doi={10.1103/PhysRevLett.129.261302}, eprint={2207.13716}, archivePrefix={arXiv}, primaryClass={hep-ph}
}

@article{CarenzaGlueball2023,
  author={Carenza, Pierluca and Ferreira, Tassia and Pasechnik, Roman and Wang, Zhi-Wei},
  title={Glueball dark matter, precisely},
  journal={Phys. Rev. D}, volume={108}, pages={123027}, year={2023},
  doi={10.1103/PhysRevD.108.123027}, eprint={2306.09510}, archivePrefix={arXiv}, primaryClass={hep-ph}
}

@article{McKeenGlueball2025,
  author={McKeen, David and Mizuta, Riku and Morrissey, David E. and Shamma, Michael},
  title={Dark matter from dark glueball dominance},
  journal={Phys. Rev. D}, volume={111}, pages={015044}, year={2025},
  doi={10.1103/PhysRevD.111.015044}, eprint={2406.18635}, archivePrefix={arXiv}, primaryClass={hep-ph}
}

@article{YamadaYonekura2023,
  author={Yamada, Masaki and Yonekura, Kazuya},
  title={Dark baryon from pure Yang--Mills theory and its {GW} signature from cosmic strings},
  journal={JHEP}, volume={09}, pages={197}, year={2023},
  doi={10.1007/JHEP09(2023)197}, eprint={2307.06586}, archivePrefix={arXiv}, primaryClass={hep-ph}
}

@article{BiondiniNonAbelian2024,
  author={Biondini, Simone and Kole{\v{s}}ov{\'a}, Helena and Procacci, Simona},
  title={Smooth reheating and dark matter via non-Abelian gauge theory},
  journal={Phys. Lett. B}, volume={857}, pages={138995}, year={2024},
  doi={10.1016/j.physletb.2024.138995}, eprint={2406.10345}, archivePrefix={arXiv}, primaryClass={hep-ph}
}

@article{MaasMarklMuller2023,
  author={Maas, Axel and Markl, Markus and Muller, Michael},
  title={Exploratory applications of the Fr{\"o}hlich--Morchio--Strocchi mechanism in quantum gravity},
  journal={Phys. Rev. D}, volume={107}, pages={025013}, year={2023},
  doi={10.1103/PhysRevD.107.025013}, eprint={2202.05117}, archivePrefix={arXiv}, primaryClass={hep-th}
}

@article{LiRenXiaoYuZheng2022,
  author={Li, Hao-Lin and Ren, Zhe and Xiao, Ming-Lei and Yu, Jiang-Hao and Zheng, Yu-Hui},
  title={Operators for generic effective field theory at any dimension: on-shell amplitude basis construction},
  journal={JHEP}, volume={04}, pages={140}, year={2022},
  doi={10.1007/JHEP04(2022)140}, eprint={2201.04639}, archivePrefix={arXiv}, primaryClass={hep-ph}
}

@article{DongMaShu2023,
  author={Dong, Zi-Yu and Ma, Teng and Shu, Jing},
  title={Constructing on-shell operator basis for all masses and spins},
  journal={Phys. Rev. D}, volume={107}, pages={L111901}, year={2023},
  doi={10.1103/PhysRevD.107.L111901}, eprint={2103.15837}, archivePrefix={arXiv}, primaryClass={hep-ph}
}

@article{GrafHenningLuMeliaMurayama2023,
  author={Gr{\'a}f, Luk{\'a}s and Henning, Brian and Lu, Xiaochuan and Melia, Tom and Murayama, Hitoshi},
  title={Hilbert series, the Higgs mechanism, and HEFT},
  journal={JHEP}, volume={02}, pages={064}, year={2023},
  doi={10.1007/JHEP02(2023)064}, eprint={2211.06275}, archivePrefix={arXiv}, primaryClass={hep-ph}
}

@article{HeidenreichChernWeil2021,
  author={Heidenreich, Ben and McNamara, Jacob and Montero, Miguel and Reece, Matthew and Rudelius, Tom and Valenzuela, Irene},
  title={Chern--Weil Global Symmetries and How Quantum Gravity Avoids Them},
  journal={JHEP}, volume={11}, pages={053}, year={2021},
  doi={10.1007/JHEP11(2021)053}, eprint={2012.00009}, archivePrefix={arXiv}, primaryClass={hep-th}
}

@article{vanBeestSwampland2022,
  author={van Beest, Marieke and Calder{\'o}n-Infante, Jos{\'e} and Mirfendereski, Delaram and Valenzuela, Irene},
  title={Lectures on the Swampland Program in String Compactifications},
  journal={Phys. Rept.}, volume={989}, pages={1--50}, year={2022},
  doi={10.1016/j.physrep.2022.09.002}, eprint={2102.01111}, archivePrefix={arXiv}, primaryClass={hep-th}
}

@article{WangWanYou2022,
  author        = {Wang, Juven and Wan, Zheyan and You, Yi-Zhuang},
  title         = {Proton Stability: From the Standard Model to Beyond Grand Unification},
  journal       = {Phys. Rev. D},
  volume        = {106},
  number        = {2},
  pages         = {025016},
  year          = {2022},
  doi           = {10.1103/PhysRevD.106.025016},
  eprint        = {2204.08393},
  archivePrefix = {arXiv},
  primaryClass  = {hep-ph}
}

@article{DevBNV2024,
  author        = {Dev, P. S. Bhupal and Koerner, L. W. and Saad, S. and others},
  title         = {Searches for Baryon Number Violation in Neutrino Experiments: A White Paper},
  journal       = {J. Phys. G},
  volume        = {51},
  number        = {3},
  pages         = {033001},
  year          = {2024},
  doi           = {10.1088/1361-6471/ad1658},
  eprint        = {2203.08771},
  archivePrefix = {arXiv},
  primaryClass  = {hep-ex}
}

@article{FileviezWise2010,
  author        = {Fileviez P{\'e}rez, Pavel and Wise, Mark B.},
  title         = {Baryon and Lepton Number as Local Gauge Symmetries},
  journal       = {Phys. Rev. D},
  volume        = {82},
  number        = {1},
  pages         = {011901},
  year          = {2010},
  doi           = {10.1103/PhysRevD.82.011901},
  eprint        = {1002.1754},
  archivePrefix = {arXiv},
  primaryClass  = {hep-ph},
  note          = {Erratum: Phys. Rev. D 82, 079901 (2010)}
}

@article{AlvesMatterParity2017,
  author        = {Alves, Alexandre and Arcadi, Giorgio and Dong, P. V. and Duarte, Laura and Queiroz, Farinaldo S. and Valle, Jos{\'e} W. F.},
  title         = {Matter-parity as a residual gauge symmetry: Probing a theory of cosmological dark matter},
  journal       = {Phys. Lett. B},
  volume        = {772},
  pages         = {825--831},
  year          = {2017},
  doi           = {10.1016/j.physletb.2017.07.056}
}

@article{HaradaE6BL2003,
  author        = {Harada, Junpei},
  title         = {Hypercharge and baryon minus lepton number in $E_6$},
  journal       = {JHEP},
  volume        = {04},
  pages         = {011},
  year          = {2003},
  doi           = {10.1088/1126-6708/2003/04/011},
  eprint        = {hep-ph/0305015},
  archivePrefix = {arXiv}
}

@article{ManganoParke1991,
  author  = {Mangano, Michelangelo L. and Parke, Stephen J.},
  title   = {Multiparton Amplitudes in Gauge Theories},
  journal = {Phys. Rept.},
  volume  = {200},
  pages   = {301--367},
  year    = {1991},
  doi     = {10.1016/0370-1573(91)90091-Y}
}

@article{BodwinBraatenLepage1995,
  author        = {Bodwin, Geoffrey T. and Braaten, Eric and Lepage, G. Peter},
  title         = {Rigorous QCD Analysis of Inclusive Annihilation and
                   Production of Heavy Quarkonium},
  journal       = {Phys. Rev. D},
  volume        = {51},
  pages         = {1125--1171},
  year          = {1995},
  doi           = {10.1103/PhysRevD.51.1125},
  eprint        = {hep-ph/9407339},
  archivePrefix = {arXiv},
  note          = {Erratum: Phys. Rev. D 55, 5853 (1997)}
}

@article{LukeManohar1997,
  author  = {Luke, Michael and Manohar, Aneesh V.},
  title   = {Bound States and Power Counting in Effective Field Theories},
  journal = {Phys. Rev. D},
  volume  = {55},
  pages   = {4129--4140},
  year    = {1997},
  doi     = {10.1103/PhysRevD.55.4129}
}

@article{ManoharGeorgi1984,
  author  = {Manohar, Aneesh and Georgi, Howard},
  title   = {Chiral Quarks and the Nonrelativistic Quark Model},
  journal = {Nucl. Phys. B},
  volume  = {234},
  pages   = {189--212},
  year    = {1984},
  doi     = {10.1016/0550-3213(84)90231-1}
}

@article{Feshbach1958,
  author  = {Feshbach, Herman},
  title   = {Unified Theory of Nuclear Reactions},
  journal = {Annals Phys.},
  volume  = {5},
  pages   = {357--390},
  year    = {1958},
  doi     = {10.1016/0003-4916(58)90007-1}
}

@article{Feshbach1962,
  author  = {Feshbach, Herman},
  title   = {A Unified Theory of Nuclear Reactions. II},
  journal = {Annals Phys.},
  volume  = {19},
  pages   = {287--313},
  year    = {1962},
  doi     = {10.1016/0003-4916(62)90221-X}
}

@book{BourbakiLie46,
  author    = {Bourbaki, Nicolas},
  title     = {Lie Groups and Lie Algebras: Chapters 4--6},
  publisher = {Springer},
  address   = {Berlin, Heidelberg},
  year      = {2002},
  series    = {Elements of Mathematics},
  isbn      = {978-3-540-42650-9}
}

@book{Hartshorne1977,
  author    = {Hartshorne, Robin},
  title     = {Algebraic Geometry},
  series    = {Graduate Texts in Mathematics},
  volume    = {52},
  publisher = {Springer},
  address   = {New York},
  year      = {1977},
  doi       = {10.1007/978-1-4757-3849-0}
}

@book{HuybrechtsLehn2010,
  author    = {Huybrechts, Daniel and Lehn, Manfred},
  title     = {The Geometry of Moduli Spaces of Sheaves},
  edition   = {2},
  publisher = {Cambridge University Press},
  year      = {2010},
  doi       = {10.1017/CBO9780511711985}
}

@book{FultonHarris1991,
  author    = {Fulton, William and Harris, Joe},
  title     = {Representation Theory: A First Course},
  series    = {Graduate Texts in Mathematics},
  volume    = {129},
  publisher = {Springer},
  address   = {New York},
  year      = {1991},
  doi       = {10.1007/978-1-4612-0979-9}
}

@book{Macdonald1995,
  author    = {Macdonald, I. G.},
  title     = {Symmetric Functions and Hall Polynomials},
  edition   = {2},
  publisher = {Oxford University Press},
  address   = {Oxford},
  year      = {1995},
  doi       = {10.1093/oso/9780198534891.001.0001}
}

@article{MachacekVaughn1983,
  author  = {Machacek, Marie E. and Vaughn, Michael T.},
  title   = {Two-Loop Renormalization Group Equations in a General Quantum Field Theory. I. Wave Function Renormalization},
  journal = {Nucl. Phys. B},
  volume  = {222},
  pages   = {83--103},
  year    = {1983},
  doi     = {10.1016/0550-3213(83)90610-7}
}

@article{Yang1950,
  author  = {Yang, Chen Ning},
  title   = {Selection Rules for the Dematerialization of a Particle into Two Photons},
  journal = {Phys. Rev.},
  volume  = {77},
  pages   = {242--245},
  year    = {1950},
  doi     = {10.1103/PhysRev.77.242}
}

@article{Landau1948,
  author  = {Landau, L. D.},
  title   = {On the Angular Momentum of a System of Two Photons},
  journal = {Dokl. Akad. Nauk SSSR},
  volume  = {60},
  pages   = {207--209},
  year    = {1948}
}

@article{HenningLuMurayama2018,
  author        = {Henning, Brian and Lu, Xiaochuan and Murayama, Hitoshi},
  title         = {One-loop Matching and Running with Covariant Derivative Expansion},
  journal       = {JHEP},
  volume        = {01},
  pages         = {123},
  year          = {2018},
  doi           = {10.1007/JHEP01(2018)123},
  eprint        = {1604.01019},
  archivePrefix = {arXiv},
  primaryClass  = {hep-ph}
}

@CONTROL{apsrev42Control,
  author="08",
  editor="1",
  pages="0",
  title="0",
  year="1"
}

\end{document}